\documentclass[12pt]{article}
\usepackage{amsfonts}
\usepackage{amsmath}
\usepackage{amssymb}
\usepackage{array}
\usepackage{bigints}
\usepackage{booktabs}
\usepackage[nosort]{cite}
\usepackage{color}
\usepackage{dsfont}
\usepackage{float}
\usepackage{framed}
\usepackage{graphicx}
\usepackage{indentfirst}
\usepackage{mathrsfs}
\usepackage{multirow}
\usepackage{pdflscape}
\usepackage{setspace}
\usepackage{subdepth}
\usepackage{subfig}
\usepackage{titlesec}
\usepackage[dotinlabels]{titletoc}
\usepackage{wrapfig}
\usepackage[all]{xy}
\usepackage{young}
\usepackage[vcentermath]{youngtab}
\usepackage{relsize}
\usepackage{stackengine}
\usepackage{verbatim}

\usepackage{hyperref}
\hypersetup{colorlinks=true}
\hypersetup{linkcolor=black}
\hypersetup{citecolor=black}
\hypersetup{urlcolor=black}

\numberwithin{equation}{section}

\usepackage[left=2.5cm,right=2.5cm,top=2.5cm,bottom=3cm]{geometry}
\titleformat{\section}{\normalfont\bfseries}{\thesection.}{4pt}{}
\titlespacing{\section}{0pt}{20pt}{6pt}

\titleformat{\subsection}{\normalfont\itshape}{\thesubsection.}{4pt}{}
\titlespacing{\subsection}{0pt}{15pt}{6pt}

\titleformat{\subsubsection}{\normalfont}{\thesubsubsection.}{4pt}{}
\titlespacing{\subsubsection}{0pt}{15pt}{6pt}

\def\ie{\begin{equation}\begin{aligned}}
\def\fe{\end{aligned}\end{equation}}

\def\tilde{\widetilde}
\def\t{\tilde}
\def\hat{\widehat}

\def\bar{\overline}
\def\b{\bar}

\def\half{{1 \over 2}}
\def\d{\partial}
\def\jbar{{\b \jmath}}

\def\1{{\mathds 1}}

\newcommand{\Z}{{\mathbb Z}}

\newcommand{\R}{{\mathbb R}}

\def\SL{{\mathscr L}}

\def\CA{{\mathcal A}}
\def\CB{{\mathcal B}}
\def\CC{{\mathcal C}}
\def\CD{{\mathcal D}}
\def\CE{{\mathcal E}}

\def\CJ{{\mathcal J}}
\def\CK{{\mathcal K}}

\def\CM{{\mathcal M}}
\def\CN{{\mathcal N}}
\def\CO{{\mathcal O}}

\def\CR{{\mathcal R}}

\def\CT{{\mathcal T}}

\def\CV{{\mathcal V}}
\def\CW{{\mathcal W}}

\DeclareFontShape{OT1}{cmr}{mx}{n}%
    {<->cmr10}{}
\newcommand{\mytitlefont}{\fontseries{mx}\selectfont}
\DeclareMathAlphabet{\titlemath}{OT1}{cmr}{mx}{n}

\begin{document}


\begin{titlepage}

\begin{center}

~\\[2cm]

{\fontsize{26pt}{0pt} \mytitlefont Multiplets of Superconformal \\[1pt] Symmetry in Diverse Dimensions}

~\\[0.5cm]

Clay C\'{o}rdova,$^{1}$ Thomas T.~Dumitrescu,$^2$ and Kenneth Intriligator\,$^3$

~\\[0.1cm]

$^1$~{\it School of Natural Sciences, Institute for Advanced Study, Princeton, NJ 08540, USA}

$^2$~{\it Department of Physics, Harvard University, Cambridge, MA 02138, USA}

$^3$ {\it Department of Physics, University of California, San Diego, La Jolla, CA 92093, USA}

~\\[1cm]

\end{center}

\noindent We systematically analyze the operator content of unitary superconformal multiplets in~$d \geq 3$ spacetime dimensions. We present a simple, general, and efficient algorithm that generates all of these multiplets by correctly eliminating possible null states. The algorithm is conjectural, but passes a vast web of consistency checks. We apply it to tabulate a large variety of superconformal multiplets. In particular, we classify and construct all multiplets that contain conserved currents or free fields, which play an important role in superconformal field theories (SCFTs). Some currents that are allowed in conformal field theories cannot be embedded in superconformal multiplets, and hence they are absent in SCFTs. We use the structure of superconformal stress tensor multiplets to show that SCFTs with more than~$16$ Poincar\'e supercharges cannot arise in~$d \geq 4$, even when the corresponding superconformal algebras exist. We also show that such theories do arise in~$d = 3$, but are necessarily free. 

\vfill 

\begin{flushleft}
December 2016
\end{flushleft}

\end{titlepage}


\tableofcontents


\section{Introduction}

\label{sec:intro}

In this paper we revisit the problem of constructing unitary multiplets of a superconformal algebra~${\mathfrak S}(d, \CN)$ in~$d \geq 3$ spacetime dimensions,\footnote{~Superconformal algebras also exist for~$d = 1,2$. In two dimensions, they arise as subalgebras of super-Virasoro algebras, whose representation theory is substantially richer. Since there is no corresponding phenomenon in higher dimensions and we would like to keep the discussion uniform, we restrict to~$d \geq 3$.} with~$\CN$-extended supersymmetry. These multiplets play an essential role in the study of the corresponding superconformal field theories (SCFTs) and their deformations. A unitary multiplet~$\CM$ of~$\mathfrak S(d, \CN)$ is conveniently described by presenting its decomposition into a finite number of unitary, irreducible representations~$\big\{\CC_a\big\}$ of the bosonic subalgebra~$\frak{so}(d,2) \times {\mathfrak R} \subset \frak S(d,\CN)$,
\begin{equation}\label{basicdecomp}
\CM = \bigoplus_a \, \CC_a \quad (\text{finite sum})~.
\end{equation}
The non-compact Lie algebra~$\frak{so}(d,2)$ is the conformal algebra, while the~$R$-symmetry algebra~$\mathfrak R$ is a compact Lie algebra; their unitary representations are well understood. There are various equivalent ways to think about~\eqref{basicdecomp}, e.g.~as the decomposition of a superfield~$\CM$ into its component fields~$\big\{\CC_a\big\}$, or as the expansion of an~$\frak S(d, \CN)$ character~$\chi_\CM$ in terms of~$\frak{so}(d,2) \times \frak R$ characters~$\chi_{\CC_a}$.

Our main result is a simple and general algorithm that outputs the decomposition~\eqref{basicdecomp} for any unitary superconformal multiplet (section~\ref{sec:alg}). Along the way, we review various aspects of unitarity (super-) conformal representations (sections~\ref{sec:intro} and~\ref{sec:usm}). We use our algorithm to tabulate  a wide variety of superconformal multiplets (section~\ref{sec:tables}); a {\tt Mathematica} package that implements the algorithm and can be used to generate any superconformal multiplet will appear in~\cite{tdmm}. We also make a detailed survey of multiplets that contain conserved currents and explore some of the implications for unitary SCFTs (section~\ref{sec:currents}). A more detailed summary of applications appears in section~\ref{sec:introapp}. 

\subsection{Conformal Field Theories}

\label{sec:introcft}

Conformal field theory (CFT) is a subject of enduring interest, with myriad applications, see for instance~\cite{Rychkov:2016iqz} for a recent introduction with references. CFTs are powerfully constrained by the~$\frak{so}(d,2)$ conformal algebra, and all local operators must reside in representations -- also called multiplets -- of this algebra. Throughout this paper, we will only discuss unitary theories, and hence unitary multiplets. The unitary, irreducible representations of~$\frak{so}(d,2)$ are well understood (see for instance~\cite{Mack:1975je,Minwalla:1997ka,Dolan:2005wy,Kinney:2005ej,Rychkov:2016iqz} and references therein). Every conformal multiplet~$\CC$ consists of (typically infinitely many) local operators~$\CO$, which can be taken to transform irreducibly under the maximal compact subalgebra~$\frak{so}(d) \times \frak{so}(2)$ generated by the (Wick-rotated) Lorentz transformations~$M_{[\mu\nu]}$ and the dilatation~$D$. (The other~$\frak{so}(d,2)$ generators are the momenta~$P_\mu$ and the special conformal transformations~$K_\mu$.) Thus, $\CO$ has definite~$\frak{so}(d)$ Lorentz weights~$L_\CO$ and a definite scaling dimension~$\Delta_\CO$ related to its~$\frak{so}(2)$ weight. Throughout, we indicate these quantum numbers as\footnote{~Unless stated otherwise, we will use integer-valued Dynkin labels to specify weights and representations such as~$L_\CO$. For instance~$[n]$ (with~$n \in \Z_{\geq 0}$) is the~$(n+1)$-dimensional spin-${n \over 2}$ representation of~$\frak{su}(2)$; it contains the weights~$[n], [n-2], \ldots, [-n]$. See appendix~\ref{app:liealg} for a summary of our Lie algebra conventions.}
\begin{equation}\label{notation}
\CO = [L_\CO]_{\Delta_\CO}~.
\end{equation} 
The full Lorentz representation containing~$\CO$ is specified by the~$\frak{so}(d)$ quantum numbers of its highest weight operator~$\CO_\text{h.w.}$. Depending on the context, we will interchangeably use the notation~\eqref{notation} to refer to full Lorentz representations, via their highest weights, or to the weights of individual operators that reside in such representations. 

The state-operator correspondence identifies every local operator~$\CO$ with a unique state $|\CO\rangle$ in radial quantization. The conformally invariant CFT vacuum corresponds to the unit operator~$\CO = \1$. We will frequently abuse notation and write~$\CO$ when we mean~$|\CO\rangle$. Every unitary, irreducible conformal representation~$\CC$ possesses a unique operator~$\CV$ of lowest scaling dimension~$\Delta_\CV$, which is known as the conformal primary (CP).  As such, it is annihilated by the special conformal generators~$K_\mu$, whose scaling dimension~$\Delta_{K_\mu} = -1$ is negative. All other states~$\CO$ in the multiplet~$\CC$ are obtained by acting on~$\CV$ with the translation generators~$P_\mu$, whose scaling dimension is~$\Delta_{P_\mu} =1$. The states~$\CO$ are referred to as conformal descendants (CDs) of the CP~$\CV$. In the operator language, the CDs~$\CO$ are simply given by spacetime derivatives~$P_\mu \sim \d_\mu$ of the CP~$\CV$. By contrast, $\CV$ cannot be written as a derivative of a well-defined, local operator. 

It follows that the structure of the entire multiplet~$\CC$ is completely determined by the quantum numbers of the conformal primary~$\CV$, and we will often use~$\CV$ to refer to the multiplet~$\CC$ generated by the CP~$\CV$ and its CDs.  In particular, there is a natural inner product on the Hilbert space of states in radial quantization, which descends from the two-point function of local operators, and the norm of all states in $\CC$ is completely determined by the quantum numbers of the CP~$\CV$ (see~\cite{Minwalla:1997ka,Rychkov:2016iqz} and references therein for additional details).\footnote{\label{fn:bosnorm}~To compute the norm of a ~CD $\CO$ one proceeds as follows. Express~$|\CO\rangle \sim P_{\mu_1} \cdots P_{\mu_n} |\CV\rangle$. In radial quantization, $P_\mu^\dagger = K_\mu$, so that~$\langle \CO |\CO\rangle \sim \langle \CV | K_{\mu_n} \cdots K_{\mu_1} P_{\mu_1} \cdots P_{\mu_n} |\CV\rangle$. This can be evaluated using~$K_\mu |\CV\rangle = 0$ and the commutation relation~$[P_\mu, K_\nu] \sim \eta_{\mu\nu} D + M_{\mu\nu}$. The result is a polynomial in the dimension~$\Delta_\CV$, with coefficients that depend on the Lorentz weights~$L_\CV$ of~$\CV$, multiplied by the norm~$\langle \CV |\CV\rangle$.}  In unitary theories, all primary and descendant operators in a conformal multiplet~$\CC$ must have non-negative norm with respect to this inner product. This results in unitarity bounds on the scaling dimension~$\Delta_\CV$ of~$\CV$ in terms of its Lorentz representation~$L_\CV$. Schematically,
\begin{equation}\label{confub}
\Delta_\CV \geq f(L_\CV)~.
\end{equation} 
This leads to the important distinction between long and short conformal multiplets:
\begin{itemize}
\item {\it Long Multiplets:} If the inequality~\eqref{confub} is strict, then all states have positive norm and we refer to~$\CC$ as a long multiplet. Given a Lorentz representation~$L_\CV$, we can always construct a long multiplet based on a CP~$\CV$ with these quantum numbers by choosing its scaling dimension~$\Delta_\CV$ to be sufficiently large.  

\item  {\it Short Multiplets:}  If~\eqref{confub} is saturated, then some states in~$\CC$ -- called null states -- have zero norm. The null states form a closed subrepresentation of~$\CC$ (also unitary unless~$\CC = \1$ is the identity), and hence they can be consistently removed from~$\CC$. The resulting conformal multiplet contains fewer states than a long multiplet based on a CP with Lorentz quantum numbers~$L_\CV$. Consequently, we refer to it as a short multiplet. 
\end{itemize}

\noindent The most extreme example of a short conformal multiplet is the unit operator~$\1$, which is annihilated by all derivatives and has $\Delta_{\1}=0.$  A less trivial example of a short conformal multiplet is a conserved flavor current~$j_\mu$. The conservation equation~$\d^\mu j_\mu = 0$ and its CDs are null states that fix the scaling dimension of~$j_\mu$ to be~$\Delta_{j_\mu} = d-1$. Yet another elementary example is a free scalar field~$\phi$ of dimension~$\Delta_\phi = {d -2 \over 2}$, whose null states are given by the equation of motion~$\d^{2} \phi = 0$ and its descendants. In general, the CP of any short conformal multiplet is annihilated by a first- or second-order differential operator.  The unitarity constraints on short conformal multiplets were worked out in~\cite{Minwalla:1997ka}. A detailed discussion can be found in section~\ref{sec:currents}.

\subsection{Superconformal Field Theories}

\label{sec:introscft}

The symmetry algebra of SCFTs contains both the conformal algebra~$\frak{so}(d,2)$ and a Poincar\'e supersymmetry (SUSY) algebra of the schematic form~$\{Q, Q\} \sim P_\mu$. Since~$P_\mu$ has scaling dimension~$\Delta_{P_\mu} = 1$, the Poincar\'e~$Q$-supercharges have~$\Delta_Q = \half$. These symmetries combine into a larger superconformal algebra~$\frak S$ that also contains superconformal~$S$-supersymmetries of scaling dimension~$\Delta_S = -\half$. They anticommute to the special conformal generators, $\{S, S\} \sim K_\mu$. Both~$Q$ and~$S$ are fundamental Lorentz spinors, which combine into a spinor of the~$\frak{so}(d,2)$ conformal algebra. Typically~$\frak S$ also contains a bosonic~$R$-symmetry subalgebra~$\frak R \subset \frak S$. It commutes with all~$\frak{so}(d,2)$ conformal generators, but~$Q$ and~$S$ transform in definite (and conjugate) representations of~$\frak R$. The fact that the~$R$-symmetry is part of the symmetry algebra~$\frak S$ is a hallmark of SCFTs. By contrast, non-conformal supersymmetric theories need not have an~$R$-symmetry, and if they do it only acts as an outer automorphism of the Poincar\'e SUSY algebra.\footnote{~There are nonconformal supersymmetric theories with exotic SUSY algebras of the schematic form~$\{Q, Q\} \sim P_\mu + R$, where~$R$ is an~$R$-symmetry generator. See for instance~\cite{Nahm:1977tg,Itzhaki:2005tu,Lin:2005nh, Agarwal:2008pu,Cordova:2016xhm}. }

The requirement that~$\frak S$ be a consistent superalgebra with these properties is very restrictive~\cite{Nahm:1977tg} (see also~\cite{Minwalla:1997ka}): superconformal algebras only exist in~$d \leq 6$ dimensions.  The bound~$d \leq 6$ is related to the fact that the consistency of the superconformal algebras relies on sporadic properties of~$\frak{so}(d,2)$ fundamental spinors. Such sporadic phenomena occur when $d$ is sufficiently small, but terminate with the triality automorphism of $\frak{so}(6,2).$  Moreover, the classification also implies that for~$3 \leq d \leq 6$ the superconformal algebra~$\frak S(d, \CN)$ is essentially uniquely determined by the spacetime dimension~$d$ and the amount of supersymmetry~$\CN$.\footnote{~For~$d = 3$ and~$\CN=8$, there is a choice of triality frame for the~$\frak{so}(8)_R$ symmetry representations of~$Q$ and~$S$. We take them to be vectors, since they are~$\frak{so}(\CN)_R$ vectors for all other values of~$\CN$ in three dimensions. In~$d = 6$, the superconformal algebra forces all~$Q$-supersymmetries to have the same chirality, while the~$S$-supersymmetries have the opposite chirality. For this reason we sometimes refer to~$\CN$-extended superconformal symmetry in six dimensions as~$(\CN,0)$.} Here~$\CN$ denotes the number of~$Q$-supercharges in units of a minimal spinor; we will write~$N_Q$ for the total number of~$Q$-supercharges. For instance, theories with~$N_Q = 8$ correspond to~$\CN = (1, 0)$ in~$d = 6$, $\CN = 1$ in~$d = 5$, $\CN = 2$ in~$d = 4$, and~$\CN = 4$ in~$d = 3$. 

The resulting list of superconformal algebras~$\frak S(d, \CN)$ in~$3 \leq d \leq 6$ dimensions is as follows:\footnote{~One might wonder if there are~$\CN=4$ SCFTs in~$d = 4$ whose superconformal algebra is~$\frak{su}(2,2|4)$, which is a central extension of~$\frak{psu}(2,2|4)$, by $\frak{u}(1)$. This possibility is ruled out in section~\ref{sec:maxsusy}.}
\begin{align}\label{scftalgs}
d=3 & \qquad \frak S(3, \CN) = \frak{osp}(\CN| 4) \; \supset \; \frak{so}(3,2)\times \frak{so}(\CN)_R~, \nonumber \\[4pt]
d=4 & \hskip11pt \begin{cases} \frak S(4, \CN) = 
\frak{su}(2,2|\CN) \; \supset \; \frak{so}(4,2)\times \frak{su}(\CN)_R\times \frak{u}(1)_R~,\quad \CN \neq 4~, \\ 
\frak S(4, 4) =  \frak{psu}(2,2|4)\supset \frak{so}(4,2)\times \frak{su}(4)_R~,\quad \CN =4~,
\end{cases} 
\nonumber
\\[4pt]
d=5& \qquad \frak S(5,1) = \frak{f}(4) \; \supset \; \frak{so}(5,2)\times \frak{su}(2)_R~, \quad \CN = 1~, \nonumber \\[4pt]
d=6 & \qquad \frak S(6,\CN) = \frak{osp}(6,2|\CN ) \; \supset \; \frak{so}(6,2)\times \frak{sp}(2\CN )_R~. 
\end{align}
In each case we have displayed the bosonic subalgebra; it is a direct product of the conformal algebra~$\frak{so}(d,2)$ and the~$R$-symmetry algebra~$\frak R$, indicated by a subscript~$R$ in~\eqref{scftalgs}.\footnote{~The only superconformal algebra in~\eqref{scftalgs} with trivial~$R$-symmetry algebra is~$\frak S(3,1)$, i.e.~$d = 3$, $\CN=1$. In~$d = 6$, we use the convention that~$\frak{sp}(2)_R = \frak{su}(2)_R$ (see appendix~\ref{app:liealg}).} In $d = 3,4,6$ dimensions, there is a superconformal algebra for every positive integer~$\CN$. Most of these algebras have~$N_Q > 16$ supercharges; by contrast, standard lore posits that local quantum field theories only arise for~$N_Q \leq 16$. We will discuss this issue in detail in section~\ref{sec:maxsusy}. In~$d = 5$, there is a unique, exceptional superconformal algebra~$\frak f (4)$, with minimal $\CN=1$ supersymmetry, i.e.~$N_Q = 8$. While quantum field theories with more supersymmetry do exist in~$d = 5$ (e.g.~maximally supersymmetric Yang-Mills theories, with~$\CN = 2, N_Q = 16$), they cannot be superconformal (see for instance~\cite{Seiberg:1997ax}). 

Many properties of conformal multiplets reviewed in section~\ref{sec:introcft} have close analogues in the superconformal case. (Useful background material on superconformal multiplets, with references, can be found in~\cite{Minwalla:1997ka,Dolan:2002zh,Kinney:2005ej,Bhattacharya:2008zy}.) Now all local operators must transform in unitary multiplets~$\CM$ of a superconformal algebra~$\frak S$ listed in~\eqref{scftalgs}. The structure of these multiplets is the main subject of this paper. Each local operator~$\CO$ is specified by its~$\frak{so}(d)$ Lorentz weights~$L_\CO$ and scaling dimension~$\Delta_\CO$, as well as its weights~$R_\CO$ under the~$R$-symmetry algebra~$\frak R$. Throughout, we will denote an operator with these quantum numbers as
\begin{equation}\label{snotation}
\CO = [L_\CO]_{\Delta_\CO}^{(R_\CO)}~,
\end{equation}
using square brackets~$[\,\cdots]$ for the Lorentz representation and parentheses~$(\,\cdots)$ for the~$R$-symmetry quantum numbers. As before, we will also use this notation to denote full representations, via their highest weights. For instance, we will often refer to the Lorentz and~$R$-symmetry representations~$\CR_Q$ of the~$Q$-supersymmetries,
\begin{equation}\label{qrep}
\CR_Q = Q_\text{h.w.} = [L_Q]^{(R_Q)}_{\Delta_Q = 1/2}~,
\end{equation}
where~$L_Q$ and~$R_Q$ depend on the choice of superconformal algebra~$\frak S(d, \CN)$. This description is appropriate when the~$Q$-supercharge representation~$\CR_Q$ is irreducible. For~$d = 3, \CN = 2$ or~$d = 4$ (with any~$\CN$) the Poincar\'e supercharges instead transform as a direct sum of two irreducible Lorentz and~$R$-symmetry representations. In these cases it is natural to distinguish two independent sets of supercharges~$Q$ and~$\b Q$ (here the notation~$\b Q$ does not mean complex conjugation), which transform irreducibly as~$\CR_Q$ and~$\CR_{\b Q}$ under the Lorentz and~$R$-symmetry. We will refer to~$Q$ and~$\b Q$ as left and right supercharges. (The terminology is borrowed from~$d = 4$, where~$Q$ and~$\b Q$ are Lorentz spinors of opposite chirality.) In order to streamline this introduction, we focus on the situation with one irreducible~$Q$. Several modifications are needed in the two-sided case; they will be explained in later sections. 

Just as in the bosonic case, every unitary, irreducible superconformal multiplet contains a unique operator~$\CV$ of lowest scaling dimension~$\Delta_\CV$, referred to as the superconformal primary~(SCP). The SCP is annihilated by all~$S$-supersymmetries (with~$\Delta_S = -\half$), and hence by $K_\mu \sim \{S, S\}$. All other operators~$\CO$ in the multiplet can be obtained by acting on the SCP~$\CV$ with the~$Q$-supersymmetries (with~$\Delta_Q = \half$); this includes the action of~$P_\mu \sim \{Q, Q\}$. The~$\CO$'s are referred to as superconformal descendants (SCDs) of~$\CV$. By contrast, $\CV$ cannot be written as a~$Q$-descendant of a well-defined, local operator. We will say that an operator~$\CO_\ell$ that can be obtained by acting~$\ell$ times with a~$Q$-supercharge on the SCP~$\CV$ is a level-$\ell$ SCD (consequently, the SCP~$\CV$ resides at level~$\ell = 0$),
\begin{equation}\label{leveldef}
\CO_\ell \sim Q^\ell \CV~, \qquad \ell = \text{level} \in \Z_{\geq 0}~.
\end{equation}
Here~$Q^\ell \CV$ can be thought of as the ordered action of~$\ell$ supercharges on the state~$|\CV\rangle$ in radial quantization, or as~$\ell$ nested, graded commutators of~$Q$-supercharges with the operator~$\CV$. 

It follows from the above that the superconformal multiplet~$\CM$ is completely fixed by the quantum numbers~$[L_\CV]_{\Delta_\CV}^{(R_\CV)}$ of the SCP~$\CV$.  In particular, the norm of every~SCD~$|\CO\rangle \sim Q_1 \cdots Q_n |\CV\rangle$ is completely determined by the norm of the SCP~$|\CV\rangle$. In analogy with the conformal case (see footnote~\ref{fn:bosnorm}), radial quantization leads to~$Q^\dagger = S$, so that~$\langle \CO | \CO\rangle \sim \langle \CV| S_n \cdots S_1 Q_1 \cdots Q_n |\CV\rangle$. This can be evaluated using~$S|\CV\rangle = 0$ and the superconformal anticommutator~$\{Q, S\} \sim D - R + M_{\mu\nu}$~(here~$R$ is an~$R$-symmetry generator, and $M_{\mu\nu}$ generates Lorentz transformations). The result for the descendant norm is a polynomial in the dimension~$\Delta_\CV$, with coefficients that depend on the Lorentz and~$R$-symmetry weights~$L_\CV$ and~$R_\CV$ of~$\CV$, multiplied by the norm of the primary~$\langle \CV |\CV\rangle$. Here the signs of $-R+M_{\mu \nu}$ schematically indicate that the norms decrease with increasing $R$ symmetry representation, and with decreasing Lorentz representation.  Thus, the state of smallest norm at a given level has the largest $R$ symmetry representation and/or the smallest Lorentz representation.  This is born out in the classification of unitary representations discussed below and in section \ref{sec:usm}.

\begin{figure}[h]
\centering
\includegraphics[trim=0 3.8cm 0cm 0,clip,height=8cm]{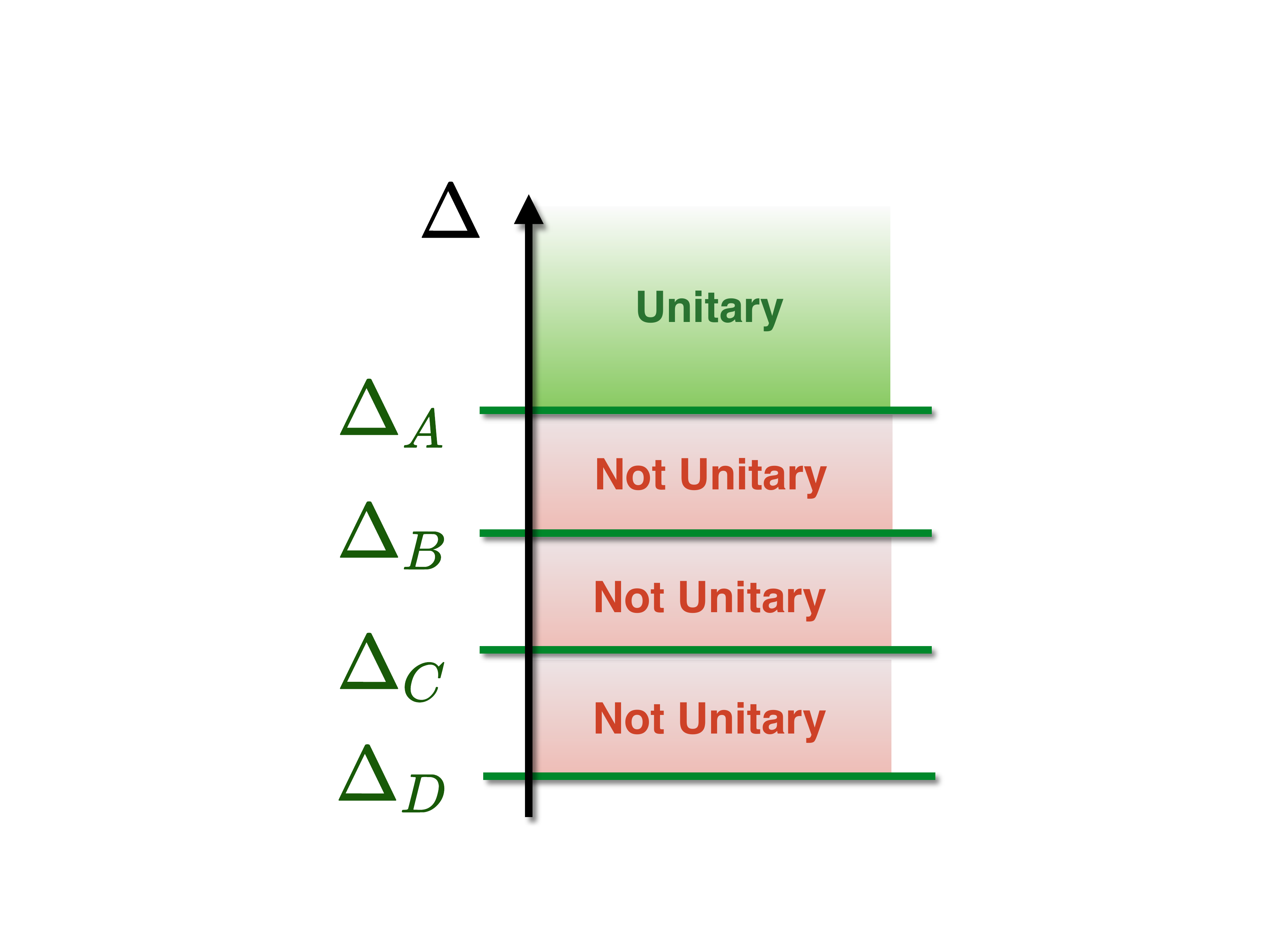}  

 \caption{Hierarchy of unitary long and short multiplets. Long multiplets exist when~$\Delta > \Delta_A$. Their continuum extends down to short~$A$-type multiplets with~$\Delta = \Delta_A$. Isolated short multiplets exist when~$\Delta = \Delta_{B,C,D}$, but not if~$\Delta$ lies in a gap between~$\Delta_A>\Delta_B >\Delta_C>\Delta_D$.} 
 \label{fig:spectrum}
\end{figure}

Unitarity dictates that all local operators in a superconformal multiplet~$\CM$ -- the SCP~$\CV$ and all SCDs~$\CO_\ell~(\ell \geq 1)$ -- must have non-negative norm. Since there are more SCDs than CDs (all~$P_\mu$-descendants are~$Q$-descendants, but not vice versa), the superconformal unitary constraints are stronger than those that follow only from conformal symmetry. These constraints were systematically worked out in~\cite{Dobrev:1985qv,Minwalla:1997ka,Ferrara:2000xg,Dobrev:2002dt,Dolan:2002zh,Kinney:2005ej,Bhattacharya:2008zy}.\footnote{~The same representations were also analyzed in the context of AdS supergravities, see for instance~\cite{Gunaydin:1984fk,Gunaydin:1985tc,Gunaydin:1984wc}.} In analogy with~\eqref{confub}, they can be expressed as a bound on the scaling dimension~$\Delta_\CV$ of the SCP~$\CV$ in terms of its Lorentz and~$R$-symmetry representations~$L_\CV$ and~$R_\CV$, 
\begin{subequations}
\begin{align}\label{scabound}
& \Delta_\CV \geq f(L_\CV)+g( R_\CV) + \delta_A \equiv \Delta_A \hskip60pt (L_\CV, R_\CV~\text{arbitrary}) \\
& \hskip86pt \text{or} \cr
& \label{scbcdbound} \Delta_\CV = f(L_\CV)+g( R_\CV) + \delta_{B, C, D} \equiv \Delta_{B,C,D} \qquad (L_\CV, R_\CV~\text{restricted})~.
\end{align}\label{scabcdboundintro}
\end{subequations}
Here~$f(L_\CV)$ differs from the function entering the bosonic unitarity bounds~\eqref{confub}. The functions $f(L_\CV)$ and $g(R_\CV)$ are the same in~\eqref{scabound} and~\eqref{scbcdbound}, while the offsets~$\delta_{A, B, C, D}$ are numerical constants that satisfy
\begin{equation}\label{gaps}
\delta_A > \delta_B > \delta_C > \delta_D \quad \Longrightarrow \quad \Delta_A > \Delta_B > \Delta_C > \Delta_D~.
\end{equation} 
This leads to a rich hierarchy\footnote{~In the discussion of long and short conformal multiplets following~\eqref{confub}, we did not distinguish between threshold and isolated short multiplets of the conformal algebra. This is because the only isolated conformal multiplet is the unit operator~$\1$ with~$\Delta_{\1} = 0$, which is separated from the continuum of Lorentz-scalar operators~$\CO$ with~$\Delta_\CO \geq {d -2 \over 2}$ by a gap if~$d \geq 3$. Short conformal multiplets with spin are always at threshold.} of long and short multiplets\footnote{~Unlike some treatments, see e.g.~\cite{Dolan:2002zh}, we do not distinguish between short and semi-short multiplets.}, also depicted in figure~\ref{fig:spectrum}:
\begin{itemize}
\item {\it Long Multiplets (L):} These are multiplets for which the inequality~\eqref{scabound} is strict, so that all states have positive norm. As in the bosonic case, we can always construct a long multiplet based on a SCP~$\CV$ with any Lorentz and~$R$-symmetry quantum numbers~$L_\CV, R_\CV$ by taking its scaling dimension~$\Delta_\CV$ to be large enough. We will use the letter~$L$, followed by the quantum numbers of its SCP~$\CV$, to denote a long multiplet,
\begin{equation}\label{lmnot}
L[L_\CV]_{\Delta_\CV}^{(R_\CV)}~, \qquad \Delta_\CV > \Delta_A~.
\end{equation}

\item {\it Short Multiplets at Threshold (A):}~These multiplets saturate the inequality~\eqref{scabound}. Hence they contain null states, which must be removed, and the scaling dimension~$\Delta_\CV$ of the SCP~$\CV$ is fixed in terms of its Lorentz and~$R$-symmetry quantum numbers. For any choice of~$L_\CV$ and~$R_\CV$, we can construct an~$A$-type short multiplet by setting~$\Delta_\CV = \Delta_A$, as in~\eqref{scabound}. Such a multiplet will be denoted as follows:
\begin{equation}\label{aelldef}
A_\ell[L_\CV]^{(R_\CV)}_{\Delta_A}~, \qquad~\ell \in \Z_{\geq 0}~.
\end{equation}
Here~$\ell$ is a positive integer that indicates the level of the first (or primary) null state. The precise range of allowed~$\ell$-values depends on the spacetime dimension~$d$. 

The null states of an~$A$-type short multiplet have the distinguishing feature that they themselves form a unitary superconformal multiplet. (More precisely, they would form a unitary multiplet if their primary had positive norm; here it has zero norm, because it is embedded as the primary null state of a parent~$A$-type short multiplet.) Generically, this null-state multiplet will also be an~$A$-type short multiplet, but for special choices of~$L_\CV$ or~$R_\CV$ it may be an isolated short multiplet of $B, C, D$-type, see below.

\item {\it Isolated Short Multiplets (B,C,D):}~These multiplets only occur for special choices of $L_\CV, R_\CV$ or spacetime dimension~$d$ (e.g.~$C$- and~$D$-type multiplets only exist when~$d \geq 5$). The scaling dimension of their SCP~$\CV$ is fixed by~\eqref{scbcdbound}, and they are isolated from all other types of multiplets (both the continuum of long or~$A$-type multiplets, as well as other isolated short multiplets) with the same Lorentz and~$R$-symmetry quantum numbers by a gap (see~\eqref{gaps} and figure~\ref{fig:spectrum}). When they exist, they will be denoted as
\begin{equation}\label{xelldef}
X_\ell[L_\CV]^{(R_\CV)}_{\Delta_{X}}~, \qquad X \in \{B, C, D\}~,\qquad~\ell \in \Z_{\geq 0}~.
\end{equation}
Here~$\ell$ (whose range depends on~$d$ and~$X$) indicates the level of the primary null state. As before, the null states of~$B, C, D$-type multiplets form closed submultiplets and must be removed, but unlike the null states of~$A$-type multiplets they do not themselves form unitary representations. A simple example of an isolated short multiplet, which exists in all SCFTs, is the unit operator~$\1$ with~$\Delta_{\1 } = 0$, which is annihilated by all supercharges.

\end{itemize}

\noindent The structure of long and short representations summarized above, and depicted in figure~\ref{fig:spectrum} has important implications for SCFTs. For instance, the fact that the scaling dimensions of short multiplets are fixed in terms of their Lorentz and~$R$-symmetry quantum numbers sometimes allows us to determine their spectrum exactly. 

This structure also leads to the notion of recombination rules: as the scaling dimension~$\Delta_\CV$ of a long multiplet~$L[L_\CV]_{\Delta_\CV}^{(R_\CV)}$ is lowered, it eventually hits the unitarity bound $\Delta_\CV = \Delta_A$ from above. At this point, the long multiplet fragments into an~$A$-type short multiplet with the same Lorentz and~$R$-symmetry quantum numbers, and another unitary short multiplet~$N$ that contains the null states of the~$A$-type multiplet. Schematically,
\begin{equation}\label{genrr}
L[L_\CV]^{(R_\CV)}_{\Delta_\CV \rightarrow \Delta_A^+} \quad \rightarrow \quad A_\ell[L_\CV]_{\Delta_A}^{(R_\CV)}~\oplus~N[L_N]^{(R_N)}_{\Delta_N = \Delta_A + {\ell/2}}~.
\end{equation}
The Lorentz- and~$R$-symmetry representations~$L_N, R_N$ of the null-state multiplet, as well as its shortening type~$N \in \{A, B, C, D\}$, depend on~$L_\CV$ and~$R_\CV$. However, its scaling dimension~$\Delta_N$ is unambiguously fixed in terms of~$\Delta_A$ and the level~$\ell$ at which the primary null state of the~$A$-type short multiplet resides. Recombination rules such as~\eqref{genrr} constitute useful consistency conditions on the structure of unitary short multiplets, and they will play an important role throughout this paper. 

The recombination phenomenon also shows that the spectrum of short multiplets need not be invariant under continuous deformations: given a family of SCFT labeled by some exactly marginal couplings (as was shown in~\cite{Cordova:2016xhm}, this can only happen for~$d = 3, \CN=1,2$ or for~$d = 4, \CN=1,2,4$), the spectrum of short multiplets can change as a function of the couplings. Some short multiplets may disappear by recombining into long ones, while some long multiplets may hit the unitarity bound and fragment into short ones. In general, the spectrum of short multiplets is only protected modulo such recombinations. Precisely this data is captured by the superconformal indices defined in~\cite{Kinney:2005ej, Bhattacharya:2008zy}. Certain special short multiplets never appear on the right-hand side of any recombination rule~\eqref{genrr}. Their spectrum is therefore preserved under exactly marginal deformations, and we will refer to them as absolutely protected short multiplets. Note that only isolated~$B, C, D$-type short multiplets can be absolutely protected. A trivial example of such a multiplet is the unit operator~$\1$, but below we will encounter more interesting examples as well.

A detailed survey of all long and short unitary superconformal multiplets in~$d \geq 3$ dimensions, including the quantum numbers of their SCPs and primary null states, as well as the resulting recombination rules, can be found in section~\ref{sec:usm}. There we also discuss the additional features that arise in two-sided situations with left and right~$Q, \b Q$ supercharges, and enumerate all absolutely protected multiplets. 

\subsection{Constructing Superconformal Multiplets}

\label{sec:introcsm}

As reviewed in the previous subsection, the operator content of a superconformal multiplet~$\CM$ is obtained by acting with all~$Q$-supercharges on the SCP~$\CV$, while setting to zero any null states. It is convenient to organize this operator content by decomposing~$\CM$ into irreducible multiplets~$\CC_a$ of the bosonic conformal and~$R$-symmetry algebra~$\frak{so}(d,2) \times \mathfrak{R}$, as in~\eqref{basicdecomp}. Every conformal multiplet~$\CC_a$ is fully specified by the Lorentz and~$R$-symmetry quantum numbers~$L_a, R_a$ and the scaling dimension~$\Delta_a$ of its CP, which we also denote by~$\CC_a$ as long as no confusion can arise. It is useful to grade the conformal multiplets~$\CC_a$ by the level of their CP within the superconformal multiplet: a CP~$\CC^{(\ell)}_{a}$ at level~$\ell$ satisfies~$\CC^{(\ell)}_{a} \sim Q^\ell \CV$. A more refined version of the decomposition~\eqref{basicdecomp} can then be written as follows:
\begin{equation}\label{basicdecompii}
\CM = \bigoplus_{a} \, \CC^{(\ell)}_{a}~~(\text{finite sum})~, \qquad \CC^{(\ell)}_{a} = [L_a]^{(R_a)}_{\Delta_a}~, \qquad \Delta_a = \Delta_\CV + {\ell \over 2}~.
\end{equation}
By focusing on the CPs~$\CC^{(\ell)}_{a}$, we can work modulo the ideal of CDs, i.e.~we can set all derivatives to zero,~$P_\mu \sim \d_\mu \sim 0$. Therefore, the~$Q$-supersymmetries effectively anticommute,
\begin{equation}\label{anticommqs}
\big\{Q_i, Q_j\big\} = 0~, \qquad i, j = 1, \ldots, N_Q~.
\end{equation}
Standard arguments about Fermi statistics then imply that CPs can only occur at levels~$\ell$ satisfying~$0 \leq \ell \leq \ell_\text{max}$ with~$\ell_\text{max} \leq N_Q$. This explains why the direct sum in~\eqref{basicdecompii} is finite. 

The constraints of unitarity on superconformal multiplets~$\CM$, expressed in terms of the quantum numbers of their SCPs~$\CV$ as in~\eqref{scabound} and~\eqref{scbcdbound}, are well understood, see for instance~\cite{Dobrev:1985qv,Minwalla:1997ka,Ferrara:2000xg,Dobrev:2002dt,Dolan:2002zh,Kinney:2005ej,Bhattacharya:2008zy} and references therein; a detailed summary appears in section~\ref{sec:usm}. By comparison, less is known about the operator content of these multiplets, although many results have been obtained in various cases of interest. This may seem surprising, because the discussion above shows that it is in principle straightforward to determine the operator content of any superconformal multiplet~$\CM$ as follows:
\begin{itemize}
\item[1.)] Find all~$Q$-descendants of the SCP~$\CV$ by acting on every state in its~$\frak{so}(d) \times \frak R$ representation with all independent combinations of~$Q$-supercharges, up to reordering using~\eqref{anticommqs}.

\item[2.)] Proceeding level by level, Clebsch-Gordon decompose all of these states into irreducible~$\frak{so}(d) \times \frak R$ representations~$\CC_a^{(\ell)}$. Every state in this decomposition is a linear combination of monomials of the schematic form~$Q^\ell \CV$, i.e.~a product of~$\ell$ supercharges acting on some particular state in the~$\frak{so}(d) \times \frak R$ multiplet of the SCP~$\CV$. This presentation makes the action of the~$Q$-supercharges on the ~$\CC^{(\ell)}_a$ completely explicit. 

\item[3.)] If~$\CM$ is a long multiplet, the~$\CC_a^{(\ell)}$ constitute the desired decomposition into conformal primaries, but if~$\CM$ is short, we must remove those~$\CC_a^{(\ell)}$ that comprise the null states of~$\CM$. These are simply given by the~$Q$-descendants of the primary null representation $N \in \{\CC_a^{(\ell)}\}$ (i.e.~the null representation with the smallest value of~$\ell$), which can be found by repeatedly acting on the states in~$N$ with the~$Q$-supersymmetries. As explained in~$2.)$ above, the action of the~$Q$-supersymmetries on states is explicitly known. 
\end{itemize}
\noindent This algorithm is simple and correct, but computationally very expensive to a degree that makes it impractical in many cases of interest. This has lead several authors to explore alternative procedures for generating superconformal multiplets that leverage the group-theoretic nature of the problem, see for instance~\cite{Dolan:2002zh, Dobrev:2004tk, Kinney:2005ej,Bianchi:2006ti,Bhattacharya:2008zy,Dolan:2008vc,Dobrev:2012me,Beem:2014kka}  and references therein. Motivated by these results, we propose a precise and general, yet reasonably efficient, algorithm that takes a superconformal multiplet~$\CM$ (characterized by the quantum numbers of its SCP~$\CV$) and outputs its operator content by enumerating the quantum numbers of all CPs~$\CC^{(\ell)}_a$ that appear in the decomposition~\eqref{basicdecompii}. A detailed description of the algorithm and its relation to previous work appears in section~\ref{sec:alg}. Here we briefly sketch some of its features. 

Our approach is strongly motivated by~\cite{Dolan:2002zh}, where the Racah-Speiser (RS) algorithm for decomposing tensor product representations of Lie algebras was applied to superconformal multiplets. The RS algorithm is briefly summarized in appendix~\ref{app:RSalg}, and further discussed in section~\ref{sec:alg}. To get some intuition for how it works, consider the decomposition of a tensor product of two~$\frak{su}(2)$ representations,\begin{equation}\label{su2dec}
j_1 \otimes j_2 = j_1+j_2 \oplus \cdots \oplus |j_1 - j_2|~.
\end{equation}
The RS algorithm reproduces this result by starting with a set of trial weights~$\CW_\text{RS}$ obtained by adding all weights of the~$j_2$ representation to the highest weight of the~$j_1$ representation:
\begin{equation}\label{su2RSintro}
\CW_\text{RS} = \{j_1 + j_2, \ldots, j_1 - j_2\}~.
\end{equation}
If~$j_1 \geq j_2$, all weights in~$\CW_\text{RS}$ are positive and the RS algorithm implies that these weights are in one-to-one correspondence with the highest weights of~$\frak{su}(2)$ representations appearing in \eqref{su2dec}. If ~$j_1 < j_2$, some trial weights~\eqref{su2RSintro} are negative and the RS algorithm  prescribes that all negative trial weights should either cancel against positive ones, or simply be removed from~$\CW_\text{RS}$, according to precise group-theoretic rules. After these cancellations, the remaining weights in~$\CW_\text{RS}$ again coincide with the highest weights that appear on the right-hand side of the decomposition~\eqref{su2dec}.  The main virtue of the RS algorithm is that it replaces the many complicated states appearing in the Clebsch-Gordon decomposition by a smaller, simple set of representative trial weights~$\CW_\text{RS}$. Conversely, the RS algorithm does not give an explicit description of the states represented by the weights in~$\CW_\text{RS}$. 

It is straightforward to construct long multiplets using the RS algorithm: since there are no null states, the operator content at level~$\ell$ is generated by acting with~$\ell$ fully antisymmetrized~$Q$-supercharges on the SCP~$\CV$. This leads to the reducible representation
\begin{equation}
\wedge^\ell \CR_Q \otimes \CV=\bigoplus_{a}\CC^{(\ell)}_{a}~,
\end{equation}
where $\CR_Q$ is the representation carried by the supercharges.  Using the RS procedure this tensor product can be decomposed into irreducible Lorentz and~$R$-symmetry representations yielding the operator content $\CC^{(\ell)}_{a}$ appearing above.
 
By contrast, it is not {\it a priori} clear how to construct short multiplets using the RS algorithm, because its use of trial weights~$\CW_\text{RS}$ rather than explicit states obscures the action of the~$Q$-supercharges. This complicates the identification and removal of null states. Following~\cite{Dolan:2002zh}, short multiplets~$X_{\ell = 1}\,,\,X \in \{A, B, C, D\}$ whose primary null representation~$N$ occurs at level~$\ell = 1$ can be constructed by simply omitting certain~$Q$-supercharges from the construction of the multiplet (more precisely, the RS trial weights~$\CW_\text{RS}$). The construction of multiplets~$X_{\ell \geq 2}$ with higher-level null states is more delicate, and various approaches have been proposed in the literature. 

In section~\ref{sec:alg} we synthesize and extend these proposals to formulate a streamlined, uniform algorithm that can be used to generate all unitary superconformal multiplets in~$d \geq 3$ dimensions. As is the case for most existing prescriptions, our proposal is conjectural, and we currently do not know of a general proof that establishes its correctness. However, we have subjected it to numerous detailed consistency checks, including an independent verification of all recombination rules~\eqref{genrr}, and a comparison with many known results. It is important to emphasize that while our algorithm is simple to state and apply for any~$d$ and~$\CN$, the resulting multiplets display a cornucopia of sporadic phenomena. This is well exemplified by multiplets that contain conserved currents, which are discussed in section~\ref{sec:currents}, and multiplets that contain supersymmetric deformations, which were systematically analyzed in\cite{Cordova:2016xhm}. We view the ability of our prescription to capture this diversity while appearing to avoid any inconsistencies as strong evidence for its correctness. 

\subsection{Applications}

\label{sec:introapp}

Here we give a brief overview of several diverse applications of our machinery:

\begin{itemize}
\item {\it Construction of Superconformal Multiplets:} Using our algorithm, we can tabulate the operator content of any superconformal multiplet. In section~\ref{sec:tables} we explicitly present such tables for all generic long and short multiplets in theories with~$N_{Q} \leq 8$ supercharges. A {\tt Mathematica} package implementing the algorithm will be made available in~\cite{tdmm}; it can be used to construct any other superconformal multiplet. Knowing the operator content of a superconformal multiplet can be used to evaluate superconformal characters and indices~\cite{Kinney:2005ej, Bhattacharya:2008zy}, or to construct superconformal blocks.

\item {\it Deformations of Superconformal Field Theories:} The structure of superconformal multiplets can be used to analyze possible deformations of SCFTs that preserve some supersymmetry. In~\cite{Cordova:2016xhm}, we have used the results of the present paper to classify such deformations. This has also lead to a streamlined understanding of the constraints of supersymmetry on moduli-space effective actions~\cite{Cordova:2015vwa, Cordova:2015fha, Cordova:2016xhm}.

\item {\it Conserved Currents in Superconformal Field Theories:}  In section~\ref{sec:currents}, we classify all superconformal multiplets that contain short representations of the conformal algebra, i.e.~conserved currents or free fields. Given~$d$ and~$\CN$, it may not be possible to embed certain currents into a superconformal multiplet, and hence they are absent in SCFTs. For instance, we show that theories with~$N_Q >8$ in~$d \geq 4$ do not admit conventional (non-$R$) flavor currents. 

\item {\it Constraints on Maximal Supersymmetry:} When $d=3, 4, 6$ the superconformal algebras in~\eqref{scftalgs} exist for arbitrary $\mathcal{N}$. Standard arguments involving massless one-particle representations show that weakly-coupled theories with~$N_Q > 16$ in~$d = 4,6$ do not admit a stress tensor, but can be well behaved in~$d = 3$. In section~\ref{sec:maxsusy}, we extend these results to arbitrary SCFTs: we show that theories with~$N_Q > 16$ in~$d = 4,6$ do not exist, because they do not admit a suitable stress tensor multiplet. By contrast, such theories do exist in~$d = 3$, but they are necessarily free. 

\item {\it Recombination Rules and Constraints on Conformal Manifolds:}

In section~\ref{sec:usm}, we present all possible recombination rules (which take the schematic form~\eqref{genrr}) in theories with~$N_Q \leq 16$ supercharges. These formulas furnish an important consistency check on our general algorithm (see section~\ref{sec:alg}). They are also a possible starting point for constructing superconformal indices~\cite{Kinney:2005ej, Bhattacharya:2008zy}. In every case, we enumerate all absolutely protected multiplets, which do not participate in any recombination rule.  The spectrum of such multiplets does not depend on exactly marginal couplings, i.e.~it is constant on the conformal manifold. (Recall that conformal manifolds can only exist when~$d = 3, \CN=1,2$ or when~$d = 4, \CN=1,2,4$, see~\cite{Cordova:2016xhm}.) For instance:
\begin{itemize}
\item[$\star$] In~$d = 4$, multiplets containing free fields are absolutely protected, i.e.~it is not possible to generate or destroy gauge-invariant free-field operators by varying exactly marginal couplings. (This statement does not apply to matter fields in a gauge theory at vanishing gauge coupling, because these are not gauge invariant.) Note that there are many examples of supersymmetric renormalization group (RG) flows, along which conformal symmetry is broken, that lead to new gauge-invariant free-field operators in the infrared. 

\item[$\star$] In~$d=4$, all multiplets containing relevant, Lorentz-scalar deformations that preserve supersymmetry are absolutely protected. The number~$\mu$ of such operators, and its monotonicity properties under RG flows, was recently studied in~\cite{Gukov:2015qea,Gukov:2016tnp}. Here we see that~$\mu$ is constant on conformal manifolds in~$d = 4$. 
\item[$\star$] The same statements hold in~$d = 3, \, \CN=2$ theories, with one exception: flavor current multiplets, which contain relevant real mass deformations (see e.g.~section~4.1 of~\cite{Cordova:2016xhm}), can recombine with a marginal chiral deformation and give rise to an irrelevant deformation residing in a long multiplet~\cite{Green:2010da}. 
\end{itemize}
\end{itemize}

\vskip0.5cm

\noindent {\it Note:~During the completion of this paper we received~\cite{Buican:2016hpb}, which has some overlap with certain subsections below.}

\section{Unitary Superconformal Multiplets}

\label{sec:usm}

In this section we review the unitary multiplets of superconformal algebras in~$3 \leq d \leq 6$ dimensions. We focus on values of~$\CN$ corresponding to~$N_Q \leq 16$ Poincar\'e supercharges. As was reviewed around~\eqref{scftalgs}, the~$d \neq 5$ superconformal algebras exist for any value of~$\CN$. (In appendix~\ref{app:genN} we briefly outline the representation theory of these algebras for generic~$\CN$.) Most of these algebras have $N_Q>16$. The (non-) existence of SCFTs with such large amounts of supersymmetry is discussed in section~\ref{sec:maxsusy}. 

The subsections below describing different values of~$d$ and~$\CN$ are largely self-contained and can be read independently. In each case we briefly summarize our conventions and review the Lorentz and~$R$-symmetry quantum numbers of the supercharges. We always use integer-valued Dynkin labels to denote Lie-algebra representations.\footnote{~Some authors (e.g.~\cite{Bhattacharya:2008zy,Minwalla:1997ka}) use orthogonal weights~$h_i$ to label representations of~$\frak{so}(n)$. The conversion between orthogonal weights and Dynkin labels is discussed around~\eqref{hdlambda} in appendix~\ref{app:liealg}.} We enumerate the possible unitarity superconformal multiplets, relying on the results of~\cite{Dobrev:1985qv,Minwalla:1997ka,Ferrara:2000xg,Dobrev:2002dt,Dolan:2002zh,Bhattacharya:2008zy}. Throughout, we follow the uniform labeling scheme for superconformal representations that was introduced around~\eqref{lmnot}, \eqref{aelldef}, and~\eqref{xelldef}: multiplets are labeled by capital letters~$L$ (long) or~$A, B, C, D$ (short). For short multiplets, we use a subscript~$\ell$ to indicate the level of the primary null state, e.g.~$A_\ell$ denotes an~$A$-type short multiplet whose primary null state resides at level~$\ell$. We also discuss the modifications that are needed when the supercharge representation~$\CR_Q$ is reducible and splits into left and right supercharges~$Q$ and~$\b Q$. (This happens for~$d = 3, \CN=2$ and in~$d = 4$.) In these theories, we denote multiplets by a pair of capital letters (one unbarred and one barred) to indicate the~$Q, \b Q$ null states, e.g.~$L \b A_2$. 

For every value of~$d$ and~$\CN$, we list the superconformal shortening conditions allowed by unitarity, the possible Lorentz and~$R$-symmetry quantum numbers of the superconformal primary, the restrictions on its scaling dimension imposed by unitarity, and the quantum numbers of the primary null state. In theories with~$Q$ and~$\b Q$ supercharges, we first list the corresponding shortening conditions independently, before combining them into consistent two-sided superconformal multiplets. 

We also spell out all possible recombination rules (or fragmentation rules) that dictate which short multiplets can combine to form long ones, and we enumerate all absolutely protected short multiplets, which do not participate in any recombination rule. When an SCFT possesses a conformal manifold of exactly marginal couplings, recombinations can in principle happen as we vary these couplings, while absolutely protected multiplets persist over the entire conformal manifold. In the absence of exactly marginal couplings, the physical meaning of the recombination rules is less clear. Nevertheless, they constitute a nontrivial and useful statement about the structure of superconformal multiplets.

\subsection{Three Dimensions}

\label{sec:3dmult}

In this section we list all unitary representations of three-dimensional SCFTs with $1 \leq \CN \leq 6$ and~$\CN=8$ supersymmetry (see~\cite{Minwalla:1997ka,Bhattacharya:2008zy} for additional details). As discussed in section~\ref{sec:maxsusy}, unitarity SCFTs with~$\CN \geq 9$ exist, but are necessarily free.  Similarly, genuine~$\CN=7$ SCFTs do not exist, see section~\ref{sec:3dn7stm} and~\cite{Bashkirov:2011fr}.

Throughout, representations of the~$\frak{so}(3) = \frak{su}(2)$ Lorentz algebra are denoted by
\begin{equation}\label{Lor3d }
[j]~, \qquad j \in \Z_{\geq 0}~.
\end{equation} 
Here~$j$ is an integer-valued~$\frak{su}(2)$ Dynkin label, so that the~$[j]$-representation is~$(j+1)$-dimensional. (The conventional half-integral~$\frak{su}(2)$ spin is~$\half j$.) We write~$[j]_\Delta$ whenever we wish to indicate the scaling dimension~$\Delta$.   

\subsubsection{$d = 3$,~$\CN=1$}
\label{sec:3dn1def}

The~$\CN=1$ superconformal algebra is~$\frak{osp}(1|4)$, which does not contain an~$R$-symmetry. The~$Q$-supersymmetries transform as \begin{equation}
Q \in [1]_\half~, \qquad N_Q = 2~.
\end{equation}
The superconformal unitarity bounds and shortening conditions are summarized in table~\ref{tab:3DN1}. 

\smallskip

\renewcommand{\arraystretch}{1.6}
\renewcommand\tabcolsep{6pt}
\begin{table}[H]
  \centering
  \begin{tabular}{ |c|lr| l|l| }
\hline
{\bf Name} &  \multicolumn{2}{c}{\bf Primary} &  \multicolumn{1}{|c|}{\bf Unitarity Bound} & \multicolumn{1}{c|}{\bf Null State } \\
\hline
\hline
$L$ & $[j]_{\Delta}~,$& $j\geq1$ &$\Delta>\frac{1}{2} \,j+1$ & \multicolumn{1}{c|}{$-$} \\
\hline
$L'$ & $[0]_{\Delta}$& $$ &$\Delta>\frac{1}{2}$ & \multicolumn{1}{c|}{$-$} \\
\hline 
\hline
$A_{1}$ & $[j]_{\Delta}~,$& $j\geq1$ &$\Delta=\frac{1}{2} \, j+1$ & $[j-1]_{\Delta+1/2}$ \\
\hline 
$A_{2}'$ & $[0]_{\Delta}$& $$ &$\Delta=\frac{1}{2}$ & $[0]_{\Delta+1}$ \\
\hline
\hline
$B_{1}$ & $[0]_{\Delta}$& $$ &$\Delta=0$ & $[1]_{\Delta+1/2}$ \\
\hline
\end{tabular}
  \caption{Shortening conditions in three-dimensional~$\CN=1$ SCFTs.}
  \label{tab:3DN1}
\end{table}

\noindent Here the notation~$L'$ and~$A_2'$ for multiplets with~$j = 0$ emphasizes the fact that their unitarity bounds do not follow from those of the~$L$ or~$A_1$ multiplets by setting~$j = 0$. 

In order to state the recombination rules, we define
\begin{equation}
\Delta_{A}=\frac{1}{2}\, j+1~, \hspace{.5in} \Delta_{A'}=\frac{1}{2}~.
\end{equation}
As~$\Delta \rightarrow \Delta_{A}^+, \Delta_{A'}^+$ we then find
\begin{equation}\label{d3n1recrules}
\begin{array}{rclcl}
L[j\geq 2]_{\Delta} & \rightarrow &A_{1}[j]_{\Delta_{A}}  &\oplus & L[j-1]_{\Delta_{A}+\half}~, \\
L[j=1]_{\Delta} & \rightarrow &A_{1}[1]_{\Delta_{A}} & \oplus  &L'[0]_{\Delta_{A}+\half}~, \\
L'[j=0]_{\Delta} & \rightarrow &A_{2}'[0]_{\Delta_{A'}}  &\oplus & L'[0]_{\Delta_{A'}+1}~. 
\end{array}
\end{equation}
These recombination rules are somewhat unusual, because the right-hand side contains long multiplets. At first sight, they even appear to be inconsistent: by examining the tables of~$d = 3, \CN=1$ multiplets in section~\ref{sec:3dn1tabs}, the right-hand side of~\eqref{d3n1recrules} naively contains more operators that the left-hand side. 

Let us examine the generic case~$j \geq 2$ in more detail: the long multiplet~$L[j]_\Delta$ contains~$2(j+1)+ 2(j+1)$ bosonic and fermionic operators, while~$A_1[j]_{\Delta_A}$ is a~$2 + 2$ multiplet of conserved currents (see the discussion around~\eqref{3dn1c}). Here we have subtracted the number of conservation laws from the operator count. Finally, the null state multiplet~$L[j-1]_{\Delta_1+\half}$ contains~$2j + 2j$ operators: some of them correspond to the conservation laws for the two current operators in~$A_1[j]_{\Delta_A}$, while the others supply the remainder of the~$L[j]_\Delta$ multiplet. This shows that one must correctly account for all conservations laws (or free-field equations of motion) when checking recombinations rules. This is particularly dramatic for the last recombination rule in~\eqref{d3n1recrules}: both the left-hand side and the right-hand side contain an~$L'[0]$ multiplet (with~$2+2$ bosonic and fermionic operators), while~$A'_2[0]_{\Delta_{A'}}$ is a multiplet of free fields, which does not contribute any net degrees of freedom (see also~\eqref{3dn1f}).  

Note that all multiplets in table~\ref{tab:3DN1} (other than the~$B_1[0]_0$ unit operator) appear in the recombination rules~\eqref{d3n1recrules}, and hence none of them are absolutely protected. 

Generic~$d = 3, \CN=1$ multiplets are tabulated in section~\ref{sec:3dn1tabs}; conserved current multiplets are studied in section~\ref{sec:3dn1curr}.

\subsubsection{$d = 3$,~$\CN=2$}

\label{sec:3dn2def}

The~$\CN=2$ superconformal algebra is~$\frak{osp}(2|4)$, hence the~$R$-symmetry is~$\frak{so}(2)_R \simeq \frak{u}(1)_R$. Operators of~$R$-charge~$r \in \R$ are denoted by~$(r)$. There are independent~$Q$ and~$\b Q$ supersymmetries, which transform as 
\begin{equation}
Q \in [1]_{\half}^{(-1)}~, \qquad \b Q \in [1]_{\half}^{(1)}~, \qquad N_Q = 4~.
\end{equation}
Superconformal multiplets obey unitarity bounds and shortening conditions with respect to both~$Q$ and~$\b Q$, which are summarized in tables~\ref{tab:3DN2C} and~\ref{tab:3DN2AC}, respectively. 

\renewcommand{\arraystretch}{1.5}
\renewcommand\tabcolsep{8pt}
\begin{table}[H]
  \centering
  \begin{tabular}{ |c|lr| l|l| }
\hline
{\bf Name} &  \multicolumn{2}{c}{\bf Primary} &  \multicolumn{1}{|c|}{\bf Unitarity Bound} & \multicolumn{1}{c|}{\bf $Q$ Null State } \\
\hline
\hline
$L$ & $[j]_{\Delta}^{(r)}$&  &$\Delta>\frac{1}{2} \,j-r+1$ & \multicolumn{1}{c|}{$-$} \\
\hline
\hline
$A_{1}$ & $[j]_{\Delta}^{(r)}~,$& $j\geq1$ &$\Delta=\frac{1}{2} \, j-r+1$ & $[j-1]_{\Delta+1/2}^{(r-1)}$ \\
\hline 
$A_{2}$ & $[0]_{\Delta}^{(r)}$& $$ &$\Delta=1-r$ & $[0]_{\Delta+1}^{(r-2)}$ \\
\hline
\hline
$B_{1}$ & $[0]_{\Delta}^{(r)}$& $$ &$\Delta=-r$ & $[1]_{\Delta+1/2}^{(r-1)}$ \\
\hline
\end{tabular}
  \caption{$Q$ shortening conditions in three-dimensional~$\CN=2$ SCFTs.}
  \label{tab:3DN2C}
\end{table}

\renewcommand{\arraystretch}{1.5}
\renewcommand\tabcolsep{8pt}
\begin{table}[H]
  \centering
  \begin{tabular}{ |c|lr| l|l| }
\hline
{\bf Name} &  \multicolumn{2}{c}{\bf Primary} &  \multicolumn{1}{|c|}{\bf Unitarity Bound} & \multicolumn{1}{c|}{\bf $\b Q$ Null State } \\
\hline
\hline
$\overline{L}$ & $[j]_{\Delta}^{(r)}$&  &$\Delta>\frac{1}{2} \,j+r+1$ & \multicolumn{1}{c|}{$-$} \\
\hline
\hline
$\overline{A}_{1}$ & $[j]_{\Delta}^{(r)}~,$& $j\geq1$ &$\Delta=\frac{1}{2} \, j+r+1$ & $[j-1]_{\Delta+1/2}^{(r+1)}$ \\
\hline 
$\overline{A}_{2}$ & $[0]_{\Delta}^{(r)}$& $$ &$\Delta=1+r$ & $[0]_{\Delta+1}^{(r+2)}$ \\
\hline
\hline
$\overline{B}_{1}$ & $[0]_{\Delta}^{(r)}$& $$ &$\Delta=r$ & $[1]_{\Delta+1/2}^{(r+1)}$ \\
\hline
\end{tabular}
  \caption{$\b Q$ shortening conditions in three-dimensional~$\CN=2$ SCFTs.}
  \label{tab:3DN2AC}
\end{table} 

\noindent Full~$\CN=2$ multiplets are two-sided: they are obtained by imposing both left~$Q$ and right~$\b Q$ unitarity bounds and shortening conditions. This can lead to restrictions on some quantum numbers, and not all left and right choices are mutually compatible. The consistent two-sided multiplets are summarized in table~\ref{tab:3DN2CAC}. 

\bigskip

\renewcommand{\arraystretch}{2}
\renewcommand\tabcolsep{2pt}
\begin{table}[H]
  \centering
  \captionsetup{justification=centering}
  \begin{tabular}{ |c||c|c|c|c|}
\hline
 &  \multicolumn{1}{c}{$ \boldsymbol{\overline{{L}}}$} &  \multicolumn{1}{|c|}{$\boldsymbol{\overline{A}_{1}}$} & \multicolumn{1}{c|}{$\boldsymbol{\overline{A}_{2}}$ }& \multicolumn{1}{c|}{$\boldsymbol{\overline{B}_{1}}$ } \\
\hline
\hline
\multirow{2}{*}{~~$\boldsymbol L$~~} &~~$[j]_{\Delta}^{(r)}$~~& $[j \geq 1]_{\Delta}^{(r>0)}$ &$[j=0]_{\Delta}^{(r>0)}$&$[j=0]_{\Delta}^{(r>\half)}$ \\[-5pt]
&~~$\Delta>\frac{1}{2} \,j+|r|+1$~~ &~~$\Delta=\frac{1}{2} \,j+r+1$ ~~&$\Delta=1+r$ &$\Delta=r$ \\
\hline
\multirow{2}{*}{~~$\boldsymbol{A_{1}}$~~} &$[j\geq 1]_{\Delta}^{(r<0)}$& $[j \geq 1]_{\Delta}^{(r=0)}$ &\multirow{2}{*}{$-$}&\multirow{2}{*}{$-$} \\[-5pt]
&~~ $\Delta=\frac{1}{2} \,j-r+1$ ~~&$\Delta=\frac{1}{2} \,j+1$ & & \\
\hline
\multirow{2}{*}{$\boldsymbol{A_{2}}$} &$[j=0]_{\Delta}^{(r<0)}$& \multirow{2}{*}{$-$} &$[j=0]_{\Delta}^{(r=0)}$&~~~$[j=0]_{\Delta}^{(r=1/2)}$~~~ \\[-5pt]
& $\Delta=1-r$ &&$\Delta=1$ &$\Delta=\half$ \\
\hline
\multirow{2}{*}{$\boldsymbol{B_{1}}$} &$[j=0]_{\Delta}^{(r<-\half)}$& \multirow{2}{*}{$-$}&~~~$[j=0]_{\Delta}^{(r=-\half)}$~~~&$[j=0]_{\Delta}^{(r=0)}$ \\[-5pt]
& $\Delta=-r$ & &$\Delta=\half$ &$\Delta=0$ \\
\hline
\end{tabular}
\caption{Consistent two-sided multiplets in three-dimensional~$\mathcal{N}=2$ theories.  A dash~($-$) indicates that the corresponding multiplet does not exist.}
  \label{tab:3DN2CAC}
 \end{table} 

\noindent In order to present the recombination rules, we define
\begin{equation}
\Delta_{A}=\frac{1}{2}\, j-r+1~, \qquad \Delta_{\overline{A}}=\frac{1}{2}\, j+r+1~.
\end{equation}
As~$\Delta \rightarrow \Delta_{A}^+$, the left-handed~$Q$ shortening conditions in table~\ref{tab:3DN2C} give rise to the following partial, left recombination rules: 
\begin{equation}
\begin{array}{rclcl}
L[j\geq 2]_{\Delta}^{(r)} & \rightarrow &A_{1}[j]_{\Delta_{A}}^{(r)}  &\oplus & A_{1}[j-1]_{\Delta_{A}+\half}^{(r-1)}~, \\
L[j=1]_{\Delta}^{(r)} & \rightarrow &A_{1}[1]_{\Delta_{A}}^{(r)}  &\oplus & A_{2}[0]_{\Delta_{A}+\half}^{(r-1)}~, \\
L[j=0]_{\Delta}^{(r)} & \rightarrow &A_{2}[0]_{\Delta_{A}}^{(r)}  &\oplus & B_{1}[0]_{\Delta_{A}+1}^{(r-2)}~.
\end{array}
\label{chiralrecomb3d}
\end{equation}
An analogous set of partial, right recombination rules follows from the~$\b Q$ shortening conditions in table~\ref{tab:3DN2AC}. 

The qualitative structure of the recombination rules for complete, two-sided multiplets is controlled by sign of the~$R$-charge~$r$, which determines the relative magnitude of~$\Delta_A$ and~$\Delta_{\b A}$. This in turn determines which unitarity bound is reached first as the dimension~$\Delta$ of a long multiplet is lowered: if~$r < 0$ then~$\Delta_A> \Delta_{\b A}$ and the left~$Q$ unitarity bound is reached first, while if~$r > 0$ the right~$\b Q$ unitarity bound is reached first, because~$\Delta_{\b A} > \Delta_A$. Finally, if~$r = 0$ and~$\Delta_A = \Delta_{\b A}$, then both unitarity bounds are reached simultaneously. Explicitly:
\begin{equation}
\begin{array}{rclclcl}
L\overline{L}[j\geq 2]_{\Delta}^{(r<0)} & \rightarrow &A_{1}\overline{L}[j]_{\Delta_{A}}^{(r)}  &\oplus & A_{1}\overline{L}[j-1]_{\Delta_{A}+\half}^{(r-1)}~, \\
L\overline{L}[j=1]_{\Delta}^{(r<0)} & \rightarrow &A_{1}\overline{L}[1]_{\Delta_{A}}^{(r)}  &\oplus & A_{2}\overline{L}[0]_{\Delta_{A}+\half}^{(r-1)}~, \\
L\overline{L}[j=0]_{\Delta}^{(r<0)} & \rightarrow &A_{2}\overline{L}[0]_{\Delta_{A}}^{(r)}  &\oplus & B_{1}\overline{L}[0]_{\Delta_{A}+1}^{(r-2)}~, \\
L\overline{L}[j\geq 2]_{\Delta}^{(r>0)} & \rightarrow &L\overline{A}_{1}[j]_{\Delta_{\overline{A}}}^{(r)}  &\oplus & L\overline{A}_{1}[j-1]_{\Delta_{\overline{A}}+\half}^{(r+1)}~, \\
L\overline{L}[j=1]_{\Delta}^{(r>0)} & \rightarrow &L\overline{A}_{1}[1]_{\Delta_{\overline{A}}}^{(r)}  &\oplus & L\overline{A}_{2}[0]_{\Delta_{\overline{A}}+\half}^{(r+1)}~, \\
L\overline{L}[j=0]_{\Delta}^{(r>0)} & \rightarrow &L\overline{A}_{2}[0]_{\Delta_{\overline{A}}}^{(r)}  &\oplus & L\overline{B}_{1}[0]_{\Delta_{\overline{A}}+1}^{(r+2)}~, \\
L\overline{L}[j\geq 2]_{\Delta}^{(r=0)} & \rightarrow &A_{1}\overline{A}_{1}[j]_{\Delta_{A}}^{(0)}  &\oplus & A_{1}\overline{L}[j-1]_{\Delta_{A}+\half}^{(-1)} & \oplus & L\overline{A}_{1}[j-1]_{\Delta_{A}+\half}^{(1)}~, \\
L\overline{L}[j=1]_{\Delta}^{(r=0)} & \rightarrow &A_{1}\overline{A}_{1}[1]_{\Delta_{A}}^{(0)}  &\oplus & A_{2}\overline{L}[0]_{\Delta_{A}+\half}^{(r-1)} & \oplus & L\overline{A}_{2}[0]_{\Delta_{A}+\half}^{(1)}~,  \\
L\overline{L}[j=0]_{\Delta}^{(r=0)} & \rightarrow &A_{2}\overline{A}_{2}[0]_{\Delta_{A}}^{(0)}  &\oplus & B_{1}\overline{L}[0]_{\Delta_{A}+1}^{(-2)} & \oplus & L\overline{B}_{1}[0]_{\Delta_{A}+1}^{(2)}~.
\end{array}
\label{3dn2recomb}
\end{equation}
Note that the recombinations rules for~$r < 0$ simply follow from the left~$Q$ recombination rules in~\ref{chiralrecomb3d} by tensoring with an~$\b L$ multiplet on the right (the case~$r > 0$ works similarly). By contrast, the recombination rules for~$r = 0$ cannot be obtained by naively tensoring left and right recombination rules, because this would result in four multiplets on the right-hand side, rather than the three multiplets that actually occur in the last three lines of~\eqref{3dn2recomb}. (This was also observed in~\cite{Minwalla:2011ma}.) In fact, examining the tables of~$d = 3, \CN=2$ multiplets in section~\ref{sec:3dn2tabs}, we might naively conclude that there are more operators on the right-hand side than on the left. As in the discussion after~\eqref{d3n1recrules} above, this apparent paradox is resolved by noting that all~$A_\ell \b A_\ell$-multiplets contain conserved currents (see section~\ref{sec:3dn2curr}), whose conservation laws must be correctly subtracted when verifying the recombination rules in~\eqref{3dn2recomb}.   

By examining the right-hand side of~\eqref{3dn2recomb} we conclude that free chiral~$A_{2}\overline{B}_{1}[0]^{(r=\half)}_{1/2}$ multiplets, as well as chiral operators of the form~$L\overline{B}_{1}[0]_r^{(r)}$ with~$\half<r<2$ (along with their complex conjugate anti-chiral multiplets) are absolutely protected. Note that the latter multiplets give rise to some of the possible relevant supersymmetric deformations of~$d = 3, \CN=2$ SCFTs (see section~3.1.2 of~\cite{Cordova:2016xhm}).

Generic~$d = 3, \CN=2$ multiplets are tabulated in section~\ref{sec:3dn2tabs}; conserved current multiplets are studied in section~\ref{sec:3dn2curr}.  

\subsubsection{$d=3$,~$\CN=3$}
\label{sec:3dn3}

The~$\CN=3$ superconformal algebra is~$\frak{osp}(3|4)$, so that there is a~$\frak{so}(3)_R \simeq \frak{su}(2)_R$ symmetry. The~$R$-charges are denoted by~$(R)$, where~$R \in \Z_{\geq 0}$ is an~$\frak{su}(2)_R$ Dynkin label. The~$Q$-supersymmetries transform in the vector representation~$\bf 3$ of~$\frak{so}(3)_R$,
 \begin{equation}
Q \in [1]_{\half}^{(2)}~, \qquad N_Q = 6~.
\end{equation}
The superconformal unitarity bounds and shortening conditions are summarized in table~\ref{tab:3DN3}. 

\renewcommand{\arraystretch}{1.7}
\renewcommand\tabcolsep{8pt}
\begin{table}[H]
  \centering
  \begin{tabular}{ |c|lr| l|l| }
\hline
{\bf Name} &  \multicolumn{2}{c}{\bf Primary} &  \multicolumn{1}{|c|}{\bf Unitarity Bound} & \multicolumn{1}{c|}{\bf Null State } \\
\hline
\hline
$L$ & $[j]_{\Delta}^{\left(R\right)}$&  &$\Delta>\frac{1}{2} \,j+\half R+1$ & \multicolumn{1}{c|}{$-$} \\
\hline
\hline
$A_{1}$ & $[j]_{\Delta}^{\left(R\right)}~,$& $j\geq1$ &$\Delta=\frac{1}{2} \, j+\half R+1$ & $[j-1]_{\Delta+1/2}^{(R+2)}$ \\
\hline 
$A_{2}$ & $[0]_{\Delta}^{\left(R\right)}$& $$ &$\Delta=\half R+1$ & $[0]_{\Delta+1}^{(R+4)}$ \\
\hline
\hline
$B_{1}$ & $[0]_{\Delta}^{\left(R\right)}$& $$ &$\Delta=\half R$ & $[1]_{\Delta+1/2}^{(R+2)}$ \\
\hline
\end{tabular}
  \caption{Shortening conditions in three-dimensional~$\CN=3$ SCFTs.}
  \label{tab:3DN3}
\end{table} 

In order to write out the recombination rules, we define
\begin{equation}
\Delta_{A}=\frac{1}{2}\, j+\frac{1}{2}R+1~.
\end{equation}
As~$\Delta \rightarrow \Delta_{A}^+$, we find that
\begin{equation}
\begin{array}{rclcl}
L[j\geq 2]^{(R)}_{\Delta} & \rightarrow &A_{1}[j]^{(R)}_{\Delta_{A}}  &\oplus & A_{1}[j-1]^{(R+2)}_{\Delta_{A}+\half}~, \\
L[j=1]^{(R)}_{\Delta} & \rightarrow &A_{1}[1]^{(R)}_{\Delta_{A}} & \oplus  &A_{2}[0]^{(R+2)}_{\Delta_{A}+\half}~, \\
L[j=0]^{(R)}_{\Delta} & \rightarrow &A_{2}[0]^{(R)}_{\Delta_{A}}  &\oplus & B_{1}[0]^{(R+4)}_{\Delta_{A}+1}~. 
\end{array}
\end{equation}
It follows that~$B_{1}[0]_{R/2}^{(R)}$ multiplets with~$R \leq 3$ are absolutely protected; the case~$R = 1$ is the free hypermultiplet in~\eqref{3dn3fh}, while~$R = 2$ corresponds to the flavor current multiplet in~\eqref{3dn3gc}. 

Generic~$d = 3, \CN=3$ multiplets are tabulated in section~\ref{sec:3dn3tabs}; conserved current multiplets are studied in section~\ref{sec:3dn3curr}.  

\subsubsection{$d=3$,~$\CN=4$}

\label{sec:3dN4defs}

The~$\CN=4$ superconformal algebra is~$\frak{osp}(4|4)$, hence the~$R$-symmetry is~$\frak{so}(4)_R \simeq \frak{su}(2)_R \times \frak{su}(2)'_R$. Its representations are denoted by~$(R\,; R')$, where~$R, R' \in \Z_{\geq 0}$ are Dynkin labels for~$\frak{su}(2)_R$ and~$\frak{su}(2)'_R$, respectively. For example, $(1;0)$ and~$(0;1)$ are the left- and right-handed spinors~$\bf 2$ and~$\bf 2'$ of~$\frak{so}(4)_R$, while~$(1;1)$ is its vector representation~$\bf 4$. Note that the~$\frak{su}(2)_R$ and~$\frak{su}(2)'_R$ factors of the~$R$-symmetry algebra are inert under complex conjugation. However, they are exchanged by the action of mirror symmetry~$M$, which is an outer automorphism of the~$\CN=4$ superconformal algebra. (It need not be a symmetry of the field theory, although it can be.) The~$Q$-supersymmetries transform as
 \begin{equation}
Q \in [1]_{\half}^{(1;1)}~~, \qquad N_Q = 8~.
\end{equation}
The superconformal unitarity bounds and shortening conditions are summarized in table~\ref{tab:3DN4}.

\renewcommand{\arraystretch}{1.5}
\renewcommand\tabcolsep{8pt}
\begin{table}[H]
  \centering
  \begin{tabular}{ |c|lr| l|l| }
\hline
{\bf Name} &  \multicolumn{2}{c|}{\bf Primary} &  \multicolumn{1}{c}{\bf Unitarity Bound} & \multicolumn{1}{|c|}{\bf Null State } \\
\hline
\hline
$L$ & $[j]_{\Delta}^{(R;R')}$&  &$\Delta>\frac{1}{2} \,j+\half \left(R+R'\right)+1$ & \multicolumn{1}{c|}{$-$} \\
\hline
\hline
$A_{1}$ & $[j]_{\Delta}^{(R;R')}~,$& $j\geq1$ &$\Delta=\frac{1}{2} \, j+\half \left(R+R'\right)+1$ & $[j-1]_{\Delta+1/2}^{(R+1; R'+1)}$ \\
\hline 
$A_{2}$ & $[0]_{\Delta}^{(R;R')}$& $$ &$\Delta=\half \left(R+R'\right)+1$ & $[0]_{\Delta+1}^{(R+2; R'+2)}$ \\
\hline
\hline
$B_{1}$ & $[0]_{\Delta}^{(R;R')}$& $$ &$\Delta=\half \left(R+R'\right)$ & $[1]_{\Delta+1/2}^{(R+1; R'+1)}$ \\
\hline
\end{tabular}
  \caption{Shortening conditions in three-dimensional~$\CN=4$ SCFTs.}
  \label{tab:3DN4}
\end{table} 
\noindent If we define
\begin{equation}
\Delta_{A}=\frac{1}{2}\, j+\frac{1}{2}(R+R')+1~,
\end{equation}
we find the following recombination rules as~$\Delta \rightarrow \Delta_{A}^+$,
\begin{equation}
\begin{array}{rclcl}
L[j\geq 2]^{(R;R')}_{\Delta} & \rightarrow &A_{1}[j]^{(R;R')}_{\Delta_{A}}  &\oplus & A_{1}[j-1]^{(R+1;R'+1)}_{\Delta_{A}+1/2}~, \\
L[j=1]^{(R;R')}_{\Delta} & \rightarrow &A_{1}[1]^{(R;R')}_{\Delta_{A}} & \oplus  &A_{2}[0]^{(R+1;R'+1)}_{\Delta_{A}+1/2}~, \\
L[j=0]^{(R;R')}_{\Delta} & \rightarrow &A_{2}[0]^{(R;R')}_{\Delta_{A}}  &\oplus & B_{1}[0]^{(R+2;R'+2)}_{\Delta_{A}+1}~. 
\end{array}
\end{equation}
We conclude that the multiplets~$B_{1}[0]_{R/2 + R'/2}^{(R;R')}$ with~$R \leq 1$ or~$R' \leq 1$ are absolutely protected. This includes the free hypermultiplets in~\eqref{3dn4fh1} and~\eqref{3dn4fh2}, the flavor current multiplets in~\eqref{3dn4gc1} and~\eqref{3dn4gc2}, as well as the extra SUSY-current multiplet in~\eqref{3dn4sc}. 

Generic~$d = 3, \CN=4$ multiplets are tabulated in section~\ref{sec:3dn4tabs}; conserved current multiplets are studied in section~\ref{sec:3dn4curr}.

\subsubsection{$d=3$,~$\CN=5$}

\label{sec:3dN5defs}

The~$\CN=5$ superconformal algebra is~$\frak{osp}(5|4)$ and therefore the~$R$-symmetry is~$\frak{so}(5)_R$. Its representations are denoted by~$(R_1,R_2)$, where~$R_1, R_2 \in \Z_{\geq 0}$ are~$\frak{so}(5)_R$ Dynkin labels. For example, $(1,0)$ is the vector representation~$\bf 5$, while~$(0,1)$ is the spinor representation~$\bf 4$.\footnote{~Note that the corresponding~$\frak{sp}(4) \simeq \frak{so}(5)$ Dynkin labels are reversed, e.g.~$(1,0)$ is the~$\bf 4$ of~$\frak{sp}(4)$, see also appendix~\ref{app:liealg}.}  The~$Q$-supersymmetries transform as
 \begin{equation}
Q \in [1]_{\half}^{(1, 0)}~, \qquad N_Q = 10~.
\end{equation}
The superconformal unitarity bounds and shortening conditions are summarized in table~\ref{tab:3DN5}. 
\renewcommand{\arraystretch}{1.7}
\renewcommand\tabcolsep{8pt}
\begin{table}[H]
  \centering
  \begin{tabular}{ |c|lr| l|l| }
\hline
{\bf Name} &  \multicolumn{2}{c}{\bf Primary} &  \multicolumn{1}{|c|}{\bf Unitarity Bound} & \multicolumn{1}{c|}{\bf Null State } \\
\hline
\hline
$L$ & $[j]_{\Delta}^{(R_1,R_2)}$&  &$\Delta>\frac{1}{2} \,j+R_1+\half R_2+1$ & \multicolumn{1}{c|}{$-$} \\
\hline
\hline
$A_{1}$ & $[j]_{\Delta}^{(R_1,R_2)}~,$& $j\geq1$ &$\Delta=\frac{1}{2} \, j+R_1+\half R_2+1$ & $[j-1]_{\Delta+1/2}^{(R_1+1,R_2)}$ \\
\hline 
$A_{2}$ & $[0]_{\Delta}^{(R_1,R_2)}$& $$ &$\Delta=R_1+\half R_2+1$ & $[0]_{\Delta+1}^{(R_1+2,R_2)}$ \\
\hline
\hline
$B_{1}$ & $[0]_{\Delta}^{(R_1,R_2)}$& $$ &$\Delta=R_1+\half R_2$ & $[1]_{\Delta+1/2}^{(R_1+1,R_2)}$ \\
\hline
\end{tabular}
  \caption{Shortening conditions in three-dimensional~$\CN=5$ SCFTs.}
  \label{tab:3DN5}
\end{table}
\noindent If we define
\begin{equation}
\Delta_{A}=\frac{1}{2}\, j+R_{1}+\frac{1}{2}R_{2}+1~,
\end{equation}
we find the following recombination rules as~$\Delta \rightarrow \Delta_{A}^+$,
\begin{equation}
\begin{array}{rclcl}
L[j\geq 2]^{(R_{1},R_{2})}_{\Delta} & \rightarrow &A_{1}[j]^{(R_{1},R_{2})}_{\Delta_{A}}  &\oplus & A_{1}[j-1]^{(R_{1}+1,R_{2})}_{\Delta_{A}+\half}~, \\
L[j=1]^{(R_{1},R_{2})}_{\Delta} & \rightarrow &A_{1}[1]^{(R_{1},R_{2})}_{\Delta_{A}} & \oplus  &A_{2}[0]^{(R_{1}+1,R_{2})}_{\Delta_{A}+\half}~, \\
L[j=0]^{(R_{1},R_{2})}_{\Delta} & \rightarrow &A_{2}[0]^{(R_{1},R_{2})}_{\Delta_{A}}  &\oplus & B_{1}[0]^{(R_{1}+2,R_{2})}_{\Delta_{A}+1}~. 
\end{array}
\end{equation}
We conclude that the multiplets~$B_1[0]^{(R_1, R_2)}_{R_1 + R_2/2}$ with~$R_1  = 0,1$ are absolutely protected. This includes the free hypermultiplet in~\eqref{3dn5free}, the extra SUSY-current multiplet in~\eqref{3dn5sc}, and the stress tensor multiplet in~\eqref{3dn5em}. 

Conserved current multiplets in~$d = 3, \CN=5$ theories are studied in section~\ref{sec:3dn5curr}.

\subsubsection{$d=3$,~$\CN=6$}

\label{sec:3dn6defs}

The~$\CN=6$ superconformal algebra is~$\frak{osp}(6|4)$ and thus the~$R$-symmetry is~$\frak{so}(6)_R$. The~$R$-symmetry representations are denoted by~$(R_1,R_2,R_3)$, where~$R_1, R_2,R_3 \in \Z_{\geq 0}$ are $\frak{so}(6)_R$ Dynkin labels. Therefore~$(1,0,0)$ is the vector representation~$\bf 6$, while $(0,1,0)$ and $(0,0,1)$ are the two chiral spinor representations~$\bf 4$ and~$\bf \b 4$, which are related by complex conjugation.\footnote{~Note that the Dynkin labels of the isomorphic~$\frak{so}(6)$ and~$\frak{su}(4)$ algebras are related by a permutation. For instance, the~$(0,1,0)$ and~$(0,0,1)$ chiral spinor representations of~$\frak{so}(6)$ correspond to the fundamental~$(1,0,0)$ and anti-fundamental~$(0,0,1)$ of~$\frak{su}(4)$, while the vector~$(1,0,0)$ of~$\frak{so}(6)$ is the~$(0,1,0)$ of~$\frak{su}(4)$. See also appendix~\ref{app:liealg}.} The~$Q$-supersymmetries transform as
 \begin{equation}
Q \in [1]_{\half}^{(1,0,0)}~, \qquad N_Q = 12~.
\end{equation}
The superconformal unitarity bounds and shortening conditions are summarized in table~\ref{tab:3DN6}. 
\renewcommand{\arraystretch}{1.7}
\renewcommand\tabcolsep{8pt}
\begin{table}[H]
  \centering
  \begin{tabular}{ |c|lr| l|l| }
\hline
{\bf Name} &  \multicolumn{2}{c}{\bf Primary} &  \multicolumn{1}{|c|}{\bf Unitarity Bound} & \multicolumn{1}{c|}{\bf Null State } \\
\hline
\hline
$L$ & $[j]_{\Delta}^{(R_1,R_2,R_3)}$&  &$\Delta>\frac{1}{2} \,j+R_1+\half \left(R_2+R_3\right)+1$ & \multicolumn{1}{c|}{$-$} \\
\hline
\hline
$A_{1}$ & $[j]_{\Delta}^{(R_1,R_2,R_3)}~,$& $j\geq1$ &$\Delta=\frac{1}{2} \, j+R_1+\half \left(R_2+R_3\right)+1$ & $[j-1]_{\Delta+1/2}^{(R_1+1,R_2,R_3)}$ \\
\hline 
$A_{2}$ & $[0]_{\Delta}^{(R_1,R_2,R_3)}$& $$ &$\Delta=R_1+\half \left(R_2+R_3\right)+1$ & $[0]_{\Delta+1}^{(R_1+2,R_2,R_3)}$ \\
\hline
\hline
$B_{1}$ & $[0]_{\Delta}^{(R_1,R_2,R_3)}$& $$ &$\Delta=R_1+\half\left(R_2+R_3\right)$ & $[1]_{\Delta+1/2}^{(R_1+1,R_2,R_3)}$ \\
\hline
\end{tabular}
  \caption{Shortening conditions in three-dimensional~$\CN=6$ SCFTs.}
  \label{tab:3DN6}
\end{table}
\noindent Defining 
\begin{equation}
\Delta_{A}=\frac{1}{2}\, j+R_{1}+\frac{1}{2}(R_{2}+R_{3})+1~,
\end{equation}
we find the following recombination rules as~$\Delta \rightarrow \Delta_{A}^+$,
\begin{equation}
\begin{array}{rclcl}
L[j\geq 2]^{(R_{1},R_{2},R_{3})}_{\Delta} & \rightarrow &A_{1}[j]^{(R_{1},R_{2},R_{3})}_{\Delta_{A}}  &\oplus & A_{1}[j-1]^{(R_{1}+1,R_{2},R_{3})}_{\Delta_{A}+\half}~, \\
L[j=1]^{(R_{1},R_{2},R_{3})}_{\Delta} & \rightarrow &A_{1}[1]^{(R_{1},R_{2},R_{3})}_{\Delta_{A}} & \oplus  &A_{2}[0]^{(R_{1}+1,R_{2},R_{3})}_{\Delta_{A}+\half}~, \\
L[j=0]^{(R_{1},R_{2},R_{3})}_{\Delta} & \rightarrow &A_{2}[0]^{(R_{1},R_{2},R_{3})}_{\Delta_{A}}  &\oplus & B_{1}[0]^{(R_{1}+2,R_{2},R_{3})}_{\Delta_{A}+1}~. 
\end{array}
\end{equation}
It follows that multiplets of the form~$B_1[0]_{R_1 + R_2/2 + R_3/2}^{(R_1, R_2, R_3)}$ with~$R_1 = 0,1$ are absolutely protected. This includes the free hypermultiplets in~\eqref{3dn6free1} and~\eqref{3dn6free2}, the extra SUSY-current multiplets in~\eqref{3dn6super1} and~\eqref{3dn6super2}, the stress tensor multiplet in~\eqref{3dn6em}, and the higher-spin current multiplet in~\eqref{3dn6hs1}.

Conserved current multiplets in~$d = 3, \CN=6$ theories are studied in section~\ref{sec:3dn6curr}.

\subsubsection{$d=3$,~$\CN=8$}

\label{sec:d3n8defs}

The~$\CN=8$ superconformal algebra is~$\frak{osp}(8|4)$ and thus the~$R$-symmetry is~$\frak{so}(8)_R$. The~$R$-symmetry representations are denoted by~$(R_1,R_2,R_3,R_4)$, where~$R_1, R_2,R_3,R_4 \in \Z_{\geq 0}$ are $\frak{so}(8)_R$ Dynkin labels. For instance, $(1,0,0,0)$ is the vector representation~$\bf 8_\text{v}$, while~$(0,0,1,0)$ and~$(0,0,0,1)$ are the two chiral spinor representations~$\bf 8_\text{s}$ and~$\bf 8_{c}$. All three representations are real (i.e.~the spinors~$\bf 8_\text{s}$, $\bf 8_\text{c}$ are Majorana-Weyl), and they are permuted by the~$S_3$ triality group, which is an outer automorphism of~$\frak{so}(8)$. We choose a triality frame in which the~$Q$-supersymmetries transform in the vector representation~$\bf 8_\text{v}$,
 \begin{equation}\label{eq:d3n8scs}
Q_\alpha \in [1]_{\half}^{(1,0,0,0)}~, \qquad N_Q = 16~.
\end{equation}
This choice preserves a~$\Z_2 \subset S_3$ triality subgroup~$T$ that exchanges~$\bf 8_\text{s} \leftrightarrow \bf 8_\text{c}$ and is similar to the mirror automorphism~$M$ of three-dimensional~$\CN=4$ theories discussed in section~\ref{sec:3dN4defs}. The superconformal unitarity bounds and shortening conditions are summarized in table~\ref{tab:3DN8}. 

\renewcommand{\arraystretch}{1.7}
\renewcommand\tabcolsep{6pt}
\begin{table}[H]
  \centering
  \begin{tabular}{ |c|lr| l|l| }
\hline
{\bf Name} &  \multicolumn{2}{c}{\bf Primary} &  \multicolumn{1}{|c|}{\bf Unitarity Bound} & \multicolumn{1}{c|}{\bf Null State } \\
\hline
\hline
$L$ & $[j]_{\Delta}^{(R_1,R_2,R_3,R_4)}$&  &$\Delta>\frac{1}{2} \,j+R_1+R_2+\half \left(R_3+R_4\right)+1$ & \multicolumn{1}{c|}{$-$} \\
\hline
\hline
$A_{1}$ & $[j]_{\Delta}^{(R_1,R_2,R_3,R_4)}~,$& $j\geq1$ &$\Delta=\frac{1}{2} \, j+R_1+R_2+\half \left(R_3+R_4\right)+1$ & $[j-1]_{\Delta+1/2}^{(R_1+1,R_2,R_3,R_4)}$ \\
\hline 
$A_{2}$ & $[0]_{\Delta}^{(R_1,R_2,R_3,R_4)}$& $$ &$\Delta=R_1+R_2+\half \left(R_3+R_4\right)+1$ & $[0]_{\Delta+1}^{(R_1+2,R_2,R_3,R_4)}$ \\
\hline
\hline
$B_{1}$ & $[0]_{\Delta}^{(R_1,R_2,R_3,R_4)}$& $$ &$\Delta=R_1+R_2+\half \left(R_3+R_4\right)$ & $[1]_{\Delta+1/2}^{(R_1+1,R_2,R_3,R_4)}$ \\
\hline
\end{tabular}
  \caption{Shortening conditions in three-dimensional~$\CN=8$ SCFTs.}
  \label{tab:3DN8}
\end{table}
\noindent In order to state the recombination rules, we define 
\begin{equation}
\Delta_{A}=\frac{1}{2}\, j+R_{1}+R_{2}+\frac{1}{2}(R_{3}+R_{4})+1~.
\end{equation}
As~$\Delta \rightarrow \Delta_{A}^+$ we then find that
\begin{equation}
\begin{array}{rclcl}
L[j\geq 2]^{(R_{1},R_{2},R_{3},R_{4})}_{\Delta} & \rightarrow &A_{1}[j]^{(R_{1},R_{2},R_{3},R_{4})}_{\Delta_{A}}  &\oplus & A_{1}[j-1]^{(R_{1}+1,R_{2},R_{3},R_{4})}_{\Delta_{A}+\half}~, \\
L[j=1]^{(R_{1},R_{2},R_{3},R_{4})}_{\Delta} & \rightarrow &A_{1}[1]^{(R_{1},R_{2},R_{3},R_{4})}_{\Delta_{A}} & \oplus  &A_{2}[0]^{(R_{1}+1,R_{2},R_{3},R_{4})}_{\Delta_{A}+\half}~, \\
L[j=0]^{(R_{1},R_{2},R_{3},R_{4})}_{\Delta} & \rightarrow &A_{2}[0]^{(R_{1},R_{2},R_{3},R_{4})}_{\Delta_{A}}  &\oplus & B_{1}[0]^{(R_{1}+2,R_{2},R_{3},R_{4})}_{\Delta_{A}+1}~. 
\end{array}
\end{equation}
This implies that~$B_{1}[0]^{(R_{1},R_{2},R_{3},R_{4})}_{R_1+R_2 + R_3/2 + R_4/2}$ multiplets with~$R_1 = 0,1$ are absolutely protected. This includes the free hypermultiplets in~\eqref{3dn8free1} and~\eqref{3dn8free2}, the stress tensor multiplets in~\eqref{3dn8em1} and~\eqref{3dn8em2}, and the higher-spin current multiplets in~\eqref{3dn8hs1}, \eqref{3dn8hs2}, and~\eqref{3dn8hs3}.

Conserved current multiplets in~$d = 3, \CN=8$ theories are studied in section~\ref{sec:3dn8curr}.

\subsection{Four Dimensions}
\label{sec:4dmults}

In this section we list all unitary multiplets of four-dimensional SCFTs with $1 \leq \CN \leq 4$ supersymmetry (see~\cite{Dobrev:1985qv,Minwalla:1997ka,Dolan:2002zh} and references therein). As discussed in section~\ref{sec:maxsusy}, unitary SCFTs with~$\CN \geq 5$ do not exist. 

Representations of the~$\frak{so}(4) = \frak{su}(2) \times \b{\frak{su}(2)}$ Lorentz algebra will be denoted by
\begin{equation}\label{Lor4d}
[j; \b j]~, \qquad j, \b j \in \Z_{\geq 0}~.
\end{equation} 
Here~$j, \b j$ are integer-valued~$\frak{su}(2)$ Dynkin labels, so that the representation in~\eqref{Lor4d} has dimension~$(j+1)(\b j+1)$. We use~$[j; \b j]_\Delta$ to indicate the Lorentz quantum numbers of an operator with scaling dimension~$\Delta$.

\subsubsection{$d=4$,~$\CN=1$}

\label{sec:4dn1defs}

The~$\CN=1$ superconformal algebra is~$\frak{su}(2,2|1)$, so that there is a~$\frak{u}(1)_R$ symmetry. Operators of~$R$-charge~$r \in \R$ are denoted by~$(r)$. The~$Q$-supersymmetries transform as \begin{equation}
Q \in [1;0]_{\half}^{(-1)}~, \qquad \b Q \in [0;1]_{\half}^{(1)}~, \qquad N_Q = 4~.
\end{equation}
Superconformal multiplets obey unitarity bounds and shortening conditions with respect to both~$Q$ and~$\b Q$, which are summarized in tables~\ref{tab:4DN1C} and~\ref{4DN1AC}, respectively. Consequently, they are labeled by a pair of capital letters. 
\renewcommand{\arraystretch}{1.5}
\renewcommand\tabcolsep{8pt}
\begin{table}[H]
  \centering
  \begin{tabular}{ |c|lr| l|l| }
\hline
{\bf Name} &  \multicolumn{2}{c}{\bf Primary} &  \multicolumn{1}{|c|}{\bf Unitarity Bound} & \multicolumn{1}{c|}{\bf $Q$ Null State } \\
\hline
\hline
$L$ & $[j;\overline{j}]_{\Delta}^{(r)}$&  &$\Delta>2 + j-{3 \over 2} r$ & \multicolumn{1}{c|}{$-$} \\
\hline
\hline
$A_{1}$ & $[j;\overline{j}]_{\Delta}^{(r)}~,$& $j\geq1$ &$\Delta= 2+j-{3 \over 2} r$ & $[j-1;\overline{j}]_{\Delta+1/2}^{(r-1)}$ \\
\hline 
$A_{2}$ & $[0;\overline{j}]_{\Delta}^{(r)}$& $$ &$\Delta=2-{3 \over 2} r$ & $[0;\overline{j}]_{\Delta+1}^{(r-2)}$ \\
\hline
\hline
$B_{1}$ & $[0;\overline{j}]_{\Delta}^{(r)}$& $$ &$\Delta=-{3 \over 2} r$ & $[1;\overline{j}]_{\Delta+1/2}^{(r-1)}$ \\
\hline
\end{tabular}
  \caption{$Q$ shortening conditions in four-dimensional~$\CN=1$ SCFTs.}
  \label{tab:4DN1C}
\end{table} 

\renewcommand{\arraystretch}{1.5}
\renewcommand\tabcolsep{8pt}
\begin{table}[H]
  \centering
  \begin{tabular}{ |c|lr| l|l| }
\hline
{\bf Name} &  \multicolumn{2}{c}{\bf Primary} &  \multicolumn{1}{|c|}{\bf Unitarity Bound} & \multicolumn{1}{c|}{\bf $\b  Q$ Null State } \\
\hline
\hline
$\overline{L}$ & $[j;\overline{j}]_{\Delta}^{(r)}$&  &$\Delta>2+ \overline{j}+{3 \over 2}r$ & \multicolumn{1}{c|}{$-$} \\
\hline
\hline
$\overline{A}_{1}$ & $[j;\overline{j}]_{\Delta}^{(r)}~,$& $\overline{j}\geq1$ &$\Delta=2+ \overline{j}+{3 \over 2}r$ & $[j;\overline{j}-1]_{\Delta+1/2}^{(r+1)}$ \\
\hline 
$\overline{A}_{2}$ & $[j;0]_{\Delta}^{(r)}$& $$ &$\Delta=2+{3 \over 2}r$ & $[j;0]_{\Delta+1}^{(r+2)}$ \\
\hline
\hline
$\overline{B}_{1}$ & $[j;0]_{\Delta}^{(r)}$& $$ &$\Delta={3 \over 2}r$ & $[j;1]_{\Delta+1/2}^{(r+1)}$ \\
\hline
\end{tabular}
  \caption{$\b Q$ shortening conditions in four-dimensional~$\CN=1$ SCFTs.}
  \label{4DN1AC}
\end{table}

\noindent Full~$\CN=1$ multiplets are two-sided: they are obtained by imposing both left~$Q$ and right~$\b Q$ unitarity bounds and shortening conditions. This can lead to restrictions on some quantum numbers. The consistent two-sided multiplets are summarized in table~\ref{tab:4DN1CAC}. 

\medskip
\renewcommand{\arraystretch}{2}
\renewcommand\tabcolsep{2pt}
\begin{table}[H]
  \centering
  \captionsetup{justification=centering}
  \begin{tabular}{ |c||c|c|c|c|}
\hline
 &  \multicolumn{1}{c}{$ \boldsymbol{\overline{{L}}}$} &  \multicolumn{1}{|c|}{$\boldsymbol{\overline{A}_{1}}$} & \multicolumn{1}{c|}{$\boldsymbol{\overline{A}_{2}}$ }& \multicolumn{1}{c|}{$\boldsymbol{\overline{B}_{1}}$ } \\
\hline
\hline
\multirow{2}{*}{~$\boldsymbol L$~} &~~$[j;\overline{j}]_{\Delta}^{(r)}$~~& ~~$[j;\overline{j}\geq 1]_{\Delta}^{(r>{1 \over 3} (j-\overline{j}))}$ ~~&~$[j;\overline{j}=0]_{\Delta}^{(r>{1 \over 3} j)}$~&~~$[j;\overline{j}=0]_{\Delta}^{(r>{1 \over 3} (j+2))}$ ~~ \\[-5pt]
& ~$\scriptstyle \Delta\,>\, 2+\max\left\{j-\frac{3}{2}r,\overline{j}+\frac{3}{2}r\right\}$~ &$\Delta=2+\overline{j}+\frac{3}{2}r$ &$\Delta=2+\frac{3}{2}r$ &$\Delta=\frac{3}{2}r$ \\
\hline
\multirow{2}{*}{~$\boldsymbol{A_{1}}$~} &$[j\geq 1;\overline{j}]_{\Delta}^{(r<{1\over 3}(j-\overline{j}))}$&~ $[j\geq 1; \overline{j}\geq 1]_{\Delta}^{(r={1 \over 3}(j-\overline{j}))}$ ~&~ $[j\geq 1; \overline{j}=0]_{\Delta}^{(r={1 \over 3}j)}$~ & ~$[j \geq 1; \overline{j}=0]_{\Delta}^{(r={1 \over 3} (j+2))}$ ~\\[-5pt]
& $\Delta=2+j-\frac{3}{2}r$ &$\Delta=2+\frac{1}{2}(j+\overline{j})$ &$\Delta=2+\frac{1}{2}\, j$ & $\Delta=1+\frac{1}{2}\, j$ \\
\hline
\multirow{2}{*}{$\boldsymbol{A_{2}}$} &$[j=0;\overline{j}]_{\Delta}^{(r<-{1 \over 3} \overline{j})}$& $[j=0; \overline{j}\geq 1]_{\Delta}^{(r=-{1 \over 3} \overline{j})}$ &$[j=0; \overline{j}=0]_{\Delta}^{(r=0)}$&~~~$[j=0; \overline{j}=0]_{\Delta}^{(r={2 \over 3})}$~~~ \\[-5pt]
& $\Delta=2-\frac{3}{2}r$ &$\Delta=2+\frac{1}{2}\, \overline{j}$&$\Delta=2$ &$\Delta=1$ \\
\hline
\multirow{2}{*}{$\boldsymbol{B_{1}}$} &$[j=0;\overline{j}]_{\Delta}^{(r<-{1 \over 3}(\overline{j}+2))}$& $[j=0; \overline{j} \geq 1]_{\Delta}^{(r=-{1 \over 3}(\overline{j}+2))}$ &~$[j=0; \overline{j}=0]_{\Delta}^{(r=-{2\over 3})}$~&$[j=0; \overline{j}=0]_{\Delta}^{(r=0)}$ \\[-5pt]
& $\Delta=-\frac{3}{2}r$ &$\Delta=1+\frac{1}{2}\, \overline{j} $&$\Delta=1$ &$\Delta=0$ \\
\hline
\end{tabular}
\caption{Consistent two-sided multiplets in four-dimensional~$\mathcal{N}=1$ theories.}
  \label{tab:4DN1CAC}
 \end{table} 

\noindent  In order to state the recombination rules, we define
\begin{equation}
\Delta_{A}=2+j-\frac{3}{2}r~, \hspace{.5in} \Delta_{\overline{A}}=2+\overline{j}+\frac{3}{2}r~.
\end{equation}
As~$\Delta \rightarrow \Delta_{A}^+$, we find the following partial, left recombination rules, 
\begin{equation}
\begin{array}{rclcl}
L[j\geq 2]_{\Delta}^{(r)} & \rightarrow &A_{1}[j]_{\Delta_{A}}^{(r)}  &\oplus & A_{1}[j-1]_{\Delta_{A}+1/2}^{(r-1)}~, \\
L[j=1]_{\Delta}^{(r)} & \rightarrow &A_{1}[1]_{\Delta_{A}}^{(r)}  &\oplus & A_{2}[0]_{\Delta_{A}+1/2}^{(r-1)}~, \\
L[j=0]_{\Delta}^{(r)} & \rightarrow &A_{2}[0]_{\Delta_{A}}^{(r)}  &\oplus & B_{1}[0]_{\Delta_{A}+1}^{(r-2)}~,
\end{array}
\label{chiralrecomb4dn1}
\end{equation}
and similarly for the partial, right recombination rules. As in the discussion around~\eqref{chiralrecomb3d}, the recombination rules for complete, two-sided multiplets are controlled by the~$R$-charge~$r$, which determines the unitarity bound that is saturated first:
\begin{itemize}
\item If~$r<{1 \over 3}(j-\overline{j})$, then~$\Delta_A> \Delta_{\b A}$ and the left~$Q$ unitarity bound is reached first, so that
\begin{equation}\label{d4n1rr1}
\begin{array}{rclclcl}
L\overline{L}[j\geq 2;\overline{j}]_{\Delta}^{(r<{1 \over 3}(j-\overline{j}))} & \rightarrow &A_{1}\overline{L}[j;\overline{j}]_{\Delta_{A}}^{(r)}  &\oplus & A_{1}\overline{L}[j-1;\overline{j}]_{\Delta_{A}+\half}^{(r-1)}~, \\
L\overline{L}[j=1;\overline{j}]_{\Delta}^{(r<{1 \over 3}(j-\overline{j}))} & \rightarrow &A_{1}\overline{L}[1;\overline{j}]_{\Delta_{A}}^{(r)}  &\oplus & A_{2}\overline{L}[0;\overline{j}]_{\Delta_{A}+\half}^{(r-1)}~, \\
L\overline{L}[j=0;\overline{j}]_{\Delta}^{(r<{1 \over 3}(j-\overline{j}))} & \rightarrow &A_{2}\overline{L}[0;\overline{j}]_{\Delta_{A}}^{(r)}  &\oplus & B_{1}\overline{L}[0;\overline{j}]_{\Delta_{A}+1}^{(r-2)}~.
\end{array}
\end{equation}
\item If~$r>{1 \over 3} (j-\overline{j})$, then~$\Delta_{\b A} > \Delta_A$ and the right~$\b Q$ unitarity bound is saturated first, 
\begin{equation}\label{d4n1rr2}
\begin{array}{rclclcl}
L\overline{L}[j;\overline{j}\geq 2]_{\Delta}^{(r>{1 \over 3} (j-\overline{j}))} & \rightarrow &L\overline{A}_{1}[j;\overline{j}]_{\Delta_{\overline{A}}}^{(r)}  &\oplus & L\overline{A}_{1}[j;\overline{j}-1]_{\Delta_{\overline{A}}+\half}^{(r+1)}~, \\
L\overline{L}[j;\overline{j}=1]_{\Delta}^{(r>{1 \over 3} (j-\overline{j}))} & \rightarrow &L\overline{A}_{1}[j;1]_{\Delta_{\overline{A}}}^{(r)}  &\oplus & L\overline{A}_{2}[j;0]_{\Delta_{\overline{A}}+\half}^{(r+1)}~, \\
L\overline{L}[j;\overline{j}=0]_{\Delta}^{(r>{1 \over 3} (j-\overline{j}))} & \rightarrow &L\overline{A}_{2}[j;0]_{\Delta_{\overline{A}}}^{(r)}  &\oplus & L\overline{B}_{1}[j;0]_{\Delta_{\overline{A}}+1}^{(r+2)}~.
\end{array}
\end{equation}
\item When $r={1 \over 3} (j-\overline{j})$, the left and right unitarity bounds are reached simultaneously, because~$\Delta_A = \Delta_{\b A}$. This gives rise to the following recombination rules: 
\end{itemize}
\begin{equation}
\begin{array}{rclclcl}
L\overline{L}[j\geq 2; \overline{j}\geq 2]_{\Delta}^{(r={1 \over 3} (j-\overline{j}))} & \rightarrow &A_{1}\overline{A}_{1}[j; \overline{j}]_{\Delta_{A}}^{(r)}  &\oplus & A_{1}\overline{L}[j-1; \overline{j}]_{\Delta_{A}+\half}^{(r-1)} & \oplus & L\overline{A}_{1}[j;\overline{j}-1]_{\Delta_{A}+\half}^{(r+1)}~, \\
L\overline{L}[j\geq 2; \overline{j}=1]_{\Delta}^{(r={1 \over 3} (j-\overline{j}))} & \rightarrow &A_{1}\overline{A}_{1}[j; 1]_{\Delta_{A}}^{(r)}  &\oplus & A_{1}\overline{L}[j-1; 1]_{\Delta_{A}+\half}^{(r-1)} & \oplus & L\overline{A}_{2}[j;0]_{\Delta_{A}+\half}^{(r+1)}~, \\
L\overline{L}[j\geq 2; \overline{j}=0]_{\Delta}^{(r={1 \over 3} (j-\overline{j}))} & \rightarrow &A_{1}\overline{A}_{2}[j;0]_{\Delta_{A}}^{(r)}  &\oplus & A_{1}\overline{L}[j-1; 0]_{\Delta_{A}+\half}^{(r-1)} & \oplus & L\overline{B}_{1}[j;0]_{\Delta_{A}+1}^{(r+2)}~, \\
L\overline{L}[j=1; \overline{j}\geq 2]_{\Delta}^{(r={1 \over 3} (j-\overline{j}))}& \rightarrow &A_{1}\overline{A}_{1}[1; \overline{j}]_{\Delta_{A}}^{(r)}  &\oplus & A_{2}\overline{L}[0; \overline{j}]_{\Delta_{A}+\half}^{(r-1)} & \oplus & L\overline{A}_{1}[1;\overline{j}-1]_{\Delta_{A}+\half}^{(r+1)}~, \\
L\overline{L}[j=0; \overline{j}\geq 2]_{\Delta}^{(r={1 \over 3} (j-\overline{j}))} & \rightarrow &A_{2}\overline{A}_{1}[0; \overline{j}]_{\Delta_{A}}^{(r)}  &\oplus & B_{1}\overline{L}[0; \overline{j}]_{\Delta_{A}+1}^{(r-2)} & \oplus & L\overline{A}_{1}[0;\overline{j}-1]_{\Delta_{A}+\half}^{(r+1)}~, \\
L\overline{L}[j=1; \overline{j}=1]_{\Delta}^{(r={1 \over 3} (j-\overline{j}))} & \rightarrow &A_{1}\overline{A}_{1}[1; 1]_{\Delta_{A}}^{(r)}  &\oplus & A_{2}\overline{L}[0; 1]_{\Delta_{A}+\half}^{(r-1)} & \oplus & L\overline{A}_{2}[1;0]_{\Delta_{A}+\half}^{(r+1)}~, \\
L\overline{L}[j=1; \overline{j}=0]_{\Delta}^{(r={1 \over 3} (j-\overline{j}))} & \rightarrow &A_{1}\overline{A}_{2}[1; 0]_{\Delta_{A}}^{(r)}  &\oplus & A_{2}\overline{L}[0; 0]_{\Delta_{A}+\half}^{(r-1)} & \oplus & L\overline{B}_{1}[1;0]_{\Delta_{A}+1}^{(r+2)}~, \\
L\overline{L}[j=0; \overline{j}=1]_{\Delta}^{(r={1 \over 3} (j-\overline{j}))} & \rightarrow &A_{2}\overline{A}_{1}[0; 1]_{\Delta_{A}}^{(r)}  &\oplus & B_{1}\overline{L}[0; 1]_{\Delta_{A}+1}^{(r-2)} & \oplus & L\overline{A}_{2}[0;0]_{\Delta_{A}+\half}^{(r+1)}~,\\
L\overline{L}[j=0; \overline{j}=0]_{\Delta}^{(r={1 \over 3} (j-\overline{j}))} & \rightarrow &A_{2}\overline{A}_{2}[0; 0]_{\Delta_{A}}^{(r)}  &\oplus & B_{1}\overline{L}[0; 0]_{\Delta_{A}+1}^{(r-2)} & \oplus & L\overline{B}_{1}[0;0]_{\Delta_{A}+1}^{(r+2)}~.\\[5pt]
\end{array}
\label{4dn1recomb}
\end{equation}
\noindent As for~$d = 3, \CN=2$ (see the discussion around~\eqref{3dn2recomb}), the right-hand side of these recombination rules consists of three, rather than four terms and cannot be obtained by naively tensoring left and right recombination rules. As was the case there, the peculiarities of~\eqref{4dn1recomb} can be traced back to the fact that all~$A_\ell \b A_{\b \ell}$-multiplets contain currents (see section~\ref{sec:d4n1curr}), whose conservation laws must be taken into account.  

By examining the right-hand sides of~\eqref{d4n1rr1}, \eqref{d4n1rr2}, and~\eqref{4dn1recomb}, we conclude that the chiral free field multiplets~$A_{\ell}\overline{B}_{1}[j;0]^{(r = {1 \over 3} (j+2))}_{\Delta=r}$ in~\eqref{4dn1freev}, as well as chiral operators of the form $L\overline{B}_{1}[j;0]_{\Delta=r}^{(r)}$ with ${1 \over 3} j + {2 \over 3} < r < {1 \over 3} j + 2$ (together with their complex conjugate anti-chiral multiplets) are absolutely protected. Note that the latter multiplets give rise to all possible relevant supersymmetric deformations of~$d = 4, \CN=1$ SCFTs (see~\cite{Green:2010da} and section~3.2.1 of~\cite{Cordova:2016xhm}). 

Generic~$d = 4, \CN=1$ multiplets are tabulated in section~\ref{sec:4dn1tabs}; conserved current multiplets are studied in section~\ref{sec:d4n1curr}.

\subsubsection{$d=4$,~$\CN=2$}

\label{sec:d4n2defs}

The~$\CN=2$ superconformal algebra is~$\frak{su}(2,2|2)$, so that there is a~$\frak{su}(2)_R \times \frak{u}(1)_R$ symmetry. The~$R$-charges of an operator are denoted by~$(R; r)$, where~$R \in \Z_{\geq 0}$ is an~$\frak{su}(2)_R$ Dynkin label, while~$r \in \R$ is the~$\frak{u}(1)_R$ charge. The~$Q$-supersymmetries transform as 
\begin{equation}
Q \in [1;0]_{\half}^{(1;-1)}~, \qquad \b Q \in [0;1]_{\half}^{(1 ; 1)}~, \qquad N_Q = 8~.
\end{equation}
Superconformal multiplets obey unitarity bounds and shortening conditions with respect to both~$Q$ and~$\b Q$, which are summarized in tables~\ref{tab:4DN2C} and~\ref{4DN2AC}, respectively. Therefore, they are labeled by a pair of capital letters.
\renewcommand{\arraystretch}{1.5}
\renewcommand\tabcolsep{8pt}
\begin{table}[H]
  \centering
  \begin{tabular}{ |c|lr| l|l| }
\hline
{\bf Name} &  \multicolumn{2}{c}{\bf Primary} &  \multicolumn{1}{|c|}{\bf Unitarity Bound} & \multicolumn{1}{c|}{\bf $Q$ Null State } \\
\hline
\hline
$L$ & $[j;\overline{j}]_{\Delta}^{(R ; r)}$&  &$\Delta>2 + j+R-\half r$ & \multicolumn{1}{c|}{$-$} \\
\hline
\hline
$A_{1}$ & $[ j;\overline{j}]_{\Delta}^{(R ; r)}~,$& $j\geq1$ &$\Delta= 2+j+R-\half r$ & $[j-1;\overline{j}]_{\Delta+1/2}^{(R+1 ; r-1)}$ \\
\hline 
$A_{2}$ & $[0;\overline{j}]_{\Delta}^{(R;r)}$& $$ &$\Delta=2+R-\half r$ & $[0;\overline{j}]_{\Delta+1}^{(R+2; r-2)}$ \\
\hline
\hline
$B_{1}$ & $[0;\overline{j}]_{\Delta}^{(R;r)}$& $$ &$\Delta=R-\half r$ & $[1;\overline{j}]_{\Delta+1/2}^{(R+1; r-1)}$ \\
\hline
\end{tabular}
  \caption{$Q$ shortening conditions in four-dimensional~$\CN=2$ SCFTs.}

  \label{tab:4DN2C}
\end{table}

\renewcommand{\arraystretch}{1.5}
\renewcommand\tabcolsep{8pt}
\begin{table}[H]
  \centering
  \begin{tabular}{ |c|lr| l|l| }
\hline
{\bf Name} &  \multicolumn{2}{c}{\bf Primary} &  \multicolumn{1}{|c|}{\bf Unitarity Bound} & \multicolumn{1}{c|}{\bf $\b  Q$ Null State } \\
\hline
\hline
$\overline{L}$ & $[j;\overline{j}]_{\Delta}^{(R ; r)}$&  &$\Delta>2+ \overline{j}+R+\half r$ & \multicolumn{1}{c|}{$-$} \\
\hline
\hline
$\overline{A}_{1}$ & $[j;\overline{j}]_{\Delta}^{(R ; r)}~,$& $\overline{j}\geq1$ &$\Delta=2+ \overline{j}+R+\half r$ & $[ j;\overline{j}-1 ]_{\Delta+1/2}^{(R+1 ; r+1)}$ \\
\hline 
$\overline{A}_{2}$ & $[j;0 ]_{\Delta}^{(R ; r)}$& $$ &$\Delta=2+R+\half r$ & $[ j;0 ]_{\Delta+1}^{(R+2 ; r+2)}$ \\
\hline
\hline
$\overline{B}_{1}$ & $[j;0]_{\Delta}^{(R ; r)}$& $$ &$\Delta=R+\half r$ & $[ j;1]_{\Delta+1/2}^{(R+1 ; r+1)}$ \\
\hline
\end{tabular}
  \caption{$\b Q$ shortening conditions in four-dimensional~$\CN=2$ SCFTs.}
  \label{4DN2AC}
\end{table}

\noindent Full~$\CN=2$ multiplets are two-sided: they are obtained by imposing both left~$Q$ and right~$\b Q$ unitarity bounds and shortening conditions. This can lead to restrictions on some quantum numbers. The consistent two-sided multiplets are summarized in table~\ref{tab:4DN2CAC}. 

\begin{landscape}
\renewcommand{\arraystretch}{2}
\renewcommand\tabcolsep{2pt}
\begin{table}
  \centering
  \captionsetup{justification=centering}
  \begin{tabular}{ |c||c|c|c|c|}
\hline
 &  \multicolumn{1}{c}{$ \boldsymbol{\overline{{L}}}$} &  \multicolumn{1}{|c|}{$\boldsymbol{\overline{A}_{1}}$} & \multicolumn{1}{c|}{$\boldsymbol{\overline{A}_{2}}$ }& \multicolumn{1}{c|}{$\boldsymbol{\overline{B}_{1}}$ } \\
\hline
\hline
\multirow{2}{*}{~~$\boldsymbol L$~~} &~~$\begin{aligned} \\[-5pt] [j;\overline{j}]_{\Delta}^{(R;r)} \end{aligned} $~~& ~~$\begin{aligned} \\[-5pt] [j;\overline{j}\geq 1]_{\Delta}^{(R;r>j-\overline{j})} \end{aligned} $ ~~&~$\begin{aligned} \\[-5pt] [j;\overline{j}=0]_{\Delta}^{(R;r>j)} \end{aligned}$~&~~$\begin{aligned} \\[-5pt] [j;\overline{j}=0]_{\Delta}^{(R;r>j+2)} \end{aligned}$ ~~ \\
&~~ $\Delta>2+R+\max\big\{j-\frac{1}{2}r,\overline{j}+\frac{1}{2}r\big\}$ ~~&~~$\Delta=2+R+\overline{j}+\frac{1}{2}r$~~ &~~$\Delta=2+R+\frac{1}{2}r$~~ &$~~~\Delta=R+\frac{1}{2}r$~~~ \\
\hline
\multirow{2}{*}{~~$\boldsymbol{A_{1}}$~~} &$\begin{aligned} \\[-5pt] [j\geq 1;\overline{j}]_{\Delta}^{(R;r< j-\overline{j} )} \end{aligned} $&~ $\begin{aligned} \\[-5pt] [j\geq 1; \overline{j}\geq 1]_{\Delta}^{(R; r= j-\overline{j} )} \end{aligned}$ ~&~ $\begin{aligned} \\[-5pt] [j\geq 1; \overline{j}=0]_{\Delta}^{(R;r=j)} \end{aligned}$~ & ~$\begin{aligned} \\[-5pt] [j \geq 1; \overline{j}=0]_{\Delta}^{(R;r=j+2)} \end{aligned}$ ~\\
& $\Delta=2+R+j-\frac{1}{2}r$ &$\Delta=2+R+\frac{1}{2}(j+\overline{j})$ &$\Delta=2+R+\frac{1}{2}\, j$ & $\Delta=1+R+\frac{1}{2}\, j$ \\
\hline
\multirow{2}{*}{$\boldsymbol{A_{2}}$} &$\begin{aligned} \\[-5pt] [j=0;\overline{j}]_{\Delta}^{(R;r<- \overline{j})} \end{aligned}$& $\begin{aligned} \\[-5pt] [j=0; \overline{j}\geq 1]_{\Delta}^{(R;r=- \overline{j})} \end{aligned}$ &$\begin{aligned} \\[-5pt] [j=0; \overline{j}=0]_{\Delta}^{(R;r=0)} \end{aligned}$&~~~$\begin{aligned} \\[-5pt] [j=0; \overline{j}=0]_{\Delta}^{(R;r=2)} \end{aligned}$~~~ \\
& $\Delta=2+R-\frac{1}{2}r$ &$\Delta=2+R+\frac{1}{2}\, \overline{j}$&$\Delta=2+R$ &$\Delta=1+R$ \\
\hline
\multirow{2}{*}{$\boldsymbol{B_{1}}$} &$\begin{aligned} \\[-5pt] [j=0;\overline{j}]_{\Delta}^{(R;r<-(\overline{j}+2))} \end{aligned} $& $\begin{aligned} \\[-5pt] [j=0; \overline{j} \geq 1]_{\Delta}^{(R;r=-(\overline{j}+2))} \end{aligned} $ &~$\begin{aligned} \\[-5pt] [j=0; \overline{j}=0]_{\Delta}^{(R;r=-2)} \end{aligned}$~&$\begin{aligned} \\[-5pt] [j=0; \overline{j}=0]_{\Delta}^{(R;r=0)} \end{aligned}$ \\
& $\Delta=R-\frac{1}{2}r$ &$\Delta=1+R+\frac{1}{2}\, \overline{j} $&$\Delta=1+R$ &$\Delta=R$ \\
\hline
\end{tabular}
\caption{Consistent two-sided multiplets in four-dimensional~$\mathcal{N}=2$ theories.}
\label{tab:4DN2CAC}
 \end{table} 
 
\end{landscape}

The relation between our labeling scheme for multiplets and the notation of~\cite{Dolan:2002zh} is as follows: $ \CA^\Delta_{R, r(j, \b \jmath)} = L\b L[j; \b \jmath]^{(R ; r)}_\Delta$ is a long multiplet, and the short multiplets are given by
\begin{align}\label{doourconv}
& \CC_{R, r(j, \b \jmath)} = A_{\ell}\b L[j; \b \jmath]^{(R ;  r)}~, & &  \CB_{R, r(0,\b \jmath)} = B_1\b L[0; \b \jmath]^{(R>0 ; r)}~,\cr
& \CE_{r(0,\b \jmath)} = B_1\b L[0; \b \jmath]^{(0 ; r)}~, & & \hat \CC_{R(j, \b \jmath)} = A_{\ell} \b A_{\b \ell}[j; \b \jmath]^{(R ; j - \b \jmath)}~,\cr
& \CD_{R(0, \b \jmath)} = B_1 \b A_{\b \ell}[0;\b \jmath]^{(R ; -\b \jmath-2)}~, & & \hat \CB_R = B_1\b B_1[0;0]^{(R ; 0)}~,
\end{align}
where~$\ell, \b \ell = 1,2$ as dictated by the quantum numbers. Analogous relations for the multiplets~$\b \CC_{R, r(j, \b \jmath)}$, $\b \CB_{R, r(j, 0)}$, $\b \CE_{r(j, 0)}$, and~$\b \CD_{R(j, 0)}$ can be obtained by complex conjugation.

In order to state the recombination rules, we define
\begin{equation}
\Delta_{A}=2+R+j-\frac{1}{2}r~, \hspace{.5in} \Delta_{\overline{A}}=2+R+\overline{j}+\frac{1}{2}r~.
\end{equation}
As~$\Delta \rightarrow \Delta_{A}^+$, we find the following partial, left (or chiral) recombination rules
\begin{equation}
\begin{array}{rclcl}
L[j\geq 2]_{\Delta}^{(R ; r)} & \rightarrow &A_{1}[j]_{\Delta_{A}}^{(R ; r)}  &\oplus & A_{1}[j-1]_{\Delta_{A}+\half}^{(R+1 ; r-1)}~, \\
L[j=1]_{\Delta}^{(R ; r)} & \rightarrow &A_{1}[1]_{\Delta_{A}}^{(R ; r)}  &\oplus & A_{2}[0]_{\Delta_{A}+\half}^{(R+1 ; r-1)}~, \\
L[j=0]_{\Delta}^{(R ; r)} & \rightarrow &A_{2}[0]_{\Delta_{A}}^{(R ; r)}  &\oplus & B_{1}[0]_{\Delta_{A}+1}^{(R+2 ; r-2)}~,
\end{array}
\label{chiralrecomb4dn2}
\end{equation}
and similarly for partial, right (or antichiral) recombination rules. As in~$\CN=1$ theories (see the discussion around~\eqref{chiralrecomb4dn1}), the structure of the full, two-sided recombination rules is controlled by the~$\frak{u}(1)_R$ charge~$r$: 
\begin{itemize}
\item When~$r<j-\overline{j}$, then~$\Delta_A > \Delta_{\b A}$ and the chiral unitarity bound is saturated first,
\begin{equation}\label{d4n2rr1}
\setstretch{1.2}
\begin{array}{rclclcl}
L\overline{L}[j\geq 2;\overline{j}]_{\Delta}^{(R ;  r<j-\overline{j})} & \rightarrow &A_{1}\overline{L}[j;\overline{j}]_{\Delta_{A}}^{(R ; r)}  &\oplus & A_{1}\overline{L}[j-1;\overline{j}]_{\Delta_{A}+\half}^{(R+1 ; r-1)}~, \\
L\overline{L}[j=1;\overline{j}]_{\Delta}^{(R ; r<j-\overline{j}))} & \rightarrow &A_{1}\overline{L}[1;\overline{j}]_{\Delta_{A}}^{(R ; r)}  &\oplus & A_{2}\overline{L}[0;\overline{j}]_{\Delta_{A}+\half}^{(R+1 ; r-1)}~, \\
L\overline{L}[j=0;\overline{j}]_{\Delta}^{(R ; r<j-\overline{j})} & \rightarrow &A_{2}\overline{L}[0;\overline{j}]_{\Delta_{A}}^{(R ; r)}  &\oplus & B_{1}\overline{L}[0;\overline{j}]_{\Delta_{A}+1}^{(R+2 ; r-2)}~.
\end{array}
\end{equation}
\item When $r>j-\overline{j}$, then~$\Delta_{\b A} > \Delta_A$ and the antichiral unitarity bound is saturated first,  
\begin{equation}\label{d4n2rr2}
\setstretch{1.2}
\begin{array}{rclclcl}
L\overline{L}[j;\overline{j}\geq 2]_{\Delta}^{(R ;  r>j-\overline{j})} & \rightarrow &L\overline{A}_{1}[j;\overline{j}]_{\Delta_{\overline{A}}}^{(R ;  r)}  &\oplus & L\overline{A}_{1}[j;\overline{j}-1]_{\Delta_{\overline{A}}+\half}^{(R+1 ; r+1)}~, \\
L\overline{L}[j;\overline{j}=1]_{\Delta}^{(R ;  r>j-\overline{j})} & \rightarrow &L\overline{A}_{1}[j;1]_{\Delta_{\overline{A}}}^{(R ; r)}  &\oplus & L\overline{A}_{2}[j;0]_{\Delta_{\overline{A}}+\half}^{(R+1 ; r+1)}~, \\
L\overline{L}[j;\overline{j}=0]_{\Delta}^{(R ; r>j-\overline{j})} & \rightarrow &L\overline{A}_{2}[j;0]_{\Delta_{\overline{A}}}^{(R ; r)}  &\oplus & L\overline{B}_{1}[j;0]_{\Delta_{\overline{A}}+1}^{(R+2 ; r+2)}~.
\end{array}
\end{equation}
\item If~$r=j-\overline{j}$, then~$\Delta_A = \Delta_{\b A}$; chiral and antichiral unitarity bounds are both saturated:
\end{itemize}
\begin{equation}
\setstretch{1.2}
\begin{array}{rclclcl}
L\overline{L}[j\geq 2; \overline{j}\geq 2]_{\Delta}^{(R ; r=j-\overline{j})} & \rightarrow &A_{1}\overline{A}_{1}[j; \overline{j}]_{\Delta_{A}}^{(R ; r)}  &\oplus & A_{1}\overline{A}_{1}[j-1; \overline{j}]_{\Delta_{A}+\half}^{(R+1 ; r-1)} \\
& \oplus & A_{1}\overline{A}_{1}[j;\overline{j}-1]_{\Delta_{A}+\half}^{(R+1 ; r+1)}& \oplus & A_{1}\overline{A}_{1}[j-1;\overline{j}-1]_{\Delta_{A}+1}^{(R+2 ; r)}~, \\
L\overline{L}[j\geq 2; \overline{j}=1]_{\Delta}^{(R ; r=j-\overline{j})} & \rightarrow &A_{1}\overline{A}_{1}[j; 1]_{\Delta_{A}}^{(R ; r)}  &\oplus & A_{1}\overline{A}_{1}[j-1; 1]_{\Delta_{A}+\half}^{(R+1 ; r-1)} \\
& \oplus & A_{1}\overline{A}_{2}[j;0]_{\Delta_{A}+\half}^{(R+1 ; r+1)}& \oplus & A_{1}\overline{A}_{2}[j-1;0]_{\Delta_{A}+1}^{(R+2 ; r)}~, \\
L\overline{L}[j\geq 2; \overline{j}=0]_{\Delta}^{(R ; r=j-\overline{j})} & \rightarrow &A_{1}\overline{A}_{2}[j;0]_{\Delta_{A}}^{(R ; r)}  &\oplus & A_{1}\overline{A}_{2}[j-1; 0]_{\Delta_{A}+\half}^{(R+1 ; r-1)}\\
 & \oplus & A_{1}\overline{B}_{1}[j;0]_{\Delta_{A}+1}^{(R+2 ; r+2)}& \oplus & A_{1}\overline{B}_{1}[j-1;0]_{\Delta_{A}+{3 \over 2}}^{(R+3 ; r+1)}~, \\
L\overline{L}[j=1; \overline{j}\geq 2]_{\Delta}^{(R ; r=j-\overline{j})} & \rightarrow &A_{1}\overline{A}_{1}[1; \overline{j}]_{\Delta_{A}}^{(R ; r)}  &\oplus & A_{2}\overline{A}_{1}[0; \overline{j}]_{\Delta_{A}+\half}^{(R+1 ; r-1)} \\
& \oplus & A_{1}\overline{A}_{1}[1;\overline{j}-1]_{\Delta_{A}+\half}^{(R+1 ; r+1)}& \oplus & A_{2}\overline{A}_{1}[0;\overline{j}-1]_{\Delta_{A}+1}^{(R+2 ; r)}~, \\
L\overline{L}[j=0; \overline{j}\geq 2]_{\Delta}^{(R ; r=j-\overline{j})} & \rightarrow &A_{2}\overline{A}_{1}[0; \overline{j}]_{\Delta_{A}}^{(R ; r)}  &\oplus & B_{1}\overline{A}_{1}[0; \overline{j}]_{\Delta_{A}+1}^{(R+2 ; r-2)} \\
& \oplus & A_{2}\overline{A}_{1}[0;\overline{j}-1]_{\Delta_{A}+\half}^{(R+1 ; r+1)}& \oplus & B_{1}\overline{A}_{1}[0;\overline{j}-1]_{\Delta_{A}+{3 \over 2}}^{(R+3 ;  r-1)}~, \\
L\overline{L}[j=1; \overline{j}=1]_{\Delta}^{(R ;  r=j-\overline{j})} & \rightarrow &A_{1}\overline{A}_{1}[1; 1]_{\Delta_{A}}^{(R ; r)}  &\oplus & A_{2}\overline{A}_{1}[0; 1]_{\Delta_{A}+\half}^{(R+1 ; r-1)}\\
 & \oplus & A_{1}\overline{A}_{2}[1;0]_{\Delta_{A}+\half}^{(R+1 ;  r+1)}& \oplus & A_{2}\overline{A}_{2}[0;0]_{\Delta_{A}+1}^{(R+2 ; r)}~, \\
L\overline{L}[j=1; \overline{j}=0]_{\Delta}^{(R ; r=j-\overline{j})} & \rightarrow &A_{1}\overline{A}_{2}[1; 0]_{\Delta_{A}}^{(R ; r)}  &\oplus & A_{2}\overline{A}_{2}[0; 0]_{\Delta_{A}+\half}^{(R+1 ; r-1)} \\
& \oplus & A_{1}\overline{B}_{1}[1;0]_{\Delta_{A}+1}^{(R+2 ; r+2)}& \oplus & A_{2}\overline{B}_{1}[0;0]_{\Delta_{A}+{3 \over 2}}^{(R+3 ; r+1)}~, \\
L\overline{L}[j=0; \overline{j}=1]_{\Delta}^{(R ;  r=j-\overline{j})} & \rightarrow &A_{2}\overline{A}_{1}[0; 1]_{\Delta_{A}}^{(R ; r)}  &\oplus & B_{1}\overline{A}_{1}[0; 1]_{\Delta_{A}+1}^{(R+2 ; r-2)}\\
 & \oplus & A_{2}\overline{A}_{2}[0;0]_{\Delta_{A}+\half}^{(R+1 ; r+1)} & \oplus & B_{1}\overline{A}_{2}[0;0]_{\Delta_{A}+{3 \over 2}}^{(R+3 ; r-1)}~,\\
L\overline{L}[j=0; \overline{j}=0]_{\Delta}^{(R ;  r=j-\overline{j})} & \rightarrow &A_{2}\overline{A}_{2}[0; 0]_{\Delta_{A}}^{(R ; r)}  &\oplus & B_{1}\overline{A}_{2}[0; 0]_{\Delta_{A}+1}^{(R+2 ; r-2)} \\
& \oplus & A_{2}\overline{B}_{1}[0;0]_{\Delta_{A}+1}^{(R+2 ; r+2)}& \oplus & B_{1}\overline{B}_{1}[0;0]_{\Delta_{A}+2}^{(R+4 ; r)}~.
\end{array}
\label{4dn2recomb}
\end{equation}
\noindent Note that unlike the~$d = 4, \CN=1$ recombination rules in~\eqref{4dn1recomb}, which contain three multiplets on the right-hand side, those in~\eqref{4dn2recomb} contain four.  

It follows from~\eqref{d4n2rr1}, \eqref{d4n2rr2}, and~\eqref{4dn2recomb} that the following multiplets (as well as their complex conjugates) are absolutely protected: 
\begin{equation}
L\overline{B}_{1}[j;0]^{(R ; r)}~~(R \leq 1)~, \quad A_{\ell}\overline{B}_{1}[j;0]^{(R ; r)}~~(R \leq 1)~, \quad B_{1}\overline{B}_{1}[0;0]^{(R ; r)}~~(R \leq 3)~.
\end{equation}
This includes the flavor current multiplet in~\eqref{4dn2flavcurr}.  

Generic~$d = 4, \CN=2$ multiplets are tabulated in section~\ref{sec:4dn2tabs}; conserved current multiplets are studied in section~\ref{sec:d4n2curr}.

\subsubsection{$d=4$,~$\CN=3$}

\label{sec:d4n3defs}

The~$\CN=3$ superconformal algebra is~$\frak{su}(2,2|3)$, with~$R$-symmetry~$\frak{su}(3)_R \times \frak{u}(1)_R$. The~$R$-charges of an operator are denoted by~$(R_1, R_2 ; r)$. Here~$R_1, R_2 \in \Z_{\geq 0}$ are~$\frak{su}(3)_R$ Dynkin labels, e.g.~$(1,0)$ denotes the fundamental~$\bf 3$ and~$(0,1)$ the anti-fundamental~$\b {\bf 3}$. The~$\frak{u}(1)_R$ charge is given by~$r \in \R$. The~$Q$-supersymmetries transform as \begin{equation}
Q \in [1;0]_{\half}^{(1,0 ; -1)}~, \qquad \b Q \in [0;1]_{\half}^{(0,1 ; 1)}~, \qquad N_Q = 12~.
\end{equation}
Superconformal multiplets obey unitarity bounds and shortening conditions with respect to both~$Q$ and~$\b Q$, summarized in tables~\ref{tab:4DN3C} and~\ref{4DN3AC}. 
\renewcommand{\arraystretch}{1.5}
\renewcommand\tabcolsep{8pt}
\begin{table}[H]
  \centering
  \begin{tabular}{ |c|lr| l|l| }
\hline
{\bf Name} &  \multicolumn{2}{c}{\bf Primary} &  \multicolumn{1}{|c|}{\bf Unitarity Bound} & \multicolumn{1}{c|}{\bf $Q$~Null State } \\
\hline
\hline
$L$ & $[ j;\overline{j} ]_{\Delta}^{(R_1,R_2 ; r)}$&  &$\Delta>2 + j+{2 \over 3} \left(2R_1+R_2\right) - {1 \over 6}r$ & \multicolumn{1}{c|}{$-$} \\
\hline
\hline
$A_{1}$ & $[ j;\overline{j} ]_{\Delta}^{(R_1,R_2 ; r)}~,$& $j\geq1$ &$\Delta= 2+j+{2 \over 3} \left(2R_1+R_2\right) - {1 \over 6}r$ & $[ j-1;\overline{j} ]_{\Delta+1/2}^{(R_1+1,R_2 ; r-1)}$ \\
\hline 
$A_{2}$ & $[ 0;\overline{j} ]_{\Delta}^{(R_1,R_2 ; r)}$& $$ &$\Delta=2+{2 \over 3} \left(2R_1+R_2\right) - {1 \over 6}r$ & $[ 0;\overline{j} ]_{\Delta+1}^{(R_1+2,R_2 ;  r-2)}$ \\
\hline
\hline
$B_{1}$ & $[ 0;\overline{j} ]_{\Delta}^{(R_1,R_2 ; r)}$& $$ &$\Delta={2 \over 3} \left(2R_1+R_2\right) - {1 \over 6}r$ & $[ 1;\overline{j} ]_{\Delta+1/2}^{(R_1+1, R_2 ;  r-1)}$ \\
\hline
\end{tabular}
  \caption{$Q$ shortening conditions in four-dimensional~$\CN=3$ SCFTs.}
  \label{tab:4DN3C}
\end{table} 

\renewcommand{\arraystretch}{1.5}
\renewcommand\tabcolsep{8pt}
\begin{table}[H]
  \centering
  \begin{tabular}{ |c|lr| l|l| }
\hline
{\bf Name} &  \multicolumn{2}{c}{\bf Primary} &  \multicolumn{1}{|c|}{\bf Unitarity Bound} & \multicolumn{1}{c|}{\bf $\b Q$ Null State } \\
\hline
\hline
$\overline{L}$ & $[ j;\overline{j} ]_{\Delta}^{(R_1,R_2 ; r)}$&  &$\Delta>2+ \overline{j}+{2 \over 3}\left(R_1+2R_2\right)+{1 \over 6} r$ & \multicolumn{1}{c|}{$-$} \\
\hline
\hline
$\overline{A}_{1}$ & $[ j;\overline{j} ]_{\Delta}^{(R_1,R_2 ; r)}~,$& $\overline{j}\geq1$ &$\Delta=2+ \overline{j}+{2 \over 3}\left(R_1+2R_2\right)+{1 \over 6} r$ & $[ j;\overline{j}-1 ]_{\Delta+1/2}^{(R_1,R_2+1 ;  r+1)}$ \\
\hline 
$\overline{A}_{2}$ & $[ j;0 ]_{\Delta}^{(R_1,R_2 ; r)}$& $$ &$\Delta=2+{2 \over 3}\left(R_1+2R_2\right)+{1 \over 6} r$ & $[ j;0 ]_{\Delta+1}^{(R_1,R_2+2 ; r+2)}$ \\
\hline
\hline
$\overline{B}_{1}$ & $[ j;0 ]_{\Delta}^{(R_1,R_2 ; r)}$& $$ &$\Delta={2 \over 3}\left(R_1+2R_2\right)+{1 \over 6} r$ & $[ j;1 ]_{\Delta+1/2}^{(R_1,R_2+1 ; r+1)}$ \\
\hline
\end{tabular}
  \caption{$\b Q$ shortening conditions in four-dimensional~$\CN=3$ SCFTs.}
  \label{4DN3AC}
\end{table}

\noindent Full~$\CN=3$ multiplets are two-sided: they are obtained by imposing both left~$Q$ and right~$\b Q$ unitarity bounds and shortening conditions. This can lead to restrictions on some quantum numbers. The consistent two-sided multiplets are summarized in table~\ref{tab:4DN3CAC}. 

\begin{landscape}
\renewcommand{\arraystretch}{2}
\renewcommand\tabcolsep{2pt}
\begin{table}
  \centering
  \captionsetup{justification=centering}
  \begin{tabular}{ |c||c|c|c|c|}
\hline
 &  \multicolumn{1}{c}{$ \boldsymbol{\overline{{L}}}$} &  \multicolumn{1}{|c|}{$\boldsymbol{\overline{A}_{1}}$} & \multicolumn{1}{c|}{$\boldsymbol{\overline{A}_{2}}$ }& \multicolumn{1}{c|}{$\boldsymbol{\overline{B}_{1}}$ } \\
\hline
\hline
\multirow{2}{*}{~~$\boldsymbol L$~~} &~~$\begin{aligned} \\[-5pt] [j;\overline{j}]_{\Delta}^{(R_1,R_2;r)} \end{aligned} $~~& ~~$\begin{aligned} \\[-5pt] [j;\overline{j}\geq 1]_{\Delta}^{(R_1,R_2;r>r_*)} \end{aligned} $ ~~&~$\begin{aligned} \\[-5pt] [j;\overline{j}=0]_{\Delta}^{(R_1,R_2;r>r_*)} \end{aligned}$~&~~$\begin{aligned} \\[-5pt] [j;\overline{j}=0]_{\Delta}^{(R_1,R_2;r>r_* + 6)} \end{aligned}$ ~~ \\[10pt]
&~ $\begin{aligned} \scriptstyle \Delta\, >\, 2+\max\big\{ & \scriptstyle j + {2 \over 3} (2 R_1 + R_2) - {1 \over 6} r \, , \\
&  \scriptstyle \overline{j} + {2 \over 3} (R_1 + 2 R_2) + {1 \over 6} r\big\} \\[-8pt] & \end{aligned}$ ~&~$\Delta=2+ \overline{j} + {2 \over 3} (R_1 + 2 R_2) + {1 \over 6}r$~ &~$\Delta=2+ {2 \over 3} (R_1 + 2 R_2) + {1 \over 6}r$~ &$~\Delta={2 \over 3} (R_1 + 2 R_2) + {1 \over 6}r$~ \\
\hline
\multirow{2}{*}{~~$\boldsymbol{A_{1}}$~~} &$\begin{aligned} \\[-5pt] [j\geq 1;\overline{j}]_{\Delta}^{(R_1,R_2;r< r_*)} \end{aligned} $&~ $\begin{aligned} \\[-5pt] [j\geq 1; \overline{j}\geq 1]_{\Delta}^{(R_1,R_2; r= r_* )} \end{aligned}$ ~&~ $\begin{aligned} \\[-5pt] [j\geq 1; \overline{j}=0]_{\Delta}^{(R_1,R_2;r=r_*)} \end{aligned}$~ & ~$\begin{aligned} \\[-5pt] [j \geq 1; \overline{j}=0]_{\Delta}^{(R_1, R_2;r=r_*+6)} \end{aligned}$ ~\\[5pt]
& ~$\Delta=2+j+{2 \over 3}(2 R_1 + R_2) -\frac{1}{6}r$~ &$\Delta=2+\frac{1}{2}(j+\overline{j}) + R_1 + R_2$ &$\Delta=2+\half j +R_1 + R_2$ & $\Delta=1+ \half j + R_1 + R_2$ \\
\hline
\multirow{2}{*}{$\boldsymbol{A_{2}}$} &$\begin{aligned} \\[-5pt] [j=0;\overline{j}]_{\Delta}^{(R_1, R_2;r<r_*)} \end{aligned}$& $\begin{aligned} \\[-5pt] [j=0; \overline{j}\geq 1]_{\Delta}^{(R_1, R_2;r=r_*)} \end{aligned}$ &$\begin{aligned} \\[-5pt] [j=0; \overline{j}=0]_{\Delta}^{(R_1, R_2;r=r_*)} \end{aligned}$&~$\begin{aligned} \\[-5pt] [j=0; \overline{j}=0]_{\Delta}^{(R_1, R_2;r=r_* + 6)} \end{aligned}$~ \\
& $\Delta=2+{2 \over 3}(2R_1 + R_2) -  {1 \over 6} r$ &$\Delta=2+\half \b j + R_1 + R_2$&$\Delta=2+R_1+ R_2$ &$\Delta=1+R_1+ R_2$ \\
\hline
\multirow{2}{*}{$\boldsymbol{B_{1}}$} &$\begin{aligned} \\[-5pt] [j=0;\overline{j}]_{\Delta}^{(R_1, R_2;r<r_*-6)} \end{aligned} $& $\begin{aligned} \\[-5pt] [j=0; \overline{j} \geq 1]_{\Delta}^{(R_1, R_2;r=r_*-6)} \end{aligned} $ &~$\begin{aligned} \\[-5pt] [j=0; \overline{j}=0]_{\Delta}^{(R_1, R_2;r=r_*-6)} \end{aligned}$~&$\begin{aligned} \\[-5pt] [j=0; \overline{j}=0]_{\Delta}^{(R_1, R_2;r=r_*)} \end{aligned}$ \\
& $\Delta={2 \over 3} (2R_1 + R_2) - {1 \over 6}r $ &$\Delta=1+\half \b{j} +R_1 + R_2 $&$\Delta=1+R_1+R_2$ &$\Delta=R_1+R_2$ \\
\hline
\end{tabular}
\caption{ \label{tab:4DN3CAC} Consistent two-sided multiplets in~$d = 4, \mathcal{N}=3$ theories. The critical $\frak{u}(1)_{r}$ charge $r_{*}$ is defined as in \eqref{rstardef}: 
  \\[4pt]
  $r_{*}(j,\overline{j},R_{1},R_{2})=3(j-\overline{j})+2(R_{1}-R_{2})$~.}
 \end{table} 
\end{landscape}

In order to write down the recombination rules, we define 
\begin{equation}
\Delta_{A}=2+j+\frac{2}{3}(2R_{1}+R_{2})-\frac{1}{6}r~, \qquad \Delta_{\overline{A}}=2+\overline{j}+\frac{2}{3}(R_{1}+2R_{2})+\frac{1}{6}r~.
\end{equation}
As~$\Delta \rightarrow \Delta_{A}^+$, we find the following partial chiral recombination rules,
\begin{equation}
\begin{array}{rclcl}
L[j\geq 2]_{\Delta}^{(R_{1}, R_{2} ; r)} & \rightarrow &A_{1}[j]_{\Delta_{A}}^{(R_{1}, R_{2} ; r)}  &\oplus & A_{1}[j-1]_{\Delta_{A}+\half}^{(R_{1}+1, R_{2} ; r-1)}~, \\
L[j=1]_{\Delta}^{(R_{1}, R_{2} ; r)} & \rightarrow &A_{1}[1]_{\Delta_{A}}^{(R_{1}, R_{2} ; r)}  &\oplus & A_{2}[0]_{\Delta_{A}+\half}^{(R_{1}+1, R_{2} ; r-1)}~, \\
L[j=0]_{\Delta}^{(R_{1}, R_{2} ; r)} & \rightarrow &A_{2}[0]_{\Delta_{A}}^{(R_{1}, R_{2} ; r)}  &\oplus & B_{1}[0]_{\Delta_{A}+1}^{(R_{1}+2, R_{2} ; r-2)}~,
\end{array}
\label{chiralrecomb4dn3}
\end{equation}
and similarly for the antichiral sector. The structure of the full, two-sided recombination rules is controlled by the critical~$\frak{u}(1)_{r}$ charge
\begin{equation}
r_{*}(j,\overline{j},R_{1},R_{2})=3(j-\overline{j})+2(R_{1}-R_{2})~. \label{rstardef}
\end{equation}
The sign of~$r-r_{*}$ determines whether the chiral or the antichiral unitarity bound is saturated first as the scaling dimension~$\Delta$ of a long multiplet is lowered:
\begin{itemize}
\item If~$r<r_{*}$, then~$\Delta_A > \Delta_{\b A}$ and the chiral unitarity bounds are saturated first:
\begin{equation}\label{d4n3rr1}
\setstretch{1.2}
\begin{array}{rclclcl}
L\overline{L}[j\geq 2;\overline{j}]_{\Delta}^{(R_{1},R_{2} ;  r<r_{*})} & \rightarrow &A_{1}\overline{L}[j;\overline{j}]_{\Delta_{A}}^{(R_{1}, R_{2} ; r)}  &\oplus & A_{1}\overline{L}[j-1;\overline{j}]_{\Delta_{A}+\half}^{(R_{1}+1, R_{2} ; r-1)}~, \\
L\overline{L}[j=1;\overline{j}]_{\Delta}^{(R_{1},R_{2} ;  r<r_{*})} & \rightarrow &A_{1}\overline{L}[1;\overline{j}]_{\Delta_{A}}^{(R_{1}, R_{2} ;  r)}  &\oplus & A_{2}\overline{L}[0;\overline{j}]_{\Delta_{A}+\half}^{(R_{1}+1, R_{2} ; r-1)}~, \\
L\overline{L}[j=0;\overline{j}]_{\Delta}^{(R_{1},R_{2} ; r<r_{*})} & \rightarrow &A_{2}\overline{L}[0;\overline{j}]_{\Delta_{A}}^{(R_{1},R_{2} ; r)}  &\oplus & B_{1}\overline{L}[0;\overline{j}]_{\Delta_{A}+1}^{(R_{1}+2,R_2{} ;  r-2)}~.
\end{array}
\end{equation}
\item If~$r>r_{*}$, then~$\Delta_{\b A}> \Delta_A$ and the antichiral unitarity bounds are saturated first:
\begin{equation}\label{d4n3rr2}
\setstretch{1.2}
\begin{array}{rclclcl}
L\overline{L}[j;\overline{j}\geq 2]_{\Delta}^{(R_{1},R_{2} ; r>r_{*})} & \rightarrow &L\overline{A}_{1}[j;\overline{j}]_{\Delta_{\overline{A}}}^{(R_{1},R_{2} ; r)}  &\oplus & L\overline{A}_{1}[j;\overline{j}-1]_{\Delta_{\overline{A}}+\half}^{(R_{1},R_{2}+1 ; r+1)}~, \\
L\overline{L}[j;\overline{j}=1]_{\Delta}^{(R_{1},R_{2} ; r>r_{*})} & \rightarrow &L\overline{A}_{1}[j;1]_{\Delta_{\overline{A}}}^{(R_{1},R_{2} ; r)}  &\oplus & L\overline{A}_{2}[j;0]_{\Delta_{\overline{A}}+\half}^{(R_{1},R_{2}+1 ; r+1)}~, \\
L\overline{L}[j;\overline{j}=0]_{\Delta}^{(R_{1},R_{2} ; r>r_{*})} & \rightarrow &L\overline{A}_{2}[j;0]_{\Delta_{\overline{A}}}^{(R_{1},R_{2} ;  r)}  &\oplus & L\overline{B}_{1}[j;0]_{\Delta_{\overline{A}}+1}^{(R_{1},R_{2}+2 ; r+2)}~.
\end{array}
\end{equation}
\item When $r=r_{*}$, then~$\Delta_A = \Delta_{\b A}$ and the chiral and antichiral unitarity bounds are saturated simultaneously:
\end{itemize}
\noindent \begin{equation}
\setstretch{1.2}
\begin{array}{rclclcl}
L\overline{L}[j\geq 2; \overline{j}\geq 2]_{\Delta}^{(R_{1},R_{2} ;  r=r_{*})} & \rightarrow &A_{1}\overline{A}_{1}[j; \overline{j}]_{\Delta_{A}}^{(R_{1},R_{2} ; r_{*})}  &\oplus & A_{1}\overline{A}_{1}[j-1; \overline{j}]_{\Delta_{A}+\half}^{(R_{1}+1,R_{2} ; r_{*}-1)} \\
& \oplus & A_{1}\overline{A}_{1}[j;\overline{j}-1]_{\Delta_{A}+\half}^{(R_{1},R_{2}+1 ; r_{*}+1)}& \oplus & A_{1}\overline{A}_{1}[j-1;\overline{j}-1]_{\Delta_{A}+1}^{(R_{1}+1,R_{2}+1 ; r_{*})}~, \\
L\overline{L}[j\geq 2; \overline{j}=1]_{\Delta}^{(R_{1},R_{2} ;  r=r_{*})} & \rightarrow &A_{1}\overline{A}_{1}[j; 1]_{\Delta_{A}}^{(R_{1},R_{2} ; r_{*})}  &\oplus & A_{1}\overline{A}_{1}[j-1; 1]_{\Delta_{A}+\half}^{(R_{1}+1,R_{2} ; r_{*}-1)} \\
& \oplus & A_{1}\overline{A}_{2}[j;0]_{\Delta_{A}+\half}^{(R_{1},R_{2}+1 ; r_{*}+1)}& \oplus & A_{1}\overline{A}_{2}[j-1;0]_{\Delta_{A}+1}^{(R_{1}+1,R_{2}+1 ; r_{*})}~, \\
L\overline{L}[j\geq 2; \overline{j}=0]_{\Delta}^{(R_{1},R_{2} ;  r=r_{*})} & \rightarrow &A_{1}\overline{A}_{2}[j;0]_{\Delta_{A}}^{(R_{1},R_{2} ; r_{*})}  &\oplus & A_{1}\overline{A}_{2}[j-1; 0]_{\Delta_{A}+\half}^{(R_{1}+1,R_{2} ; r_{*}-1)}\\
 & \oplus & A_{1}\overline{B}_{1}[j;0]_{\Delta_{A}+1}^{(R_{1},R_{2}+2 ; r_{*}+2)}& \oplus & A_{1}\overline{B}_{1}[j-1;0]_{\Delta_{A}+{3 \over 2}}^{(R_{1}+1,R_{2}+2 ; r_{*}+1)}~, \\
L\overline{L}[j=1; \overline{j}\geq 2]_{\Delta}^{(R_{1},R_{2} ; r=r_{*})} & \rightarrow &A_{1}\overline{A}_{1}[1; \overline{j}]_{\Delta_{A}}^{(R_{1},R_{2} ; r_{*})}  &\oplus & A_{2}\overline{A}_{1}[0; \overline{j}]_{\Delta_{A}+\half}^{(R_{1}+1,R_{2} ;  r_{*}-1)} \\
& \oplus & A_{1}\overline{A}_{1}[1;\overline{j}-1]_{\Delta_{A}+\half}^{(R_{1},R_{2}+1 ; r_{*}+1)}& \oplus & A_{2}\overline{A}_{1}[0;\overline{j}-1]_{\Delta_{A}+1}^{(R_{1}+1,R_{2}+1 ; r_{*})}~, \\
L\overline{L}[j=0; \overline{j}\geq 2]_{\Delta}^{(R_{1},R_{2} ; r=r_{*})} & \rightarrow &A_{2}\overline{A}_{1}[0; \overline{j}]_{\Delta_{A}}^{(R_{1},R_{2} ; r_{*})}  &\oplus & B_{1}\overline{A}_{1}[0; \overline{j}]_{\Delta_{A}+1}^{(R_{1}+2,R_{2} ; r_{*}-2)} \\
& \oplus & A_{2}\overline{A}_{1}[0;\overline{j}-1]_{\Delta_{A}+\half}^{(R_{1},R_{2}+1 ; r_{*}+1)}& \oplus & B_{1}\overline{A}_{1}[0;\overline{j}-1]_{\Delta_{A}+{3 \over 2}}^{(R_{1}+2,R_{2}+1 ; r_{*}-1)}~, \\
L\overline{L}[j=1; \overline{j}=1]_{\Delta}^{(R_{1},R_{2} ; r=r_{*})} & \rightarrow &A_{1}\overline{A}_{1}[1; 1]_{\Delta_{A}}^{(R_{1},R_{2} ; r_{*})}  &\oplus & A_{2}\overline{A}_{1}[0; 1]_{\Delta_{A}+\half}^{(R_{1}+1,R_{2} ; r_{*}-1)}\\
 & \oplus & A_{1}\overline{A}_{2}[1;0]_{\Delta_{A}+\half}^{(R_{1},R_{2}+1 ; r_{*}+1)}& \oplus & A_{2}\overline{A}_{2}[0;0]_{\Delta_{A}+1}^{(R_{1}+1,R_{2}+1 ; r_{*})}~, \\
L\overline{L}[j=1; \overline{j}=0]_{\Delta}^{(R_{1},R_{2} ; r=r_{*})} & \rightarrow &A_{1}\overline{A}_{2}[1; 0]_{\Delta_{A}}^{(R_{1},R_{2} ; r_{*})}  &\oplus & A_{2}\overline{A}_{2}[0; 0]_{\Delta_{A}+\half}^{(R_{1}+1,R_{2} ; r_{*}-1)} \\
& \oplus & A_{1}\overline{B}_{1}[1;0]_{\Delta_{A}+1}^{(R_{1},R_{2}+2 ; r_{*}+2)}& \oplus & A_{2}\overline{B}_{1}[0;0]_{\Delta_{A}+{3\over 2}}^{(R_{1}+1,R_{2}+2 ; r_{*}+1)}~, \\
L\overline{L}[j=0; \overline{j}=1]_{\Delta}^{(R_{1},R_{2} ; r=r_{*})} & \rightarrow &A_{2}\overline{A}_{1}[0; 1]_{\Delta_{A}}^{(R_{1},R_{2} ; r_{*})}  &\oplus & B_{1}\overline{A}_{1}[0; 1]_{\Delta_{A}+1}^{(R_{1}+2,R_{2} ; r_{*}-2)}\\
 & \oplus & A_{2}\overline{A}_{2}[0;0]_{\Delta_{A}+\half}^{(R_{1},R_{2}+1 ; r_{*}+1)} & \oplus & B_{1}\overline{A}_{2}[0;0]_{\Delta_{A}+{3\over 2}}^{(R_{1}+2,R_{2}+1 ; r_{*}-1)}~,\\
L\overline{L}[j=0; \overline{j}=0]_{\Delta}^{(R_{1},R_{2} ;  r=r_{*})} & \rightarrow &A_{2}\overline{A}_{2}[0; 0]_{\Delta_{A}}^{(R_{1},R_{2} ; r_{*})}  &\oplus & B_{1}\overline{A}_{2}[0; 0]_{\Delta_{A}+1}^{(R_{1}+2,R_{2} ; r_{*}-2)} \\
& \oplus & A_{2}\overline{B}_{1}[0;0]_{\Delta_{A}+1}^{(R_{1},R_{2}+2 ;  r_{*}+2)}& \oplus & B_{1}\overline{B}_{1}[0;0]_{\Delta_{A}+2}^{(R_{1}+2,R_{2}+2 ; r_{*})}~.
\end{array}
\label{4dn3recomb}
\end{equation}
\noindent By examining~\eqref{d4n3rr1}, \eqref{d4n3rr2}, and~\eqref{4dn3recomb}, we conclude that the following multiplets (as well as their complex conjugates) are absolutely protected: 
\begin{align}
& L\overline{B}_{1}[j;0]^{(R_{1}, R_{2} ; r>r_{*}+2)}~~(R_{2} \leq 1)~, &&A_{\ell}\overline{B}_{1}[j;0]^{(R_{1}, R_{2} ; r_{*}+2)}~~(R_{2} \leq 1)~,\cr
& B_{1}\overline{B}_{1}[0;0]^{(R_{1}, R_{2} ; r_{*})}~~(R_1 \leq 1~\text{or}~R_2 \leq 1)~. &&
\end{align}
These include the extra SUSY-current~\eqref{d4n3escm} and the stress tensor multiplet~\eqref{d4n3stm}. 

Conserved current multiplets in~$d = 4, \CN=3$ theories are studied in section~\ref{sec:d4n3curr}. 

\subsubsection{$d = 4$,~$\CN=4$} \label{sec:4dn4defs}

The~$\CN=4$ superconformal algebra is~$ \frak{psu}(2,2|4)$, with~$R$-symmetry~$\frak{su}(4)_R \simeq \frak{so}(6)_R$. The~$R$-charges are denoted by~$\frak{su}(4)_R$ Dynkin labels~$(R_1, R_2, R_3)$ with~$R_1, R_2, R_3 \in \Z_{\geq 0}\,$. For instance, $(1,0,0)$ and~$(0,0,1)$ are the fundamental~$\bf 4$ and the anti-fundamental~$\b {\bf 4}$ of~$\frak{su}(4)_R$, while~$(0,1,0)$ is the fundamental vector representation~$\bf 6$ of~$\frak{so}(6)_R$. The~$Q$-supercharges are 
\begin{equation}
Q \in [1;0]_{\half}^{(1,0,0)}~, \qquad \b Q \in [0;1]_{\half}^{(0,0,1)}~, \qquad N_Q = 16~.
\end{equation}
Superconformal multiplets obey unitarity bounds and shortening conditions with respect to both~$Q$ and~$\b Q$, summarized in tables~\ref{tab:4DN4C} and~\ref{4DN4AC}, and are labeled by a pair of capital letters. 

\renewcommand{\arraystretch}{1.5}
\renewcommand\tabcolsep{8pt}
\begin{table}[H]
  \centering
  \begin{tabular}{ |c|lr| l|l| }
\hline
{\bf Name} &  \multicolumn{2}{c}{\bf Primary} &  \multicolumn{1}{|c|}{\bf Unitarity Bound} & \multicolumn{1}{c|}{\bf $Q$~Null State } \\
\hline
\hline
$L$ & $[ j;\overline{j} ]_{\Delta}^{(R_1,R_2,R_3)}$&  &$\Delta>2 + j+\half\left(3R_1 +2R_2 +R_3\right)$ & \multicolumn{1}{c|}{$-$} \\
\hline
\hline
$A_{1}$ & $[j;\overline{j} ]_{\Delta}^{(R_1,R_2,R_3)}~,$& $j\geq1$ &$\Delta= 2+j+\half\left(3R_1 +2R_2 +R_3\right)$ & $[j-1;\overline{j} ]_{\Delta+1/2}^{(R_1+1,R_2,R_3)}$ \\
\hline 
$A_{2}$ & $[ 0;\overline{j} ]_{\Delta}^{(R_1,R_2,R_3)}$& $$ &$\Delta=2+\half\left(3R_1 +2R_2 +R_3\right)$ & $[0;\overline{j}]_{\Delta+1}^{(R_1+2,R_2,R_3)}$ \\
\hline
\hline
$B_{1}$ & $[0;\overline{j}]_{\Delta}^{(R_1,R_2,R_3)}$& $$ &$\Delta=\half\left(3R_1 +2R_2 +R_3\right)$ & $[1;\overline{j}]_{\Delta+1/2}^{(R_1+1, R_2,R_3)}$ \\
\hline
\end{tabular}
  \caption{$Q$ shortening conditions in four-dimensional~$\CN=4$ SCFTs.}
  \label{tab:4DN4C}
\end{table}

\renewcommand{\arraystretch}{1.5}
\renewcommand\tabcolsep{8pt}
\begin{table}[H]
  \centering
  \begin{tabular}{ |c|lr| l|l| }
\hline
{\bf Name} &  \multicolumn{2}{c}{\bf Primary} &  \multicolumn{1}{|c|}{\bf Unitarity Bound} & \multicolumn{1}{c|}{\bf $\b Q$~Null State } \\
\hline
\hline
$\overline{L}$ & $[j;\overline{j}]_{\Delta}^{(R_1,R_2,R_3)}$&  &$\Delta>2+ \overline{j}+\half\left(R_1+2R_2 +3 R_3\right)$ & \multicolumn{1}{c|}{$-$} \\
\hline
\hline
$\overline{A}_{1}$ & $[j;\overline{j}]_{\Delta}^{(R_1,R_2,R_3)}~,$& $\overline{j}\geq1$ &$\Delta=2+ \overline{j}+\half\left(R_1+2R_2 +3 R_3\right)$ & $[j;\overline{j}-1]_{\Delta+1/2}^{(R_1,R_2,R_3+1)}$ \\
\hline 
$\overline{A}_{2}$ & $[j;0]_{\Delta}^{(R_1,R_2,R_3)}$& $$ &$\Delta=2+\half\left(R_1+2R_2 +3 R_3\right)$ & $[j;0]_{\Delta+1}^{(R_1,R_2,R_3+2)}$ \\
\hline
\hline
$\overline{B}_{1}$ & $[j;0]_{\Delta}^{(R_1,R_2,R_3)}$& $$ &$\Delta=\half\left(R_1+2R_2+3R_3\right)$ & $[j;1]_{\Delta+1/2}^{(R_1,R_2,R_3+1)}$ \\
\hline
\end{tabular}
  \caption{$\b Q$ shortening conditions in four-dimensional~$\CN=4$ SCFTs.}
  \label{4DN4AC}
\end{table}

\noindent Full~$\CN=4$ multiplets are two-sided: they are obtained by imposing both left~$Q$ and right~$\b Q$ unitarity bounds and shortening conditions. This can lead to restrictions on some quantum numbers. The consistent two-sided multiplets are summarized in table~\ref{tab:4DN4CAC}. 

\begin{landscape}
\renewcommand{\arraystretch}{2}
\renewcommand\tabcolsep{2pt}
\begin{table}
  \centering
  \captionsetup{justification=centering}
  \begin{tabular}{ |c||c|c|c|c|}
\hline
 &  \multicolumn{1}{c}{$ \boldsymbol{\overline{{L}}}$} &  \multicolumn{1}{|c|}{$\boldsymbol{\overline{A}_{1}}$} & \multicolumn{1}{c|}{$\boldsymbol{\overline{A}_{2}}$ }& \multicolumn{1}{c|}{$\boldsymbol{\overline{B}_{1}}$ } \\
\hline
\hline
\multirow{3}{*}{~~$\boldsymbol L$~~} &$\begin{aligned} \\[-5pt] [j;\overline{j}]_{\Delta}^{(R_{1},R_{2}, R_{3})} \end{aligned} $& $\begin{aligned} \\[-5pt] [j;\overline{j}\geq 1]_{\Delta}^{(R_{1},R_{2}, R_{3})} \end{aligned} $ &$\begin{aligned} \\[-5pt] [j;\overline{j}=0]_{\Delta}^{(R_{1},R_{2}, R_{3})} \end{aligned} $&$\begin{aligned} \\[-5pt] [j; \overline{j}=0]_{\Delta}^{(R_{1},R_{2}, R_{3})} \end{aligned}$ \\
& ~~$\begin{aligned}  \\[-8pt] \scriptstyle \Delta \, > \, 2+\max\big\{& \scriptstyle j+\frac{1}{2}(3R_{1}+2R_{2}+R_{3})~, \\[-2pt] & \scriptstyle  \overline{j}+\frac{1}{2}(R_{1}+2R_{2}+3R_{3})\big\} 
\end{aligned} $~~ &$\scriptstyle \Delta\, =\, 2+\overline{j}+\frac{1}{2}(R_{1}+2R_{2}+3R_{3})$ &$\scriptstyle \Delta \, = \, 2+\frac{1}{2}(R_{1}+2R_{2}+3R_{3})$ &$\scriptstyle \Delta \, = \, \frac{1}{2}(R_{1}+2R_{2}+3R_{3})$ \\[-22pt]
& &$\scriptstyle R_{3}\, >\, j-\overline{j}+R_{1}$&$\scriptstyle R_{3}\, >\, j+R_{1}$&$\scriptstyle R_{3}\, >\, 2+j+R_{1}$\\
\hline
\multirow{3}{*}{~~$\boldsymbol{A_{1}}$~~} &$\begin{aligned} \\[-5pt] [j\, \geq \, 1 ; \overline{j}]_{\Delta}^{(R_{1},R_{2}, R_{3})} \end{aligned} $& $\begin{aligned} \\[-5pt] [j \, \geq \, 1 ; \overline{j} \, \geq \, 1]_{\Delta}^{(R_{1},R_{2}, R_{3})} \end{aligned} $ & $ \begin{aligned} \\[-5pt] ~~[j\, \geq \, 1; \overline{j}=0]_{\Delta}^{(R_{1},R_{2}, R_{3})}~~ \end{aligned} $ & ~~$\begin{aligned} \\[-5pt] [j \, \geq \, 1; \overline{j}=0]_{\Delta}^{(R_{1},R_{2}, R_{3})} \end{aligned} $~~ \\
& $\scriptstyle \Delta\, =\, 2+j+\frac{1}{2}(3R_{1}+2R_{2}+R_{3})$ &$\scriptstyle \Delta \, =\, 2+\frac{1}{2}(j+\overline{j})+R_{1}+R_{2}+R_{3}$ &$\scriptstyle \Delta \, = \, 2+\frac{1}{2}\, j+R_{1}+R_{2}+R_{3}$ & $\scriptstyle \Delta \, = \, 1+\frac{1}{2}\, j+R_{1}+R_{2}+R_{3}$ \\[-15pt]
&$\scriptstyle R_{1} \, > \, \overline{j}-j+R_{3}$&$\scriptstyle j+R_{1} \, = \, \overline{j}+R_{3}$&$\scriptstyle j+R_{1} \, = \, R_{3}$&$\scriptstyle R_{3} \, = \, 2+j+R_{1}$\\
\hline
\multirow{3}{*}{$\boldsymbol{A_{2}}$} &$\begin{aligned} \\[-5pt]  [j\, =\, 0;\overline{j}]_{\Delta}^{(R_{1},R_{2}, R_{3})} \end{aligned} $& $\begin{aligned} \\[-5pt]  [j\, =\, 0; \overline{j}\,  \geq \, 1]_{\Delta}^{(R_{1},R_{2}, R_{3})} \end{aligned} $ &$\begin{aligned} \\[-5pt] [j\, =\, 0; \overline{j}=0]_{\Delta}^{(R_{1},R_{2}, R_{3})}\end{aligned} $&$\begin{aligned} \\[-5pt]  [j\, =\, 0; \overline{j}=0]_{\Delta}^{(R_{1},R_{2}, R_{3})}\end{aligned} $ \\
& $\scriptstyle \Delta\, =\, 2+\frac{1}{2}(3R_{1}+2R_{2}+R_{3})$ &$\scriptstyle \Delta\, =\, 2+\frac{1}{2}\, \overline{j}+R_{1}+R_{2}+R_{3}$&$\scriptstyle \Delta\, =\, 2+R_{1}+R_{2}+R_{3}$ &$\scriptstyle \Delta\, =\, 1+R_{1}+R_{2}+R_{3}$ \\[-15pt]
&$\scriptstyle R_{1}\,>\,\overline{j}+R_{3}$&$\scriptstyle R_{1}\, =\, \overline{j}+R_{3}$&$\scriptstyle R_{1}\, =\, R_{3}$&$\scriptstyle R_{3}\, =\, 2+R_{1}$\\
\hline
\multirow{3}{*}{$\boldsymbol{B_{1}}$} &$\begin{aligned} \\[-5pt] [j\,=\,0;\overline{j}]_{\Delta}^{(R_{1},R_{2}, R_{3})} \end{aligned} $&$\begin{aligned} \\[-5pt] [j\,=\,0; \overline{j}\,\geq \, 1]_{\Delta}^{(R_{1},R_{2}, R_{3})}\end{aligned} $&$\begin{aligned} \\[-5pt] [j\,=\,0; \overline{j}\,=\,0]_{\Delta}^{(R_{1},R_{2}, R_{3})}\end{aligned} $&$\begin{aligned} \\[-5pt] [j\,=\,0; \overline{j}\,=\,0]_{\Delta}^{(R_{1},R_{2}, R_{3})}\end{aligned}$ \\
& $\scriptstyle \Delta\,=\,\frac{1}{2}(3R_{1}+2R_{2}+R_{3})$ &$\scriptstyle \Delta\,=\,1+\frac{1}{2}\, \overline{j} +R_{1}+R_{2}+R_{3}$&$\scriptstyle \Delta\,=\,1+R_{1}+R_{2}+R_{3}$ &$\scriptstyle \Delta\,=\,R_{1}+R_{2}+R_{3}$ \\[-15pt]
&$\scriptstyle R_{1}\,>\,2+\overline{j}+R_{3}$&$\scriptstyle R_{1}\, =\, 2+\overline{j}+R_{3}$&$\scriptstyle R_{1}\,=\, 2+R_{3}$&$\scriptstyle R_{1}\,=\,R_{3}$\\
\hline
\end{tabular}
  \caption{ \label{tab:4DN4CAC} Consistent two-sided multiplets in four-dimensional~$\CN = 4$ theories.   }
 \end{table} 
\end{landscape}

In order to summarize the recombination rules, we define 
\begin{equation}
\Delta_{A}=2+j+\frac{1}{2}(3R_{1}+2R_{2}+R_{3})~, \hspace{.5in} \Delta_{\overline{A}}=2+\overline{j}+\frac{1}{2}(R_{1}+2R_{2}+3R_{3})~.
\end{equation}
As~$\Delta \rightarrow \Delta_{A}^+$, we find the following chiral recombination rules
\begin{equation}
\begin{array}{rclcl}
L[j\geq 2]_{\Delta}^{(R_{1}, R_{2}, R_{3})} & \rightarrow &A_{1}[j]_{\Delta_{A}}^{(R_{1}, R_{2}, R_{3})}  &\oplus & A_{1}[j-1]_{\Delta_{A}+\half}^{(R_{1}+1, R_{2}, R_{3})}~, \\
L[j=1]_{\Delta}^{(R_{1}, R_{2}, R_{3})} & \rightarrow &A_{1}[1]_{\Delta_{A}}^{(R_{1}, R_{2}, R_{3})}  &\oplus & A_{2}[0]_{\Delta_{A}+\half}^{(R_{1}+1, R_{2}, R_{3})}~, \\
L[j=0]_{\Delta}^{(R_{1}, R_{2}, R_{3})} & \rightarrow &A_{2}[0]_{\Delta_{A}}^{(R_{1}, R_{2}, R_{3})}  &\oplus & B_{1}[0]_{\Delta_{A}+1}^{(R_{1}+2, R_{2}, R_{3})}~,
\end{array}
\label{chiralrecomb4dn4}
\end{equation}
and similarly in the antichiral sector. For~$\CN\neq 4$, the structure of fully two-sided recombination rules is determined by the~$\frak{u}(1)_R$ charge; when~$\CN=4$ it is instead controlled by 
\begin{equation} \label{chidef}
\chi=j-\overline{j} +R_{1} -{R}_{3}~.
\end{equation}
The sign of~$\chi$ determines whether the chiral or antichiral unitarity bounds are saturated first as the dimension~$\Delta$ of a long multiplet is lowered:
\begin{itemize}
\item If~$\chi>0$, then~$\Delta_A > \Delta_{\b A}$ and the chiral unitarity bounds are saturated first:
\begin{equation}\label{d4n4rr1}
\setstretch{1.2}
\begin{array}{rclclcl}
L\overline{L}[j\geq 2;\overline{j}]_{\Delta}^{(R_{1},R_{2},R_{3})} & \rightarrow &A_{1}\overline{L}[j;\overline{j}]_{\Delta_{A}}^{(R_{1}, R_{2},R_{3})}  &\oplus & A_{1}\overline{L}[j-1;\overline{j}]_{\Delta_{A}+\half}^{(R_{1}+1, R_{2},R_{3})}~, \\
L\overline{L}[j=1;\overline{j}]_{\Delta}^{(R_{1},R_{2},R_{3})} & \rightarrow &A_{1}\overline{L}[1;\overline{j}]_{\Delta_{A}}^{(R_{1},R_{2},R_{3})}  &\oplus & A_{2}\overline{L}[0;\overline{j}]_{\Delta_{A}+\half}^{(R_{1}+1,R_{2},R_{3})}~, \\
L\overline{L}[j=0;\overline{j}]_{\Delta}^{(R_{1},R_{2},R_{3})} & \rightarrow &A_{2}\overline{L}[0;\overline{j}]_{\Delta_{A}}^{(R_{1},R_{2},R_{3})}  &\oplus & B_{1}\overline{L}[0;\overline{j}]_{\Delta_{A}+1}^{(R_{1}+2,R_{2},R_{3})}~.
\end{array}
\end{equation}
\item When~$\chi<0$, then~$\Delta_{\b A} > \Delta_A$ and the antichiral unitarity bounds are saturated first:
\begin{equation}\label{d4n4rr2}
\setstretch{1.2}
\begin{array}{rclclcl}
L\overline{L}[j;\overline{j}\geq 2]_{\Delta}^{(R_{1},R_{2},R_{3})} & \rightarrow &L\overline{A}_{1}[j;\overline{j}]_{\Delta_{\overline{A}}}^{(R_{1},R_{2},R_{3})}  &\oplus & L\overline{A}_{1}[j; \overline{j}-1]_{\Delta_{\overline{A}}+\half}^{(R_{1},R_{2},R_{3}+1)}~, \\
L\overline{L}[j;\overline{j}=1]_{\Delta}^{(R_{1},R_{2},R_{3})} & \rightarrow &L\overline{A}_{1}[j;1]_{\Delta_{\overline{A}}}^{(R_{1},R_{2},R_{3})}  &\oplus & L\overline{A}_{2}[j;0]_{\Delta_{\overline{A}}+\half}^{(R_{1},R_{2},R_{3}+1)}~, \\
L\overline{L}[j;\overline{j}=0]_{\Delta}^{(R_{1},R_{2},R_{3})} & \rightarrow &L\overline{A}_{2}[j;0]_{\Delta_{\overline{A}}}^{(R_{1},R_{2},R_{3})}  &\oplus & L\overline{B}_{1}[j;0]_{\Delta_{\overline{A}}+1}^{(R_{1},R_{2},R_{3}+2)}~.
\end{array}
\end{equation}
\item If~$\chi=0$, then~$\Delta_A = \Delta_{\b A}$ and the chiral and antichiral unitarity bounds are saturated simultaneously:
\end{itemize}
\begin{equation}
\setstretch{1.2}
\begin{array}{rclclcl}
L\overline{L}[j\geq 2; \overline{j}\geq 2]_{\Delta}^{(R_{1},R_{2},R_{3})} & \rightarrow &A_{1}\overline{A}_{1}[j; \overline{j}]_{\Delta_{A}}^{(R_{1},R_{2},R_{3})}  &\oplus & A_{1}\overline{A}_{1}[j-1; \overline{j}]_{\Delta_{A}+\half}^{(R_{1}+1,R_{2},R_{3})} \\
& \oplus & A_{1}\overline{A}_{1}[j;\overline{j}-1]_{\Delta_{A}+\half}^{(R_{1},R_{2},R_{3}+1)}& \oplus & A_{1}\overline{A}_{1}[j-1;\overline{j}-1]_{\Delta_{A}+1}^{(R_{1}+1,R_{2},R_{3}+1)}~, \\
L\overline{L}[j\geq 2; \overline{j}=1]_{\Delta}^{(R_{1},R_{2},R_{3})} & \rightarrow &A_{1}\overline{A}_{1}[j; 1]_{\Delta_{A}}^{(R_{1},R_{2},R_{3})}  &\oplus & A_{1}\overline{A}_{1}[j-1; 1]_{\Delta_{A}+\half}^{(R_{1}+1,R_{2},R_{3})} \\
& \oplus & A_{1}\overline{A}_{2}[j;0]_{\Delta_{A}+\half}^{(R_{1},R_{2},R_{3}+1)}& \oplus & A_{1}\overline{A}_{2}[j-1;0]_{\Delta_{A}+1}^{(R_{1}+1,R_{2},R_{3}+1)}~, \\
L\overline{L}[j\geq 2; \overline{j}=0]_{\Delta}^{(R_{1},R_{2},R_{3})} & \rightarrow &A_{1}\overline{A}_{2}[j;0]_{\Delta_{A}}^{(R_{1},R_{2},R_{3})}  &\oplus & A_{1}\overline{A}_{2}[j-1; 0]_{\Delta_{A}+\half}^{(R_{1}+1,R_{2},R_{3})}\\
 & \oplus & A_{1}\overline{B}_{1}[j;0]_{\Delta_{A}+1}^{(R_{1},R_{2},R_{3}+2)}& \oplus & A_{1}\overline{B}_{1}[j-1;0]_{\Delta_{A}+{3 \over 2}}^{(R_{1}+1,R_{2},R_{3}+2)}~, \\
L\overline{L}[j=1; \overline{j}\geq 2]_{\Delta}^{(R_{1},R_{2},R_{3})} & \rightarrow &A_{1}\overline{A}_{1}[1; \overline{j}]_{\Delta_{A}}^{(R_{1},R_{2},R_{3})}  &\oplus & A_{2}\overline{A}_{1}[0; \overline{j}]_{\Delta_{A}+\half}^{(R_{1}+1,R_{2},R_{3})} \\
& \oplus & A_{1}\overline{A}_{1}[1;\overline{j}-1]_{\Delta_{A}+\half}^{(R_{1},R_{2},R_{3}+1)}& \oplus & A_{2}\overline{A}_{1}[0;\overline{j}-1]_{\Delta_{A}+1}^{(R_{1}+1,R_{2},R_{3}+1)}~, \\
L\overline{L}[j=0; \overline{j}\geq 2]_{\Delta}^{(R_{1},R_{2},R_{3})} & \rightarrow &A_{2}\overline{A}_{1}[0; \overline{j}]_{\Delta_{A}}^{(R_{1},R_{2},R_{3})}  &\oplus & B_{1}\overline{A}_{1}[0; \overline{j}]_{\Delta_{A}+1}^{(R_{1}+2,R_{2},R_{3})} \\
& \oplus & A_{2}\overline{A}_{1}[0;\overline{j}-1]_{\Delta_{A}+\half}^{(R_{1},R_{2},R_{3}+1)}& \oplus & B_{1}\overline{A}_{1}[0;\overline{j}-1]_{\Delta_{A}+{3 \over 2}}^{(R_{1}+2,R_{2},R_{3}+1)}~, \\
L\overline{L}[j=1; \overline{j}=1]_{\Delta}^{(R_{1},R_{2},R_{3})} & \rightarrow &A_{1}\overline{A}_{1}[1; 1]_{\Delta_{A}}^{(R_{1},R_{2},R_{3})}  &\oplus & A_{2}\overline{A}_{1}[0; 1]_{\Delta_{A}+\half}^{(R_{1}+1,R_{2},R_{3})}\\
 & \oplus & A_{1}\overline{A}_{2}[1;0]_{\Delta_{A}+\half}^{(R_{1},R_{2},R_{3}+1)}& \oplus & A_{2}\overline{A}_{2}[0;0]_{\Delta_{A}+1}^{(R_{1}+1,R_{2},R_{3}+1)}~, \\
L\overline{L}[j=1; \overline{j}=0]_{\Delta}^{(R_{1},R_{2},R_{3})} & \rightarrow &A_{1}\overline{A}_{2}[1; 0]_{\Delta_{A}}^{(R_{1},R_{2},R_{3})}  &\oplus & A_{2}\overline{A}_{2}[0; 0]_{\Delta_{A}+\half}^{(R_{1}+1,R_{2},R_{3})} \\
& \oplus & A_{1}\overline{B}_{1}[1;0]_{\Delta_{A}+1}^{(R_{1},R_{2},R_{3}+2)}& \oplus & A_{2}\overline{B}_{1}[0;0]_{\Delta_{A}+{3 \over 2}}^{(R_{1}+1,R_{2},R_{3}+2)}~, \\
L\overline{L}[j=0; \overline{j}=1]_{\Delta}^{(R_{1},R_{2},R_{3})} & \rightarrow &A_{2}\overline{A}_{1}[0; 1]_{\Delta_{A}}^{(R_{1},R_{2},R_{3})}  &\oplus & B_{1}\overline{A}_{1}[0; 1]_{\Delta_{A}+1}^{(R_{1}+2,R_{2},R_{3})}\\
 & \oplus & A_{2}\overline{A}_{2}[0;0]_{\Delta_{A}+\half}^{(R_{1},R_{2},R_{3}+1)} & \oplus & B_{1}\overline{A}_{2}[0;0]_{\Delta_{A}+{3 \over 2}}^{(R_{1}+2,R_{2},R_{3}+1)}~,\\
L\overline{L}[j=0; \overline{j}=0]_{\Delta}^{(R_{1},R_{2},R_{3})} & \rightarrow &A_{2}\overline{A}_{2}[0; 0]_{\Delta_{A}}^{(R_{1},R_{2},R_{3})}  &\oplus & B_{1}\overline{A}_{2}[0; 0]_{\Delta_{A}+1}^{(R_{1}+2,R_{2},R_{3})} \\
& \oplus & A_{2}\overline{B}_{1}[0;0]_{\Delta_{A}+1}^{(R_{1},R_{2},R_{3}+2)}& \oplus & B_{1}\overline{B}_{1}[0;0]_{\Delta_{A}+2}^{(R_{1}+2,R_{2},R_{3}+2)}~.
\end{array}
\label{4dn4recomb}
\end{equation}
\noindent It follows from the recombination rules~\eqref{d4n4rr1}, \eqref{d4n4rr2}, and~\eqref{4dn4recomb} that the following multiplets are absolutely protected:
\begin{equation}
B_{1}\overline{B}_{1}[0;0]^{(R_{1}, R_{2}, R_3)}\quad (R_1 = R_3 = 0,1)~.
\end{equation}
Multiplets with~$R_1 = R_3 =  0$ are~$\half$-BPS; an example is the stress tensor multiplet in~\eqref{d4n4stm}. 

Conserved current multiplets in~$d = 4, \CN=4$ theories are studied in section~\ref{sec:d4n4curr}.

\subsection{Five Dimensions}
\label{sec:5dusm}  

Here we list all unitary multiplets of five-dimensional SCFTs. The unique superconformal algebra in~$d = 5$ is the exceptional superalgebra~$\frak{f}(4)$, which corresponds to~$\CN=1$ supersymmetry (see~\cite{Minwalla:1997ka,Bhattacharya:2008zy} and references therein for additional details). 

The Lorentz algebra is~$\frak{so}(5) = \frak{sp}(4)$ and the~$R$-symmetry is~$\frak{su}(2)_R$. Lorentz representations are denoted by~$\frak{sp}(4)$ Dynkin labels~$j_1, j_2 \in \Z_{\geq 0}$, e.g.~$[1,0]$ and~$[0,1]$ are the spinor~$\bf 4$ and the vector~$\bf 5$ representations of~$\frak{so}(5)$. The~$R$-charges are denoted by~$(R)$, where~$R \in \Z_{\geq 0}$ is an~$\frak{su}(2)_R$ Dynkin label. The quantum numbers of an operator with scaling dimension~$\Delta$ are indicated as follows,
\begin{equation}\label{5dlabeling}
[j_1, j_2]^{(R)}_\Delta~, \qquad j_1, j_2, R \in \Z_{\geq 0}~.
\end{equation} 
The~$Q$-supersymmetries transform as \begin{equation}
Q \in [1,0]_{\half}^{(1)}~, \qquad N_Q = 8~.
\end{equation}
The superconformal unitarity bounds and shortening conditions are summarized in table~\ref{tab:5Dmult}.  

\renewcommand{\arraystretch}{1.7}
\renewcommand\tabcolsep{10pt}
\begin{table}[H]
  \centering
  \begin{tabular}{ |c|lr| l|l| }
\hline
{\bf Name} &  \multicolumn{2}{c}{\bf Primary} &  \multicolumn{1}{|c|}{\bf Unitarity Bound} & \multicolumn{1}{c|}{\bf Null State } \\
\hline
\hline
$L$ & $[j_{1},j_{2} ]_{\Delta}^{(R)}$&  &$\Delta>j_{1}+j_{2}+\frac{3}{2}\,R+4$ & \multicolumn{1}{c|}{$-$} \\
\hline
\hline
$A_{1}$ &$[j_{1}, j_{2}]_{\Delta}^{(R)}~,$& $j_{1}\geq1$ &$\Delta=j_{1}+j_{2}+\frac{3}{2}\,R+4$ & $[j_{1}-1,j_{2}]_{\Delta+1/2}^{(R+1)}$ \\
\hline 
$A_{2}$ & $[0,j_{2}]_{\Delta}^{(R)}~,$& $j_{2}\geq1$ &$\Delta=j_{2}+\frac{3}{2}\,R+4$ & $[0,j_{2}-1]_{\Delta+1}^{(R+2)}$ \\
\hline
$A_{4}$ & $[0,0]_{\Delta}^{(R)}$& $$ &$\Delta=\frac{3}{2}\,R+4$ & $[0,0]_{\Delta+2}^{(R+4)}$ \\
\hline
\hline
$B_{1}$ & $[0,j_{2}]_{\Delta}^{(R)}~$& $j_2\geq 1$&$\Delta=j_{2}+\frac{3}{2}\,R+3$ & $[1,j_{2}-1]_{\Delta+1/2}^{(R+1)}$ \\
\hline
$B_{2}$ & $[0,0]_{\Delta}^{(R)}$& &$\Delta=\frac{3}{2}\,R+3$ & $[0,0]_{\Delta+1}^{(R+2)}$ \\
\hline
\hline
$C_{1}$ & $[0,0]_{\Delta}^{(R)}$& &$\Delta=\frac{3}{2}\,R$ & $[1,0]_{\Delta+1/2}^{(R+1)}$ \\
\hline
\end{tabular}
  \caption{Shortening conditions in five-dimensional~$\CN=1$ SCFTs.}
  \label{tab:5Dmult}
\end{table} 

\smallskip

\noindent In order to state the recombination rules, we define 
\begin{equation}
\Delta_A = j_1 + j_{2}  + \frac{3}{2} R + 4~.
\end{equation}
As~$\Delta \rightarrow \Delta_A^+$, we find that 
\begin{equation}
\begin{array}{rclcl}
  L[j_1\geq 2, j_2]_\Delta^{(R)} &  \longrightarrow &A_1[j_1, j_2]_{\Delta_A}^{(R)} &\oplus & A_1[j_1-1, j_2]_{\Delta_A+\half}^{(R+1)}~,\\
 L[j_1=1, j_2\geq 1]_\Delta^{(R)} &  \longrightarrow &A_1[1, j_2]_{\Delta_A}^{(R)} &\oplus & A_2[0, j_2]_{\Delta_A+\half}^{(R+1)}~,\\
  L[j_1=1, j_2=0]_\Delta^{(R)} &  \longrightarrow &A_1[1, 0]_{\Delta_A}^{(R)} &\oplus & A_4[0, 0]_{\Delta_A+\half}^{(R+1)}~,\\
  L[j_{1}=0, j_2\geq 2]_\Delta^{(R)} &  \longrightarrow &A_2[0, j_2]_{\Delta_A}^{(R)} &\oplus & B_1[0, j_2-1]_{\Delta_A+1}^{(R+2)}~,\\
  L[j_{1}=0, j_2=1]_\Delta^{(R)} &  \longrightarrow &A_2[0, 1]_{\Delta_A}^{(R)} &\oplus & B_2[0, 0]_{\Delta_A+1}^{(R+2)}~,\\
   L[j_{1}=0,j_{2}=0]_\Delta^{(R)} &  \longrightarrow &A_4[0, 0]_{\Delta_A}^{(R)} &\oplus & C_1[0, 0]_{\Delta_A+2}^{(R+4)}~.
\end{array}
\end{equation}
This implies the following list of absolutely protected multiplets:
\begin{equation}\label{d5absprot}
B_{\ell}[0, j_2]^{(R)}~~(R \leq 1)~, \qquad C_{1}[0,0]^{(R)}~~(R \leq 3)~, 
\end{equation}
where~$\ell = 1$ if~$j_2 \geq 1$ and~$\ell = 2$ if~$j_2 = 0$. All~$d = 5, \CN=1$ current multiplets, which are analyzed in section~\ref{sec:5dcurrents}, are examples of such absolutely protected multiplets. 

Generic~$d = 5, \CN=1$ multiplets are tabulated in section~\ref{sec:5dn1tabs}.  

\subsection{Six Dimensions}

\label{sec:6dusm}

In this section we enumerate all unitary multiplets of six-dimensional~$(\CN,0)$~SCFTs with~$\CN=1,2$ (see~\cite{Minwalla:1997ka,Ferrara:2000xg,Dobrev:2002dt,Bhattacharya:2008zy} and references therein for additional details). As discussed in section~\ref{sec:maxsusy}, unitarity~$(\CN, 0)$ SCFTs with~$\CN \geq 3$ do not exist. 

Throughout, representations of the~$\frak{so}(6) = \frak{su}(4)$ Lorentz algebra are denoted using~$\frak{su}(4)$ Dynkin labels,
\begin{equation}\label{Lor6d}
[j_1, j_2, j_3]~, \qquad j_1, j_2, j_3 \in \Z_{\geq 0}~.
\end{equation} 
For instance, $[1,0,0]$ and~$[0,0,1]$ are the left- and right-handed chiral spinor representations~${\bf 4}, {\bf 4'}$ of~$\frak{so}(6)$,\footnote{~As representations of~$\frak{su}(4)$, the~$\bf 4'$ is typically denoted by~$\bf \b 4$, which is related to the~$\bf 4$ by complex conjugation. However, in six-dimensional Minkowski space, chiral spinors are not related by complex conjugation.} while~$[0,1,0]$ is the vector representation~$\bf 6$ of~$\frak{so}(6)$. Operators of scaling dimension~$\Delta$ are denoted by~$[j_1,j_2,j_3]_\Delta$.

\subsubsection{$d=6$,~$\CN=(1,0)$}

\label{sec:d6n1defs}

The~$\CN=(1,0)$ superconformal algebra is~$\frak{osp}(8|2)$, with~$R$-symmetry~$\frak{sp}(2)_R \simeq \frak{su}(2)_R$. Its representations are denoted by~$(R)$, where~$R \in \Z_{\geq 0}$ is an~$\frak{su}(2)_R$ Dynkin label. The~$Q$-supersymmetries transform as
 \begin{equation}
Q \in [1,0,0]_{\half}^{(1)}~~, \qquad N_Q = 8~.
\end{equation}
The superconformal unitarity bounds and shortening conditions are summarized in table~\ref{tab:6DN1}. 

\renewcommand{\arraystretch}{1.7}
\renewcommand{\tabcolsep}{8pt}
\begin{table}[H]
  \centering
  \begin{tabular}{ |c|lr| l|l| }
\hline
{\bf Name} &  \multicolumn{2}{c}{\bf Primary} &  \multicolumn{1}{|c|}{\bf Unitarity Bound} & \multicolumn{1}{c|}{\bf Null State } \\
\hline
\hline
$L$ & $[j_1,j_2,j_3]_{\Delta}^{(R)}$&  &$\Delta>\frac{1}{2}\left(j_1+2j_2+3j_3\right)+2R+6$ & \multicolumn{1}{c|}{$-$} \\
\hline
\hline
$A_{1}$ &$[j_1,j_2,j_3]_{\Delta}^{(R)}~,$& $j_3\geq1$ &$\Delta=\frac{1}{2}\left(j_1+2j_2+3j_3\right)+2R+6$ & $[j_1,j_2,j_3-1]_{\Delta+1/2}^{(R+1)}$ \\
\hline 
$A_{2}$ &$[j_1,j_2,0]_{\Delta}^{(R)}~,$& $j_2\geq1$ &$\Delta=\frac{1}{2}\left(j_1+2j_2\right)+2R+6$ & $[j_1,j_2-1,0]_{\Delta+1}^{(R+2)}$ \\
\hline
$A_{3}$ &$[j_1,0,0]_{\Delta}^{(R)}~,$& $j_1\geq1$ &$\Delta=\frac{1}{2}\, j_1+2R+6$ & $[j_1-1,0,0]_{\Delta+3/2}^{(R+3)}$ \\
\hline
$A_{4}$ &$[0,0,0]_{\Delta}^{(R)}$& $$ &$\Delta=2R+6$ & $[0,0,0]_{\Delta+2}^{(R+4)}$ \\
\hline
\hline
$B_{1}$ &$[j_1,j_2,0]_{\Delta}^{(R)}~,$& $j_2\geq1$ &$\Delta=\frac{1}{2}\left(j_1+2j_2\right)+2R+4$ & $[j_1,j_2-1,1]_{\Delta+1/2}^{(R+1)}$ \\
\hline
$B_{2}$ &$[j_1,0,0]_{\Delta}^{(R)}~,$& $j_1\geq1$ &$\Delta=\frac{1}{2}\, j_1+2R+4$ & $[j_1-1,0,1]_{\Delta+1}^{(R+2)}$ \\
\hline
$B_{3}$ &$[0,0,0]_{\Delta}^{(R)}$& $$ &$\Delta=2R+4$ & $[0,0,1]_{\Delta+3/2}^{(R+3)}$ \\
\hline
\hline
$C_{1}$ &$[j_1,0,0]_{\Delta}^{(R)}~,$& $j_1 \geq 1$ &$\Delta=\frac{1}{2} \, j_1+2R+2$ & $[j_1-1,1,0]_{\Delta+1/2}^{(R+1)}$ \\
\hline
$C_{2}$ &$[0,0,0]_{\Delta}^{(R)}$& $$ &$\Delta=2R+2$ & $[0,1,0]_{\Delta+1}^{(R+2)}$ \\
\hline
\hline
$D_{1}$ &$[0,0,0]_{\Delta}^{(R)}$& $$ &$\Delta=2R$ & $[1,0,0]_{\Delta+1/2}^{(R+1)}$ \\
\hline
\end{tabular}
  \caption{Shortening conditions in six-dimensional~$\CN=(1,0)$ theories.}
  \label{tab:6DN1}
\end{table} 

\noindent If we define
\begin{equation}
\Delta_A = \half \left(j_1 + 2 j_2 + 3 j_3 \right) + 2 R + 6~,
\end{equation}
then the following recombination rules apply as~$\Delta \rightarrow \Delta_A^+$:
\begin{equation}
\begin{array}{rclcl}
  L[j_1, j_2, j_3 \geq 2]_\Delta^{(R)} &  \longrightarrow &A_1[j_1, j_2, j_3]_{\Delta_A}^{(R)} &\oplus & A_1[j_1, j_2, j_3-1]_{\Delta_A+\half}^{(R+1)}~,\\
 L[j_1, j_2 \geq 1, j_3  = 1]_\Delta^{(R)} & \longrightarrow &A_1[j_1, j_2, 1]_{\Delta_A}^{(R)} & \oplus & A_2[j_1, j_2, 0]_{\Delta_A+\half}^{(R+1)}~,\\
  L[j_1 \geq 1, j_2 = 0, j_3  = 1]_\Delta^{(R)} & \longrightarrow &A_1[j_1, 0, 1]_{\Delta_A}^{(R)} &\oplus & A_3[j_1, 0, 0]_{\Delta_A+\half}^{(R+1)}~,\\
  L[j_1 =0, j_2 = 0, j_3  = 1]_\Delta^{(R)} & \longrightarrow & A_1[0, 0, 1]_{\Delta_A}^{(R)} &\oplus & A_4[0, 0, 0]_{\Delta_A+\half}^{(R+1)}~,\\
  L[j_1, j_2 \geq 2, j_3  = 0]_\Delta^{(R)} & \longrightarrow & A_2[j_1, j_2, 0]_{\Delta_A}^{(R)} &\oplus & B_1[j_1, j_2-1, 0]_{\Delta_A+1}^{(R+2)}~,\\
 L[j_1 \geq 1, j_2 = 1, j_3  = 0]_\Delta^{(R)} & \longrightarrow & A_2[j_1, 1, 0]_{\Delta_A}^{(R)} &\oplus & B_2[j_1, 0, 0]_{\Delta_A+1}^{(R+2)}~,\\
L[j_1 = 0, j_2 = 1, j_3  = 0]_\Delta^{(R)} & \longrightarrow & A_2[0, 1, 0]_{\Delta_A}^{(R)} &\oplus & B_3[0, 0, 0]_{\Delta_A+1}^{(R+2)}~,\\
L[j_1 \geq 2, j_2 = 0, j_3  = 0]_\Delta^{(R)} & \longrightarrow & A_3[j_1, 0, 0]_{\Delta_A}^{(R)} &\oplus & C_1[j_1-1, 0, 0]_{\Delta_A+{3 \over 2}}^{(R+3)}~,\\
 L[j_1 = 1, j_2 = 0, j_3  = 0]_\Delta^{(R)} & \longrightarrow & A_3[1, 0, 0]_{\Delta_A}^{(R)} &\oplus & C_2[0, 0, 0]_{\Delta_A+{3 \over 2}}^{(R+3)}~,\\
  L[j_1 = 0, j_2 = 0, j_3  = 0]_\Delta^{(R)} & \longrightarrow & A_4[0, 0, 0]_{\Delta_A}^{(R)} &\oplus & D_1[0, 0, 0]_{\Delta_A+2}^{(R+4)}~.
\end{array}
\end{equation}
This leads to the following list of absolutely protected multiplets:
\begin{equation}
B_{\ell}[j_1, j_2, 0]^{(R)}~~(R \leq 1)~, \qquad C_{\ell}[j_1, 0,0]^{(R)}~~(R \leq 2)~, \qquad D_1[0,0,0]^{(R)}~~(R \leq 3)~.
\end{equation}
Here the value of~$\ell$ depends on~$j_1, j_2$ (see table~\ref{tab:6DN1}). All~$d = 6, \CN=(1,0)$ current multiplets, which are discussed in section~\ref{sec:d6n1curr}, are examples of such absolutely protected multiplets. 

Generic~$d = 6, \CN=(1,0)$ multiplets are tabulated in section~\ref{sec:6dn1tabs}.

\subsubsection{$d=6$,~$\CN=(2,0)$}

\label{sec:d6n2defs}

The~$\CN=(2,0)$ superconformal algebra is~$\frak{osp}(8|4)$, so that the~$R$-symmetry is~$\frak{sp}(4)_R$. Its representations are denoted by~$(R_1,R_2)$, where~$R_1,R_2 \in \Z_{\geq 0}$ are~$\frak{sp}(4)_R$ Dynkin labels, e.g.~$(1,0)$ and~$(0,1)$ denote the~$\bf 4$ and~$\bf 5$, respectively. The~$Q$-supersymmetries transform as 
 \begin{equation}
Q \in [1,0,0]_{\half}^{(1,0)}~~, \qquad N_Q = 16~.
\end{equation}
The superconformal unitarity bounds and shortening conditions are summarized in table~\ref{tab:6DN2}. 

\renewcommand{\arraystretch}{1.7}
\renewcommand\tabcolsep{3.5pt}
\begin{table}[H]
  \centering
  \begin{tabular}{ |c|lr| l|l| }
\hline
{\bf Name} &  \multicolumn{2}{c}{\bf Primary} &  \multicolumn{1}{|c|}{\bf Unitarity Bound} & \multicolumn{1}{c|}{\bf Null State } \\
\hline
\hline
$L$ & $[j_1,j_2,j_3]_{\Delta}^{(R_1,R_2)}$&  &$\Delta>\frac{1}{2}\left(j_1+2j_2+3j_3\right)+2(R_1+R_2)+6$ & \multicolumn{1}{c|}{$-$} \\
\hline
\hline
$A_{1}$ &$[j_1,j_2,j_3]_{\Delta}^{(R_1,R_2)}~,$& $j_3\geq1$ &$\Delta=\frac{1}{2}\left(j_1+2j_2+3j_3\right)+2(R_1+R_2)+6$ & $[j_1,j_2,j_3-1]_{\Delta+1/2}^{(R_1+1,R_2)}$ \\
\hline 
$A_{2}$ &$[j_1,j_2,0]_{\Delta}^{(R_1,R_2)}~,$& $j_2\geq1$ &$\Delta=\frac{1}{2}\left(j_1+2j_2\right)+2(R_1+R_2)+6$ & $[j_1,j_2-1,0]_{\Delta+1}^{(R_1+2,R_2)}$ \\
\hline
$A_{3}$ &$[j_1,0,0]_{\Delta}^{(R_1,R_2)}~,$& $j_1\geq1$ &$\Delta=\frac{1}{2}\, j_1+2(R_1+R_2)+6$ & $[j_1-1,0,0]_{\Delta+3/2}^{(R_1+3,R_2)}$ \\
\hline
$A_{4}$ &$[0,0,0]_{\Delta}^{(R_1,R_2)}$& $$ &$\Delta=2(R_1+R_2)+6$ & $[0,0,0]_{\Delta+2}^{(R_1+4,R_2)}$ \\
\hline
\hline
$B_{1}$ &$[j_1,j_2,0]_{\Delta}^{(R_1,R_2)}~,$& $j_2\geq1$ &$\Delta=\frac{1}{2}\left(j_1+2j_2\right)+2(R_1+R_2)+4$ & $[j_1,j_2-1,1]_{\Delta+1/2}^{(R_1+1,R_2)}$ \\
\hline
$B_{2}$ &$[j_1,0,0]_{\Delta}^{(R_1,R_2)}~,$& $j_1\geq1$ &$\Delta=\frac{1}{2}\, j_1+2(R_1+R_2)+4$ & $[j_1-1,0,1]_{\Delta+1}^{(R_1+2,R_2)}$ \\
\hline
$B_{3}$ &$[0,0,0]_{\Delta}^{(R_1,R_2)}$& $$ &$\Delta=2(R_1+R_2)+4$ & $[0,0,1]_{\Delta+3/2}^{(R_1+3,R_2)}$ \\
\hline
\hline
$C_{1}$ &$[j_1,0,0]_{\Delta}^{(R_1,R_2)}~,$& $j_1 \geq 1$ &$\Delta=\frac{1}{2} \, j_1+2(R_1+R_2)+2$ & $[j_1-1,1,0]_{\Delta+1/2}^{(R_1+1,R_2)}$ \\
\hline
$C_{2}$ &$[0,0,0]_{\Delta}^{(R_1,R_2)}$& $$ &$\Delta=2(R_1+R_2)+2$ & $[0,1,0]_{\Delta+1}^{(R_1+2,R_2)}$ \\
\hline
\hline
$D_{1}$ &$[0,0,0]_{\Delta}^{(R_1,R_2)}$& $$ &$\Delta=2(R_1+R_2)$ & $[1,0,0]_{\Delta+1/2}^{(R_1+1,R_2)}$ \\
\hline
\end{tabular}
  \caption{Shortening conditions in six-dimensional~$\CN=(2,0)$ theories.}
  \label{tab:6DN2}
\end{table} 

\noindent In order to state the recombination rules, we define 
\begin{equation}
\Delta_A = \half \left(j_1 + 2 j_2 + 3 j_3 \right) + 2 (R_{1}+R_{2}) + 6~.
\end{equation}
As~$\Delta \rightarrow \Delta_A^+$, we find that 
\begin{equation}
\begin{array}{rclcl}
  L[j_1, j_2, j_3 \geq 2]_\Delta^{(R_{1},R_{2})} &  \longrightarrow &A_1[j_1, j_2, j_3]_{\Delta_A}^{(R_{1},R_{2})} &\oplus & A_1[j_1, j_2, j_3-1]_{\Delta_A+\half}^{(R_{1}+1,R_{2})}~,\\
 L[j_1, j_2 \geq 1, j_3  = 1]_\Delta^{(R_{1},R_{2})} & \longrightarrow &A_1[j_1, j_2, 1]_{\Delta_A}^{(R_{1},R_{2})} & \oplus & A_2[j_1, j_2, 0]_{\Delta_A+\half}^{(R_{1}+1,R_{2})}~,\\
  L[j_1 \geq 1, j_2 = 0, j_3  = 1]_\Delta^{(R_{1},R_{2})} & \longrightarrow &A_1[j_1, 0, 1]_{\Delta_A}^{(R_{1},R_{2})} &\oplus & A_3[j_1, 0, 0]_{\Delta_A+\half}^{(R_{1}+1,R_{2})}~,\\
  L[j_1 =0, j_2 = 0, j_3  = 1]_\Delta^{(R_{1},R_{2})} & \longrightarrow & A_1[0, 0, 1]_{\Delta_A}^{(R_{1},R_{2})} &\oplus & A_4[0, 0, 0]_{\Delta_A+\half}^{(R_{1}+1,R_{2})}~,\\
  L[j_1, j_2 \geq 2, j_3  = 0]_\Delta^{(R_{1},R_{2})} & \longrightarrow & A_2[j_1, j_2, 0]_{\Delta_A}^{(R_{1},R_{2})} &\oplus & B_1[j_1, j_2-1, 0]_{\Delta_A+1}^{(R_{1}+2,R_{2})}~,\\
 L[j_1 \geq 1, j_2 = 1, j_3  = 0]_\Delta^{(R_{1},R_{2})} & \longrightarrow & A_2[j_1, 1, 0]_{\Delta_A}^{(R_{1},R_{2})} &\oplus & B_2[j_1, 0, 0]_{\Delta_A+1}^{(R_{1}+2,R_{2})}~,\\
L[j_1 = 0, j_2 = 1, j_3  = 0]_\Delta^{(R_{1},R_{2})} & \longrightarrow & A_2[0, 1, 0]_{\Delta_A}^{(R_{1},R_{2})} &\oplus & B_3[0, 0, 0]_{\Delta_A+1}^{(R_{1}+2,R_{2})}~,\\
L[j_1 \geq 2, j_2 = 0, j_3  = 0]_\Delta^{(R_{1},R_{2})} & \longrightarrow & A_3[j_1, 0, 0]_{\Delta_A}^{(R_{1},R_{2})} &\oplus & C_1[j_1-1, 0, 0]_{\Delta_A+{3\over 2}}^{(R_{1}+3,R_{2})}~,\\
 L[j_1 = 1, j_2 = 0, j_3  = 0]_\Delta^{(R_{1},R_{2})} & \longrightarrow & A_3[1, 0, 0]_{\Delta_A}^{(R_{1},R_{2})} &\oplus & C_2[0, 0, 0]_{\Delta_A+{3\over 2}}^{(R_{1}+3,R_{2})}~,\\
  L[j_1 = 0, j_2 = 0, j_3  = 0]_\Delta^{(R_{1},R_{2})} & \longrightarrow & A_4[0, 0, 0]_{\Delta_A}^{(R_{1},R_{2})} &\oplus & D_1[0, 0, 0]_{\Delta_A+2}^{(R_{1}+4,R_{2})}~.
\end{array}
\end{equation}
This leads to the following list of absolutely protected multiplets:
\begin{equation}
B_{\ell}[j_1, j_2, 0]^{(R_1, R_2)}~(R_1 \leq 1)~, ~~C_{\ell}[j_1, 0,0]^{(R_1, R_2)}~(R_1 \leq 2)~, ~~D_1[0,0,0]^{(R_1, R_2)}~(R_1 \leq 3)~.
\end{equation}
Here the value of~$\ell$ depends on~$j_1, j_2$ (see table~\ref{tab:6DN2}). All~$d = 6, \CN=(2,0)$ current multiplets, which are analyzed in section~\ref{sec:d6n2curr}, are examples of such absolutely protected multiplets.

\section{Algorithms for Constructing Superconformal Multiplets} 

\label{sec:alg}

In this section we motivate, and precisely formulate, the algorithm that we will use to generate all unitary superconformal multiplets in~$d \geq 3$ dimensions. As we will explain, the algorithm is conjectural but passes a variety of non-trivial consistency checks.  Along the way, we compare and contrast our proposal with several existing approaches in the literature. 

\subsection{Input and Output}

\label{sec:inoutalg}

The algorithm takes the following data as input:
\begin{itemize}
\item[1.)] A superconformal algebra~$\frak S(d,\CN)$ chosen from the list in~\eqref{scftalgs}. Its bosonic subalgebra consists of the conformal algebra~$\frak{so}(d,2)$, which itself contains the~$\frak{so}(d)$ Lorentz algebra, and the compact~$R$-symmetry algebra~$\frak R$. We denote the Lorentz and~$R$-symmetry representation of the~$Q$-supercharges by~$\CR_Q$. In~$d = 3, \,\CN=2$ and in~$d = 4$ this representation is reducible and splits into a direct sum~$\CR_Q \oplus \CR_{\b Q}$ corresponding to left~$Q$-supercharges and right~$\b Q$-supercharges. 

\item[2.)] A unitary superconformal multiplet~$\CM$ of the algebra~$\frak S(d, \CN)$. As reviewed in sections~\ref{sec:introscft} and~\ref{sec:usm}, the structure of~$\CM$ is completely determined by the Lorentz and~$R$-symmetry representations~$L_\CV$ and~$R_\CV$, as well as the scaling dimension~$\Delta_\CV$, of its SCP~$\CV$, 
\begin{equation}\label{primhwdef}
\CV = \big[L_\CV\big]_{\Delta_\CV}^{(R_\CV)}~.
\end{equation}
\noindent The restrictions of unitarity on~$\CV$ were summarized in section~\ref{sec:usm}. We continue to use a labeling scheme for~$\CM$ that conveys some information about its null states, e.g.~$L$ denotes a long multiplet without null states, while~$A_\ell$ is an~$A$-type short multiplet whose primary null state resides at level~$\ell$. We therefore write
\begin{equation}\label{mlabel}
\CM = X_\ell[L_\CV]_{\Delta_\CV}^{(R_\CV)}~, \qquad X \in \{L, A, B, C, D\}~,
\end{equation}
where~$\ell$ denotes the level of the primary null representation~$N$ within~$\CM$.\footnote{~Long multiplets do not possess null states and we will omit the subscript~$\ell$ when~$X = L$.} As discussed around~\eqref{aelldef} and~\eqref{xelldef}, as well as in section~\ref{sec:usm}, the allowed values of~$\ell$ and~$X$ depend on the spacetime dimension~$d$ and the amount of supersymmetry~$\CN$. 

\medskip

In cases with left~$Q$- and right~$\b Q$-supercharges, i.e.~$d = 3,\, \CN =2$ or~$d = 4$, multiplets can shorten both from the left and from the right, so that so that~\eqref{mlabel} is modified to 
\begin{equation}
\CM = X_\ell \b X_{\b \ell} [L_\CV]_{\Delta_\CV}^{(R_\CV)}~, \qquad X \in \{L, A, B\}~, \qquad \b X \in \{ \b L, \b A, \b B\}~.
\end{equation}
Note that only~$A$- and~$B$-type short multiplets (and their barred counterparts) can arise in the two-sided case, see sections~\ref{sec:3dn2def} and~\ref{sec:4dmults} for a detailed discussion. Now~$\ell$ and~$\b \ell$ denote the levels of the left and right primary null representations~$N$ and~$\b N$ within~$\CM$. 

\end{itemize}

\bigskip

The output of the algorithm consists of the operator content of the superconformal multiplet~$\CM$. As explained in section~\ref{sec:introcsm}, it suffices to enumerate the Lorentz and~$R$-symmetry representations~$L_a, R_a$ and scaling dimensions~$\Delta_a$ of the CPs~$\CC^{(\ell)}_a$ that appear in the decomposition of~$\CM$ into a finite number of~$\frak{so}(d, 2) \times \frak R$ multiplets. (Here~$\ell$ denotes the level of the CP within~$\CM$.) This decomposition takes the form~\eqref{basicdecompii}, which we repeat here
\begin{equation}\label{basicdecompiii}
\CM = \bigoplus_{a} \, \CC^{(\ell)}_{a}~~(\text{finite sum})~, \qquad \CC^{(\ell)}_{a} = [L_a]^{(R_a)}_{\Delta_a}~, \qquad \Delta_a = \Delta_\CV + {\ell \over 2}~.
\end{equation}
By focusing on the CPs, we effectively set all spacetime derivatives to zero, so that the~$Q$-supercharges anticommute as in~\eqref{anticommqs},
\begin{equation}\label{anticommqsii}
\big\{Q_i, Q_j\big\} = 0~, \qquad i, j = 1, \ldots, N_Q = \dim \CR_Q~.
\end{equation}
More generally, there can be~$N_Q = \dim \CR_Q$ left supercharges~$Q_i$ and~$N_{\b Q} = \dim \CR_{\b Q}$ right supercharges~$\b Q_{\b i}$. In that case we can distinguish left and right levels~$\ell, \b \ell$ as well as the total level~$\ell_\text{tot} = \ell + \b \ell$. Fermi statistics imply that CPs can only occur at levels~$0 \leq \ell \leq N_Q$ and~$0 \leq \b \ell \leq N_{\b Q}$. In what follows, we will always first discuss situations with a single irreducible~$\CR_Q$ before pointing out the necessary modifications in the two-sided cases.

\subsection{Long Multiplets: Clebsch-Gordon and Racah-Speiser Algorithms}

\label{sec:rslm}

Long multiplets do not possess null states, and hence the action of the~$Q$-supercharges on the superconformal primary~$\CV$ is only restricted by Fermi statistics~\eqref{anticommqsii}. Products of~$\ell$ supercharges transform in the totally antisymmetric wedge power~$\wedge^\ell \CR_Q$, with~$1 \leq \ell \leq N_Q$. When acting on~$\CV$, they produce the following, generally reducible, representation, 
\begin{equation}\label{wedgerqv}
\wedge^\ell \CR_Q \otimes \CV~.
\end{equation} 
At every level~$\ell$, the components~$\CC_a^{(\ell)}$ appearing in~\eqref{basicdecompiii} can therefore be obtained by decomposing~\eqref{wedgerqv} into irreducible Lorentz and~$R$-symmetry representations. 

This is a well-posed group theory problem. As explained in section~\ref{sec:introcsm}, it can in principle be solved by applying the Clebsch-Gordon algorithm to decompose the space of CPs at a given level~$\ell$ into irreducible representations of the Lorentz and~$R$-symmetry. This space is spanned by all independent monomials of the schematic form~$Q^\ell \CV$, i.e.~a product of~$\ell$~distinct $Q$-supercharges acting on some state in the SCP representation.\footnote{~More precisely, the CPs take this form up to some CDs, i.e.~derivatives, of lower-level CPs, but as explained around~\eqref{anticommqsii} we are not keeping track of such terms.} The Clebsch-Gordon states are explicit, typically complicated, linear combinations of such monomials. The need to identify and keep track of them makes the Clebsch-Gordan approach computationally costly. 

A much more efficient method for decomposing the states~$\wedge^\ell \CR_Q \otimes \CV$ at level~$\ell$ into irreducible representations of the Lorentz and~$R$-symmetry utilizes the Racah-Speiser (RS) algorithm for decomposing tensor product representations of simple Lie algebras. This algorithm can be derived from the Weyl character formula; see~\cite{Fuchs:1997jv} for a textbook discussion, and~\cite{Dolan:2002zh,Kinney:2005ej,Bhattacharya:2008zy} for previous applications to superconformal multiplets. A short review can be found in appendix~\ref{app:RSalg}. Here we will explain the essential ingredients in the context of long multiplets (the adaptation to short multiplets will be discussed in section~\ref{rsshortmult} below).  As was already briefly explained around~\eqref{su2dec}, the RS algorithm offers two significant simplifications over the Clebsch-Gordon algorithm:
\begin{itemize}
\item It only utilizes the highest-weight state~$\CV_\text{h.w.} = \big[L_\CV\big]_{\Delta_\CV}^{(R_\CV)} \in \CV$ of the superconformal primary representation~$\CV$, rather than all states in~$\CV$. 
\item It suffices to consider simple, monomial states of the form~$Q_{i_1} Q_{i_2} \cdots Q_{i_\ell} \CV_\text{h.w.}$, rather than the more complicated Clebsch-Gordon states.
\end{itemize}
In general, the state~$Q_{i_1} Q_{i_2} \cdots Q_{i_\ell} \CV_\text{h.w.}$ is not a highest-weight state of any Lorentz or~$R$-symmetry representation. Nevertheless, the weights of this state (obtained by adding the weights of~$Q_{i_1}$ through~$Q_{i_\ell}$ to those of~$\CV_\text{h.w.}$) can generically be interpreted as highest weights of irreducible Lorentz and~$R$-symmetry representations. We will refer to the weights~$\CW_\text{RS}^{(\ell)}$ generated by acting with all sequences of~$\ell$ supercharges, distinct up to rearrangements using the anticommutators in~\eqref{anticommqsii}, as the RS trial weights at level~$\ell$. The RS algorithm guarantees that the trial weights~$\CW^{(\ell)}_\text{RS}$ coincide with the highest weights of the irreducible representations occurring in~\eqref{wedgerqv}, as long as the highest weights~$\CV_\text{h.w.}$ of the superconformal primary (more precisely, all of its Dynkin labels) are sufficiently large. Roughly, this is because the RS trial state~$Q_{i_1} Q_{i_2} \cdots Q_{i_\ell} \CV_\text{h.w.}$ is -- to a certain, limited extent -- a proxy for the true highest-weight state with the same quantum numbers as long as the representation~$\CV$ is sufficiently large. 

The RS algorithm also applies when the representation~$\CV$ is small, or non-generic, even though the RS trial states~$Q_{i_1} Q_{i_2} \cdots Q_{i_\ell} \CV_\text{h.w.}$ are no longer good proxies for true highest-weight states. As we will see below, this is the regime in which some of the most interesting, sporadic aspects of superconformal representation theory arise. When the representation~$\CV$ is too small, it may happen that a trial weight~$w \in \CW^{(\ell)}_\text{RS}$ can no longer be interpreted as a highest weight, because some of its Dynkin labels are negative. In this case the RS algorithm specifies that~$w$ should be removed from~$\CW^{(\ell)}_\text{RS}$, perhaps to be replaced by a different weight~$\t w$ with non-negative Dynkin labels. The precise group-theoretic rules that determine the map~$w \rightarrow \t w$ are reviewed in appendix~\ref{app:RSalg}; some examples appear below. The map~$w \rightarrow \t w$ also determines an overall sign factor~$\sigma = \pm 1, 0$. If~$\sigma = +1$, then~$\t w$ is simply added to~$\CW_\text{RS}^{(\ell)}$, while~$\sigma = -1$ indicates that~$\t w$ cancels against another weight in~$\CW^{(\ell)}_\text{RS}$ with coefficient~$+1$ and the same Dynkin labels. Finally, if~$\sigma =  0$ then the original trial weight~$w$ should be removed from~$\CW_\text{RS}^{(\ell)}$ without adding~$\t w$. The RS algorithm guarantees that, once all cancellations have been carried out, the remaining trial weights in~$\CW_\text{RS}^{(\ell)}$ have positive multiplicity and correctly capture the decomposition of~\eqref{wedgerqv} into irreducible Lorentz and~$R$-symmetry representations.  

We will now illustrate this procedure by constructing the first few levels of a long multiplet in three-dimensional~$\CN=3$ theories. As discussed in section \ref{sec:3dn3}, the~$R$-symmetry is~$\frak{R} = \frak{su}(2)$ and the Lorentz algebra is~$\frak{su}(2)$. The supercharges~$Q_\alpha^{(ij)}$ (with~$\alpha, i, j = \pm$) transform as a Lorentz doublet and an~$R$-symmetry triplet, so that~$\CR_Q = [1]^{(2)}_{1/2}$. We will write the quantum numbers of the SCP as~$\CV = [j]^{(R)}_\Delta$, with~$j, R \in \Z_{\geq 0}$. The scaling dimension~$\Delta$ must satisfy the unitarity bound~$\Delta > \half j + \half R + 1$ for a long multiplet. Then, the multiplet we construct is denoted by
\begin{equation}\label{tdn3longmult}
\CM = L[j]_\Delta^{(R)}~.
\end{equation}
The weights of the~$N_Q = \dim \CR_Q = 6$ individual supercharges in~$\CR_Q$ are
\begin{equation}
Q_\pm^{++} = [\pm 1]^{(2)}~, \qquad Q_\pm^{+-} = [\pm 1]^{(0)}~, \qquad Q_\pm^{--} = [\pm 1]^{(-2)}~.
\end{equation}
The RS trial states and their weights at levels~$\ell = 0,1,2$ are given by
\begin{align}\label{tdnthreeex}
\ell = 0: & \qquad \CV_\text{h.w.}~, \qquad \CW^{(0)}_\text{RS} = \left\{[j]_\Delta^{(R)} \right\}~,\cr
\ell = 1: & \qquad Q_\pm^{++}\CV_\text{h.w.}~,~Q_\pm^{+-}\CV_\text{h.w.}~,~Q_\pm^{--}\CV_\text{h.w.}~,\qquad  \CW^{(1)}_\text{RS} = \left\{[j\pm1]_{\Delta+1/2}^{(R+2)\oplus (R) \oplus (R-2)}\right\}~, \cr
\ell = 2: & \qquad Q^{++}_{\pm}Q^{+-}_{\pm} \CV_\text{h.w.}~,~Q^{++}_{\pm}Q^{--}_{\pm} \CV_\text{h.w.}~,~Q^{+-}_{\pm}Q^{--}_{\pm} \CV_\text{h.w.}~,~Q^{\pm\pm}_{+}Q^{\pm\pm}_{-} \CV_\text{h.w.}~,\cr
& \qquad Q^{\pm\pm}_{+}Q^{+-}_{-} \CV_\text{h.w.} ~,~Q^{\pm\pm}_{+}Q^{\mp\mp}_{-} \CV_\text{h.w.}~,~Q^{+-}_{+}Q^{\pm\pm}_{-} \CV_\text{h.w.}~,~Q^{+-}_{+}Q^{+-}_{-} \CV_\text{h.w.}\cr
& \qquad \CW^{(2)}_\text{RS} = \left\{[j\pm2]_{\Delta+1}^{(R+2) \oplus (R) \oplus (R-2)}~, \quad [j]_{\Delta+1}^{(R\pm4) \oplus 2 (R\pm2) \oplus 3(R)} \right\}~.
\end{align}
As long as the Dynkin labels~$j, R$ of the SCP are sufficiently large (more precisely, if~$j \geq 2$ and~$R \geq 4$), all trial weights in~$\CW^{(\ell = 0,1,2)}_\text{RS}$ have non-negative Dynkin labels and correctly specify the decomposition of levels~$\ell=0,1,2$ of the multiplet~$\CM$ into conformal primaries. 

We now examine what happens as we lower the values of~$j$ and~$R$. For instance, if~$j = 1$ but~$R \geq 4$ then the trial weights at levels~$\ell = 0,1$ are not modified but~$\CW^{(2)}_\text{RS}$ now contains~$[-1]_{\Delta+1}^{(R\pm 2) \oplus (R)}$. Since the Lorentz weight~$w = [-1]$ is negative, it must be mapped to a non-negative weight~$\t w$ according to the rules of the RS algorithm. For~$\frak{su}(2)$ representations the map is particularly simple (see appendix~\ref{app:RSalg}),
\begin{equation}\label{rsreflsu2}
w = [-j] \qquad \longrightarrow \qquad \begin{cases} 
 \t w= -[j-2] \quad (\sigma = -1) \quad \text{if} \quad j \in \Z_{\geq 2} \\
\t w\,~\text{removed}~\,~\quad(\sigma = 0) \quad~~ \,\text{if} \quad j = 1
\end{cases}
\end{equation}
Therefore the correct RS trial weights at level~$\ell=2$, for~$j =1$ but generic~$R$, are given by 
\begin{equation}
\CW^{(2)}_\text{RS} = \left\{[3]_{\Delta+1}^{(R+2) \oplus (R) \oplus (R-2)}~, \quad [1]_{\Delta+1}^{(R\pm4) \oplus 2 (R\pm2) \oplus 3(R)} \right\}~.
\end{equation}
If we lower~$j$ further and consider~$j = 0$, the weights~$[-1]^{(R \pm 2) \oplus (R)}$ at level~$\ell=1$ must also be dropped and there are genuine cancellations at level~$\ell=2$,
\begin{align}\label{jzerorstrial}
 \CW^{(1)}_\text{RS} & = \left\{[1]_{\Delta+1/2}^{(R\pm2)\oplus (R)}\right\}~,\cr
 \CW^{(2)}_\text{RS} & = \left\{[2]_{\Delta+1}^{(R\pm 2) \oplus (R)}~, \quad [0]_{\Delta+1}^{(R\pm4) \oplus 2 (R\pm2) \oplus 3(R)}, \quad [-2]_{\Delta+1}^{(R\pm 2) \oplus (R)} = - [0]_{\Delta+1}^{(R\pm 2) \oplus (R)}\right\} \cr
& =  \left\{[2]_{\Delta+1}^{(R\pm 2) \oplus (R)}~, \quad [0]_{\Delta+1}^{(R\pm4) \oplus  (R\pm2) \oplus 2(R)} \right\}
\end{align}
Finally, we can take both~$j$ and~$R$ to be small, e.g.~$j = R = 0$. Now we must apply~\eqref{rsreflsu2} to both negative Lorentz and negative~$R$-symmetry weights, multiplying the resulting~$\sigma$-factors. Performing the ensuing cancellations leads to
\begin{equation}
\CW^{(1)}_\text{RS}  = \left\{[1]_{\Delta+1/2}^{(2)}\right\}~, \qquad \CW^{(2)}_\text{RS}  =  \left\{[2]_{\Delta+1}^{(2)}~, \quad [0]_{\Delta+1}^{(4) \oplus (0))} \right\}~,
\end{equation} 
which correctly captures the operator content of the~$L[j = 0]^{(R = 0)}_\Delta$ multiplet at levels~$\ell = 1,2$. 

It is straightforward to generate long multiplets for any~$d$ and~$\CN$ using the RS algorithm. The appropriate generalization of~\eqref{rsreflsu2} to other Lorentz and~$R$-symmetry algebras is spelled out in appendix~\ref{app:RSalg}.

\subsection{Aspects of Short Multiplets}
\label{rsshortmult}

As was already emphasized in the previous subsection, a key advantage of the RS algorithm is that it only uses simple trial states of the form~$Q_{i_1} \cdots Q_{i_\ell} \CV_\text{h.w.}$. If we denote all states in the SCP representation by~$\CV = \big\{\CV_A\big\}_{A = 1}^{\dim \CV}$ with~$\CV_1 = \CV_\text{h.w.}$ the highest-weight state, then only~$\CV_1$ appears in the RS algorithm. By contrast, the Clebsch-Gordon procedure involves all states in~$\CV$. The drawback of working with RS trial states is that we lose some information: even though they can be thought of as proxies for true highest-weight states for some purposes, they do not capture all properties of the latter. Crucially, the RS trial states may fail to correctly indicate the action of the~$Q$-supersymmetries. 

As a simple example, consider the~$L[j]_\Delta^{(R)}$ multiplets for~$d = \CN = 3$ discussed in the previous subsection. Here we focus on the case~$j = 1$ with generic~$R$. We will need the following two states from the representation of the SCP~$\CV$,
\begin{equation}
\CV_1 = \CV_\text{h.w.} = [1]^{(R)}~, \qquad \CV_2 = [-1]^{(R)}~. 
\end{equation}
The RS trial state corresponding to the representation~$[0]^{(R+2)}$ at level~$\ell = 1$ in~\eqref{jzerorstrial} is
\begin{equation}\label{fakehwex}
[0]^{(R+2)}\big|_\text{RS trial} = Q_-^{++} \CV_1~, \end{equation}
while the true highest-weight state of that representation is actually
\begin{equation}\label{truehwex}
[0]^{(R+2)} \big|_\text{true} = Q_-^{++} \CV_1 - Q_+^{(++)}\CV_2~.
\end{equation}
Note that the behavior of the two states in~\eqref{fakehwex} and~\eqref{truehwex} under the action of the~$Q$-supercharges is different, e.g.~$Q_-^{++}$ annihilates the trial state~$Q_-^{++} \CV_1$ in~\eqref{fakehwex}, but it does not annihilate the true highest-weight state in~\eqref{truehwex}. 

The fact that the trial states of the RS algorithm obscure the action of the~$Q$-supercharges complicates the construction of short superconformal multiplets~$\CM$.\footnote{~As discussed in~\cite{Cordova:2016xhm}, it also presents a challenge to the identification of supersymmetric deformations, i.e.~CPs annihilated by all~$Q$-supercharges, up to total derivatives.} Such multiplets possess null states, which can be identified and removed by finding all~$Q$-descendants of the primary null representation~$N \subset \CM$. This is straightforward in the Clebsch-Gordan approach, where the action of the~$Q$-supercharges is manifest. By contrast, it is not {\it a priori} clear how to correctly implement the removal of null states in an approach based on the RS algorithm. A variety of proposals for dealing with this problem has appeared in the literature. Our discussion is most directly inspired by~\cite{Dolan:2002zh}, as well as closely related discussions in~\cite{Dobrev:2004tk,Bianchi:2006ti,Dolan:2008vc,Dobrev:2012me,Beem:2014kka}.

In the remainder of this subsection, we comment on various issues that arise when one tries to apply or adapt the existing proposals to construct different short multiplets of increasing complexity. A synthesis appears in section~\ref{sec:ouralg} below, where we spell out a uniform algorithm for constructing all unitary superconformal multiplets in~$d \geq 3$ dimensions.

\bigskip

\subsubsection{Generic Multiplets with One Primary Null Representation at Level One} 

\label{sec:gennone}

Consider a short multiplet~$\CM = X_{\ell = 1}\CV$ (with~$X \in \{A, B, C, D\}$) whose primary null representation~$N$ resides at level~$\ell = 1$. It can be checked that there is always a single~$Q$-supercharge~$Q_*$ such that~$Q_* \CV_\text{h.w.}$ has the same Lorentz and~$R$-symmetry weights as the highest-weight state of~$N$. In this case it is natural to use the RS trial state~$Q_* \CV_\text{h.w.}$ as a proxy for the true highest-weight state of~$N$. This leads to the prescription found in~\cite{Dolan:2002zh}: to remove the null states from~$\CM$, discard all RS trial states of the form~$Q_* Q^n \CV_\text{h.w.}~(n \in \Z_{\geq 0})$. Here the RS trial states~$Q_* Q^n \CV_\text{h.w.}$ with~$n \geq 1$ serve as proxies for the true descendants of~$N$. The remaining RS trial states are used to determine the RS trial weights~$\CW_\text{RS}^{(\ell)}$, which are manipulated as in section~\ref{sec:rslm}. As long as all trial weights with negative coefficients cancel (unlike for long multiplets, this is no longer guaranteed; see section~\ref{sec:leftovernegs} below), we obtain a list of highest weights that describe the decomposition of~$\CM$ into CPs~$\CC_a^{(\ell)}$. 

As an example, we will construct the first few levels of a generic~$A_1$-type short multiplet~$\CM$ when~$d = \CN = 3$. Recalling the unitarity conditions from table~\ref{tab:3DN3}, we have
\begin{equation}\label{ma13dn3mult}
\CM = A_1[j]^{(R)}_{\Delta}~, \qquad \Delta = \half j + \half R + 1~,
\end{equation}
with~$j$ and~$R$ sufficiently large. Its primary null multiplet is given by (see table~\ref{tab:3DN3})
\begin{equation}\label{nmultaonetdn3}
N = [j-1]_{\Delta + 1/2}^{(R+2)}~.
\end{equation}
The weights of the supercharges, and the RS trial states and weights for a long multiplet in~$d = \CN = 3$ theories can be found in~\eqref{tdn3longmult} and~\eqref{tdnthreeex}. We see that the only trial state at level~$\ell = 1$ in~\eqref{tdnthreeex} whose quantum numbers match those of~$N$ in~\eqref{nmultaonetdn3} is~$Q_-^{++} \CV_\text{h.w.}$, so that~$Q_* = Q_-^{++}$. Omitting all RS trial states in~\eqref{tdnthreeex} that involve this supercharge, we obtain
\begin{align}\label{tdnthreea1}
\ell = 0: & \qquad \CV_\text{h.w.}~, \qquad \CW^{(0)}_\text{RS} = \left\{[j]_\Delta^{(R)} \right\}~,\cr
\ell = 1: & \qquad Q_+^{++}\CV_\text{h.w.}~,~Q_\pm^{+-}\CV_\text{h.w.}~,~Q_\pm^{--}\CV_\text{h.w.}~,\cr
&\qquad \CW^{(1)}_\text{RS} = \left\{[j+1]_{\Delta+1/2}^{(R+2)\oplus (R) \oplus (R-2)}~,~[j-1]_{\Delta+1/2}^{(R) \oplus (R-2)}\right\}~, \cr
\ell = 2: & \qquad Q^{++}_{+}Q^{+-}_{+} \CV_\text{h.w.}~,~Q^{++}_{+}Q^{--}_{+} \CV_\text{h.w.}~,~Q^{+-}_{\pm}Q^{--}_{\pm} \CV_\text{h.w.}~,~Q^{--}_{+}Q^{--}_{-} \CV_\text{h.w.}~,\cr
& \qquad Q^{\pm\pm}_{+}Q^{+-}_{-} \CV_\text{h.w.} ~,~Q^{++}_{+}Q^{--}_{-} \CV_\text{h.w.}~,~Q^{+-}_{+}Q^{--}_{-} \CV_\text{h.w.}~,~Q^{+-}_{+}Q^{+-}_{-} \CV_\text{h.w.}\cr
& \qquad \CW^{(2)}_\text{RS} = \Big\{[j+2]_{\Delta+1}^{(R+2) \oplus (R) \oplus (R-2)}~,\cr
& \qquad \qquad\qquad\,  [j]_{\Delta+1}^{(R+2) \oplus 2 (R) \oplus 2 (R-2) \oplus (R-4)}~,~[j-2]_{\Delta+1}^{(R-2)} \Big\}~.
\end{align}
For generic~$j$ and~$R$, the weights in~$\CW_\text{RS}^{(\ell = 0,1,2)}$ correctly capture the operator content of the short multiplet~$\CM$ in~\eqref{ma13dn3mult} at levels~$\ell = 0,1,2$. 

The prescription summarized above can be used to generate all short multiplets~$\CM$ whose primary null representation~$N$ resides at level~$\ell = 1$, as long as the quantum numbers of the SCP~$\CV$ are suitably generic (i.e.~its Dynkin labels must be sufficiently large). However, as was already discussed in the context of long multiplets, using RS trial states as proxies for true highest-weight states can lead to wrong results, especially if the quantum numbers of~$\CV$ are sufficiently small. Indeed, the prescription above can fail in such cases and must therefore be modified. This is explained in section~\ref{sec:nongennone} below. 

\subsubsection{Generic Multiplets with Two Primary Null Representation at Level One} 

\label{sec:genmultwoprimlone}

For~$d = 3,~\CN=2$ and~$d = 4$ there are~$Q$- and~$\b Q$-supercharges, which can lead to distinct left and right shortening conditions (see sections~\ref{sec:3dn2def} and~\ref{sec:4dmults}). The procedure described in section~\ref{sec:gennone} above then applies to multiplets that are either short on the left and long on the right, i.e.~multiplets of the form~$X_1 \b L$ with~$X \in \{A, B\}$, or multiplets that are long on the left and short on the right, i.e.~multiplets of the form~$L \b X_1$ with~$\b X \in \{\b A, \b B\}$. 

It is straightforward to adapt the procedure to multiplets of the form~$\CM = X_1 \b X_1$ that are short on both the left and the right, with the corresponding primary null representations~$N$ and~$\b N$ at levels~$\ell = 1\,,~\b \ell = 0$ and~$\ell = 0\,,~\b \ell = 1$, respectively. In this case one can always find unique left and right supercharges~$Q_*$ and~$\b Q_{\b *}$, such that the trial states~$Q_* \CV_\text{h.w.}$ and~$\b Q_{\b *} \CV_\text{h.w.}$ have the same quantum numbers as~$N$ and~$\b N$. We then drop all independent trial states of the form~$Q_* Q^m \b Q^n \CV_\text{h.w.}$ or~$\b Q_{\b *} Q^m \b Q^n \CV_\text{h.w.}$ with~$m, n \in \Z_{\geq 0}$. As before, the remaining RS trial states are used to determine the set of trial weights~$\CW^{(\ell)}_\text{RS}$, which are subjected to the RS cancellation procedure described in sections~\ref{sec:rslm} and~\ref{sec:gennone}. This procedure for generating short multiplets with two primary null representations at level~$\ell_\text{tot} = 1$ was thoroughly explored for~$d = 4$ and~$\CN=2, 4$ in~\cite{Dolan:2002zh}. 

Here we consider an example in~$d = 3$, with~$\CN=2$. As reviewed in section~\ref{sec:3dn2def}, there is a~$\frak{u}(1)_{r}$ symmetry and the Lorentz algebra is~$\frak{su}(2)$. There are left supercharges~$Q_\alpha~(\alpha = \pm)$ transforming as a Lorentz doublet of~$R$-charge~$-1$, i.e.~$\CR_Q = [1]_{1/2}^{(-1)}$, and right supercharges~$\b Q_\alpha~(\alpha = \pm)$ transforming as~$\CR_{\b Q} = [1]^{(1)}_{1/2}$. The multiplet we consider is (see table~\ref{tab:3DN2CAC}) 
\begin{equation}\label{a1a1b3dn2ex}
\CM = A_1 \b A_1[j \geq 1]_{\Delta}^{(r = 0)}~, \qquad \Delta = \half j +1~.
\end{equation}
This multiplet has a left primary null representation~$N$ at~$\ell = 1$ with respect to~$Q_\alpha$ and a right null representation~$\b N$ at~$\b \ell = 1$ with respect to~$\b Q_\alpha$ (see tables~\ref{tab:3DN2C} and~\ref{tab:3DN2AC}),
\begin{equation}\label{twsdnsex}
N = [j-1]^{(-1)}_{\Delta + 1/2}~, \qquad \b N = [j-1]^{(1)}_{\Delta + 1/2}~.
\end{equation}
The list of all possible RS trial states at levels~$\ell_\text{tot.} = 0,1,2$ and their weights are given by
\begin{align}\label{3dn2a1a1trial}
\ell_\text{tot.} = 0~, \qquad & \CV_\text{h.w.} = [j]^{(0)}_{\Delta}~,\cr
\ell_\text{tot.} = 1~, \qquad & Q_\pm \CV_\text{h.w.} = [j \pm 1]^{(1)}_{\Delta + 1/2}~, \quad\b Q_\pm \CV_\text{h.w.} = [j \pm 1]^{(-1)}_{\Delta + 1/2}~,\cr
 \ell_\text{tot.} = 2~, \qquad & Q_+ Q_- \CV_\text{h.w.} = [j]^{(-2)}_{\Delta+1}~, \quad \b Q_+ \b Q_- \CV_\text{h.w.} = [j]_{\Delta+1}^{(2)}~,\cr
&  Q_\pm \b Q_\pm \CV_\text{h.w.} = [j \pm 2]_{\Delta+1}^{(0)}~, \quad  Q_\pm \b Q_\mp \CV_\text{h.w.} = 2 [j]_{\Delta+1}^{(0)}~.
\end{align}
The only trial states at level~$\ell_\text{tot.} = 1$ with the quantum numbers of~$N$ and~$\b N$ in~\eqref{twsdnsex} are~$Q_- \CV_\text{h.w.}$ and~$\b Q_- \CV_\text{h.w.}$. We thus drop all trial states involving~$Q_-$ or~$\b Q_-$, leaving only
\begin{align}\label{a1a1barmult}
\ell_\text{tot.} = 0~, \qquad & \CV_\text{h.w.} = [j]^{(0)}_{\Delta}~,\cr
\ell_\text{tot.} = 1~, \qquad & Q_+ \CV_\text{h.w.} = [j +1]^{(1)}_{\Delta + 1/2}~, \quad \b Q_+ \CV_\text{h.w.} = [j + 1]^{(-1)}_{\Delta + 1/2}~,\cr
 \ell_\text{tot.} = 2~, \qquad & Q_+ \b Q_+ \CV_\text{h.w.} = [j + 2]_{\Delta+1}^{(0)}~.
\end{align}
Note that in this example there can be no additional trial states at levels~$\ell_\text{tot} \geq 3$, because any such state must contain either~$Q_-$ or~$\b Q_-$. Therefore~\eqref{a1a1barmult} represents the full operator content of the multiplet~$\CM$ in~\eqref{a1a1b3dn2ex}. Every CP in~$\CM$ is a conserved current; this multiplet of currents is also discussed in section~\ref{sec:currents}.

\subsubsection{Non-Generic Multiplets with Primary Null Representations at Level One} 

\label{sec:nongennone}

The prescription explained in sections~\ref{sec:gennone} and~\ref{sec:genmultwoprimlone} above is appropriate for short multipelts~$\CM$ whose SCP~$\CV$ is suitably generic, i.e.~all of its Dynkin labels are sufficiently large. Roughly speaking, this is because the prescription used RS trial states as proxies for true highest-weight states. As we already explained in the context of long multiplets,  around~\eqref{fakehwex} and~\eqref{truehwex}, this can be misleading, especially if some of the Dynkin labels of~$\CV$ are sufficiently small. In such cases, the algorithm described above may fail, even if the primary null representations reside at level one; simple examples arise for~$B_1$-type multiplets when~$d = 3\,,~\CN = 6$ or for~$B_1 \b B_1$-type multiplets when~$d = 4\,,~\CN=4$. 

In order to address this problem, the authors of~\cite{Dolan:2002zh} proposed to remove additional supercharges from the construction of RS trial states, beyond the supercharge~$Q_*$ (as well as~$\b Q_{\b *}$, in two-sided cases) connecting the SCP~$\CV$ and the primary null representation~$N$ (as well as~$\b N$, in two-sided cases). The precise rule is as follows: if there is a Lorentz or~$R$-symmetry lowering generator~$E_{-\alpha}$, with~$\alpha$ a simple root, that annihilates the highest weight state of the SCP, $E_{-\alpha} \CV_\text{h.w.} = 0$, then we should also omit the supercharge~$E_{-\alpha} Q_*$ from the construction of RS trial states. Intuitively, this prescription reflects the fact that~$Q_* \CV_\text{h.w.}$ is the RS trial state representing the primary null representation. Setting this trial state to zero and acting with~$E_{-\alpha}$ then leads to
\begin{equation}\label{moreqstars}
Q_* \CV_\text{h.w.} = 0 \quad \text{and} \quad E_{-\alpha} \CV_\text{h.w.} = 0 \qquad \Longrightarrow \qquad \left(E_{-\alpha} Q_*\right) \CV_\text{h.w.} = 0~.
\end{equation}
More generally, if the SCP is annihilated by several Lorentz or~$R$-symmetry lowering generators, we should omit all~$Q$-supercharges that can be obtained from~$Q_*$ by acting with any combination of these lowering generators. This leads to a finite list~$Q_*\,,\, Q_{**}\,,\,\ldots\,,\,Q_{*^k}$ (with~$k \leq N_Q$), of supercharges that should not be used in the construction of RS trial states. 

Note that there is a simple criterion for when a highest weight like~$\CV_\text{h.w.}$ is annihilated by a lowering generator~$E_{-\alpha_i}$, for some simple root~$\alpha_i$. As reviewed in appendix~\ref{app:liealg}, there is a one-to-one correspondence between the simple roots~$\alpha_j$ and the Dynkin labels~$\lambda_j \in \Z_{\geq 0}$ of~$\CV_\text{h.w.}$. Then~$E_{-\alpha_i} \CV_\text{h.w.} = 0$ if and only if~$\lambda_i = 0$. 

The fact that the procedure for constructing short multiplets must be modified when the quantum numbers of the SCP~$\CV$ are sufficiently small shows that it is in general not possible to construct such non-generic short multiplets by starting with more generic ones, substituting the small quantum numbers, and applying the RS algorithm to cancel any trial weights with negative Dynkin labels. This can be illustrated using~$B_1 \b B_1[0;0]^{(R_1, R_2, R_1)}$ short multiplets for~$d = 4\,, \CN=4$: specializing from generic~$R_1, R_2$ to~$R_1 = 0$ and applying the RS algorithm does not correctly reproduce a~$B_1 \b B_1[0;0]^{(0, R_2, 0)}$ short multiplet. As explained above, the correct procedure for constructing the latter involves removing extra supercharges (in addition to~$Q_*$ and~$\b Q_{\b *}$), because the~$R_1$ Dynkin label vanishes.  

Some authors have augmented the prescription described above by effectively removing even more RS trial states, which do not involve any of the supercharges~$Q_*\,,\, Q_{**}\,,\,\ldots\,,\,Q_{*^k}$ identified above. An example appears in appendix~C of~\cite{Beem:2014kka}, where the construction of the short multiplet~$\CM = D_1[0,0,0]^{(R_1, R_2)}_{\Delta = 2(R_1+R_2)}$ for~$d = 6\,,~\CN=(2,0)$ (see table~\ref{tab:6DN2}) is discussed. The primary null representation at level~$\ell = 1$ is given by~$N = [1,0,0]^{(R_1 + 1, R_2)}_{\Delta+1/2}$, so that~$Q_* = [1,0,0]^{(1,0)}_{1/2}$. As long as~$R_1, R_2 \geq 1$, the highest-weight state~$\CV_\text{h.w.} = [0,0,0]^{(R_1, R_2)}_\Delta$ is annihilated by all Lorentz lowering operators, but none corresponding to the~$R$-symmetry. The prescription described around~\eqref{moreqstars} then instructs us to omit the entire Lorentz-multiplet of~$Q_*$ from the construction of RS trial states. Explicitly, 
\begin{equation}\label{qstarsex}
Q_* = [1,0,0]^{(1,0)}_{1/2}~,~Q_{*^2} = [-1,1,0]^{(1,0)}_{1/2}~,~Q_{*^3} = [0,-1,1]^{(1,0)}_{1/2}~,~Q_{*^4} = [0,0,-1]^{(1,0)}_{1/2}~.
\end{equation} 
In our language, the prescription of~\cite{Beem:2014kka} also involves dropping certain RS trial states that are not constructed using any of the supercharges in~\eqref{qstarsex}. For instance, when~$R_1 = 3$, the proposal of~\cite{Beem:2014kka} also involves omitting the following RS trial state,
\begin{align}
Q^4 \cdot \CV_\text{h.w.}  & =  [1,0,0]^{(-1,1)}_{1/2} [-1,1,0]^{(-1,1)}_{1/2} [0,-1,1]^{(-1,1)}_{1/2}[0,0,-1]^{(-1,1)}_{1/2}~\cdot~ [0,0,0]^{(R_1 = 1, R_2)}_\Delta \cr
&  = [0,0,0]^{(-1,4)}_{\Delta + 2}~.
\end{align}
However, this trial state is automatically set to zero by the RS algorithm (see appendix~\ref{app:RSalg}), since the first~$R$-symmetry Dynkin label is~$-1$. Although we have not checked every example, we expect that all additional RS trial states removed in~\cite{Beem:2014kka}, which do not involve any of the supercharges~$Q_*\,,\, Q_{**}\,,\,\ldots\,,\,Q_{*^k}$, can similarly be accounted for by the RS algorithm.

\subsubsection{Leftover Trial Weights with Negative Coefficients }

\label{sec:leftovernegs}

At intermediate stages, the RS algorithm can lead to trial weights with negative coefficients. When applied to long multiplets (see section~\ref{sec:rslm}), it is guaranteed that all negative coefficients cancel in the end. This is no longer the case when we apply the algorithm to short multiplets, because we remove various RS trial weights along the way to account for null states. 
This phenomenon was noted in~\cite{Dolan:2002zh}, in the context of 4d $\CN =2$ and $\CN =4$, and for the various cases explored there the leftover negative-coefficient weights could be interpreted as null states associated with short representations of the conformal group, i.e.~conserved currents or free fields.\footnote{~For instance, the conservation law~$\d^\mu j_\mu = 0$ for a dimension~$\Delta = d-1$ vector current was interpreted as a Lorentz-singlet CP of dimension~$\Delta_{\d j} = d$ with coefficient~$-1$.} However, there are examples where this interpretation does not hold. For instance, a~$D_1[0,0,0]^{(1,2)}_6$ multiplet for~$d = 6\,,~\CN=(2,0)$ does not contain any conserved currents or free fields. Nevertheless, the RS procedure used to construct this multiplet results in many trial states with negative coefficients that do not cancel. Rather than attempting to interpret these leftover negative-coefficient weights, we simply propose to eliminate them at the end of the procedure. A detailed discussion of multiplets that contain conserved currents or free fields and their null states appears in section~\ref{sec:currents}.

\subsubsection{Multiplets with Higher-Level Primary Null Representations}

\label{sec:hlns}

So far we have only discussed short multiplets of~$ X_{\ell = 1}$-type whose primary null representations reside at level one. Here we will discuss how to treat multiplets of the form~$\CM = X_{\ell \geq 2}$, whose primary null representation~$N$ resides at level two or higher. In many cases, there is a unique monomial~$Q_*^\ell = Q_* \cdots Q_{*^\ell}$ in the supercharges such that~$Q^\ell_* \CV_\text{h.w.}$ has the same quantum numbers as~$N$. In this case it is natural to generalize the procedure discussed above by simply omitting all RS trial states that involve the monomial~$Q_*^\ell$. A simple example of this sort arises for~$d = 6\,,~\CN=(1,0)$ (see section~\ref{sec:d6n1defs}): consider a short multiplet~$\CM = A_2[j_1, j_2 \geq 1, 0]^{(R)}_{\Delta}$ with~$\Delta = \half (j_1 + 2 j_2) + 2 R + 6$ (see table~\ref{tab:6DN1}), whose primary null representation~$N = [j_1, j_2-1, 1]^{(R+2)}_{\Delta + 1}$ resides at level~$\ell = 2$. It can be checked that there are only two supercharges, 
\begin{equation}
Q_* = [0,-1,1]^{(1)}_{1/2}~, \qquad Q_{**} = [0,0,1]_{1/2}^{(1)}~,
\end{equation}
such that~$Q_* Q_{**} \CV_\text{h.w.}$ has the same quantum numbers as~$N$. In the construction of this multiplet, we therefore discard all RS trial states of the form~$Q_* Q_{**} Q^n \CV_\text{h.w.}$, where~$Q^n$ is any product of~$n \in \Z_{\geq 0}$ other supercharges. 

However, for some short multiplets~$\CM = X_{\ell\geq 2}$ there are several distinct products of~$\ell$ supercharges whose quantum numbers make up the difference between~$\CV_\text{h.w.}$ and~$N$. Consider, for instance, the multiplet~$\CM =  B_2[0,0]^{(R)}_{\Delta}$ with~$\Delta = {3 \over 2} R + 3$ in~$d = 5$, whose primary null representation has quantum numbers~$N = [0,0]^{(R+2)}_{\Delta +1}$ and resides at level~$\ell = 2$ (see table~\ref{tab:5Dmult}). The weights of the supercharges are
\begin{equation}
Q^\pm_1 = [1,0]_{1/2}^{(\pm 1)}~, \quad Q^\pm_2 = [-1,1]_{1/2}^{(\pm 1)}~, \quad Q^\pm_3 = [1,-1]_{1/2}^{(\pm 1)}~, \quad Q^\pm_4 = [-1,0]_{1/2}^{(\pm 1)}~.
\end{equation}
Hence there are two distinct monomials in the~$Q$'s that can be applied to the SCP~$\CV_\text{h.w.}$ to produce a state with the quantum numbers of~$N$, \begin{equation}\label{5dqmonoex}
Q_1^+ Q_4^+ \sim Q_2^+ Q_3^+ \sim [0,0]^{(2)}_1~.
\end{equation}
It is therefore not clear which linear combination of these monomials to omit from the construction of RS trial states, and for some choices the resulting multiplet is incorrect. 

We propose to resolve this ambiguity by examining the true highest-weight state of~$N$. For the~$B_2[0,0]^{(R)}$ example, if we denote the full SCP representation by~$\CV^{(i_1 \cdots i_R)}$, where $i_1, \ldots, i_R = \pm$ are~$\frak{su}(2)_R$ doublet indices, then the full primary null representation~$N^{(i_1 \cdots i_{R+ 2})}$ is given by 
\begin{equation}
N^{(i_1 \cdots i_{R+ 2})} = \Omega^{\alpha\beta}Q^{(i_1}_\alpha Q^{i_2}_\beta \CV^{i_3 \cdots i_{R+2})}~.
\end{equation}
Here~$\Omega$ is a~$4 \times 4$ symplectic matrix used to raise and lower fundamental~$\frak{sp}(4)$ indices (i.e.~spinor indices of the~$\frak{so}(5)$ Lorentz group). Setting~$i_1 = \cdots = i_{R + 2} = +$, we obtain the highest-weight state 
\begin{equation}\label{primnspoly}
N_\text{h.w.} = \left(Q_1^+ Q_4^+ - Q_2^+ Q_3^+\right) \CV^{+ \cdots +}~.
\end{equation}
Therefore~$N_\text{h.w.}$ selects a particular linear combination of the monomials in~\eqref{5dqmonoex}, and we propose to omit this particular linear combination from the construction of RS trial states. 

As before, the prescription above applies to a generic~$B_2[0,0]_\Delta^{(R)}$ multiplet. When~$R = 0$, we follow the logic of section~\ref{sec:nongennone} and omit additional combination of supercharges that can be reached from~$Q_1^+ Q_4^+ - Q_2^+ Q_3^+$ in~\eqref{primnspoly} using the~$R$-symmetry lowering generators. Here, this leads to an entire Lorentz and~$R$-symmetry multiplet of supercharge combinations that should be removed from the construction of RS trial states:
\begin{equation}
Q_1^\pm Q_4^\pm - Q_2^\pm Q_3^\pm~, \qquad Q_1^+ Q_4^- + Q_1^- Q_4^+ - Q_2^+ Q_3^- - Q_2^- Q_3^+~.
\end{equation}

\subsection{Proposal for a Uniform Algorithm}

\label{sec:ouralg}

We will now synthesize the discussion in sections~\ref{sec:gennone} through~\ref{sec:hlns} into a precise algorithm for generating all unitary superconformal multiplets in dimensions~$d \geq 3$ and for all allowed values of~$\CN$. As is the case for most existing prescriptions in the literature, we do not know of a general proof that establishes the {\it a priori} correctness of our algorithm. However, it passes a large variety of detailed consistency checks, which are summarized below. 

The input and output of the algorithm are spelled out in section~\ref{sec:inoutalg}. The precise rules for constructing a short multiplet~$\CM$ are as follows:\footnote{~The construction of long multiplets was already discussed in section~\ref{sec:rslm}.} 

\begin{itemize}
\item[1.)] Identify the SCP~$\CV$ and the primary null representations: if the supercharge representation~$\CR_Q$ is irreducible, there is a single primary null representation~$N$ at level~$\ell$; if there are left and right supercharges~$Q$ and~$\b Q$ (which occurs when~$d = 3\,,~\CN=2$ or when~$d = 4$ for all~$\CN$), then~$\CM$ may have a primary null representation~$N$ at level~$\ell$ on the left, or a primary null representation~$\b N$ at level~$\b \ell$ on the right, or both.

\item[2.)]  Let~$Q_i~(i = 1, \ldots, N_Q)$ be the (left) supercharges; in the two-sided cases there are also right supercharges~$\b Q_{\b i}~(\b i = 1, \ldots, \b N_{\b Q})$. Denote all states in the SCP representation by~$\CV = \{\CV_A\}_{A = 1}^{\dim \CV}$, with~$\CV_1 = \CV_\text{h.w.}$ being the highest-weight state. In the one-sided case, the true highest-weight state of the primary null representation~$N$ takes the form
\begin{equation}\label{exactnhw}
N_\text{h.w.} = \sum_{A = 1}^{\dim \CV} P^{(\ell)}_A(Q_i) \CV_A~.
\end{equation}
Here the~$P_A^{(\ell)}(Q_i)$ are degree-$\ell$ polynomials in the~$Q$-supercharges. In the two-sided case there may also be a primary null representation~$\b N$ on the right, whose highest-weight state can be written using degree-$\b \ell$ polynomials~${\b P}^{(\b \ell)}_A(\b Q_{\b i})$ in the~$\b Q$-supercharges,
\begin{equation}\label{exactnbarhw}
\b N_\text{h.w.} = \sum_{A = 1}^{\dim \CV} {\b P}^{(\b \ell)}_A(\b Q_{\b i}) \CV_A~.
\end{equation}
We now project~\eqref{exactnhw} and~\eqref{exactnbarhw} onto the highest-weight state~$\CV_1 = \CV_\text{h.w.}$ to obtain constraints that should be imposed on RS trial states (see point 4.) below): 
\begin{equation}\label{RSconstraints}
P_1^{(\ell)}(Q_i) \CV_\text{h.w.} = 0~, \qquad \b P_1^{(\b \ell)}( \b Q_{\b i}) \CV_\text{h.w.} = 0~.
\end{equation}
Here it is understood that the second constraint is absent in the one-sided case. 

\item[3.)] If the quantum numbers of~$\CV$ are not suitably generic, additional constraints need to be imposed to supplement those in~\eqref{RSconstraints}. They are determined by acting on~\eqref{RSconstraints} with all Lorentz or~$R$-symmetry lowering operators~$E_{-\alpha}$ (here~$\alpha$ is a simple root) that annihilate the highest-weight state~$\CV_\text{h.w.}$ of the SCP representation. This leads to a set of constraints of the form
\begin{equation}\label{RSconstraintsaug}
p^{(\ell)}_k (Q_i) \CV_\text{h.w.} = 0~, \qquad \b p^{(\b \ell)}_{\b k} ( \b Q_{\b i}) \CV_\text{h.w.} = 0~.
\end{equation}
Here~$p^{(\ell)}_k (Q_i)$ and~$\b p^{(\b \ell)}_{\b k}(\b Q_{\b i})$ are degree~$\ell$ and~$\b \ell$ polynomials in the left and right supercharges. The labels~$k$ and~$\b k$ index the different left and right constraints, with~$k = \b k = 1$ corresponding to the original constraints in~\eqref{RSconstraints}, i.e.~$p_1^{(\ell)} = P_1^{(\ell)}$ and~$\b p_{\b 1}^{(\b \ell)} = \b P_1^{(\b \ell)}$. As before, it is understood that the~$\b p_{\b k}^{(\b \ell)}$-constraints are absent in the one-sided case. 

\item[4.)] Construct all RS trial states~$Q^n \CV_\text{h.w.}$ with~$n \in \Z_{\geq 0}$ in the one-sided case, or~$Q^n \b Q^{\b n}\CV_\text{h.w.}$ with~$n, \b n \in \Z_{\geq 0}$ in the two-sided case. Then impose all constraints in~\eqref{RSconstraintsaug}. The remaining trial states are used to determine the set of RS trial weights.

\item[5.)] If all RS trial weights constructed in point 4.) above are valid highest weights, i.e.~if all of their Dynkin labels are nonnegative, the algorithm terminates. Otherwise, apply the RS algorithm (see section~\ref{sec:rslm} and appendix~\ref{app:RSalg}) to the set of RS trial weights. The result is a list of valid highest weights, some of which may have negative coefficients.\footnote{~Weights that are assigned vanishing coefficients by the RS algorithm are simply eliminated.} Finally, drop all negative-coefficient weights.

\end{itemize}

\noindent The set of valid highest weights with positive multiplicities that remains after the algorithm terminates describes the quantum numbers of the CPs~$\CC^{(\ell)}_a$ in the decomposition~\eqref{basicdecompiii}, which (together with their CDs) constitute the operator content of the short multiplet~$\CM$.

It is an important feature of our algorithm that the constraints on short multiplets due to null states are determined once and for all by examining the primary null representations~$N$ and (possibly)~$\b N$ (see points 2.) and 3.) above). These constraints are then applied at all levels to eliminate RS trial states. It is not necessary to further manipulate the trial states, e.g.~by eliminating additional states according to some level-by-level prescription, before applying the RS algorithm at the end.

As was already emphasized above, our proposed algorithm is conjectural, but satisfies a large number of consistency checks that support its correctness:
\begin{itemize}
\item[(i)] It produces multiplets that agree with many examples that have already been constructed in the literature, see for instance~\cite{Dolan:2002zh,Bianchi:2006ti,Dolan:2008vc,Minwalla:2011ma,Beem:2014kka} and references therein.

\item[(ii)] It produces multiplets that contain the same number of bosonic and fermionic operators.\footnote{~More generally, every operator representation of the Poincar\'e supersymmetry algebra must also satisfy this sum rule, see for instance~\cite{Weinberg:2000cr}.} For multiplets containing conserved currents, this is only true if we correctly account for conservation laws (see also~(iv) below).  For instance, a conserved current~$j_\mu$ in~$d$ dimensions has~$d -1$ bosonic degrees of freedom, because~$\d^\mu j_\mu = 0$. 

\item[(iii)] The short multiplets constructed using our algorithm satisfy all recombination rules. As explained around~\eqref{genrr} and in section~\ref{sec:usm}, these rules constitute an infinite number of nontrivial {\it a priori} consistency conditions on short multiplets. As in~(ii) above, the recombination rules involving multiplets containing conserved currents are only satisfied if we properly account for conservation laws (see also~(iv) below). 

\item[(iv)] Despite its uniformity and simplicity, our algorithm correctly describes the rich variety of sporadic short multiplets that exist for different values of~$d$ and~$\CN$. This is well illustrated by multiplets containing conserved currents, which are discussed in section~\ref{sec:currents}. As was already emphasized above, the consistency conditions in~(ii) and~(iii) are only satisfied if we properly account for the conservation laws of any conserved currents, as well as other null states of the conformal group, such as free-field equations of motion. Since we did not explicitly remove these extra null states by hand, we view the fact that they are correctly captured by our algorithm as further evidence for its correctness. 
\end{itemize}

\section{Tables of Select Superconformal Multiplets}

\label{sec:tables}

In this section we tabulate the full operator content of all generic superconformal multiplets in~$3\leq d \leq 6$ dimensions with~$N_Q \leq 8$ supercharges. (Typical multiplets in theories with~$N_Q > 8$ are increasingly difficult to display explicitly, since the number of operators they contain grows exponentially with~$N_Q$.) We call a multiplet generic if all nonzero Dynkin labels of its SCP are sufficiently large. Additional cancellations occur when some of the Dynkin labels take small values. Copious examples of such non-generic multiplets are furnished by the superconformal current multiplets discussed in section~\ref{sec:currents}. 

Throughout, we will not explicitly indicate the scaling dimension of every operator in a given multiplet, since this is easily inferred from the level of the operator within the multiplet and the dimension~$\Delta$ of the SCP. In situations with left and right supercharges (i.e.~$d = 3, \CN=2$ and~$d = 4$), complex conjugation relates multiplets of the schematic form~$X \b Y$ and~$Y \b X$ (where~$X, Y \in \{L, A, B\}$). Whenever possible, we will use this fact to reduce the number of multiplets that need to be tabulated. 

\subsection{$d = 3,~ \CN=1$}
\label{sec:3dn1tabs}

The unitary superconformal multiplets are summarized in table~\ref{tab:3DN1}. We present them for generic values of the~$\frak{su}(2)$ Lorentz Dynkin label~$j$, whenever it is non-vanishing:

\begin{align*}
 \xymatrix{*++[F-,]{\boldsymbol{L}}} & \qquad \xymatrix{*++[F=]{{ [j ]_\Delta}~,~\Delta > \half j + 1} \ar[r]^-Q & *++[F]{[j \pm 1]} \ar[r]^-{Q} & *++[F]{[j]}}  \\[8pt]
  \xymatrix{*++[F-,]{\boldsymbol{ L'}}}   & \qquad \xymatrix{*++[F=]{{[0]_\Delta}~,~\Delta > \half} \ar[r]^-Q & *++[F]{[1]} \ar[r]^-{Q} & *++[F]{[0]} 
} \\[8pt]
\xymatrix{*++[F-,]{\boldsymbol{ A_1}}} & \qquad \xymatrix{*++[F=]{{ [j]_\Delta}~,~\Delta = \half j + 1~
} \ar[r]^-Q & *++[F]{[j + 1]} }  \\[8pt]
\xymatrix{*++[F-,]{\boldsymbol{ A'_2}}} & \qquad \xymatrix{*++[F=]{{[0]_\Delta}~,~\Delta = \half~
} \ar[r]^-Q & *++[F]{[1]} }  \\[8pt]
\xymatrix{*++[F-,]{\boldsymbol{ B_1}}} & \qquad \xymatrix{*++[F=]{{~[0]_0}~}}
\end{align*}

\subsection{$d = 3,~ \CN=2$}
\label{sec:3dn2tabs}

The unitary superconformal multiplets are summarized in table~\ref{tab:3DN2CAC}. We present their operator content for generic values of the~$\frak{su}(2)$ Lorentz Dynkin label~$j$, whenever it is non-vanishing (see also~\cite{Minwalla:2011ma}). 

\begin{align*}
\xymatrix{
*++[F-,]{{\boldsymbol{ L \b L}}}} & \xymatrix @C=7pc @R=7pc @!0 @dr {
*++[F=]{[j]^{(r)}_\Delta~,~\Delta > \frac{1}{2}\, j+|r|+1} \ar[r]|--{{~\b Q~}} \ar[d]|--{~Q~} 
& *++[F]{[j \pm 1 ]^{(r+1)}} \ar[r]|--{{~\b Q~}} \ar[d]|--{~Q~} 
& *++[F]{[j]^{(r+2)}} \ar[d]|--{~Q~}\\
*++[F]{[j \pm 1]^{(r-1)}} 
\ar[r]|--{{~\b Q~}} \ar[d]|--{~Q~} 
& *++[F]{ [j\pm 2]^{(r)} \oplus [j]^{2(r)}}\ar[r]|--{{~\b Q~}}\ar[d]|--{~Q~}
 & *++[F]{[ j \pm 1]^{(r+1)}} \ar[d]|--{~Q~}\\
*++[F]{[j]^{(r-2)}} \ar[r]|--{{~\b Q~}}
& *++[F]{[ j \pm 1]^{(r-1)}} \ar[r]|--{{~\b Q~}}  
& *++[F]{[j]^{(r)}} 
}
\end{align*}

\begin{align*}
\xymatrix{
*++[F-,]{{\boldsymbol{ L \b A_1}}}} & \xymatrix @C=7pc @R=7pc @!0 @dr {
*++[F=]{[j]_\Delta^{(r)}~,~\Delta = \frac{1}{2}\, j + r + 1~,~r >0} \ar[r]|--{{~\b Q~}} \ar[d]|--{~Q~} 
& *++[F]{[j+ 1]^{(r+1)}}  
\ar[d]|--{~Q~}
\\
*++[F]{[j \pm 1]^{(r-1)}} 
\ar[r]|--{~\b Q~}
\ar[d]|--{~Q~} 
& *++[F]{[j+2]^{(r)} \oplus [j]^{(r)}}\ar[d]|--{~Q~}
\\
*++[F]{[j]^{(r-2)}} \ar[r]|--{{~\b Q~}}
& *++[F]{[j+ 1]^{(r-1)}} 
}
\end{align*}

\begin{align*}
\xymatrix{
*++[F-,]{{\boldsymbol{ L \b A_2}}}} & \xymatrix @C=7pc @R=7pc @!0 @dr {
*++[F=]{[0]_\Delta^{(r)}~,~\Delta =  r + 1~,~r >0} \ar[r]|--{{~\b Q~}} \ar[d]|--{~Q~} 
& *++[F]{[ 1]^{(r+1)}}  
\ar[d]|--{~Q~}
\\
*++[F]{[1]^{(r-1)}} 
\ar[r]|--{~\b Q~}
\ar[d]|--{~Q~} 
& *++[F]{[2]^{(r)} \oplus [0]^{(r)}}\ar[d]|--{~Q~}
\\
*++[F]{[0]^{(r-2)}} \ar[r]|--{{~\b Q~}}
& *++[F]{[1]^{(r-1)}} 
}
\end{align*}

\begin{align*}
\xymatrix{
*++[F-,]{{\boldsymbol{ L \b B_1}}}} & \xymatrix @C=7pc @R=7pc @!0 @dr {
*++[F=]{[ 0]_\Delta^{(r)}~,~\Delta = r ~,~r >0 }  \ar[d]|--{~Q~} \\
*++[F]{[1]^{(r-1)}} 
\ar[d]|--{~Q~} 
\\
*++[F]{[0]^{(r-2)}}
}
\end{align*}

\begin{align*}
\xymatrix{
*++[F-,]{{\boldsymbol{ A_1 \b A_1}}}} &~~~~~~~~~~ \xymatrix @C=7pc @R=7pc @!0 @dr {
*++[F=]{[j]_\Delta^{(0)}~,~\Delta = {1 \over 2}\,j + 1~} \ar[r]|--{{~\b Q~}} \ar[d]|--{~Q~} 
& *++[F]{[j + 1]^{(1)}}  
\ar[d]|--{~Q~}
\\
*++[F]{[j + 1]^{(-1)}} 
\ar[r]|--{~\b Q~}
& *++[F]{[j+2]^{(0)}}
}
\end{align*}

\begin{align*}
\xymatrix{
*++[F-,]{{\boldsymbol{ A_2 \b A_2}}}} & \xymatrix @C=7pc @R=7pc @!0 @dr {
*++[F=]{[0]^{(0)}_\Delta~,~\Delta = 1~} \ar[r]|--{{~\b Q~}} \ar[d]|--{~Q~} 
& *++[F]{[1]^{(1)}}  
\ar[d]|--{~Q~}
\\
*++[F]{[1]^{(-1)}} 
\ar[r]|--{~\b Q~}
& *++[F]{[2]^{(0)} \oplus  [0]^{(0)}}
}
\end{align*}

\begin{align*}
\xymatrix{
*++[F-,]{{\boldsymbol{ A_2 \b B_1}}}} &~~~~~~~~~~ \xymatrix @C=7pc @R=7pc @!0 @dr {
*++[F=]{[0]^{(\half)}_\Delta~,~\Delta=\frac{1}{2} ~ }  \ar[d]|--{~Q~} \\
*++[F]{[1]^{(-\half)}} 
}
\end{align*}

\begin{align*}
\xymatrix{
*++[F-,]{{\boldsymbol{ B_1 \b B_1}}}} &~~~~~~~~~~ \xymatrix @C=7pc @R=7pc @!0 @dr {
*++[F=]{~[0]^{(0)}_0~}  
}
\end{align*}

\subsection{$d = 3, ~\CN=3$}
\label{sec:3dn3tabs}

The unitary superconformal multiplets are summarized in table~\ref{tab:3DN3}. We present their operator content for generic values of the~$\frak{su}(2)$ Lorentz and~$\frak{su}(2)_R$ Dynkin labels~$j$ and~$R$, whenever they are non-vanishing (see also~\cite{Minwalla:2011ma}). We used a condensed notation in which any~$\pm$ offsets for the two~$\frak{su}(2)$'s are independent, e.g.~$[j \pm 1]^{(R \pm 2)}$ denotes four operators. 

\begin{align*}
{\xymatrix{*++[F-,]{\boldsymbol{ L}}}}  \quad &{\xymatrix @R=1.4pc@C=2pc{*++[F=]{[j]_\Delta^{(R)}~,~\Delta > \frac{1}{2}\,j+ \frac{1}{2}R + 1}}} \\[5pt]
Q: \quad & \boxed{[  j\pm1 ]^{(R\pm2)\oplus (R)}~
}\\[5pt]
Q^2: \quad & \boxed{
 [ j \pm 2]^{(R\pm2)\oplus(R)}~,~[j]^{(R\pm4) \oplus 2(R\pm 2)\oplus 3(R)}~}\\[5pt]
Q^3: \quad & \boxed{ 
[j\pm 3]^{(R)}~,~[j\pm1]^{(R\pm 4)\oplus 2(R\pm2)\oplus 3(R)}~
}\\[5pt]
Q^4:  \quad & \boxed{
 [j \pm 2 ]^{(R\pm2)\oplus(R)}~,~[ j ]^{(R\pm4) \oplus 2(R\pm 2)\oplus 3(R)}~}\\[5pt]
Q^5: \quad & \boxed{[ j\pm1 ]^{(R\pm2)\oplus (R)}~
}\\[5pt]
Q^6: \quad & \boxed{[ j ]^{(R)}~
}\\[20pt]
{\xymatrix{*++[F-,]{\boldsymbol{ A_{1}}}}}  \quad &{\xymatrix @R=1.4pc@C=2pc{*++[F=]{[j]_\Delta^{(R)}~,~\Delta = \frac{1}{2}\,j+ \frac{1}{2}R + 1}}} \\[5pt]
Q: \quad & \boxed{[ j+1 ]^{(R\pm2)\oplus (R)}~,~[  j-1 ]^{(R-2)\oplus (R)}~
}\\[5pt]
Q^2: \quad & \boxed{
 [ j + 2 ]^{(R\pm2)\oplus(R)}~,~[ j ]^{(R+2) \oplus 2(R)\oplus 2(R-2)\oplus(R-4)}~,~[ j-2 ]^{(R-2)}~}\\[5pt]
Q^3: \quad & \boxed{ 
[ j+ 3 ]^{(R)}~,~[ j+1 ]^{(R+2)\oplus 2(R)\oplus 2(R-2)\oplus(R-4)}~,~[ j-1 ]^{(R)\oplus (R-2)\oplus (R-4)}~
}\\[5pt]
Q^4:  \quad & \boxed{
 [ j + 2 ]^{(R)\oplus(R-2)}~,~[ j ]^{(R) \oplus (R-2)\oplus (R-4)}~}\\[5pt]
Q^5: \quad & \boxed{[ j+1 ]^{(R-2)}~
}\\[20pt]
{\xymatrix{*++[F-,]{\boldsymbol{ A_{2}}}}}  \quad &{\xymatrix @R=1.4pc@C=2pc{*++[F=]{[ 0 ]_\Delta^{(R)}~,~\Delta =  \frac{1}{2}R + 1}}} \\[5pt]
Q: \quad & \boxed{[1]^{(R\pm2)\oplus (R)}~
}\\[5pt]
Q^2: \quad & \boxed{
 [2]^{(R\pm2)\oplus(R)}~,~[0]^{(R+2) \oplus 2(R)\oplus (R-2)\oplus(R-4)}~}\\[5pt]
Q^3: \quad & \boxed{ 
[3]^{(R)}~,~[1]^{(R+2)\oplus 2(R)\oplus 2(R-2)\oplus(R-4)}~
}\\[5pt]
Q^4:  \quad & \boxed{
 [2]^{(R)\oplus(R-2)}~,~[0]^{(R) \oplus (R-2)\oplus (R-4)}~}\\[5pt]
Q^5: \quad & \boxed{[1]^{(R-2)}~}
\end{align*}

\begin{align*}
{\xymatrix{*++[F-,]{\boldsymbol{ B_{1}}}}}  \quad &{\xymatrix @R=1.4pc@C=2pc{*++[F=]{[0]_\Delta^{(R)}~,~\Delta =  \frac{1}{2}R}}} \\[5pt]
Q: \quad & \boxed{[1]^{(R)\oplus (R-2)}~
}\\[5pt]
Q^2: \quad & \boxed{
 [2]^{(R-2)}~,~[0]^{(R) \oplus (R-2)\oplus (R-4)}~}\\[5pt]
Q^3: \quad & \boxed{ 
[1]^{(R-2)\oplus(R-4)}~
}\\[5pt]
Q^4:  \quad & \boxed{
[0]^{(R-4)}~}\\[5pt]
\end{align*}

\subsection{$d = 3, ~\CN=4$}
\label{sec:3dn4tabs}

The unitary superconformal multiplets are summarized in table~\ref{tab:3DN4}. We present their operator content for generic values of the~$\frak{su}(2)$ Lorentz Dynkin label~$j$ and the~$\frak{su}(2)_R \oplus \frak{su}(2)'_{R}$ Dynkin labels~$(R; R')$, whenever these are non-vanishing. We used a condensed notation in which any~$\pm$ offsets for the three~$\frak{su}(2)$'s are independent, e.g.~$[j \pm 1]^{(R \pm 1; R' \pm 1)}$ denotes eight operators. Moreover, we define a transpose operation~$T$ on numerical~$R$-symmetry offsets~$x, y$,
\begin{equation}
T\{(R+x ; R'+y)\}=(R+y ; R'+x)~,
\end{equation}
and use it to introduce the following shorthand: whenever we write~$\{ \cdots\} \oplus T\{\cdots\}$, we mean the sum of all representations in~$\{ \cdots\}$ and their~$T$-transposes. For instance, 
\begin{equation}
\{(R\pm2 ; R')\}\oplus T\{\cdots\}=(R\pm2 ; R')\oplus (R ; R'\pm2)~.
\end{equation}
With these conventions, the generic~$d = 3, \CN=4$ multiplets take the following form: 

\begin{align*}
{\xymatrix{*++[F-,]{\boldsymbol{ L}}}}  \quad &{\xymatrix @R=1.4pc@C=2pc{*++[F=]{[ j ]_\Delta^{(R ; R')}~,~\Delta > \frac{1}{2}\,j+ \frac{1}{2}(R+R') + 1}}} \\[5pt]
Q: \quad & \boxed{[ j\pm1 ]^{(R\pm1 ; R'\pm1)}~
}\\[5pt]
Q^2: \quad & \boxed{
 [ j \pm 2 ]^{\{(R\pm2 ; R')\}\oplus T\{\cdots\}  \oplus 2(R ; R')}~,~[ j ]^{
  (R\pm2 ; R'\pm2) \oplus 4 (R ; R')  \oplus \{2(R\pm 2 ; R')\}  \oplus T\{\cdots\} }~}\\[5pt]
Q^3: \quad & \boxed{ 
[j\pm 3]^{(R\pm1 ; R'\pm1)}~,~[j\pm1]^{ \{(R\pm3 ; R'\pm1)\}\oplus T\{\cdots\} \oplus 4(R\pm1 ; R'\pm1) }~
}\\[5pt]
Q^4:  \quad & \boxed{ \begin{aligned} & 
 [ j \pm 4 ]^{(R ; R')}~,~[ j \pm 2]^{(R\pm2 ; R'\pm2)\oplus 4(R ; R') \oplus \{2(R\pm2 ;  R')\} \oplus T\{\cdots\}} \\
 &  [ j ]^{ \{(R\pm4 ; R')\oplus 4(R\pm2 ; R')\} \oplus T\{\cdots\} \oplus 2(R\pm2 ; R'\pm2)\oplus 8(R ; R')}
 \end{aligned}~}\\[5pt]
Q^5: \quad & \boxed{ 
[j\pm 3]^{(R\pm1 ; R'\pm1)}~,~[j\pm1]^{ \{(R\pm3 ; R'\pm1)\}\oplus T\{\cdots\} \oplus 4(R\pm1 ; R'\pm1) }~
}\\[5pt]
Q^6:\quad & \boxed{
  [ j \pm 2 ]^{\{(R\pm2 ; R')\}\oplus T\{\cdots\}  \oplus 2(R ; R')}~,~[ j]^{
  (R\pm2 ; R'\pm2) \oplus 4 (R ; R') \oplus \{2(R\pm 2 ; R')\}  \oplus T\{\cdots\} }~}\\[5pt]
 Q^{7}: \quad & \boxed{[  j\pm1 ]^{(R\pm1 ;  R'\pm1)}~
}\\[5pt]
Q^{8}: \quad & \boxed{[ j]^{(R ; R')}~
}\\[5pt]
\end{align*}

\begin{align*}
{\xymatrix{*++[F-,]{\boldsymbol{ A_{1}}}}}  \quad &{\xymatrix @R=1.4pc@C=2pc{*++[F=]{[ j ]_\Delta^{(R ; R')}~,~\Delta = \frac{1}{2}\,j+ \frac{1}{2}(R+R') + 1}}} \\[5pt]
Q: \quad & \boxed{[  j+1 ]^{(R\pm1 ;  R'\pm1)}~,~[  j-1 ]^{\{(R+1 ; R'-1)\}\oplus T\{\cdots \} \oplus(R-1 ; R'-1) }~
}\\[5pt]
Q^2: \quad & \boxed{\begin{aligned} &
 [ j +2 ]^{\{(R\pm2 ; R')\}\oplus T\{\cdots\}  \oplus 2(R ; R')}~,~ [ j -2 ]^{\{(R-2 ; R')\}\oplus T\{\cdots\}  \oplus (R ; R')} \\
 & [ j]^{
  \{(R+2 ; R'-2)\oplus 2(R-2 ; R')\oplus (R+2 ; R')\}  \oplus T\{\cdots\} \oplus 3 (R ; R') \oplus (R-2 ; R'-2) }\end{aligned}~}\\[5pt]
Q^3: \quad & \boxed{ \begin{aligned}
&[j+3]^{(R\pm1 ; R'\pm1)}~,~[j+1]^{ \{(R+3 ; R'-1)\oplus (R-3 ; R'\pm1)\oplus 3(R+1 ; R'-1)\}\atop \oplus T\{\cdots\} \oplus 2(R+1 ; R'+1)\oplus 4(R-1 ; R'-1)}~\\
&[j-1]^{ \{(R-3 ; R'\pm1)\oplus 2(R+1 ; R'-1)\} \oplus T\{\cdots\} \atop \oplus (R+1 ; R'+1)\oplus 3(R-1 ; R'-1)}~,~[j-3]^{(R-1 ; R'-1)}~
\end{aligned}}\\[5pt]
Q^4:  \quad & \boxed{\begin{aligned}
 &[ j +4 ]^{(R ; R')}~,~[ j + 2 ]^{\{2(R ;R'-2)\oplus(R+2 ; R')\oplus(R+2 ; R'-2)\} \atop \oplus T\{\cdots\} \oplus 3(R ; R')\oplus(R-2 ; R'-2)}~\\
 & [ j ]^{ \{(R-4 ; R')\oplus (R+2 ; R')\oplus(R+2 ; R'-2)\oplus 3(R ; R'-2)\}\atop \oplus T\{\cdots\} \oplus 4(R ; R')\oplus 2(R-2 ; R'-2)}~,~[j-2]^{(R ; R')\oplus(R-2 ; R'-2)\atop \{(R ; R'-2)\}\oplus T\{\cdots\}}~
 \end{aligned}}\\[5pt]
Q^5: \quad & \boxed{ \begin{aligned} & 
[j+ 3]^{\{(R+1 ; R'-1)\} \oplus T\{\cdots\} \oplus (R-1 ; R'-1)} \\
& [j+1]^{ \{(R\pm1 ; R'-3)\oplus 2(R+1 ; R'-1)\}\oplus T\{\cdots\} \oplus (R+1 ; R'+1)\oplus 3(R-1 ; R'-1) } \\
& [j-1]^{\{(R+1 ; R'-1)\oplus (R-1 ; R'-3)\} \oplus T\{\cdots\}\oplus 2(R-1; R'-1)}~\end{aligned}
}\\[5pt]
Q^6:\quad & \boxed{
  [j + 2 ]^{\{(R ; R'-2)\}\oplus T\{\cdots\}  \oplus (R ; R')}~,~[ j]^{
  \{(R ; R'-2)\} \oplus T\{\cdots\}  (R ; R') \oplus (R- 2 ; R'-2) }~}\\[5pt]
 Q^{7}: \quad & \boxed{[ j+1 ]^{(R-1 ;  R'-1)}~
}\\[5pt]
\end{align*}

\begin{align*}
{\xymatrix{*++[F-,]{\boldsymbol{ A_{2}}}}}  \quad &{\xymatrix @R=1.4pc@C=2pc{*++[F=]{[ 0 ]_\Delta^{(R ; R')}~,~\Delta =  \frac{1}{2}(R+R') + 1}}} \\[5pt]
Q: \quad & \boxed{[ 1 ]^{(R\pm1 ;  R'\pm1)}~
}\\[5pt]
Q^2: \quad & \boxed{
 [2]^{\{(R\pm2 ; R')\}\oplus T\{\cdots\}  \oplus 2(R ; R')}~,~[0]^{
  \{(R+2 ; R'-2)\oplus (R\pm2 ; R')\} \oplus T\{\cdots\} \atop \oplus 2 (R ; R') \oplus (R-2 ; R'-2) }~}\\[5pt]
Q^3: \quad & \boxed{
[3]^{(R\pm1 ; R'\pm1)}~,~[1]^{ \{(R+3 ; R'-1)\oplus (R-3 ; R'\pm1)\oplus 3(R+1 ; R'-1)\}\atop \oplus T\{\cdots\} \oplus 2(R+1 ; R'+1)\oplus 3(R-1 ; R'-1)}~}\\[5pt]
Q^4:  \quad & \boxed{\begin{aligned} &
 [  4 ]^{(R ; R')}~,~[2]^{\{2(R  ;  R'-2)\oplus(R+2 ; R')\oplus(R+2 ; R'-2)\}  \oplus T\{\cdots\} \oplus 3(R ; R')\oplus(R-2 ; R'-2)} \\
& [0]^{ \{(R-4 ; R')\oplus (R+2 ; R')\oplus(R+2 ; R'-2)\oplus 2(R ; R'-2)\} \oplus T\{\cdots\} \oplus 3(R ; R')\oplus (R-2 ; R'-2)}~\end{aligned}
 }\\[5pt]
Q^5: \quad & \boxed{ 
[3]^{\{(R+1 ; R'-1)\} \oplus T\{\cdots\} \oplus (R-1 ;R'-1)}~,~[1]^{ \{(R\pm1 ; R'-3)\oplus 2(R+1 ; R'-1)\}\oplus T\{\cdots\}\atop \oplus (R+1 ; R'+1)\oplus 3(R-1 ; R'-1) }~
}\\[5pt]
Q^6:\quad & \boxed{
  [ 2 ]^{\{(R ; R'-2)\}\oplus T\{\cdots\}  \oplus (R ; R')}~,~[0]^{
  \{(R ; R'-2)\} \oplus T\{\cdots\}  (R ; R') \oplus (R- 2 ; R'-2) }~}\\[5pt]
 Q^{7}: \quad & \boxed{[1]^{(R-1 ;  R'-1)}~
}\\[40pt]
{\xymatrix{*++[F-,]{\boldsymbol{ B_{1}}}}}  \quad &{\xymatrix @R=1.4pc@C=2pc{*++[F=]{[ 0 ]_\Delta^{(R ; R')}~,~\Delta =  \frac{1}{2}(R+R') }}} \\[5pt]
Q: \quad & \boxed{[1]^{\{(R+1 ;  R'-1)\}\oplus T\{\cdots\} \oplus (R-1 ; R'-1)}~
}\\[5pt]
Q^2: \quad & \boxed{
 [2]^{\{(R ; R'-2)\}\oplus T\{\cdots\}  \oplus (R ; R')}~,~[0]^{
  \{(R+2 ; R'-2)\oplus (R ; R'-2)\} \oplus T\{\cdots\} \oplus  (R ; R') \oplus (R-2 ; R'-2) }~}\\[5pt]
Q^3: \quad & \boxed{
[3]^{(R-1 ; R'-1)}~,~[1]^{ \{ (R-3 ; R'\pm1)\oplus (R+1 ; R'-1)\} \oplus T\{\cdots\} \oplus 2(R-1 ; R'-1)}~}\\[5pt]
Q^4:  \quad & \boxed{
[2]^{\{(R  ;  R'-2)\oplus T\{\cdots\} \oplus(R-2 ; R'-2)}~,~[0]^{ \{(R ; R'-4)\oplus (R ; R'-2)\} \oplus T\{\cdots\}  \oplus (R ; R')\oplus (R-2 ; R'-2)}~
 }\\[5pt]
Q^5: \quad & \boxed{ 
[1]^{ \{(R-1 ; R'-3)\}\oplus T\{\cdots\} \oplus (R-1 ; R'-1) }~
}\\[5pt]
Q^6:\quad & \boxed{
 [0]^{
  (R- 2 ; R'-2) }~}
\end{align*}

\subsection{$d = 4, ~\CN=1$}
\label{sec:4dn1tabs}

The unitary superconformal multiplets are summarized in table~\ref{tab:4DN1CAC}.  We present their operator content for generic values of their non-vanishing~$\frak{su}(2) \oplus \b {\frak{su}(2)}$ Lorentz Dynkin labels. We condense the notation by declaring any~$\pm$ offsets for the two~$\frak{su}(2)$'s to be independent, e.g.~$[j\pm1; \b j \pm 1]$ denotes four operators.  

\begin{align*}
\xymatrix{
*++[F-,]{{\boldsymbol{ L \b L}}}} & \xymatrix @C=7pc @R=7pc @!0 @dr {
*++[F=]{[ j;\b j]^{(r)}_\Delta~,~\Delta > \max\left\{2 +  j - {3 \over 2} r,~ 2+ \b j + {3 \over 2} r \right\}} \ar[r]|--{{~\b Q~}} \ar[d]|--{~Q~} 
& *++[F]{[j ; \b j \pm 1]^{(r+1)}} \ar[r]|--{{~\b Q~}} \ar[d]|--{~Q~} 
& *++[F]{[ j ; \b j]^{(r+2)}} \ar[d]|--{~Q~}\\
*++[F]{[j \pm 1; \b j]^{(r-1)}} \ar[r]|--{{~\b Q~}} \ar[d]|--{~Q~} 
& *++[F]{ [ j\pm 1; \b j \pm 1]^{(r)}}\ar[r]|--{{~\b Q~}}\ar[d]|--{~Q~}
 & *++[F]{[ j \pm 1; \jbar ]^{(r+1)}} \ar[d]|--{~Q~}\\
*++[F]{[ j; \b j]^{(r-2)}} \ar[r]|--{{~\b Q~}}
& *++[F]{[j; \b j \pm 1]^{(r-1)}} \ar[r]|--{{~\b Q~}}  
& *++[F]{[j; \jbar]^{(r)}} 
} \\[20pt]
\xymatrix{
*++[F-,]{{\boldsymbol{ L \b A_{\bar{\ell}}}}}} & \xymatrix @C=7pc @R=7pc @!0 @dr {
*++[F=]{[j; \b j]_\Delta^{(r)}~,~\Delta = 2 + \b j + {3 \over 2} r~,~r > {1 \over3} \left(j - \b j\right)} \ar[r]|--{{~\b Q~}} \ar[d]|--{~Q~} 
& *++[F]{[j; \b j + 1]^{(r+1)}}  
\ar[d]|--{~Q~}
\\
*++[F]{[j \pm 1; \b j]^{(r-1)}} 
\ar[r]|--{~\b Q~}
\ar[d]|--{~Q~} 
& *++[F]{[j\pm1; \b j + 1]^{(r)}}\ar[d]|--{~Q~}
\\
*++[F]{[j ;\b j]^{(r-2)}} \ar[r]|--{{~\b Q~}}
& *++[F]{[ j; \b j + 1\,]^{(r-1)}} 
}
\end{align*}
\noindent In the preceding table~$\b \ell = 1$ if~$\b j \geq 1$ and~$\b \ell = 2$ if~$\b j = 0$.

\begin{align*}
\xymatrix{
*++[F-,]{{\boldsymbol{ L \b B_1}}}} & \xymatrix @C=7pc @R=7pc @!0 @dr {
*++[F=]{[j; 0]_\Delta^{(r)}~,~\Delta = {3 \over 2} r ~,~r > {1 \over3} \left(j+2\right) }  \ar[d]|--{~Q~} \\
*++[F]{[j \pm 1; 0]^{(r-1)}} 
\ar[d]|--{~Q~} 
\\
*++[F]{[j; 0]^{(r-2)}}
}
\end{align*}

\begin{align*}
\xymatrix{
*++[F-,]{{\boldsymbol{ A_{\ell} \b A_{\bar{\ell}}}}}} & ~~~~~~~~~~\xymatrix @C=7pc @R=7pc @!0 @dr {
*++[F=]{[j; \b j]^{(r)}_\Delta~,~\Delta = {1 \over 2}\left(j + \jbar\right) + 2~,~r = {1 \over3} \left(j - \jbar\right)} \ar[r]|--{{~\b Q~}} \ar[d]|--{~Q~} 
& *++[F]{[j; \b j + 1]^{(r+1)}}  
\ar[d]|--{~Q~}
\\
*++[F]{[j + 1; \b j]^{(r-1)}} 
\ar[r]|--{~\b Q~}
& *++[F]{[j+1; \b j + 1]^{(r)}}
}
\end{align*}
In the preceding table~$\ell = 1$ if~$j \geq 1$ and~$\ell = 2$ if~$j = 0$, and similarly for~$\b \ell$ and~$\b j$.

\begin{align*}
\xymatrix{
*++[F-,]{{\boldsymbol{ A_{\ell} \b B_1}}}} &~~~~~~~~~~ \xymatrix @C=7pc @R=7pc @!0 @dr {
*++[F=]{[j; 0]^{(r)}_\Delta~,~\Delta = {3 \over 2} r ~,~r = {1 \over3} \left(j+2\right) }  \ar[d]|--{~Q~} \\
*++[F]{[j + 1; 0]^{(r-1)}} 
}
\end{align*}
In the preceding table~$\ell = 1$ if~$j \geq 1$ and~$\ell = 2$ if~$j = 0$. 

\begin{align*}
\xymatrix{
*++[F-,]{{\boldsymbol{ B_1 \b B_1}}}} & ~~~~~~~~~~\xymatrix @C=7pc @R=7pc @!0 @dr {
*++[F=]{[0; 0]^{(0)}_0}  
}
\end{align*}

\subsection{$d=4$, $\mathcal{N}=2$}
\label{sec:4dn2tabs}

The unitary superconformal multiplets are summarized in table~\ref{tab:4DN2CAC}.  We present their operator content for generic values of their non-vanishing~$\frak{su}(2) \oplus \b {\frak{su}(2)}$ Lorentz and~$\frak{su}(2)_R$ Dynkin labels (see~\cite{Dolan:2002zh}; our conventions are related via~\eqref{doourconv}). We condense the notation by declaring any~$\pm$ offsets for the three different~$\frak{su}(2)$'s to be independent, e.g.~$[j\pm1; \b j\pm 1]^{(R \pm 2)}$ denotes eight operators. We also do not indicate the~$\frak{u}(1)_R$ charges of the operators, since these are easily determined by the left and right levels~$\ell, \b \ell$ and the~$\frak{u}(1)_R$ charge of the SCP. 
\begin{align*}
\xymatrix{
*++[F-,]{{\boldsymbol{ L \b L}}}} & \hskip-30pt
\xymatrix @C=6.6pc @R=6.6pc @!0 @dr {
*++[F=]{ [j; \b j]^{(R;r)}_{\Delta}~,~\Delta>2+R+\max\big\{j-\frac{1}{2}r,\overline{j}+\frac{1}{2}r\big\}} 
\ar[d]|--{~Q~} \ar[r]|--{{~\b Q~}}
& *+[F]{\scriptstyle [j;\b j\pm1]^{(R \pm 1)} } 
\ar[d]|--{~Q~} \ar[r]|--{{~\b Q~}}
& *+[F]{ \scriptstyle [j;\b j\pm2]^{(R)} \oplus [j; \b j]^{(R\pm 2) \oplus 2(R)} }\ar[d]|--{~Q~} \ar[r]|--{{~\b Q~}} 
& *+[F]{ \scriptstyle [j;\b j\pm1]^{(R \pm 1)} }  \ar[d]|--{{~ Q~}} \ar[r]|--{{~\b Q~}} 
& *+[F]{\scriptstyle [j;\b j]^{(R)}}\ar[d]|--{{~ Q~}} 
\\
*+[F]{ \scriptstyle [j\pm1;\b j]^{(R\pm 1)}} \ar[d]|--{~Q~} \ar[r]|--{{~\b Q~}} & *+[F]{ \scriptstyle [j\pm1;\b j\pm1]^{(R \pm 2) \oplus 2(R) } } \ar[d]|--{~Q~} \ar[r]|--{{~\b Q~}} 
& *[F]{ \begin{aligned} &~ \scriptstyle [j \pm 1; \b j \pm 2 ]^{(R \pm 1)} \\[-8pt] 
&~ \scriptstyle [j \pm1 ; \b j]^{(R \pm 3) \oplus 3(R \pm 1)}~ \end{aligned}}  \ar[d]|--{~Q~} \ar[r]|--{~\b Q~}
&  *+[F]{ \scriptstyle [j \pm 1; \b j \pm 1]^{(R \pm 2) \oplus 2(R)} }\ar[d]|--{~Q~} \ar[r]|--{~\b Q~} & *+[F]{ \scriptstyle [j\pm1;\b j]^{(R\pm 1)}} \ar[d]|--{~Q~} 
\\
*+[F]{\scriptstyle [j; \b j]^{(R \pm 2) \oplus 2(R)} \oplus [j\pm2; \b j]^{(R)}} \ar[r]|--{{~\b Q~}} \ar[d]|--{~Q~} 
& *[F]{ \begin{aligned} &~ \scriptstyle [j \pm 2; \b j \pm 1 ]^{(R \pm 1)} \\[-8pt] 
&~ \scriptstyle [j; \b j \pm1 ]^{(R \pm 3) \oplus 3(R \pm 1)}~ \end{aligned}}  \ar[d]|--{~Q~} \ar[r]|--{{~\b Q~}} 
& *[F]{ \begin{aligned} &~ \scriptstyle [j \pm 2; \b j \pm 2]^{(R)} \\[-8pt] 
&~ \scriptstyle \left\{ [j \pm 2; \b j] \oplus [j ; \b j\pm 2]\right\}^{(R \pm 2) \oplus 2(R)}~ \\[-8pt] 
& ~\scriptstyle [j; \b j]^{(R \pm 4) \oplus 4(R \pm 2) \oplus 6(R)} ~ \end{aligned}}  \ar[d]|--{~Q~} \ar[r]|--{{~\b Q~}} 
& *[F]{ \begin{aligned} &~ \scriptstyle [j \pm 2; \b j \pm 1 ]^{(R \pm 1)} \\[-8pt] 
&~ \scriptstyle [j; \b j \pm1 ]^{(R \pm 3) \oplus 3(R \pm 1)}~ \end{aligned}} \ar[d]|--{~Q~} \ar[r]|--{~\b Q~} & 
*+[F]{ \scriptstyle [j; \b j]^{(R \pm 2) \oplus 2(R)} \oplus [j\pm2; \b j]^{(R)}} \ar[d]|--{~Q~}
\\
*+[F]{ \scriptstyle [j\pm1;\b j]^{(R\pm 1)}} \ar[d]|--{~Q~} \ar[r]|--{{~\b Q~}} & *+[F]{ \scriptstyle [j\pm1;\b j\pm1]^{(R \pm 2) \oplus 2(R) } } \ar[d]|--{~Q~} \ar[r]|--{{~\b Q~}} 
& *[F]{ \begin{aligned} &~ \scriptstyle [j \pm 1; \b j \pm 2 ]^{(R \pm 1)} \\[-8pt] 
&~ \scriptstyle [j \pm1 ; \b j]^{(R \pm 3) \oplus 3(R \pm 1)}~ \end{aligned}}  \ar[d]|--{~Q~} \ar[r]|--{~\b Q~}
&  *+[F]{ \scriptstyle [j \pm 1; \b j \pm 1]^{(R \pm 2) \oplus 2(R)} }\ar[d]|--{~Q~} \ar[r]|--{~\b Q~} & *+[F]{ \scriptstyle [j\pm1;\b j]^{(R\pm 1)}} \ar[d]|--{~Q~} 
\\
*+[F]{\scriptstyle  [j;\b j]^{(R)}}  \ar[r]|--{{~\b Q~}}  & *+[F]{\scriptstyle [j;\b j\pm1]^{(R \pm 1)} } 
\ar[r]|--{{~\b Q~}}
& *+[F]{ \scriptstyle [j;\b j\pm2]^{(R)} \oplus [j; \b j]^{(R\pm 2) \oplus 2(R)} } \ar[r]|--{{~\b Q~}} 
& *+[F]{ \scriptstyle [j;\b j\pm1]^{(R \pm 1)} }   \ar[r]|--{{~\b Q~}} 
& *+[F]{\scriptstyle [j;\b j]^{(R)}}
}
\end{align*}

\begin{align*}
\xymatrix{
*++[F-,]{{\boldsymbol{ L \b A_{1} }}}} & \hskip-40pt
\xymatrix @C=7.6pc @R=7.6pc @!0 @dr {
*++[F=]{  [j; \b j]^{(R;r)}_{\Delta}~,~\Delta=2+R+\overline{j}+\frac{1}{2}r~,~ r > j - \b j} 
\ar[d]|--{~Q~} \ar[r]|--{{~\b Q~}}
& *+[F]{\scriptstyle  [j;\b j-1]^{(R - 1)} \oplus[j;\b j+1]^{(R \pm 1)} } 
\ar[d]|--{~Q~} \ar[r]|--{{~\b Q~}}
& *+[F]{ \scriptstyle [j;\b j+2]^{(R)} \oplus [j; \b j]^{(R) \oplus (R-2)} }\ar[d]|--{~Q~} \ar[r]|--{{~\b Q~}} 
& *+[F]{ \scriptstyle [j;\b j+1]^{(R - 1)} }  \ar[d]|--{{~ Q~}}  
\\
*+[F]{ \scriptstyle [j\pm1;\b j]^{(R\pm 1)}} \ar[d]|--{~Q~} \ar[r]|--{{~\b Q~}} & *[F]{  \begin{aligned} &~ \scriptstyle [j \pm 1; \b j - 1 ]^{(R) \oplus (R-2)} \\[-8pt] 
&~ \scriptstyle [j \pm 1; \b j + 1 ]^{(R \pm 2) \oplus 2(R)}~ \end{aligned} } \ar[d]|--{~Q~} \ar[r]|--{{~\b Q~}} 
& *[F]{ \begin{aligned} &~ \scriptstyle [j \pm 1; \b j + 2 ]^{(R \pm 1)} \\[-8pt] 
&~ \scriptstyle [j \pm1 ; \b j]^{(R+1) \oplus 2(R-1) \oplus (R-3)}~ \end{aligned}}  \ar[d]|--{~Q~} \ar[r]|--{~\b Q~}
&  *+[F]{ \scriptstyle [j \pm 1; \b j + 1]^{(R) \oplus (R-2)} }\ar[d]|--{~Q~} 
\\
*+[F]{\scriptstyle [j; \b j]^{(R \pm 2) \oplus 2(R)} \oplus [j\pm2; \b j]^{(R)}} \ar[r]|--{{~\b Q~}} \ar[d]|--{~Q~} 
& *[F]{ \begin{aligned} &~ \scriptstyle [j \pm 2; \b j - 1 ]^{(R - 1)} , \, [j \pm 2; \b j + 1 ]^{(R \pm 1)} ~ \\[-8pt] 
&~ \scriptstyle [j; \b j - 1 ]^{(R + 1) \oplus 2(R-1) \oplus (R-3)}~ \\[-8pt]
&~ \scriptstyle [j; \b j + 1 ]^{(R \pm 3) \oplus 3(R \pm 1)}~ \end{aligned}}  \ar[d]|--{~Q~} \ar[r]|--{{~\b Q~}} 
& *[F]{ \begin{aligned} &~ \scriptstyle [j \pm 2; \b j + 2]^{(R)} \oplus  [j \pm 2; \b j]^{ (R) \oplus (R-2)}~ \\[-8pt]
&~\scriptstyle  [j ; \b j\pm 2]^{(R \pm 2) \oplus 2(R)} \\[-8pt] 
& ~\scriptstyle [j; \b j]^{(R+2) \oplus 3(R) \oplus 3(R-2) \oplus (R-4)} ~ \end{aligned}}  \ar[d]|--{~Q~} \ar[r]|--{{~\b Q~}} 
& *[F]{ \begin{aligned} &~ \scriptstyle [j \pm 2; \b j + 1 ]^{(R - 1)} \\[-8pt] 
&~ \scriptstyle [j; \b j + 1 ]^{(R+1) \oplus 2(R-1) \oplus (R-3)}~ \end{aligned}} \ar[d]|--{~Q~} 
\\
*+[F]{ \scriptstyle [j\pm1;\b j]^{(R\pm 1)}} \ar[d]|--{~Q~} \ar[r]|--{{~\b Q~}} & *[F]{  \begin{aligned} &~ \scriptstyle [j \pm 1; \b j - 1 ]^{(R) \oplus (R-2)} \\[-8pt] 
&~ \scriptstyle [j \pm 1; \b j + 1 ]^{(R \pm 2) \oplus 2(R)}~ \end{aligned} } \ar[d]|--{~Q~} \ar[r]|--{{~\b Q~}} 
& *[F]{ \begin{aligned} &~ \scriptstyle [j \pm 1; \b j + 2 ]^{(R \pm 1)} \\[-8pt] 
&~ \scriptstyle [j \pm1 ; \b j]^{(R+1) \oplus 2(R-1) \oplus (R-3)}~ \end{aligned}}  \ar[d]|--{~Q~} \ar[r]|--{~\b Q~}
&  *+[F]{ \scriptstyle [j \pm 1; \b j + 1]^{(R) \oplus (R-2)} }\ar[d]|--{~Q~} 
\\
*+[F]{\scriptstyle  [j;\b j]^{(R)}}  \ar[r]|--{{~\b Q~}}  & *+[F]{\scriptstyle  [j;\b j-1]^{(R - 1)} \oplus[j;\b j+1]^{(R \pm 1)} } \ar[r]|--{{~\b Q~}}
& *+[F]{ \scriptstyle [j;\b j+2]^{(R)} \oplus [j; \b j]^{(R) \oplus (R-2)} }  \ar[r]|--{{~\b Q~}} 
& *+[F]{ \scriptstyle [j;\b j+1]^{(R - 1)} } 
}
\end{align*}

\begin{align*}
\xymatrix{
*++[F-,]{{\boldsymbol{ L \b A_{2} }}}} & \hskip-30pt
\xymatrix @C=7.4pc @R=7.4pc @!0 @dr {
*++[F=]{  [j; 0]^{(R;r)}_{\Delta}~,~\Delta=2+R+\frac{1}{2}r~,~ r > j} 
\ar[d]|--{~Q~} \ar[r]|--{{~\b Q~}}
& *+[F]{\scriptstyle  [j; 1]^{(R \pm 1)} } 
\ar[d]|--{~Q~} \ar[r]|--{{~\b Q~}}
& *+[F]{ \scriptstyle [j; 2]^{(R)} \oplus [j; 0]^{(R) \oplus (R-2)} }\ar[d]|--{~Q~} \ar[r]|--{{~\b Q~}} 
& *+[F]{ \scriptstyle [j; 1]^{(R - 1)} }  \ar[d]|--{{~ Q~}}  
\\
*+[F]{ \scriptstyle [j\pm1;0]^{(R\pm 1)}} \ar[d]|--{~Q~} \ar[r]|--{{~\b Q~}} & *+[F]{  \scriptstyle [j \pm 1;  1 ]^{(R \pm 2) \oplus 2(R)}~  } \ar[d]|--{~Q~} \ar[r]|--{{~\b Q~}} 
& *[F]{ \begin{aligned} &~ \scriptstyle [j \pm 1;  2 ]^{(R \pm 1)} \\[-8pt] 
&~ \scriptstyle [j \pm1 ; 0]^{(R+1) \oplus 2(R-1) \oplus (R-3)}~ \end{aligned}}  \ar[d]|--{~Q~} \ar[r]|--{~\b Q~}
&  *+[F]{ \scriptstyle [j \pm 1; 1]^{(R) \oplus (R-2)} }\ar[d]|--{~Q~} 
\\
*+[F]{\scriptstyle [j; 0]^{(R \pm 2) \oplus 2(R)} \oplus [j\pm2; 0]^{(R)}} \ar[r]|--{{~\b Q~}} \ar[d]|--{~Q~} 
& *[F]{ \begin{aligned} &~ \scriptstyle [j \pm 2; 1 ]^{(R \pm 1)} ~ \\[-8pt] 
&~ \scriptstyle [j; 1 ]^{(R \pm 3) \oplus 3(R \pm 1)}~ \end{aligned}}  \ar[d]|--{~Q~} \ar[r]|--{{~\b Q~}} 
& *[F]{ \begin{aligned} &~ \scriptstyle [j \pm 2; 2]^{(R)} \oplus  [j \pm 2; 0]^{ (R) \oplus (R-2)}~ \\[-8pt]
&~\scriptstyle  [j ; 2]^{(R \pm 2) \oplus 2(R)} \\[-8pt] 
& ~\scriptstyle [j; 0]^{(R) \oplus 2(R-2) \oplus (R-4)} ~ \end{aligned}}  \ar[d]|--{~Q~} \ar[r]|--{{~\b Q~}} 
& *[F]{ \begin{aligned} &~ \scriptstyle [j \pm 2; 1 ]^{(R - 1)} \\[-8pt] 
&~ \scriptstyle [j; 1 ]^{(R+1) \oplus 2(R-1) \oplus (R-3)}~ \end{aligned}} \ar[d]|--{~Q~} 
\\
*+[F]{ \scriptstyle [j\pm1;0]^{(R\pm 1)}} \ar[d]|--{~Q~} \ar[r]|--{{~\b Q~}} & *+[F]{  \scriptstyle [j \pm 1;  1 ]^{(R \pm 2) \oplus 2(R)}~  } \ar[d]|--{~Q~} \ar[r]|--{{~\b Q~}} 
& *[F]{ \begin{aligned} &~ \scriptstyle [j \pm 1;  2 ]^{(R \pm 1)} \\[-8pt] 
&~ \scriptstyle [j \pm1 ; 0]^{(R+1) \oplus 2(R-1) \oplus (R-3)}~ \end{aligned}}  \ar[d]|--{~Q~} \ar[r]|--{~\b Q~}
&  *+[F]{ \scriptstyle [j \pm 1; 1]^{(R) \oplus (R-2)} }\ar[d]|--{~Q~} 
\\
*+[F]{\scriptstyle  [j; 0]^{(R)}}  \ar[r]|--{{~\b Q~}}  & *+[F]{\scriptstyle [j; 1]^{(R \pm 1)} } \ar[r]|--{{~\b Q~}}
& *+[F]{ \scriptstyle [j; 2]^{(R)} \oplus [j; 0]^{(R) \oplus (R-2)} }  \ar[r]|--{{~\b Q~}} 
& *+[F]{ \scriptstyle [j;1]^{(R - 1)} } 
}
\end{align*}

\begin{align*}
\xymatrix{
*++[F-,]{{\boldsymbol{ L \b B_1}}}} & \hskip-30pt \xymatrix @C=8pc @R=8pc @!0 @dr {
*++[F=]{[ j ; 0]^{(R ; r)}_\Delta~,~\Delta = R + \half r~,~r > j + 2} \ar[r]|--{{~\b Q~}} \ar[d]|--{~Q~} 
& *++[F]{ [ j ;1]^{(R-1)}} \ar[r]|--{{~\b Q~}} \ar[d]|--{~Q~}& *++[F]{[ j ; 0]^{(R-2)}} \ar[d]|--{~Q~}\\
*++[F]{ [ j\pm 1;0]^{(R \pm1)} } \ar[r]|--{{~\b Q~}}\ar[d]|--{~Q~} 
& *++[F]{  [ j\pm1;1]^{(R) \oplus (R-2)} }\ar[r]|--{{~\b Q~}}\ar[d]|--{~Q~}
 & *++[F]{  [ j\pm1;0]^{(R-1) \oplus (R-3)} } \ar[d]|--{~Q~}\\
*+[F]{\begin{aligned} & [ j \pm 2; 0]^{(R)} \\[-2pt]
& [ j; 0]^{(R \pm 2) \oplus 2(R)}
\end{aligned}} \ar[r]|--{{~\b Q~}} \ar[d]|--{~Q~} 
& *+[F]{\begin{aligned} & [ j \pm 2 ; 1]^{(R-1)} \\[-2pt]
& [ j ; 1]^{(R +1) \oplus 2(R-1) \oplus (R-3)}
\end{aligned}} \ar[r]|--{{~\b Q~}} \ar[d]|--{~Q~} 
& *+[F]{\begin{aligned} & [ j \pm 2; 0]^{(R-2)} \\[-2pt]
& [ j; 0]^{(R) \oplus 2(R-2) \oplus (R-4)}
\end{aligned}} \ar[d]|--{~Q~} \\
*++[F]{ [ j\pm 1;0]^{(R \pm1)} } \ar[r]|--{{~\b Q~}}\ar[d]|--{~Q~} 
& *++[F]{  [ j\pm1;1]^{(R) \oplus (R-2)} }\ar[r]|--{{~\b Q~}}\ar[d]|--{~Q~}
 & *++[F]{  [ j\pm1;0]^{(R-1) \oplus (R-3)} } \ar[d]|--{~Q~}\\
*++[F]{[ j ; 0]^{(R)}} \ar[r]|--{{~\b Q~}}
& *++[F]{ [ j;1]^{(R-1)}} \ar[r]|--{{~\b Q~}} & *++[F]{[ j;0]^{(R-2)}} 
}
\end{align*}

\begin{align*}
\xymatrix{
*++[F-,]{{\boldsymbol{ A_\ell \b A_{\b \ell}}}}}  & \hskip-60pt
\xymatrix @C=9.1pc @R=9.1pc @!0 @dr {
*++[F=]{[ j; \b j]^{(R; r)}_\Delta~,~\Delta = 2 + R + \half \left(j + \b j \right)~,~r = j - \b j} \ar[d]|--{~Q~} \ar[r]|--{{~\b Q~}} 
& *+[F]{\begin{aligned} & \scriptstyle [ j ; \b j+ 1]^{(R \pm 1)}~\\[-6pt] 
& \scriptstyle [ j ; \b j-1]^{(R-1)}
\end{aligned}} 
\ar[d]|--{~Q~} \ar[r]|--{{~\b Q~}}
& *+[F]{\begin{aligned} & \scriptstyle [ j ; \b j + 2]^{(R)}~\\[-6pt]
& \scriptstyle [  j ; \b j]^{(R) \oplus (R-2)}
~\end{aligned}} \ar[d]|--{~Q~} \ar[r]|--{{~\b Q~}} 
& *+[F]{\scriptstyle [ j ; \b j+1]^{(R-1)}} \ar[d]|--{~Q~} 
\\
*+[F]{\begin{aligned} & \scriptstyle [ j+1 ; \b j ]^{(R \pm 1)}~\\[-6pt]
& \scriptstyle [ j-1 ; \b j]^{(R-1)}
\end{aligned}} \ar[d]|--{~Q~} \ar[r]|--{{~\b Q~}} & *+[F]{\begin{aligned} & \scriptstyle [ j+1; \b j+1 ]^{(R \pm 2) \oplus 2(R)}~\\[-6pt]
& \scriptstyle \left\{ [ j+1; \b j -1 ] \oplus [ j-1; \b j +1 ] \right\}^{(R) \oplus (R-2)}~\\[-6pt]
& \scriptstyle [ j-1; \b j-1 ]^{(R-2)}\end{aligned}} \ar[d]|--{~Q~} \ar[r]|--{{~\b Q~}} 
& *+[F]{ \begin{aligned} & \scriptstyle [ j+1; \b j+2]^{(R \pm 1)}~\\[-6pt]
& \scriptstyle [ j+1; \b j]^{(R+1) \oplus 2(R-1) \oplus (R-3)}~\\[-6pt]
& \scriptstyle [ j-1; \b j+2 ]^{(R-1)}~\\[-6pt]
& \scriptstyle [ j-1; \b j]^{(R-1) \oplus (R-3)}
\end{aligned}}  \ar[d]|--{~Q~} \ar[r]|--{{~\b Q~}} 
& *+[F]{\begin{aligned} & \scriptstyle [ j+1; \b j+1]^{(R) \oplus (R-2)}~\\[-6pt]
&\scriptstyle  [ j-1; \b j +1]^{(R-2)}
\end{aligned}} \ar[d]|--{~Q~}
\\
*+[F]{\begin{aligned} & \scriptstyle [ j+2; \b j]^{(R)}~\\[-6pt]
& \scriptstyle [ j; \b j]^{(R) \oplus (R-2)}~
\end{aligned}} \ar[r]|--{{~\b Q~}} \ar[d]|--{~Q~} 
& *+[F]{\begin{aligned} & \scriptstyle [ j+2; \b j+1 ]^{(R\pm1)}~\\[-6pt]
& \scriptstyle [ j+2; \b j-1 ]^{(R-1)}~\\[-6pt] 
& \scriptstyle [j; \b j+1 ]^{(R+1) \oplus 2(R-1) \oplus (R-3)}~\\[-6pt]
& \scriptstyle [ j; \b j-1]^{(R-1) \oplus (R-3)}
\end{aligned}} \ar[d]|--{~Q~}\ar[r]|--{{~\b Q~}} 
& *+[F]{\begin{aligned} & \scriptstyle [ j+2; \b j+2]^{(R)}~\\[-6pt]
& \scriptstyle [ j+2; \b j]^{(R) \oplus (R-2)}~\\[-6pt]
& \scriptstyle [ j; \b j+2 ]^{(R) \oplus (R-2)}~\\[-6pt]
& \scriptstyle [ j; \b j]^{(R) \oplus 2(R-2) \oplus (R-4)}~
\end{aligned}} \ar[d]|--{~Q~} \ar[r]|--{{~\b Q~}} 
& *+[F]{\begin{aligned} & \scriptstyle [ j+2; \b j+1]^{(R-1)}~\\[-6pt]
& \scriptstyle [ j; \b j+1]^{(R-1) \oplus (R-3)}~
\end{aligned}} \ar[d]|--{~Q~}
\\
*+[F]{\scriptstyle [ j+1; \b j]^{(R-1)}} \ar[r]|--{{~\b Q~}} 
& *+[F]{\begin{aligned} & \scriptstyle [ j+1;
\b j + 1]^{(R) \oplus (R-2)}~\\[-6pt] 
& \scriptstyle [ j+1; \b j-1]^{(R-2)~}
\end{aligned}} \ar[r]|--{{~\b Q~}}
& *+[F]{\begin{aligned} & \scriptstyle [ j+1; \b j + 2]^{(R-1)}~\\[-6pt]
& \scriptstyle [  j+1; \b j]^{(R-1) \oplus (R-3)}
~\end{aligned}} \ar[r]|--{{~\b Q~}} 
& *+[F]{\scriptstyle [j+1; \b j+1]^{(R-2)}} 
}
\end{align*}

\medskip

\noindent In the preceding table~$\ell = \b \ell = 1$ if ~$j, \b j \geq 1$, while~$\ell = 2$ if~$j = 0$ and~$\b \ell = 2$ if~$\b j = 0$. In the latter two cases, we must delete all operators of the form~$[j-1; \ldots]$ and~$[\ldots;\b j-1]$, respectively. 
\vskip-2cm
\begin{align*}
\xymatrix{
*++[F-,]{{\boldsymbol{ A_\ell \b B_1}}}} & \hskip-30pt \xymatrix @C=7.2pc @R=7.2pc @!0 @dr {
*++[F=]{[j; 0]^{(R;r)}_\Delta~,~\Delta = R + \half r~,~r = j + 2} \ar[r]|--{{~\b Q~}} \ar[d]|--{~Q~} 
& *+[F]{ [j;1]^{(R-1)}} \ar[r]|--{{~\b Q~}} \ar[d]|--{~Q~}& *+[F]{[j;0]^{(R-2)}} \ar[d]|--{~Q~}\\
*+[F]{\begin{aligned} & [j+1;0]^{(R \pm1)}~\\[-2pt]
& [j-1;0]^{(R - 1)} 
\end{aligned}} \ar[r]|--{{~\b Q~}}\ar[d]|--{~Q~} 
& *+[F]{\begin{aligned} & [j+1;1]^{(R) \oplus (R-2)}~\\[-2pt]
& [j-1;1]^{(R-2)} 
\end{aligned}}\ar[r]|--{{~\b Q~}}\ar[d]|--{~Q~}
 & *+[F]{\begin{aligned} & [j+1;0]^{(R-1) \oplus (R-3)}~\\[-2pt]
& [j-1;0]^{(R-3)}
\end{aligned}} \ar[d]|--{~Q~}\\
*+[F]{\begin{aligned} & [j + 2; 0]^{(R)}~\\[-2pt]
& [j; 0]^{(R) \oplus (R-2)}
\end{aligned}} \ar[r]|--{{~\b Q~}} \ar[d]|--{~Q~} 
& *+[F]{\begin{aligned} & [j + 2;1]^{(R-1)}~\\[-2pt]
& [j; 1]^{(R-1) \oplus (R-3)}
\end{aligned}} \ar[r]|--{{~\b Q~}} \ar[d]|--{~Q~} 
& *+[F]{\begin{aligned} & [j + 2; 0]^{(R-2)}~\\[-2pt]
& [j; 0]^{(R-2) \oplus (R-4)}
\end{aligned}} \ar[d]|--{~Q~} \\
*+[F]{[j+1;0]^{(R-1)}} \ar[r]|--{{~\b Q~}}
& *+[F]{ [j+1;1]^{(R-2)}}\ar[r]|--{{~\b Q~}} 
 & *+[F]{[j+1;0]^{(R-3)}} 
}
\end{align*}
\vskip-5pt \noindent In the preceding table~$\ell = 1$ if~$j \geq 1$. If~$j = 0$, then~$\ell = 2$ and we drop all operators~$[j-1;\cdots]$. 
\vskip-10pt 
\begin{align*}
\xymatrix{
*++[F-,]{{\boldsymbol{ B_1 \b B_1}}}} & 
\xymatrix @C=6pc @R=6pc @!0 @dr {
*++[F=]{[0; 0]^{(R; 0)}_\Delta~,~\Delta = R} \ar[r]|--{{~\b Q~}} \ar[d]|--{~Q~} & *++[F]{ [0;1]^{(R-1)}} \ar[r]|--{{~\b Q~}} \ar[d]|--{~Q~}& *++[F]{[0;0]^{(R-2)}} \ar[d]|--{~Q~}\\
*++[F]{[1;0]^{(R-1)}} \ar[r]|--{{~\b Q~}}\ar[d]|--{~Q~} & *++[F]{[1;1]^{(R-2)}}\ar[r]|--{{~\b Q~}}\ar[d]|--{~Q~} & *++[F]{ [1;0]^{(R-3)}} \ar[d]|--{~Q~}\\
*++[F]{[0;0]^{(R-2)}} \ar[r]|--{{~\b Q~}} & *++[F]{[0;1]^{(R-3)}} \ar[r]|--{{~\b Q~}}& *++[F]{ [0;0]^{(R-4)}} 
}
\end{align*}

\subsection{$d=5,$ $\mathcal{N}=1$}
\label{sec:5dn1tabs}

The unitary superconformal multiplets are summarized in table~\ref{tab:5Dmult}.  We present their operator content for generic values of their non-vanishing~$\frak{sp}(4)$ Lorentz Dynkin labels~$j_1, j_2$ and~$\frak{su}(2)_R$ Dynkin label~$R$. We condense the notation by declaring that~$\pm$ offsets for the Lorentz and the~$R$-symmetry are independent. However, the offsets for the two Lorentz Dynkin labels~$j_1, j_2$ are correlated. This means that~$[j_1 \pm1, j_2]^{(R \pm 1)}$ denotes four operators, while~$[j_1 \pm 1, j_2 \mp 1]$ only denotes two operators, $[j_1 + 1, j_2 - 1]$ and~$[j_1 - 1, j_2 + 1]$.

\begin{align*}
{\xymatrix{*++[F-,]{\boldsymbol{ L}}}}  \quad &{\xymatrix @R=1.4pc@C=2pc{*++[F=]{[j_1,j_2]_\Delta^{(R)}~,~\Delta > j_1 + j_2 + {3 \over 2} R + 4}}} \\[5pt]
Q: \quad & \boxed{[j_1\pm1,j_2]^{(R\pm1)}~,~[j_1\pm1,j_2\mp1]^{(R\pm1)}
}\\[5pt]
Q^2: \quad & \boxed{\begin{aligned} 
& [j_1\pm2,j_2]^{(R)}~,~[j_1\pm2,j_2\mp1]^{(R\pm2) \oplus 2(R)}~,~[j_1\pm2,j_2\mp2]^{(R)}~\\[1pt]
& [j_1,j_2\pm1]^{(R\pm2) \oplus 2(R)}~,~[j_1,j_2]^{2 (R \pm 2) \oplus 4(R)}
\end{aligned}
}\\[5pt]
Q^3: \quad & \boxed{\begin{aligned} 
& [j_1\pm3,j_2\mp1]^{(R\pm1)}~,~[j_1\pm3,j_2\mp2]^{(R\pm1)}~,~[j_1\pm1,j_2\pm1]^{(R\pm1)}~\\[1pt]
& [j_1\pm1,j_2]^{(R\pm3) \oplus 4 (R \pm1)}~,~[j_1\pm1, j_2\mp1]^{(R \pm 3) \oplus 4(R \pm1)}~,~[j_1\pm1, j_2\mp2]^{(R \pm 1)}
\end{aligned}
}\\[5pt]
Q^4: \quad & \boxed{\begin{aligned}
& [j_1\pm4, j_2\mp2]^{(R)}~,~[j_1\pm2, j_2]^{(R\pm2) \oplus 2(R)}~,~[j_1\pm2, j_2\mp1]^{2 (R\pm2) \oplus 4(R)}~\\[1pt]
& [j_1\pm2, j_2\mp2]^{(R \pm 2) \oplus 2(R)}~,~[j_1, j_2\pm2]^{(R)}~,~[j_1, j_2\pm1]^{2 (R\pm2) \oplus 4(R)}~,\\[1pt]
& [j_1, j_2]^{(R \pm 4) \oplus 4(R\pm 2) \oplus 8(R)}
\end{aligned}
}\\[5pt]
Q^5: \quad & \boxed{\begin{aligned} 
& [j_1\pm3,j_2\mp1]^{(R\pm1)}~,~[j_1\pm3,j_2\mp2]^{(R\pm1)}~,~[j_1\pm1,j_2\pm1]^{(R\pm1)}~\\[1pt]
& [j_1\pm1,j_2]^{(R\pm3) \oplus 4 (R \pm1)}~,~[j_1\pm1, j_2\mp1]^{(R \pm 3) \oplus 4(R \pm1)}~,~[j_1\pm1, j_2\mp2]^{(R \pm 1)}
\end{aligned}
}\\[5pt]
Q^6: \quad & \boxed{\begin{aligned} 
& [j_1\pm2,j_2]^{(R)}~,~[j_1\pm2,j_2\mp1]^{(R\pm2) \oplus 2(R)}~,~[j_1\pm2,j_2\mp2]^{(R)}~\\[1pt]
& [j_1,j_2\pm1]^{(R\pm2) \oplus 2(R)}~,~[j_1,j_2]^{2 (R \pm 2) \oplus 4(R)}
\end{aligned}
}\\[5pt]
Q^7: \quad & \boxed{[j_1\pm1,j_2]^{(R\pm1)}~,~[j_1\pm1,j_2\mp1]^{(R\pm1)}
}\\[5pt]
Q^8: \quad & \boxed{[j_1, j_2]^{(R)}}
\end{align*}

\begin{align*}
{\xymatrix{*++[F-,]{\boldsymbol{ A_1}}}}  \quad &{\xymatrix @R=1.4pc@C=2pc{*++[F=]{[j_1,j_2]_\Delta^{(R)}~,~\Delta = j_1 + j_2 + {3 \over 2} R + 4}}} \\[5pt]
Q: \quad & \boxed{[j_1+1,j_2]^{(R\pm1)}~,~[j_1+1,j_2-1]^{(R\pm1)}~,~[j_1-1,j_2]^{(R-1)}~,~[j_1-1,j_2+1]^{(R\pm1)}
}\\[5pt]
Q^2: \quad & \boxed{\begin{aligned} 
& [j_1+2,j_2]^{(R)}~,~[j_1+2,j_2-1]^{(R\pm2) \oplus 2(R)}~,~[j_1+2,j_2-2]^{(R)}~\\[1pt]
& [j_1,j_2+1]^{(R\pm2) \oplus 2(R)}~,~[j_1,j_2]^{(R+2) \oplus  3(R) \oplus 2 (R - 2) }~,~[j_1,j_2-1]^{(R) \oplus (R-2)}~\\[1pt]
& [j_1-2,j_2+2]^{(R)}~,~[j_1-2,j_2+1]^{(R) \oplus (R-2)}
\end{aligned}
}\\[5pt]
Q^3: \quad & \boxed{\begin{aligned} 
& [j_1+3,j_2-1]^{(R\pm1)}~,~[j_1+3,j_2-2]^{(R\pm1)}~,~[j_1+1,j_2+1]^{(R\pm1)}~\\[1pt]
& [j_1+1,j_2]^{(R\pm3) \oplus 3(R+1) \oplus 4 (R -1)}~,~[j_1+1, j_2-1]^{2(R +1) \oplus 3(R-1) \oplus (R - 3)}~\\[1pt]
& [j_1+1, j_2-2]^{(R - 1)}~,~[j_1-1, j_2+2]^{(R \pm 1)}~,~[j_1-1, j_2+1]^{2(R +1) \oplus 3(R-1) \oplus (R - 3)}~\\[1pt]
& [j_1-1,j_2]^{ (R +1) \oplus 2(R-1)\oplus (R-3)}~,~[j_1-3,j_2+2]^{(R-1)}
\end{aligned}
}\\[5pt]
Q^4: \quad & \boxed{\begin{aligned}
& [j_1+4, j_2-2]^{(R)}~,~[j_1+2, j_2]^{(R\pm2) \oplus 2(R)}~,~[j_1+2, j_2-1]^{(R+2)  \oplus 3(R) \oplus 2 (R-2)}~\\[1pt]
& [j_1+2, j_2-2]^{(R) \oplus (R-2)}~,~[j_1, j_2+2]^{(R)}~,~[j_1, j_2+1]^{(R+2) \oplus 3(R)  \oplus 2 (R-2)}~\\[1pt]
& [j_1, j_2]^{ (R+ 2) \oplus  4(R) \oplus 3(R-2) \oplus (R-4) }~,~[j_1, j_2-1]^{ (R)\oplus (r-2) }~,~[j_1-2, j_2+2]^{(R) \oplus (R-2)}~\\[1pt]
& [j_1-2, j_2+1]^{(R) \oplus (R-2)}
\end{aligned}
}\\[5pt]
Q^5: \quad & \boxed{\begin{aligned} 
& [j_1+3,j_2-1]^{(R\pm1)}~,~[j_1+3,j_2-2]^{(R-1)}~,~[j_1+1,j_2+1]^{(R\pm1)}~\\[1pt]
& [j_1+1,j_2]^{2 (R +1) \oplus 3(R-1) \oplus (R-3)}~,~[j_1+1, j_2-1]^{(R+1) \oplus 2(R-1) \oplus (R-3)}~\\[1pt]
& [j_1-1, j_2+2]^{(R - 1)}~,~[j_1-1, j_2+1]^{(R +1) \oplus 2(R-1) \oplus (R - 3)}~,~[j_1-1,j_2]^{(R-1)}
\end{aligned}
}\\[5pt]
Q^6: \quad & \boxed{\begin{aligned} 
& [j_1+2,j_2]^{(R)}~,~[j_1+2,j_2-1]^{(R) \oplus (R-2)}~,~[j_1,j_2+1]^{(R) \oplus (R-2)}~,~[j_1,j_2]^{(R) \oplus (R-2)}
\end{aligned}
}\\[5pt]
Q^7: \quad & \boxed{[j_1+1,j_2]^{(R-1)}}
\end{align*}

\begin{align*}
{\xymatrix{*++[F-,]{\boldsymbol{ A_2}}}}  \quad &{\xymatrix @R=1.4pc@C=2pc{*++[F=]{[0,j_2]_\Delta^{(R)}~,~\Delta =  j_2 + {3 \over 2} R + 4}}} \\[5pt]
Q: \quad & \boxed{[1,j_2]^{(R\pm1)}~,~[1,j_2-1]^{(R\pm1)}
}\\[5pt]
Q^2: \quad & \boxed{\begin{aligned} 
& [2,j_2]^{(R)}~,~[2,j_2-1]^{(R\pm2) \oplus 2(R)}~,~[2,j_2-2]^{(R)}~\\[1pt]
& [0,j_2+1]^{(R\pm2) \oplus (R)}~,~[0,j_2]^{(R\pm2) \oplus  2(R) }~,~[0,j_2-1]^{(R) \oplus (R-2)}
\end{aligned}
}\\[5pt]
Q^3: \quad & \boxed{\begin{aligned} 
& [3,j_2-1]^{(R\pm1)}~,~[3,j_2-2]^{(R\pm1)}~,~[1,j_2+1]^{(R\pm1)}~\\[1pt]
& [1,j_2]^{(R\pm3) \oplus 3(R\pm1) }~,~[1, j_2-1]^{2(R +1) \oplus 3(R-1) \oplus (R - 3)}~,~[1, j_2-2]^{(R - 1)}
\end{aligned}
}\\[5pt]
Q^4: \quad & \boxed{\begin{aligned}
& [4, j_2-2]^{(R)}~,~[2, j_2]^{(R\pm2) \oplus 2(R)}~,~[2, j_2-1]^{(R+2)  \oplus 3(R) \oplus 2 (R-2)}~,~[2, j_2-2]^{(R) \oplus (R-2)}~\\[1pt]
&[0, j_2+2]^{(R)}~,~[0, j_2+1]^{(R\pm2) \oplus 2(R) }~,~[0, j_2]^{ (R+ 2) \oplus  3(R) \oplus 2(R-2) \oplus (R-4) }~,~[0, j_2-1]^{ (R)\oplus (R-2) }
\end{aligned}
}\\[5pt]
Q^5: \quad & \boxed{\begin{aligned} 
& [3,j_2-1]^{(R\pm1)}~,~[3,j_2-2]^{(R-1)}~,~[1,j_2+1]^{(R\pm1)}~\\[1pt]
& [1,j_2]^{2 (R +1) \oplus 3(R-1) \oplus (R-3)}~,~[1, j_2-1]^{(R+1) \oplus 2(R-1) \oplus (R-3)}
\end{aligned}
}\\[5pt]
Q^6: \quad & \boxed{
[2,j_2]^{(R)}~,~[2,j_2-1]^{(R) \oplus (R-2)}~,~[0,j_2+1]^{(R) \oplus (R-2)}~,~[0,j_2]^{(R) \oplus (R-2)}
}\\[5pt]
Q^7: \quad & \boxed{[1,j_2]^{(R-1)}}
\\[25pt]
{\xymatrix{*++[F-,]{\boldsymbol{ A_4}}}}  \quad &{\xymatrix @R=1.4pc@C=2pc{*++[F=]{[0,0]^{(R)}_\Delta~,~\Delta =   {3 \over 2} R + 4}}} \\[5pt]
Q: \quad & \boxed{[1,0]^{(R\pm1)}
}\\[5pt]
Q^2: \quad & \boxed{
[2,0]^{(R)}~,~ [0,1]^{(R\pm2) \oplus (R)}~,~[0,0]^{(R\pm2) \oplus  (R) }
}\\[5pt]
Q^3: \quad & \boxed{
[1,1]^{(R\pm1)}~,~ [1,0]^{(R\pm3) \oplus 2(R\pm1) }
}\\[5pt]
Q^4: \quad & \boxed{
 [2, 0]^{(R\pm2) \oplus (R)}~,~[0, 2]^{(R)}~,~[0, 1]^{(R\pm2) \oplus 2(R) }~,~[0, 0]^{ (R+ 2) \oplus  2(r) \oplus (R-2) \oplus (R-4) }
}\\[5pt]
Q^5: \quad & \boxed{
[1,1]^{(R\pm1)}~,~ [1,0]^{2 (R \pm1)  \oplus (R-3)}
}\\[5pt]
Q^6: \quad & \boxed{
[2,0]^{(R)}~,~[0,1]^{(R) \oplus (R-2)}~,~[0,0]^{(R) \oplus (R-2)}
}\\[5pt]
Q^7: \quad & \boxed{[1,0]^{(R-1)}}
\end{align*}

\begin{align*}
{\xymatrix{*++[F-,]{\boldsymbol{ B_1}}}}  \quad &{\xymatrix @R=1.4pc@C=2pc{*++[F=]{[0,j_2]_\Delta^{(R)}~,~\Delta = j_2 + {3 \over 2} R + 3}}} \\[5pt]
Q: \quad & \boxed{[1, j_2]^{(R\pm1)}~,~[1, j_2-1]^{(R - 1)}
}\\[5pt]
Q^2: \quad & \boxed{\begin{aligned}
& [2, j_2]^{(R)}~,~[2, j_2-1]^{(R) \oplus (R-2)}~\\[1pt]
& [0, j_2+1]^{(R\pm2) \oplus (R)}~,~[0, j_2]^{ (R) \oplus (R-2)}~,~[0, j_2-1]^{(R-2)}
\end{aligned}
}\\[5pt]
Q^3: \quad & \boxed{\begin{aligned}
& [3,j_2-1]^{(R-1)}~,~ [1,j_2+1]^{(R\pm1)}~,~[1, j_2]^{(R + 1) \oplus 2(R -1) \oplus (R-3)}~\\[1pt]
& [1, j_2-1]^{(R - 1) \oplus (R-3)}
\end{aligned}
}\\[5pt]
Q^4: \quad & \boxed{\begin{aligned} 
& [2,j_2]^{(R) \oplus (R-2)}~,~[2,j_2-1]^{(R-2)}~,~[0,j_2+2]^{(R)}~\\[1pt]
& [0,j_2+1]^{(R) \oplus (R-2)}~,~[0,j_2]^{(R) \oplus (R-2) \oplus (R-4)}
\end{aligned}
}\\[5pt]
Q^5: \quad & \boxed{[1,j_2+1]^{(R-1)}~,~[1,j_2]^{(R-1) \oplus (R-3)}
}\\[5pt]
Q^6: \quad & \boxed{[0, j_2+1]^{(R-2)}}
\\[30pt]
{\xymatrix{*++[F-,]{\boldsymbol{ B_2}}}}  \quad &{\xymatrix @R=1.4pc@C=2pc{*++[F=]{[0,0]_\Delta^{(R)}~,~\Delta = {3 \over 2} R + 3}}} \\[5pt]
Q: \quad & \boxed{[1, 0]^{(R\pm1)}
}\\[5pt]
Q^2: \quad & \boxed{[2, 0]^{(R)}~,~[0, 1]^{(r\pm2) \oplus (R)}~,~[0, 0]^{ (R) \oplus (R-2)}
}\\[5pt]
Q^3: \quad & \boxed{[1,1]^{(R\pm1)}~,~[1, 0]^{(R + 1) \oplus 2(R -1) \oplus (R-3)}
}\\[5pt]
Q^4: \quad & \boxed{\begin{aligned} 
& [2,0]^{(R) \oplus (R-2)}~,~[0,2]^{(R)}~\\[1pt]
& [0,1]^{(R) \oplus (R-2)}~,~[0,0]^{(R) \oplus (R-2) \oplus (R-4)}
\end{aligned}
}\\[5pt]
Q^5: \quad & \boxed{[1,1]^{(R-1)}~,~[1,0]^{(R-1) \oplus (R-3)}
}\\[5pt]
Q^6: \quad & \boxed{[0, 1]^{(R-2)}}
\end{align*}

\begin{align*}
{\xymatrix{*++[F-,]{\boldsymbol{ C_1}}}}  \quad &{\xymatrix @R=1.4pc@C=2pc{*++[F=]{[0,0]_\Delta^{(R)}~,~\Delta = {3 \over 2} R}}} \\[5pt]
Q: \quad & \boxed{
 [1,0]^{(R-1)}
}\\[5pt]
Q^2: \quad & \boxed{[0,1]^{(R-2)}~,~[0,0]^{(R-2)}
}\\[5pt]
Q^3: \quad & \boxed{[1,0]^{(R-3)}
}\\[5pt]
Q^4: \quad & \boxed{[0, 0]^{(R-4)}}
\end{align*}

\subsection{$d=6$, $\mathcal{N}=(1,0)$}

\label{sec:6dn1tabs}

The unitary superconformal multiplets are summarized in table~\ref{tab:6DN1}.  We present their operator content for generic values of their non-vanishing~$\frak{su}(4)$ Lorentz Dynkin labels~$j_1, j_2, j_3$ and~$\frak{su}(2)_R$ Dynkin label~$R$. We condense the notation by declaring that~$\pm$ offsets for the Lorentz and the~$R$-symmetry are independent. However, the offsets for the Lorentz Dynkin labels~$j_1, j_2, j_3$ are correlated. Therefore~$[j_1, j_2 \pm 1, j_3]^{(R \pm 2)}$ denotes four operators, while $[j_1 \pm 1, j_2 \mp 1, j_3 \pm 1]$ only denotes two operators, $[j_1 + 1, j_2 - 1, j_3 + 1]$ and~$[j_1 - 1, j_2 + 1, j_3 - 1]$.

\begin{align*}
{\xymatrix{*++[F-,]{\boldsymbol{ L}}}}  \quad &{\xymatrix @R=1.4pc@C=2pc{*++[F=]{[j_1,j_2,j_3]_\Delta^{(R)}~,~\Delta > \half (j_1 + 2j_2 + 3j_3) + 2R + 6}}} \\[5pt]
Q: \quad & \boxed{[j_1+1,j_2,j_3]^{(R \pm1)}~,~[j_1,j_2,j_3-1]^{(R \pm 1)}~,~[j_1,j_2-1,j_3+1]^{(R \pm 1)}~,~[j_1-1, j_2+1, j_3]^{(R \pm 1)}
}\\[5pt]
Q^2: \quad & \boxed{\begin{aligned} & [j_1+2,j_2,j_3]^{(R)}~,~[j_1\pm1,j_2,j_3\mp1]^{(R \pm 2) \oplus 2(R)}~,~[j_1\pm1,j_2\mp1,j_3\pm1]^{(R \pm 2) \oplus 2(R)}~\\[1pt]
& [j_1,j_2 \pm1,j_3]^{(R \pm 2) \oplus 2(R)}~,~[j_1,j_2,j_3-2]^{(R)}~,~[j_1,j_2-2,j_3+2]^{(R)}~,~[j_1-2,j_2+2,j_3]^{(R)}
\end{aligned}
}\\[5pt]
Q^3: \quad & \boxed{\begin{aligned} & [j_1+2,j_2,j_3-1]^{(R \pm 1)}~,~[j_1+2,j_2-1,j_3+1]^{(R \pm 1)}~,~[j_1+1,j_2+1,j_3]^{(R\pm1)}~\\[1pt]
& [j_1+1,j_2,j_3-2]^{(R \pm 1)}~,~[j_1+1,j_2-1,j_3]^{(R \pm 3) \oplus 3 (R \pm 1)}~,~[j_1+1,j_2-2,j_3+2]^{(R \pm 1)}~\\[1pt]
& [j_1,j_2+1,j_3-1]^{(R \pm 3) \oplus 3 (R \pm 1)}~,~[j_1,j_2,j_3+1]^{(R \pm 3) \oplus 3(R \pm 1)}~,~[j_1,j_2-1,j_3-1]^{(R \pm 1)}~\\[1pt]
& [j_1,j_2-2,j_3+1]^{(R\pm 1)}~,~[j_1-1,j_2+2,j_3]^{(R \pm 1)}~,~[j_1-1,j_2+1,j_3-2]^{(R \pm 1)}~\\[1pt]
& [j_1-1,j_2,j_3]^{(R \pm 3) \oplus 3 (R \pm 1)}~,~[j_1-1,j_2-1,j_3+2]^{(R \pm 1)}~\\[1pt]
& [j_1-2,j_2+2, q-1]^{(R \pm 1)}~,~[j_1-2,j_2+1,j_3+1]^{(R \pm 1)}
\end{aligned}
}\\[5pt]
Q^4: \quad & \boxed{\begin{aligned} & [j_1\pm2,j_2,j_3\mp2]^{(R)}~,~[j_1\pm2,j_2\mp1,j_3]^{(R \pm 2) \oplus 2(R)}~,~[j_1\pm2,j_2\mp2,j_3\pm2]^{(R)}~\\[1pt]
& [j_1\pm1,j_2\pm1,j_3\mp1]^{(R \pm2) \oplus 2(R)}~,~[j_1\pm1,j_2,j_3\pm1]^{(R \pm 2) \oplus 2(R)}~\\[1pt]
& [j_1\pm1,j_2\mp1,j_3\mp1]^{(R \pm2) \oplus 2(R)}~,~[j_1\pm1,j_2\mp2,j_3\pm1]^{(R \pm 2) \oplus 2(R)}~\\[1pt]
& [j_1,j_2\pm2,j_3]^{(R)}~,~[j_1,j_2\pm1, q\mp2]^{(R \pm 2) \oplus 2(R)}~,~[j_1,j_2,j_3]^{(R \pm 4) \oplus 4 (R\pm 2) \oplus 6 (R)}
\end{aligned}
}\\[5pt]
Q^5: \quad & \boxed{\begin{aligned} & [j_1+2,j_2-1,j_3-1]^{(R \pm 1)}~,~[j_1+2,j_2-2, q+1]^{(R \pm 1)}~,~[j_1+1,j_2+1,j_3-2]^{(R \pm 1)}~\\[1pt]
& [j_1+1,j_2,j_3]^{(R \pm 3) \oplus 3 (R \pm 1)}~,~[j_1+1,j_2-1,j_3+2]^{(R \pm 1)}~,~[j_1+1,j_2-2,j_3]^{(R \pm 1)}~\\[1pt]
& [j_1,j_2+2,j_3-1]^{(R\pm 1)}~,~[j_1,j_2+1,j_3+1]^{(R \pm 1)}~,~[j_1,j_2,j_3-1]^{(R \pm 3) \oplus 3(R \pm 1)}~\\[1pt]
& [j_1,j_2-1,j_3+1]^{(R \pm 3) \oplus 3 (R \pm 1)}~,~[j_1-1,j_2+2,j_3-2]^{(R \pm 1)}~\\[1pt]
& [j_1-1,j_2+1,j_3]^{(R \pm 3) \oplus 3 (R \pm 1)}~,~[j_1-1,j_2,j_3+2]^{(R \pm 1)}~,~[j_1-1,j_2-1,j_3]^{(R\pm1)}~\\[1pt]
& [j_1-2,j_2+1,j_3-1]^{(R \pm 1)}~,~[j_1-2,j_2,j_3+1]^{(R \pm 1)}
\end{aligned}
}\\[5pt]
Q^6: \quad & \boxed{\begin{aligned} & [j_1+2,j_2-2,j_3]^{(R)}~,~[j_1\pm1,j_2,j_3\mp1]^{(R \pm 2) \oplus 2(R)}~,~[j_1\pm1,j_2\mp1,j_3\pm1]^{(R \pm 2) \oplus 2(R)}~\\[1pt]
& [j_1,j_2+2,j_3-2]^{(R)}~,~[j_1, j_2\pm1,j_3]^{(R \pm 2) \oplus 2(R)}~,~[j_1,j_2,j_3+2]^{(R)}~,~[j_1-2,j_2,j_3]^{(R)}
\end{aligned}
}\\[5pt]
Q^7: \quad & \boxed{[j_1+1,j_2-1,j_3]^{(R \pm 1)}~,~[j_1,j_2+1,j_3-1]^{(R \pm 1)}~,~[j_1,j_2,j_3+1]^{(R \pm 1)}~,~[j_1-1,j_2,j_3]^{(R \pm 1)}
}\\[5pt]
Q^8: \quad & \boxed{[j_1,j_2,j_3]^{(R)}}
\end{align*}

\begin{align*}
{\xymatrix{*++[F-,]{\boldsymbol{ A_1}}}}  \quad &{\xymatrix @R=1.4pc@C=2pc{*++[F=]{[j_1,j_2,j_3]_\Delta^{(R)}~,~\Delta = \half (j_1 + 2j_2 + 3j_3) + 2R + 6}}} \\[5pt]
Q: \quad & \boxed{[j_1+1, j_2, j_3]^{(R \pm 1)}~,~[j_1, j_2, j_3-1]^{(R-1)}~,~[j_1, j_2-1, j_3+1]^{(R \pm 1)}~,~[j_1-1, j_2+1, j_3]^{(R \pm 1)}}\\[5pt]
Q^2: \quad & \boxed{\begin{aligned} & [j_1+2,j_2,j_3]^{(R)}~,~[j_1+1,j_2,j_3-1]^{(R) \oplus (R-2)}~,~[j_1+1,j_2-1, j_3+1]^{(R \pm 2) \oplus 2 (R)}~\\[1pt]
& [j_1,j_2+1, j_3]^{(R \pm 2) \oplus 2 (R)}~,~[j_1,j_2-1,j_3]^{(R) \oplus (R-2)}~,~[j_1,j_2-2,j_3+2]^{(R)}~\\[1pt]
& [j_1-1, j_2+1, j_3-1]^{(R) \oplus (R-2)}~,~[j_1-1,j_2, j_3+1]^{(R \pm 2) \oplus 2 (R)}~,~[j_1-2, j_2+2, j_3]^{(R)}\end{aligned}}
\\[5pt]
Q^3: \quad & \boxed{\begin{aligned} & [j_1+2,j_2, j_3-1]^{(R-1)}~,~[j_1+2, j_2-1, j_3+1]^{(R \pm 1)}~,~[j_1+1,j_2+1,j_3]^{(R \pm 1)}~\\[1pt]
& [j_1+1,j_2-1,j_3]^{(R+1) \oplus 2 (R-1) \oplus (R-3)}~,~[j_1+1,j_2-2,j_3+2]^{(R\pm1)}~\\[1pt]
& [j_1,j_2+1,j_3-1]^{(R+1) \oplus 2 (R-1) \oplus (R-3)}~,~[j_1,j_2,j_3+1]^{(R\pm3) \oplus 3(R \pm1)}~\\[1pt]
& [j_1,j_2-2,j_3+1]^{(R-1)}~,~[j_1-1,j_2+2,j_3]^{(R \pm 1)}~,~[j_1-1,j_2,j_3]^{(R+1) \oplus 2 (R-1) \oplus (R-3)}~\\[1pt]
& [j_1-1,j_2-1,j_3+2]^{(R\pm1)}~,~[j_1-2,j_2+2,j_3-1]^{(R-1)}~,~[j_1-2,j_2+1,j_3+1]^{(R\pm1)}
\end{aligned}}\\[5pt]
Q^4: \quad & \boxed{\begin{aligned} & [j_1+2,j_2-1,j_3]^{(R) \oplus (R-2)}~,~[j_1+2,j_2-2,j_3+2]^{(R)}~,~[j_1+1,j_2+1,j_3-1]^{(R) \oplus (R-2)}~\\[1pt]
& [j_1+1,j_2,j_3+1]^{(R\pm2) \oplus 2 (R)}~,~[j_1+1,j_2-2,j_3+1]^{(R) \oplus (R-2)}~,~[j_1,j_2+2,j_3]^{(R)}~\\[1pt]
& [j_1,j_2,j_3]^{(R+2) \oplus 3 (R) \oplus 3(R-2) \oplus (R-4)}~,~[j_1,j_2-1,j_3+2]^{(R\pm2) \oplus 2(R)}~\\[1pt]
& [j_1-1,j_2+2,j_3-1]^{(R) \oplus (R-2)}~,~[j_1-1,j_2+1,j_3+1]^{(R\pm2) \oplus 2 (R)}~\\[1pt]
& [j_1-1,j_2-1,j_3+1]^{(R) \oplus (R-2)}~,~[j_1-2,j_2+1,j_3]^{(R) \oplus (R-2)}~,~[j_1-2,j_2,j_3+2]^{(R)}
\end{aligned}}\\[5pt]
Q^5: \quad & \boxed{\begin{aligned} & [j_1+2,j_2-2,j_3+1]^{(R-1)}~,~[j_1+1,j_2,j_3]^{(R+1)\oplus 2(R-1) \oplus (R-3)}~,~[j_1+1,j_2-1,j_3+2]^{(R\pm1)}~\\[1pt]
& [j_1,j_2+2,j_3-1]^{(R-1)}~,~[j_1,j_2+1,j_3+1]^{(R\pm1)}~,~[j_1,j_2-1,j_3+1]^{(R+1) \oplus 2 (R-1) \oplus (R-3)}~\\[1pt]
& [j_1-1,j_2+1,j_3]^{(R+1) \oplus 2(R-1) \oplus (R-3)}~,~[j_1-1,j_2,j_3+2]^{(R\pm1)}~,~[j_1-2,j_2,j_3+1]^{(R-1)}
\end{aligned}}\\[5pt]
Q^6: \quad & \boxed{\begin{aligned} & [j_1+1,j_2-1,j_3+1]^{(R) \oplus (R-2)}~,~[j_1,j_2+1,j_3]^{(R) \oplus (R-2)}~\\[1pt]
& [j_1,j_2, j_3+2]^{(R)}~,~[j_1-1,j_2,j_3+1]^{(R) \oplus (R-2)}~\end{aligned}}\\[5pt]
Q^7: \quad & \boxed{[j_1,j_2,j_3+1]^{(R-1)}}
\end{align*}

\begin{align*}
{\xymatrix{*++[F-,]{\boldsymbol{ A_2}}}}  \quad &{\xymatrix @R=1.4pc@C=2pc{*++[F=]{[j_1,j_2,0]_\Delta^{(R)}~,~\Delta = \half (j_1 + 2j_2) + 2R + 6}}} \\[5pt]
Q: \quad & \boxed{[j_1+1,j_2, 0]^{(R \pm 1)}~,~[j_1,j_2-1, 1]^{(R \pm 1)}~,~[j_1-1,j_2+1, 0]^{(R \pm 1)}
}\\[5pt]
Q^2: \quad & \boxed{\begin{aligned} & [j_1+2,j_2,0]^{(R)}~,~[j_1+1,j_2-1, 1]^{(R \pm 2) \oplus 2 (R)}~\\[1pt]
& [j_1,j_2+1, 0]^{(R \pm 2) \oplus 2 (R)}~,~[j_1,j_2-1,0]^{(R) \oplus (R-2)}~,~[j_1,j_2-2,2]^{(R)}~\\[1pt]
& [j_1-1,j_2, 1]^{(R \pm 2) \oplus 2 (R)}~,~[j_1-2,j_2+2, 0]^{(R)}
\end{aligned}
}\\[5pt]
Q^3: \quad & \boxed{\begin{aligned} & [j_1+2,j_2-1, 1]^{(R \pm 1)}~,~[j_1+1,j_2+1,0]^{(R \pm 1)}~,~[j_1+1,j_2-1,0]^{(R+1) \oplus 2 (R-1) \oplus (R-3)}~\\[1pt]
& [j_1+1,j_2-2,2]^{(R\pm1)}~,~[j_1,j_2,1]^{(R\pm3) \oplus 3(R \pm1)}~,~[j_1,j_2-2,1]^{(R-1)}~\\[1pt]
& [j_1-1,j_2+2,0]^{(R \pm 1)}~,~[j_1-1,j_2,0]^{(R+1) \oplus 2 (R-1) \oplus (R-3)}~,~[j_1-1,j_2-1,2]^{(R\pm1)}~\\[1pt]
& [j_1-2,j_2+1,1]^{(R\pm1)}
\end{aligned}
}\\[5pt]
Q^4: \quad & \boxed{\begin{aligned} & [j_1+2,j_2-1,0]^{(R) \oplus (R-2)}~,~[j_1+2,j_2-2,2]^{(R)}~,~[j_1+1,j_2,1]^{(R\pm2) \oplus 2 (R)}~\\[1pt]
& [j_1+1,j_2-2,1]^{(R) \oplus (R-2)}~,~[j_1,j_2+2,0]^{(R)}~,~[j_1,j_2,0]^{(R+2) \oplus 3 (R) \oplus 3(R-2) \oplus (R-4)}~\\[1pt]
& [j_1,j_2-1,2]^{(R\pm2) \oplus 2(R)}~,~[j_1-1,j_2+1,1]^{(R\pm2) \oplus 2 (R)}~,~[j_1-1,j_2-1,1]^{(R) \oplus (R-2)}~\\[1pt]
& [j_1-2,j_2+1,0]^{(R) \oplus (R-2)}~,~[j_1-2,j_2,2]^{(R)}
\end{aligned}
}\\[5pt]
Q^5: \quad & \boxed{\begin{aligned} & [j_1+2,j_2-2,1]^{(R-1)}~,~[j_1+1,j_2,0]^{(R+1)\oplus 2(R-1) \oplus (R-3)}~,~[j_1+1,j_2-1,2]^{(R\pm1)}~\\[1pt]
& [j_1,j_2+1,1]^{(R\pm1)}~,~[j_1,j_2-1,1]^{(R+1) \oplus 2 (R-1) \oplus (R-3)}~\\[1pt]
& [j_1-1,j_2+1,0]^{(R+1) \oplus 2(R-1) \oplus (R-3)}~,~[j_1-1,j_2,2]^{(R\pm1)}~,~[j_1-2,j_2,1]^{(R-1)}
\end{aligned}
}\\[5pt]
Q^6: \quad & \boxed{[j_1+1,j_2-1,1]^{(R) \oplus (R-2)}~,~[j_1,j_2+1,0]^{(R) \oplus (R-2)}~,~[j_1,j_2, 2]^{(R)}~,~[j_1-1,j_2,1]^{(R) \oplus (R-2)}
}\\[5pt]
Q^7: \quad & \boxed{[j_1,j_2,1]^{(R-1)}}
\end{align*}

\begin{align*}
{\xymatrix{*++[F-,]{\boldsymbol{ A_3}}}}  \quad &{\xymatrix @R=1.4pc@C=2pc{*++[F=]{[j_1,0,0]^{(R)}_\Delta~,~\Delta = \half j_1  + 2R + 6}}} \\[5pt]
Q: \quad & \boxed{[j_1+1, 0, 0]^{(R \pm 1)}~,~[j_1-1, 1, 0]^{(R \pm 1)}
}\\[5pt]
Q^2: \quad & \boxed{\begin{aligned} & [j_1+2,0,0]^{(R)}~,~[j_1,1, 0]^{(R \pm 2) \oplus 2 (R)}~,~ [j_1-1,0, 1]^{(R \pm 2) \oplus (R)}~,~[j_1-2, 2, 0]^{(R)}
\end{aligned}
}\\[5pt]
Q^3: \quad & \boxed{\begin{aligned} & [j_1+1,1,0]^{(R \pm 1)}~,~[j_1,0,1]^{(R\pm3) \oplus 2(R \pm1)}~,~[j_1-1,2,0]^{(R \pm 1)}~\\[1pt]
& [j_1-1,0,0]^{(R\pm1) \oplus (R-3)}~,~[j_1-2,1,1]^{(R\pm1)}
\end{aligned}
}\\[5pt]
Q^4: \quad & \boxed{\begin{aligned} & [j_1+1,0,1]^{(R\pm2) \oplus  (R)}~,~[j_1,2,0]^{(R)}~,~ [j_1,0,0]^{(R+2) \oplus 2 (R) \oplus 2(R-2) \oplus (R-4)} \\[1pt]
& [j_1-1,1,1]^{(R\pm2) \oplus 2 (R)}~,~[j_1-2,1,0]^{(R) \oplus (R-2)}~,~[j_1-2,0,2]^{(R)}
\end{aligned}
}\\[5pt]
Q^5: \quad & \boxed{\begin{aligned} & [j_1+1,0,0]^{(R\pm1)\oplus (R-3)}~,~[j_1,1,1]^{(R\pm1)}~,~ [j_1-2,0,1]^{(R-1)} \\[1pt]
& [j_1-1,1,0]^{(R+1) \oplus 2(R-1) \oplus (R-3)}~,~[j_1-1,0,2]^{(R\pm1)}
\end{aligned}
}\\[5pt]
Q^6: \quad & \boxed{[j_1,1,0]^{(R) \oplus (R-2)}~,~[j_1,0, 2]^{(R)}~,~[j_1-1,0,1]^{(R) \oplus (R-2)}
}\\[5pt]
Q^7: \quad & \boxed{[j_1,0,1]^{(R-1)}}
\\[25pt]
{\xymatrix{*++[F-,]{\boldsymbol{ A_4}}}}  \quad &{\xymatrix @R=1.4pc@C=2pc{*++[F=]{[0,0,0]_\Delta^{(R)}~,~\Delta = 2R+ 6}}} \\[5pt]
Q: \quad & \boxed{[1, 0, 0]^{(R \pm 1)}
}\\[5pt]
Q^2: \quad & \boxed{[2,0,0]^{(R)}~,~[0,1, 0]^{(R \pm 2) \oplus (R)}
}\\[5pt]
Q^3: \quad & \boxed{[1,1,0]^{(R \pm 1)}~,~[0,0,1]^{(R\pm3) \oplus (R \pm1)}
}\\[5pt]
Q^4: \quad & \boxed{\begin{aligned} & [1,0,1]^{(R\pm2) \oplus  (R)}~,~[0,2,0]^{(R)}~,~\\
& [0,0,0]^{(R\pm2) \oplus (R) \oplus (R-4)}~
\end{aligned}
}\\[5pt]
Q^5: \quad & \boxed{[1,0,0]^{(R\pm1)\oplus (R-3)}~,~[0,1,1]^{(R\pm1)}
}\\[5pt]
Q^6: \quad & \boxed{[0,1,0]^{(R) \oplus (R-2)}~,~[0,0, 2]^{(R)}
}\\[5pt]
Q^7: \quad & \boxed{[0,0,1]^{(R-1)}}
\end{align*}

\begin{align*}
{\xymatrix{*++[F-,]{\boldsymbol{ B_1}}}}  \quad &{\xymatrix @R=1.4pc@C=2pc{*++[F=]{[j_1,j_2,0]_\Delta^{(R)}~,~\Delta = \half (j_1 + 2 j_2) + 2R + 4}}} \\[5pt]
Q: \quad & \boxed{[j_1+1,j_2, 0]^{(R \pm 1)}~,~[j_1,j_2-1,1]^{(R-1)}~,~[j_1-1,j_2+1, 0]^{(R \pm 1)}
}\\[5pt]
Q^2: \quad & \boxed{\begin{aligned} & [j_1+2,j_2, 0]^{(R)}~,~[j_1+1,j_2-1,1]^{(R) \oplus (R-2)}~,~[j_1,j_2+1,0]^{(R\pm2) \oplus 2(R) }~\\[1pt]
& [j_1,j_2-1,0]^{(R-2)}~,~[j_1-1,j_2,1]^{(R) \oplus (R-2)}~,~[j_1-2,j_2+2,0]^{(R)}
\end{aligned}
}\\[5pt]
Q^3: \quad & \boxed{\begin{aligned} & [j_1+2,j_2-1,1]^{(R-1)}~,~[j_1+1,j_2+1, 0]^{(R\pm1)}~,~[j_1+1,j_2-1,0]^{(R-1) \oplus (R-3)}~\\[1pt]
& [j_1,j_2,1]^{(R+1) \oplus 2 (R-1) \oplus (R-3)}~,~[j_1-1,j_2+2, 0]^{(R\pm1)}~,~[j_1-1,j_2,0]^{(R-1) \oplus (R-3)}~\\[1pt]
& [j_1-2,j_2+1,1]^{(R-1)}
\end{aligned}
}\\[5pt]
Q^4: \quad & \boxed{\begin{aligned} & [j_1+2,j_2-1,0]^{(R-2)}~,~[j_1+1,j_2,1]^{(R) \oplus (R-2) }~,~[j_1,j_2+2,0]^{(R)}~,~\\[1pt]
& [j_1,j_2,0]^{(R) \oplus 2(R-2) \oplus (R-4)}~,~[j_1-1,j_2+1,1]^{(R) \oplus (R-2)}~,~[j_1-2,j_2+1,0]^{(R-2)}
\end{aligned}
}\\[5pt]
Q^5: \quad & \boxed{[j_1+1,j_2,0]^{(R-1) \oplus (R-3) }~,~[j_1,j_2+1,1]^{(R-1)}~,~[j_1-1,j_2+1,0]^{(R-1) \oplus (R-3)}
}\\[5pt]
Q^6: \quad & \boxed{[j_1,j_2+1,0]^{(R-2)}
}
\\ \\
{\xymatrix{*++[F-,]{\boldsymbol{ B_2}}}}  \quad &{\xymatrix @R=1.4pc@C=2pc{*++[F=]{[j_1,0,0]_\Delta^{(R)}~,~\Delta = \half j_1 + 2R + 4}}} \\[5pt]
Q: \quad & \boxed{[j_1+1,0,0]^{(R \pm1)}~,~[j_1-1,1,0]^{(R \pm 1)}
}\\[5pt]
Q^2: \quad & \boxed{ [j_1+2,0,0]^{(R)}~,~[j_1,1,0]^{(R\pm2) \oplus 2(R)}~,~[j_1-1,0,1]^{(R) \oplus (R-2)}~,~[j_1-2,2,0]^{(R)}
}\\[5pt]
Q^3: \quad & \boxed{\begin{aligned}& [j_1+1,1,0]^{(R\pm1)}~,~[j_1,0,1]^{(R+1) \oplus 2 (R-1) \oplus (R-3)}~,~[j_1-1,2,0]^{(R\pm1)}~\\[1pt]
& [j_1-1,0,0]^{(R-1) \oplus (R-3)}~,~[j_1-2,1,1]^{(R-1)}
\end{aligned}
}\\[5pt]
Q^4: \quad & \boxed{\begin{aligned} & [j_1+1,0,1]^{(R) \oplus (R-2)}~,~[j_1,2,0]^{(R)}~,~[j_1,0,0]^{(R) \oplus 2(R-2) \oplus (R-4)}~\\[1pt]
& [j_1-1,1,1]^{(R) \oplus (R-2)}~,~[j_1-2,1,0]^{(R-2)}
\end{aligned}
}\\[5pt]
Q^5: \quad & \boxed{[j_1+1,0,0]^{(R-1) \oplus (R-3)}~,~[j_1,1,1]^{(R-1)}~,~[j_1-1,1,0]^{(R-1) \oplus (R-3)}
}\\[5pt]
Q^6: \quad & \boxed{[j_1,1,0]^{(R-2)}
}
\end{align*}

\begin{align*}
{\xymatrix{*++[F-,]{\boldsymbol{ B_3}}}}  \quad &{\xymatrix @R=1.4pc@C=2pc{*++[F=]{[0,0,0]_\Delta^{(R)}~,~\Delta = 2R + 4}}} \\[5pt]
Q: \quad & \boxed{[1,0,0]^{(R \pm 1)}
}\\[5pt]
Q^2: \quad & \boxed{[2,0,0]^{(R)}~,~[0,1,0]^{(R \pm 2) \oplus (R)}
}\\[5pt]
Q^3: \quad & \boxed{[1,1,0]^{(R \pm 1)}~,~[0,0,1]^{(R \pm 1) \oplus (R-3)}
}\\[5pt]
Q^4: \quad & \boxed{[1,0,1]^{(R) \oplus (R-2)}~,~[0,2,0]^{(R)}~,~[0,0,0]^{(R) \oplus (R-2) \oplus  (R-4)}
}\\[5pt]
Q^5: \quad & \boxed{[1,0,0]^{(R-1) \oplus (R-3)}~,~[0,1,1]^{(R-1)}
}\\[5pt]
Q^6: \quad & \boxed{[0,1,0]^{(R-2)}
}
\end{align*}

\begin{align*}
{\xymatrix{*++[F-,]{\boldsymbol{ C_1}}}}  \quad &{\xymatrix @R=1.4pc@C=2pc{*++[F=]{[j_1,0,0]_\Delta^{(R)}~,~\Delta = \half j_1 + 2R + 2}}} \\[5pt]
Q: \quad & \boxed{[j_1+1,0,0]^{(R \pm 1)}~,~[j_1-1,1,0]^{(R-1)}
}\\[5pt]
Q^2: \quad & \boxed{[j_1+2,0,0]^{(R)}~,~[j_1,1,0]^{(R) \oplus (R-2)}~,~[j_1-1,0,1]^{(R-2)}
}\\[5pt]
Q^3: \quad & \boxed{[j_1+1,1,0]^{(R-1)}~,~[j_1,0,1]^{(R-1) \oplus (R-3)}~,~[j_1-1,0,0]^{(R-3)}
}\\[5pt]
Q^4: \quad & \boxed{[j_1+1,0,1]^{(R-2)}~,~[j_1,0,0]^{(R-2) \oplus (R-4)}
}\\[5pt]
Q^5: \quad & \boxed{[j_1+1,0,0]^{(R-3)}
}
\end{align*}

\begin{align*}
{\xymatrix{*++[F-,]{\boldsymbol{ C_2}}}}  \quad &{\xymatrix @R=1.4pc@C=2pc{*++[F=]{[0,0,0]_\Delta^{(R)}~,~\Delta = 2R + 2}}} \\[5pt]
Q: \quad & \boxed{[1,0,0]^{(R \pm 1)}
}\\[5pt]
Q^2: \quad & \boxed{[2,0,0]^{(R)}~,~[0,1,0]^{(R) \oplus (R-2)}
}\\[5pt]
Q^3: \quad & \boxed{[1,1,0]^{(R-1)}~,~[0,0,1]^{(R-1) \oplus (R-3)}
}\\[5pt]
Q^4: \quad & \boxed{[1,0,1]^{(R-2)}~,~[0,0,0]^{(R-2) \oplus (R-4)}
}\\[5pt]
Q^5: \quad & \boxed{[1,0,0]^{(R-3)}
}
\end{align*}

\begin{align*}
{\xymatrix{*++[F-,]{\boldsymbol{ D_1}}}}  \quad &{\xymatrix @R=1.4pc@C=2pc{*++[F=]{[0,0,0]_\Delta^{(R)}~,~\Delta =  2R}}} \\[5pt]
Q: \quad & \boxed{[1,0,0]^{(R-1)}
}\\[5pt]
Q^2: \quad & \boxed{[0,1,0]^{(R-2)}
}\\[5pt]
Q^3: \quad & \boxed{[0,0,1]^{(R-3)}
}\\[5pt]
Q^4: \quad & \boxed{[0,0,0]^{(R-4)}
}
\end{align*}

\bigskip

\section{Conserved Currents in Superconformal Field Theories}

\label{sec:currents}

In this section we present a complete classification of all unitary superconformal multiplets~$\CJ$ whose decomposition into CPs includes short representations of the conformal algebra~$\frak{so}(d,2)$, i.e.~conserved currents or free fields. We will refer to~$\CJ$ as a superconformal current multiplet. We also explore some physical implications of this classification. In particular, we rule out SCFTs with~$N_Q > 16$ supercharges in~$d \geq 4$. By contrast, we show that such theories exist when~$d = 3$, but are necessarily free. 

\subsection{Examples and Applications of Superconformal Current Multiplets}

\label{sec:exapcumult}

In this subsection we will explore some prototypical examples of superconformal current multiplets -- those containing flavor currents~$j_\mu$ and their superpartners, as well as those containing the stress tensor~$T_{\mu\nu}$ and its superpartners -- and some of their consequences for SCFTs.  As we will see below, $j_\mu$ and~$T_{\mu\nu}$ are annihilated by all~$Q$-supercharges, up to certain total-derivative terms (i.e.~CDs). In the terminology of~\cite{Cordova:2016xhm}, this makes them top components of their respective superconformal multiplets. Recall also that some multiplets admit several distinct top components, which may reside at different levels.

Identifying top components played a crucial role in~\cite{Cordova:2016xhm}, because they give rise to supersymmetric deformations of SCFTs. As was explained there (see especially section~2.2 of~\cite{Cordova:2016xhm}), one can distinguish between two kinds of top components:
\begin{itemize}
\item[1.)] Manifest top components are forced to map into CDs by the~$Q$-supercharges, because of selection rules on their quantum numbers. 
\item[2.)] Accidental top components are mapped into CDs by the~$Q$-supercharges, even though selection rules do not prohibit them from mapping into CPs. 
\end{itemize}
The main result of~\cite{Cordova:2016xhm} was a classification of all manifest Lorentz-scalar top components. We do not know examples of accidental Lorentz-scalar top components. 

However, there are accidental top components that are not Lorentz scalars. An example that we will encounter below is the flavor current~$[2]_2^{(0,0,0)}$ that resides at level two of the $d = 3$,~$\CN=6$ stress tensor multiplet tabulated in~\eqref{3dn6em}. Even though this operator is not prohibited from mapping to the supersymmetry current~$[3]_{5 \over 2}^{(1,0,0)}$ at level three by quantum numbers, it in fact does not, i.e.~it only maps to CDs.  While accidental top components appear to be rare, highly sporadic phenomena, the fact that they occur complicates some of our arguments. We do not know of examples in~$d \geq 4$, but we have not systematically ruled them out either.

\subsubsection{Flavor Current Multiplets}

\label{sec:fcm}

We will call a continuous global symmetry a flavor symmetry if its generator~$F$ commutes with all superconformal generators. The flavor charge~$F$ is then a Lorentz scalar of scaling dimension~$\Delta_F = 0$,  which commutes with all~$Q$-supercharges and~$R$-symmetries. Such a charge should arise by integrating an~$R$-neutral, conserved flavor current~$j_\mu$ of dimension~$\Delta_{j_\mu} = d-1$ over a spatial slice
\begin{equation}\label{fjint}
F = \int d^{d-1}x~\, j^0~.
\end{equation}
The fact that~$[Q, F] = 0$ implies that
\begin{equation}\label{qjcomm}
[Q, j_\mu] = \left(\text{total derivatives}\right)~,
\end{equation}
where the total derivative terms should be consistent with current conservation and vanish when integrated over a spatial slice. In particular, the right-hand side of~\eqref{qjcomm} cannot contain any CPs. As was reviewed at the beginning of section~\ref{sec:exapcumult}, this means that~$j_\mu$ is a top component of its superconformal multiplet. We will refer to any superconformal multiplet~$\CJ$ that contains~$j_\mu$ and satisfies~\eqref{qjcomm} as a flavor current multiplet. 

Importantly, not all conserved spin-$1$ currents are flavor currents, because the corresponding global charges may not commute with all superconformal generators. For instance, $R$-currents~$j_\mu^{(R)}$ are not flavor currents, since the~$Q$-supersymmetries transform in a representation of the~$R$-symmetry, i.e.~$[Q, R] \sim Q$. It follows that~$R$-currents reside in the same multiplet as the supersymmetry currents and the stress tensor (see section~\ref{sec:stm} below). If a spin-$1$ current is neither a flavor current nor an~$R$-current, then the commutator of~$Q$ with the corresponding global charge gives rise to new fermionic generators that extend the superconformal algebra. For instance, this can happen in free theories, where the superconformal algebra is enhanced to a higher-spin superalgebra.

\subsubsection{Stress Tensor Multiplets}

\label{sec:stm}

In an SCFT, the superconformal generators should themselves arise from local currents, whose quantum numbers are determined by those of the corresponding charges: 
\begin{itemize}
\item The conformal generators~$P_\mu, D, K_\mu$ are integrals of a conserved, traceless stress tensor~$T_{\mu\nu}$ of dimension~$\Delta_{T_{\mu\nu}} = d$, which must be~$R$-neutral.
\item The~$Q$- and~$S$-supercharges are integrals of a conserved, traceless supersymmetry current~$S_{\mu\alpha}$ of dimension~$\Delta_{S_{\mu\alpha}} = d-\half$. It carries the same~$R$-symmetry representation as the~$Q$-supercharges. 
\item Possible~$R$-symmetry generators arise from a conserved~$R$-current~$j_\mu^{(R)}$ of dimension $\Delta_{j_\mu^{(R)}} = d-1$, which must transform in the adjoint representation of the~$R$-symmetry. 
\end{itemize}
All of these currents must, in fact, reside in the same superconformal multiplet:
\begin{itemize}
\item The Poincar\'e SUSY algebra~$\{Q, Q\} \sim P_\mu$ implies that
\begin{equation} \label{QScurralg}
\{Q, S_{\mu\alpha}\} \sim T_{\mu\nu} + \left(\text{total derivatives}\right)~.
\end{equation}
As in the discussion around~\eqref{qjcomm}, the total derivative terms on the right-hand side should be consistent with current conservation and vanish when integrated over a spatial slice. It follows from~\eqref{QScurralg} that~$S_{\mu\alpha}$ and~$T_{\mu\nu}$ must reside in the same multiplet. 

\item The fact that the~$Q$-supercharges carry~$R$-charge means that~$[Q, R] \sim Q$, so that
\begin{equation}\label{qjralg}
[Q, j_\mu^{(R)}] \sim S_{\mu\alpha}  + \left(\text{total derivatives}\right)~.
\end{equation}
As in~\eqref{QScurralg}, the total derivative terms must not contaminate the integrated charge algebra. We conclude that~$j_\mu^{(R)}$ and~$S_{\mu\alpha}$ reside in the same multiplet. 
\item Since~$[Q, P_\mu] = 0$, it follows that
\begin{equation} \label{Ttop}
[Q, T_{\mu\nu}]= \left(\text{total derivatives}\right)~,
\end{equation}
so that~$T_{\mu\nu}$ is a top component of its superconformal multiplet (see the beginning of section of section~\ref{sec:exapcumult}). 
\end{itemize}
We will refer to any superconformal multiplet~$
\CT$ that contains operators~$j_\mu^{(R)}, S_{\mu\alpha}, T_{\mu\nu}$ satisfying the relations~\eqref{QScurralg}, \eqref{qjralg}, and~\eqref{Ttop} as a stress-tensor multiplet.\footnote{~Note that~$j_\mu^{(R)}$ is absent when~$d = 3,\, \CN = 1$, since these theories do not have a continuous~$R$-symmetry.}

\subsubsection{Multiplets Containing Extra Supersymmetry Currents}

\label{sec:escm}

It is sometimes convenient to view a theory with~$\CN$-supersymmetry as a special case of a theory with less supersymmetry~$\hat \CN < \CN$ (see also section~\ref{sec:subalgdec}). In this case the~$Q$-supercharges of~$\CN$-supersymmetry decompose as
\begin{equation}\label{qdecomp}
Q~\rightarrow~\hat Q \oplus Q'~.
\end{equation}
Here~$\hat Q$ denotes the~$\hat \CN$-supercharges, while the~$Q'$ are additional supercharges that enhance~$\hat \CN$ to~$\CN$. Similarly, the~$R$-symmetry algebra decomposes as
\begin{equation}\label{rdecomp}
\frak R~\rightarrow~\frak {\hat R} \oplus \frak F \oplus \CR_\text{off-diag.}~.
\end{equation}
Here~$\frak {\hat R}$ is the~$\hat R$-subalgebra corresponding to~$\hat \CN$, while~$\frak F$ is its commutant inside~$\frak R$. The remaining off-diagonal generators~$\CR_\text{off-diag.}$ of~$\frak R$ are charged under both~$\frak {\hat R}$ and~$\frak F$. The~$\hat Q$-supercharges in~\eqref{qdecomp} are charged under~$\frak{\hat R}$, but not under~$\frak F$. Hence~$\frak F$ is a flavor symmetry from the point of view of~$\hat \CN$-supersymmetry. The quantum numbers of the~$Q'$-supercharges under~$\frak {\hat R}$ and~$\frak F$ depend on the spacetime dimension~$d$ and the precise values of~$\frak \CN$ and~${\hat \CN}$.  

It is instructive to consider how a stress tensor multiplet~$\CT$ of the larger~$\CN$-supersymmetry decomposes under the~$\hat \CN$-subalgebra. Such a decomposition may include a variety of multiplets, some of which need not contain any conserved currents. However, it must certainly include the following current multiplets of~$\hat \CN$-supersymmetry:
\begin{itemize}
\item A single stress tensor multiplet~$\hat \CT$ that contains~$T_{\mu\nu}$, as well as the~$\hat \CN$-supersymmetry currents~$\hat S_{\mu\alpha}$ and the~$\hat R$-currents~$j_\mu^{(\hat R)}$. 
\item The currents that give rise to the flavor generators~$\frak F$ must reside in flavor current multiplets of~$\hat \CN$-supersymmetry.
\item The supersymmetry currents~$S'_{\mu\alpha}$ that give rise to the additional~$Q'$-supercharges must reside in a new type of superconformal current multiplet. It does not contain a stress tensor, because~$T_{\mu\nu}$ is already embedded in~$\hat \CT$. Therefore~$S'_{\mu\alpha}$ is a top component of its multiplet with respect to the~$\hat Q$-supercharges,
\begin{equation}\label{hatqsprime}
\{\hat Q, S'_{\mu\alpha}\} = \left(\text{total derivatives}\right)~.
\end{equation}
Moreover, the off-diagonal~$R$-currents~$j_\mu^{(\text{off-diag.})}$ must reside in the same multiplet, since they give rise to the~$R$-symmetry generators~$\CR_\text{off-diag.}$, which mix~$\hat Q$ and~$Q'$. Explicitly, 
\begin{equation}\label{hatqjod}
[\hat Q, j_\mu^{(\text{off-diag.})} ] \sim S'_{\mu\alpha} + \left(\text{total derivatives}\right)~.
\end{equation} 
We will refer to any multiplet with these properties as an extra SUSY-current multiplet.  
\end{itemize}

\subsubsection{Constraints on Maximal Supersymmetry}

\label{sec:maxsusy}

It is standard lore that non-gravitational quantum field theories admit at most~$N_Q = 16$ $Q$-supercharges. In~$d \geq 4$ dimensions, this is typically argued to follow from the structure of massless one-particle representations of the Poincar\'e supersymmetry algebra (see for instance~\cite{Wess:1992cp,Weinberg:2000cr}). If~$N_Q > 16$, these necessarily include particles of helicity~$h > 1$,\footnote{~In~$d \geq 4$, the massless little group is~$SO(d-2)$. Hence there is more than one helicity label if~$d \geq 6$.} which violate the bound~$h \leq 1$ that follows from the existence of a well-defined stress tensor~$T_{\mu\nu}$ in quantum field theory~\cite{Weinberg:1980kq}. (Importantly, this argument breaks down in~$d = 3$, because the massless little group is discrete and there is no notion of helicity, see below.) Since these arguments rely on the notion of one-particle states, they do not apply to strongly-coupled SCFTs.\footnote{~For example, it was recently argued that there are strongly-coupled~$\CN=3$ SCFTs in~$d = 4$ that do not enhance to~$\CN=4$~\cite{Aharony:2015oyb,Garcia-Etxebarria:2015wns,Aharony:2016kai}, even though this is the case for the corresponding one-particle representations.} We would therefore like to understand whether such theories can support~$N_Q > 16$ supercharges. Note that this question does not arise in~$d = 5$, where there is a unique superconformal algebra (corresponding to~$\CN=1$, see~\eqref{scftalgs}) with~$N_Q = 8$ supercharges.   

We will show that the superconformal algebras that can arise in quantum field theory are strongly constrained by the requirement that all generators arise from currents residing in a suitable stress tensor multiplet (see section~\ref{sec:stm}).\footnote{~This is similar to~\cite{Dumitrescu:2011iu}, where general restrictions on extended Poincar\'e supersymmetry algebras were derived from the structure of consistent stress tensor multiplets.} In particular, we show that superconformal algebras with~$N_Q > 16$ in~$d = 4,6$ do not admit a stress tensor multiplet, confirming the standard lore. We start with~$d = 4$, where the argument is simplest. As an aside, we show that the central extension~$\frak{su}(2,2|4)$ of the standard~$\CN=4$ superconformal algebra~$\frak{psu}(2,2|4)$ by~$\frak{u}(1)$ also does not admit a stress tensor multiplet, and hence cannot be realized in SCFTs. In~$d = 6$ the argument against~$N_Q > 16$ is complicated by the possibility of accidental top components. In~$d = 3$, the situation is richer: SCFTs with~$N_Q > 16$ exist, but they are necessarily free field theories. 
\begin{itemize}
\item [$d = 4$:] In order to rule out SCFTs with~$\CN \geq 5$, it suffices to show that~$\hat \CN=4$ theories do not admit an extra SUSY-current multiplet, since such a multiplet would necessarily arise by decomposing the stress tensor multiplet of~$\CN$-supersymmetry into~$\hat \CN = 4$ submultiplets (see section~\ref{sec:escm}). If we decompose the~$\CN$-supercharges~$Q$ as in~\eqref{qdecomp}, we obtain the~$\hat \CN = 4$ supercharges~$\hat Q$ in the~$(1,0,0)$ representation of the~$\frak{su}(4)_R$-symmetry, and additional supercharges~$Q'$ that are~$\frak{su}(4)_R$ singlets. Therefore, they should arise from extra SUSY-currents~$S'_{\mu\alpha}$ with quantum numbers~$[2; 1]_{7/2}^{(0,0,0)}$, which reside in an~$\hat \CN = 4$ multiplet. All~$\hat \CN = 4$ multiplets that contain conserved currents are explicitly tabulated in section~\ref{sec:d4n4curr} below. It is straightforward to check that none of them contains an operator with the quantum numbers of~$S'_{\mu\alpha}$. 

As an aside, note that all known~$\hat \CN=4$ SCFTs realize the algebra~$\frak{psu}(2,2|4)$ in~\eqref{scftalgs}. This algebra admits a central extension by~$\frak{u}(1)$, with generator~$Z$, to~$\frak{su}(2,2|4)$. Since~$Z$ commutes with all spacetime and~$R$-symmetries, it is a~$\frak{u}(1)$ flavor symmetry, which should arise from a current~$j_\mu$ with quantum numbers~$[1;1]_{3}^{(0,0,0)}$. The fact that~$j_\mu$ itself commutes with~$Z$ implies that it must reside in a flavor current multiplet (see section~\ref{sec:fcm}) of the ordinary~$\hat \CN=4$ algebra~$\frak{psu}(2,2|4)$. However, such a multiplet does not exist. To see this, note that the only candidate~$\hat \CN = 4$ multiplets in section~\ref{sec:d4n4curr} that could contain~$j_\mu$ are~${\hat B}_1 \hat  {\b B}_1[0;0]_2^{(1,0,1)}$ in~\eqref{d4n4hs1} and the Konishi multiplet~${\hat A}_2 \hat{\b A}_2[0;0]_2^{(0,0,0)}$ in~\eqref{konishi}.\footnote{~The SCP of the multiplet~$\hat {A}_1\hat{ \b A}_1[1;1]_3^{(0,0,0)}$ in~\eqref{konishi} also has the quantum numbers of~$j_\mu$, but it cannot be a flavor current, because it is manifestly not a top component of its multiplet.} However, this would require~$j_\mu$ to be an accidental top component of these multiplets. It can be checked explicitly that this is not the case.\footnote{~\label{wwfn} Alternatively, both candidate flavor current multiplets can be realized as bilinears of two free, abelian vector multiplets~\eqref{freed4n4vecmul}, whose SCPs~$X^a, Y^a~(a = 1, \ldots, 6)$ have quantum numbers~$[0;0]_1^{(0,1,0)}$. Explicitly,
\begin{equation}
\hat {B}_1\hat{ \b B}_1[0;0]_2^{(1,0,1)} \sim X^{[a} Y^{b]}~, \qquad \hat{A}_2\hat{ \b A}_2[0;0]_2^{(0,0,0)} \sim \sum_{a = 1}^6 X^a X^a~.
\end{equation} 
If these multiplets actually contained flavor currents, the abelian gauge fields~$F^{(X)}_{\mu\nu}, F^{(Y)}_{\mu\nu}$ descended from~$X^a, Y^a$ would be charged under the corresponding flavor symmetries, in contradiction with~\cite{Weinberg:1980kq}.}

The fact that~$\hat \CN=4$ SCFTs do not admit flavor current multiplets also provides an alternative argument that SCFTs with~$\CN \geq 5$ do not exist: the latter would have~$R$-symmetry~$\frak{su}(\CN) \oplus \frak{u}(1)$ and would therefore necessarily give rise to a rank-$(\CN-3)$ flavor symmetry of the~$\hat \CN = 4$ theory.  

\item[$d = 6$:] In order to rule out~$(\CN, 0)$ SCFTs with~$\CN \geq 3$, it suffices to consider~$\CN =3$. We will present two arguments that such theories do not admit a stress tensor multiplet (see section~\ref{sec:stm}). 

The first argument is based on decomposing with respect to an~$\hat \CN = 2$ subalgebra. Then the~$\CN=3$~$R$-symmetry decomposes as
\begin{equation}
\frak{sp}(6)_R~\rightarrow~ \frak{sp}(4)_{\hat R} \oplus \frak{sp}(2)_F~.
\end{equation}
Here~$\frak{sp}(4)_{\hat R}$ is the~$\hat R$-symmetry of the~$\hat \CN = 2$ subalgebra, while~$\frak{sp}(2)_F$ is a flavor symmetry that commutes with all~$\hat \CN=2$ supercharges~$\hat Q$. However, there is no $\hat \CN = 2$ flavor current multiplet that could give rise to this flavor symmetry. To see this, note that the only candidate~$\hat \CN=2$ multiplets that contain an operator with the quantum numbers~$[0,1,0]_5^{(0,0)}$ of a flavor current are~$\hat D_1[0,0,0]_4^{(2,0)}$ in~\eqref{6dn2hsd1} and~$\hat C_1[1,0,0]^{(1,0)}_{9/2}$ in~\eqref{6dn2c1r1oner2zerohs}.\footnote{~The SCP of~$\hat B_1[0,1,0]_5^{(0,0)}$ in~\eqref{6dn2b1r1r2zerohs} also has the quantum numbers of a flavor current, but it is manifestly not a top component of its multiplet.} Moreover, the candidate flavor current would have to be an accidental top component of these multiplets, but it can be shown that this is not the case.\footnote{~This can be argued by adapting the argument in footnote~\ref{wwfn} to the present case. Both candidate flavor current multiplets can be realized as bilinears of two free~$\hat \CN=2$ tensor multiplets~\eqref{6dT2}, whose SCPs~$X^a, Y^a~(a = 1, \ldots, 5)$ have quantum numbers~$[0,0,0]_2^{(0,1)}$. Explicitly,
\begin{equation}\label{6dn2candflav}
\hat D_1[0,0,0]_4^{(2,0)} \sim X^{[a} Y^{b]}~, \qquad \hat C_1[1,0,0]^{(1,0)}_{9\over 2} \sim \sum_{a = 1}^5 X^a {\left(\gamma_a\right)^i}_j \psi^{(X)j}_\alpha~.
\end{equation} 
Here~$\psi^{(X) i}_\alpha = [1,0,0]_{5/2}^{(1,0)}$ is the level-one descendant of~$X^a$, and the~${\left(\gamma_a\right)^i}_j$ are~$\frak{so}(5)_{\hat R}$ gamma matrices. If the multiplets in~\eqref{6dn2candflav} actually contained flavor currents, the self-dual three-form field strengths~$H^{(X)+}_{\mu\nu\rho}, H^{(Y)+}_{\mu\nu\rho}$ descended from~$X^a, Y^a$ would be charged under the corresponding flavor symmetries. This is ruled out by a straightforward extension of~\cite{Weinberg:1980kq} to~$d = 6$: a flavor current~$j_\mu$ can only give charge to massless particles with~$SO(4)$ little-group helicities~$(h, \t h)~(h, \t h \in \half \Z_{\geq 0})$ if~$h + \t h < 1$. In this notation, a self-dual three-form field strength creates a massless one-particle state with helicities~$(1,0)$, which violate the bound.}

The second argument against~$\CN = 3$ SCFTs involves showing more directly that the only candidate stress tensor multiplet does not in fact obey all the properties stipulated in section~\ref{sec:stm}. By applying the discussion around~\eqref{6dcurrsymbd} and the results in appendix~\ref{sec:d6app} to the case~$\CN=3$, it can be shown that the only~$\CN=3$ multiplet~$\CT$ that contains operators with the quantum numbers of the~$R$-currents~$j_\mu^{(R)}$, supersymmetry currents~$S_{\mu\alpha}$, and stress tensor~$T_{\mu\nu}$ is given by
\begin{equation}
\CT = D_{1}[0,0,0]_{4}^{(0,2,0)}~.
\end{equation}
The SCP~$\CT$ is a Lorentz scalar in the~$(0,2,0)$ representation of the~$\frak{sp}(6)_R$-symmetry.  

We claim that~$\CT$ is not a viable stress tensor multiplet, because the candidate stress tensor~$T_{\mu\nu}$, which resides at level~$\ell = 4$ (i.e.~$T_{\mu\nu} \sim Q^4 \CT$), is not a top component. To see this, we decompose with respect to an~$\hat \CN = 1$ subalgebra, so that
\begin{equation}
\frak{sp}(6)_R~\rightarrow~\frak{sp}(2)_{\hat R} \oplus \frak{sp}(4)_F~.
\end{equation}
Here~$\frak{sp}(2)_{\hat R}$ is the~$\hat R$-symmetry of the~$\hat \CN=1$ subalgebra, while~$\frak{sp}(4)_F$  is a flavor symmetry that commutes with all~$\hat \CN=1$ supercharges~$\hat Q$. We write the corresponding decomposition of~$R$-symmetry representations as~$(R_1, R_2, R_2)\rightarrow (\hat R ; F_1, F_2)$, where~$\hat R$ and~$F_{1,2}$ are~$\frak{sp}(2)_{\hat R}$ and~$\frak{sp}(4)_R$ Dynkin labels, respectively. For instance, the supercharges decompose as follows, 
\begin{equation}
Q = [1,0,0]^{(1,0,0)}  \quad \longrightarrow \quad \hat Q = [1,0,0]^{(1; 0,0)}~\oplus~Q' = [1,0,0]^{(0; 1,0)}~.
\end{equation}
Here the~$\hat Q$ are the~$\hat \CN=1$ supercharges, while the~$Q'$ make up the remainder of the~$\CN=3$ supercharges. 

The decomposition of the SCP~$\CT = D_1[0,0,0]_4^{(0,2,0)}$ at level~$\ell = 0$ gives rise to the following~$\hat \CN=1$ multiplets; all of them are Lorentz scalars, so we only indicate their~$\hat R$-symmetry and flavor quantum numbers:
\begin{equation}
\CT = D_1^{(0,2,0)} ~\longrightarrow~\hat D_{1}^{(2; 2,0)}\oplus \hat C_{2}^{(1; 1,1)}\oplus \hat C_{2}^{(1; 1,0)} \oplus  \hat B_{3}^{(0; 0,2)} \oplus  \hat B_{3}^{(0; 0,1)} \oplus  \hat B_{3}^{(0; 0,0)}~. \label{30decomp}
\end{equation}
The full~$\CN=3$ multiplet based on~$\CT$ includes additional~$\hat \CN=1$ multiplets, whose SCPs are~$Q'$ descendants of these operators and hence reside at higher levels.  The multiplet~$\hat B_{3}^{(0; 0,0)}$ that appears in~\eqref{30decomp} is the $\hat \CN=1$ stress tensor multiplet~\eqref{6dEM1}, which contains the candidate stress tensor~$T_{\mu\nu}$ at level~$\ell = 4$,  
\begin{equation}
T_{\mu\nu} \sim \hat Q^{4}\hat B_{3}^{(0; 0,0)}~.
\end{equation}
Therefore~$T_{\mu\nu}$ is an~$\hat \CN=1$ top component, i.e.~$\hat Q \, T_{\mu\nu}$ is a CD, in accord with~\eqref{Ttop}.

We will now complete the argument by showing that the action of the additional~$Q'$ supercharges on~$T_{\mu\nu}$ gives rise to new CPs, so that~\eqref{Ttop} is violated. Since
\begin{equation}
Q' \, T_{\mu\nu}\sim \widehat Q^{4} \left(Q'\hat B_{3}^{(0; 0,0)}\right)~,~\label{nonzero6dop}
\end{equation}
it suffices to consider the $\hat \CN=1$ multiplet with SCP (see section~\ref{sec:d6n1defs})
\begin{equation}
Q'\hat B_{3}^{(0; 0,0)} = \hat B_2 [1,0,0]_{9\over2}^{(0;1,0)}~.
\end{equation} 
By examining the operator content of this multiplet,\footnote{~This can be obtained by applying the full algorithm described in section~\ref{sec:ouralg}. In this case, the result can also be obtained more quickly by specializing the quantum numbers of the generic~$B_2$-multiplet tabulated in section~\ref{sec:6dn1tabs} and applying the Racah-Speiser cancellation procedure described in section~\ref{sec:ouralg}.} we find that it has a non-vanishing CP with quantum numbers~$[1,2,0]^{(0;1,0)}_{13/2}$ at level~$\ell = 4$, and hence the right-hand side of~\eqref{nonzero6dop} is not a total derivative. In other words, $T_{\mu\nu}$ is not a top component with respect to the~$Q'$ supercharges, and hence it is not a viable~$\CN=3$ stress tensor.

\item[$d = 3$:] Free, massless one-particle representations in~$d = 3$ are not characterized by a helicity quantum number; the little group is~$\Z_2$ and only distinguishes massless bosons and fermions. Consequently, the standard free-particle argument against field theories with~$N_Q>16$ supercharges reviewed at the beginning of section~\ref{sec:maxsusy} does not apply. In fact, it is straightforward to construct free SCFTs with any amount~$\CN$ of supersymmetry: let~$\phi^{i}$ and~$\psi ^i_\alpha$ be a Lorentz scalar and a spin-$\half$ fermion, where~$i$ is a spinor index of the~$\frak{so}(\mathcal{N})_R$-symmetry. Then the following supersymmetry transformations close on the free equations of motion~$\square \phi^i = \d^{\alpha\beta} \psi^i_\beta = 0$,
\begin{equation}\label{3dn9ffreal}
Q^{a}_{\alpha}\phi^{i}= {\left(\gamma^a\right)^i}_j \psi_{\alpha}^{j}~, \qquad Q_{\alpha}^{a}\psi_{\beta}^{i}= {\left(\gamma^a\right)^i}_j  \d_{\alpha \beta}\phi^{j}~.
\end{equation}
Here~$a = 1, \ldots, \CN$ is an~$\frak{so}(\mathcal{N})_R$ vector index and the~${\left(\gamma^a\right)^i}_j$ are~$\frak{so}(\mathcal{N})_R$ gamma matrices, while~$\d_{\alpha\beta} = \gamma^\mu_{\alpha\beta} \d_\mu$ is a standard spacetime derivative in bispinor notation of the~$\frak{su}(2)$ Lorentz algebra.

We will now show that all~$\CN \geq 9$ SCFTs are necessarily free. It suffices to consider the case~$\CN=9$. By applying the general constraints around~\eqref{genncurcons} and the results in appendix~\ref{sec:d3app} to the case~$\CN=9$, it can be shown that the only candidate~$\CN=9$ stress tensor multiplet consistent with the requirements of section~\ref{sec:stm} is given by 
 \begin{equation}
\xymatrix  @R=1pc {
 *++[F=]{B_{1}[0]_{1}^{(0,0,0,2)}} \ar[r]^-Q&  *++[F]{[1]_{\frac{3}{2}}^{(0,0,0,2)\oplus (0,0,1,0)}} \ar[r]^-Q&  *++[F]{[0]_{2}^{(0,0,0,2)} \oplus [2]_{2}^{(0,0,1,0)\oplus (0,1,0,0)}} \ar[dddll]_{Q}\\
 \\
 \\
 *++[F]{[3]_{\frac{5}{2}}^{(0,1,0,0) \oplus (1,0,0,0) }}\ar[r]^-Q &*++[F]{[4]_{3}^{(0,0,0,0)\oplus (1,0,0,0) }}\ar[r]^-Q& *++[F]{[5]_{\frac{7}{2}}^{(0,0,0,0) }} 
 }
 \label{3dn9em}
 \end{equation}
The operators~$[2]_2^{(0,1,0,0)}$, $[3]^{(1,0,0,0)}_{5/2}$, and~$[4]_3^{(0,0,0,0)}$ have the quantum numbers of the~$R$-currents~$j_\mu^{(R)}$ in the adjoint of~$\frak{so}(9)_R$, the supersymmetry currents~$S_{\mu\alpha}$ in the vector representation of~$\frak{so}(9)_R$, and the~$R$-symmetry neutral stress tensor~$T_{\mu\nu}$. Moreover, $R$-symmetry selection rules ensure that the action of the~$Q$-supercharges on these operators is consistent with~\eqref{QScurralg}, \eqref{qjralg}, and~\eqref{Ttop}. This shows that~\eqref{3dn9em} is in fact a stress tensor multiplet. However, the multiplet also contains other operators with the same Lorentz quantum numbers as~$j_\mu^{(R)}, S_{\mu\alpha}, T_{\mu\nu}$, but different~$R$-symmetry representations. The action of the supercharges on these operators terminates on a higher-spin current~$[5]_{7 \over 2}^{(0,0,0,0)}$. According to the results of~\cite{Maldacena:2011jn}, the presence of such a current requires that a quantum field theory be (locally) free. In terms of the free field multiplet~\eqref{3dn9ffreal}, the SCP~$[0]_{1}^{(0,0,0,2)}$ of the stress tensor multiplet in~\eqref{3dn9em} is a suitable~$R$-symmetry projection of the symmetric bilinear~$\phi^{(i} \phi^{j)}$. 
\end{itemize}

\subsection{Short Multiplets of the Conformal Algebra}

\label{sec:smca}

The basic structure of short, unitary~$\frak{so}(d,2)$ multiplets was reviewed in section~\ref{sec:introcft}. We will now discuss some of their properties in more detail (see also~\cite{Minwalla:1997ka,Rychkov:2016iqz} and references therein). Consider a conformal multiplet with CP
\begin{equation}
\CC = [L]_\Delta~.
\end{equation}
Unitarity requires the norms of all level-one CDs~$P_\mu \CC$ to be non-negative. We can decompose these descendants into irreducible Lorentz representations~$L'$,
\begin{equation}\label{lpdef}
L_\text{vec.} \otimes L = \bigoplus_{L'} L' \qquad (\text{finite sum})~,
\end{equation}
where~$L_\text{vec.}$ is the vector representation of~$\frak{so}(d)$ carried by~$P_\mu$. Up to a positive factor (indicated by~$\sim$ below), the norms of level-one CDs in the~$L'$ representation are given by 
\begin{equation}
|| P_\mu \CC \big|_{L'} ||^2 \sim \Delta + \half \Big(c_2(L') - c_2(L_\text{vec.}) - c_2(L) \Big)~.
\end{equation}
Here~$c_2(L)$ is the quadratic Casimir invariant of~$L$, normalized so that the~$d$-dimensional vector representation~$L_\text{vec.}$ has~$c_2(L_\text{vec.}) = d-1$. Clearly the states with the smallest norm have the smallest value of~$c_2(L')$, and hence they lead to the strongest unitarity bound,
\begin{equation}
\Delta \geq \Delta_* = \frac{1}{2}\left(c_2(L)+d-1-\min_{L'} c_2(L')\right)~. \label{bosonicunitarity}
\end{equation}
In this formula~$L'$ is the unique Lorentz representation on the right-hand side of~\eqref{lpdef} that minimizes~$c_2(L')$. When the  bound~\eqref{bosonicunitarity} is saturated, we obtain a short conformal multiplet of dimension~$\Delta = \Delta_*$ , with a primary null representation~$[L']_{\Delta_*+1}$ at level one. Therefore, some contraction of a spacetime derivative with the CP~$\CC$ vanishes (see below for examples).

As long as the Lorentz representation~$L$ is non-trivial, the condition in~\eqref{bosonicunitarity} is both necessary and sufficient for unitarity, i.e.~no independent unitarity bounds arise at higher levels. The only exception occurs when~$\CC$ is a Lorentz scalar, in which case~\eqref{bosonicunitarity} reduces to~$\Delta \geq 0$. This bound is saturated by the unit operator, which is annihilated by all~$P_\mu$. If~$\Delta > 0$, there is a second independent unitarity constraint,
\begin{equation}
\Delta \geq \frac{d-2}{2}~, \label{scalarsec}
\end{equation}
which arises by demanding that the level-two descendant~$\square \CC = \d^\mu \d_\mu \CC$ have positive norm. Hence the bound in~\eqref{scalarsec} is saturated if and only if~$\CC$ is a free scalar annihilated by~$\square$.

We have encountered two kinds of short conformal multiplets~$\CC$: those with non-trivial~$L$, which carry spin and vanish when contracted with a single spacetime derivative, and free Lorentz scalars, which are annihilated by~$\square = \d^\mu \d_\mu $. For some short multiplets with spin, the level-one primary null state also gives rise to a level-two descendant null state of the form~$\square \CC = 0$. Such multiplets describe conformal free fields with spin, e.g.~free Dirac fermions, which exist in any dimension~$d$. In general, we will refer to any short conformal multiplet (with or without spin) that is annihilated by~$\square$ as a free field, because it creates a normalizable single-particle state when acting on the standard Minkowski vacuum.\footnote{~A detailed discussion of which short conformal multiplets are actually free fields annihilated by~$\square$ can be found in~\cite{Siegel:1988gd} and section~2.6 of~\cite{Minwalla:1997ka}.} All other short~$\frak{so}(d,2)$ multiplets are annihilated by a first-order differential operator and will be referred to as (generalized) conserved currents. 

A simple class of conserved currents, which exists in every dimension~$d$, is furnished by symmetric traceless Lorentz tensors of rank~$s \in \Z_{\geq 1}$, 
\begin{equation}\label{rankstensors}
T_{\mu_1 \cdots \mu_s} = T_{(\mu_1 \cdots \mu_s)} - (\text{traces})~.
\end{equation}
In this case the unitarity bound~\eqref{bosonicunitarity} takes the simple form
\begin{equation}
\Delta \geq d+s-2~.
\end{equation}  
If this bound is saturated, the resulting level-one null state leads to the following conservation equation,
\begin{equation}\label{symtensconslaw}
\d^{\mu_1} T_{\mu_1 \cdots \mu_s}=0~.
\end{equation}
The special cases~$s = 1$ and~$s = 2$ correspond to a conserved flavor current with~$\Delta = d-1$ and a conserved, symmetric, traceless stress tensor with~$\Delta = d$, respectively. In~$d \geq 3$ dimensions, conserved higher-spin currents with~$s \geq 3$ are only believed to exist in CFTs that contain a locally free, decoupled subsector~\cite{Maldacena:2011jn,Alba:2015upa}.

In order to define a notion of spin for operators in more general Lorentz representations, we employ an orthogonal basis where the half-integral~$\frak{so}(d)$ weights~$h_i~(i = 1, \ldots, r)$ are eigenvalues under rotations in~$r$ mutually orthogonal planes, which define a Cartan subalgebra (see appendix~\ref{app:liealg} for more detail). A highest weight always satisfies
\begin{equation}
h_1 \geq h_2 \geq \cdots \geq |h_r| \geq 0~,
\end{equation}
so that~$h_1$ is the largest eigevalue that can arise for any rotation. For instance, the symmetric, traceless rank-$s$ tensors described around~\eqref{rankstensors} have highest weights~$h_1 = s$, $h_{i \geq 2} = 0$. More generally, we will refer to the~$h_1$ highest weight of any Lorentz representation as its spin. In~$d \geq 4$, there are increasingly complicated patterns of higher-spin conserved currents with~$h_1 > 2$, which are currently not well understood. By contrast, higher-spin free fields with~$h_1 > 1$ are ruled out by the existence of a well-defined stress tensor~\cite{Weinberg:1980kq}. 

Another useful property of the~$h_1$ spin is that it simplifies the expression for the scaling dimension~$\Delta = \Delta_*$ of a short conformal multiplet that saturates the unitarity bound~\eqref{bosonicunitarity}. For a sufficiently generic Lorentz representation~$L$,\footnote{~See section~2 of~\cite{Minwalla:1997ka} for a general discussion that applies in any spacetime dimension~$d$. A detailed summary of the cases~$3 \leq d \leq 6$ appears below.}
\begin{equation}\label{deltastar}
\Delta_* = h_1(L) + d-2\qquad (L~\text{generic})~.
\end{equation}
If~$L$ is not generic, then the offset on the right-hand side of~\eqref{deltastar} may be smaller than~$d-2$. However, the right-hand side of~\eqref{deltastar} always constitutes an upper bound,
\begin{equation}\label{deltastarbound}
\Delta_* \leq h_1(L) + d-2 \qquad (\text{any}~L).
\end{equation}
This bound will play an important role below.

\subsection{Identifying Superconformal Current Multiplets}

\label{sec:findsccm}

We will now describe how to identify unitary superconformal multiplets~$\CM$ that contain short representations of the conformal algebra, i.e.~conserved currents or free fields~$\CJ$. We will loosely refer to such~$\CM$ as superconformal current multiplets. Since~$\CM$ necessarily contains the~$\frak{so}(d,2)$ null states associated with~$\CJ$, it must be a short superconformal multiplet. Often, the SCP~$\CV$ of~$\CM$ is not a conserved current or free field, and~$\CJ$ resides at some positive level~$\ell \in \Z_{>0}$ inside~$\CM$, so that~$\CJ \sim Q^\ell \CV$. Therefore, their scaling dimensions are related as follows,
\begin{equation}
\Delta_\CV = \Delta_\CJ - {\ell \over 2}~.
\end{equation}
As explained in the previous subsection, the scaling dimension of~$\CJ$ satisfies the bound~\eqref{deltastarbound}, so that
\begin{equation}
\Delta_\CV \leq h_1(\CJ)+d-2 - \frac{\ell}{2}~. \label{currentdim}
\end{equation}

Since the multiplet~$\CM$ is labled by the quantum numbers of~$\CV$, it is desirable to turn~\eqref{currentdim} into a bound that only involves those quantum numbers. This can be done, because the orthogonal highest weight~$h_1$ is subadditive under tensor products (see appendix~\ref{app:liealg}): given two irreducible~$\frak{so}(d)$ representations~$L_1, L_2$ and a third irreducible representation~$L$ in their tensor product,~$L \in L_1 \otimes L_2$, it follows that
\begin{equation}
h_1(L)\leq h_1(L_{1})+h_1(L_{2})~.\label{subadditive}
\end{equation} 
Since the current~$\CJ \sim Q^\ell \CV$ resides at level~$\ell$, its Lorentz representation must occur in an~$\ell$-fold tensor product of~$Q$-supercharges with the SCP~$\CV$. Since the supercharges are fundamental spinors with~$h_1(Q) = \half$, applying~\eqref{subadditive} leads to 
\begin{equation}
h_1(\CJ) \leq h_1(\CV) + {\ell \over 2}~.
\end{equation} 
Substituting into~\eqref{currentdim} then leads to the bound 
\begin{equation}
\Delta_\CV \leq  h_1(\CV)+d-2~. \label{primaryh}
\end{equation}
Note that this bound only involves the quantum numbers of~$\CV$; the level~$\ell$ at which the current resides has dropped out. These properties make~\eqref{primaryh} a very effective necessary condition that can be used to identify a list of candidate superconformal current multiplets~$\CM$ that could conceivably contain a conserved current or free field. 

The remainder of this section is ordered according to increasing spacetime dimension $d$. For each value~$3 \leq d \leq 6$, we spell out the structure of short~$\frak{so}(d,2)$ multiplets that were summarized more generally in section~\ref{sec:smca}. In particular, we explicitly translate from~$\frak{so}(d)$ orthogonal weights~$h_i$ to Dynkin labels,  which are used throughout the rest of the paper. For each value of~$\CN$ with~$N_Q \leq 16$ (see section~\ref{sec:maxsusy}), we use the criterion~\eqref{primaryh} to generate a list of candidate superconformal current multiplets. We then apply the algorithm described in section~\ref{sec:ouralg} to deduce the operator content of these candidate multiplets and determine whether a conserved current or free field is actually present.

\subsection{Superconformal Current Multiplets:~$d=3$}

In this section we tabulate superconformal current multiplets in three dimensions, which contain short representations of the conformal algebra~$\frak{so}(3,2)$. The unitary representations of this algebra are summarized in table~\ref{tab:3Dcurrents} (see also page~10 of~\cite{Minwalla:1997ka}), where we specify the Lorentz quantum number of the CP, unitarity bounds on its scaling dimension, and the quantum numbers of the primary null state that results when these bounds are saturated (see section~\ref{sec:3dmult} for a summary of our conventions in~$d = 3$). We also indicate whether the CP is a free field annihilated by~$\square$. 

\smallskip

\renewcommand{\arraystretch}{1.6}
\renewcommand\tabcolsep{6pt}
\begin{table}[H]
  \centering
  \begin{tabular}{ | l  | l |l  |c| }
\hline
\multicolumn{1}{|c|}{\bf Primary} &  \multicolumn{1}{|c|}{\bf Unitarity Bound} & \multicolumn{1}{|c|}{\bf Null State } & \multicolumn{1}{|c|}{\bf Comments}\\
\hline
\hline
$ [j]_{\Delta}~,~j\geq 2 $ & $\Delta \geq \half \, j  + 1 $ & $[j-2]_{\Delta+1}$ & -\\
\hline
\hline
$ [1]_{\Delta}~~ $ & $\Delta \geq   1 $ & $[1]_{\Delta+1}$ & free field\\
\hline
$ [0]_{\Delta}~~ $ & $\Delta \geq   \half $ & $[0]_{\Delta+2}$ & free field\\
\hline
\hline
$ [0]_{\Delta}~~ $ & $\Delta = 0 $ & $[2]_{\Delta+1}$ & unit operator\\
\hline
\end{tabular}
  \caption{Unitary representations of the conformal algebra in~$d=3$. }
   \label{tab:3Dcurrents}
\end{table}

\smallskip

\noindent In general, we count currents or free fields modulo conservation laws or field equations. Applying this prescription to the operators in table~\ref{tab:3Dcurrents}, and recalling that the dimension of the~$\frak{su}(2)$ Lorentz representation~$[j]$ is~$j+1$, we conclude that the currents $[j \geq 2]_{j/2+1}$ always contain $(j+1) - (j-1) = 2$ independent operators. By contrast, the free fields~$[1]_1, [0]_{1/2}$ do not contain any operator degrees of freedom.

We now apply the method described in section~\ref{sec:findsccm} to identify all superconformal current multiplets in~$d = 3$, i.e.~multiplets that contain short conformal multiplets. To this end, we substitute~$h_1 = \half j$ (see~\eqref{hdlambda} in appendix~\ref{app:liealg}) and~$d = 3$ into the general bound~\eqref{primaryh} that must be satisfied by the SCP~$\CV$ of any superconformal current multiplet, 
\begin{equation}\label{primaryhii}
\Delta_\CV \leq {j \over 2} + 1~.
\end{equation}
We can now combine this bound with the superconformal unitarity constraints on~$\CV$ summarized in section~\ref{sec:3dmult} to identify and analyze all candidate superconformal current multiplets. Below, we will do this explicitly for~$1 \leq \CN \leq 8$; as discussed in section~\ref{sec:maxsusy}, quantum field theories with~$\CN \geq 9$ are necessarily free, and we will not discuss them further here.  It is straightforward to enumerate the candidate current multiplets for all values of~$\CN \geq 3$ (the cases~$\CN=1,2$ require a separate discussion), using the form of the superconformal unitarity bounds in appendix~\ref{app:genN}, which applies uniformly for general values of~$\CN$:
\begin{itemize}
\item $A_{1}, A_{2}$ multiplets in trivial representations of the $R$-symmetry.
\item $B_{1}$ multiplets whose~$\frak{so}(\CN)_R$ Dynkin labels~$R_i$ satisfy
\begin{equation}\label{genncurcons}
\begin{cases}
R_{1}+R_{2}+\cdots +R_{\frac{\mathcal{N}}{2}-2}+\frac{1}{2}\left(R_{\frac{\mathcal{N}}{2}-1}+R_{\frac{\mathcal{N}}{2}}\right)\leq 1~, & \mathcal{N}~\mathrm{even}\\
R_{1}+R_{2}+\cdots +R_{\lfloor\frac{\mathcal{N}}{2}\rfloor-1}+\frac{1}{2} \, R_{\lfloor\frac{\mathcal{N}}{2}\rfloor} \leq 1~, & \mathcal{N}~\mathrm{odd}~.
\end{cases}
\end{equation}
\end{itemize}

\subsubsection{$d=3$,~$\mathcal{N}=1$}

\label{sec:3dn1curr}

By comparing the bound~\eqref{primaryhii} with the superconformal unitarity restrictions summarized in section~\ref{sec:3dn1def}, we find the following superconformal current multiplets:
\begin{itemize}
\item The~$B_1[0]_{0}$ multiplet consists of the unit operator. 
\item $A'_2$-multiplets contain a free scalar and a free spin-$\half$ fermion,
\begin{equation}  \label{3dn1f}
 \xymatrix  @R=1pc {
 *++[F=]{{A'_2[0]}_\half} \ar[r]^-Q&  *++[F]{{[1]}_{1}}
}
\end{equation}

\item $A_1$-multiplets have the following operator content:
\begin{equation}
\xymatrix  @R=1pc {
 *++[F=]{{A_1[j \geq 1]}_{{j \over 2}+1}} \ar[r]^-Q&  *++[F]{{[j+1]}_{{j \over 2}+{3 \over 2}}}
}
 \label{3dn1c}
 \end{equation}
When~$j = 1$, the bottom component is a spin-$\half$ fermion of dimension~$\Delta = {3 \over 2}$, which does not satisfy any differential equation, and the top component is a conserved spin-$1$ current. Therefore~$A_1[1]_{3/2}$ is a flavor current multiplet (see section~\ref{sec:fcm}). Similarly, when~$j = 2$ we find spin-$1$ and spin-$3 \over 2$ currents in an~$A_1[2]_{2}$ extra SUSY-current multiplet (see section~\ref{sec:escm}), and when~$j = 3$ we find spin-$3 \over 2$ and spin-$2$ currents in an~$A_1[3]_{5/2}$ stress tensor multiplet (see section~\ref{sec:stm}, as well as section 2 of ~\cite{Bergshoeff:2010ui}). All~$A_1$-multiplets with~$j \geq 4$ contain higher-spin currents. After taking into account conservation laws, we find that all~$A_1$-multiplets have~$2+2$ bosonic and fermionic operators. 
\end{itemize}

\subsubsection{$d=3$,~$\mathcal{N}=2$}

\label{sec:3dn2curr}

Here we must compare the bound~\eqref{primaryhii} with the consistent~$\CN=2$ multiplets summarized in section~\ref{sec:3dn2def}. Some chiral~$L \b B_1[0]^{(r)}_{\Delta = r}$ multiplets (with~$\half < r \leq 1$) and some anti-chiral~$B_1 \b L[0]^{(r)}_{\Delta = - r}$ multiplets (with~$- 1 \leq r < - \half$) satisfy the bound, even though they do not contain any conserved currents. (The operator content of these multiplets is tabulated in section~\ref{sec:3dn2tabs}.) All other candidate multiplets that satisfy~\eqref{primaryhii} contain currents or free fields:
\begin{itemize}
\item The~$B_1 \b B_1[0]^{(0)}_{0}$ multiplet consists of the unit operator. 
\item Free chiral~$A_2 \b B_1[0]^{(1/2)}_{1/2}$ multiplets and their conjugate anti-chiral~$\b A_2 B_1[0]^{(- 1/2)}_{\Delta = 1/2}$ multiplets contain a free scalar and a free spin-$\half$ fermion:
\begin{equation}
\xymatrix  @R=1pc {
 *++[F=]{A_2 \b B_1[0]_{\frac{1}{2}}^{(\frac{1}{2})}} \ar[r]^-Q&  *++[F]{[1]_{1}^{(-\frac{1}{2})}}}
 \label{3dn2fc}
\end{equation}
\begin{equation}
\xymatrix  @R=1pc {
 *++[F=]{\b A_2 B_1[0]_{\frac{1}{2}}^{(-\frac{1}{2})}} \ar[r]^-{\overline Q}&  *++[F]{[1]_{1}^{(\frac{1}{2})}} }
 \label{3dn2fc}
\end{equation}

\item The~$A_2 \b A_2[0]_1^{(0)}$ multiplet is a~$4+4$ flavor current multiplet (see section~\ref{sec:fcm}): 
\begin{equation}
\xymatrix @C=7pc @R=7pc @!0 @dr {
*++[F=]{A_2 \b A_2[0]_1^{(0)}}  \ar[r]|--{{~\b Q~}} \ar[d]|--{~Q~} 
& *++[F]{[1]^{(+1)}_{3\over2}}  
\ar[d]|--{~Q~}
\\
*++[F]{[1]^{(-1)}_{3\over2}} 
\ar[r]|--{~\b Q~}
& *++[F]{[0]_2^{(0)} \oplus [2]_2^{(0)}}
}
\label{3dn2g}
\end{equation}

\item $A_1 \b A_1[j]_{\half j+1}^{(0)}$ multiplets have the following operator content: 
\begin{equation}
\xymatrix @C=7pc @R=7pc @!0 @dr {
*++[F=]{A_1 \b A_1[j \geq 1]_{\half j+1}^{(0)}}  \ar[r]|--{{~\b Q~}} \ar[d]|--{~Q~} 
& *++[F]{[j+1]_{\half j+{3\over2}}^{(+1)}}  
\ar[d]|--{~Q~}
\\
*++[F]{[j+1]_{\half j+{3\over2}}^{(-1)}} 
\ar[r]|--{~\b Q~}
& *++[F]{[j+2]_{\half j+2}^{(0)}}
}
\label{3dn2gen}
\end{equation}
When~$j = 1$, the bottom component is a spin-$\half$ fermion with~$\Delta = {3 \over 2}$, which does not satisfy any differential equation. The middle and top components are conserved spin-$1$ and spin-$3 \over 2$ currents, making~$A_1 \b A_1[1]_{3/2}^{(0)}$ an extra SUSY-current multiplet (see section~\ref{sec:escm}). If~$j = 1$, the bottom component is a spin-$1$~$\frak{u}(1)_R$-current, the middle components are spin-$3 \over 2$ supersymmetry currents, and the top component is a spin-$2$ stress tensor, making~$A_1 \b A_1[2]_{2}^{(0)}$ a stress tensor multiplet (see section~\ref{sec:stm}, as well as section 2 of~\cite{Bergshoeff:2010ui}). All~$A_1 \b A_1[j]_{\half j+1}^{(0)}$ multiplets with~$j \geq 3$ contain higher-spin currents. After taking into account conservation laws, all of these multiplets contain~$4+4$ bosonic and fermionic operators.
\end{itemize}

\subsubsection{$d=3$,  $\mathcal{N}=3$}

\label{sec:3dn3curr}

Here we can use the results around~\eqref{genncurcons} for~$\CN=3$ (see section~\ref{sec:3dn3}), to conclude that currents can only reside in~$A_{1,2}$-multiplets with~$R = 0$ and~$B_1$-multiplets with~$R = 0,1,2$. We now examine these multiplets in turn:

\begin{itemize}
\item The~$B_1[0]_0^{(0)}$ representation consists of the unit operator. 
\item The~$B_{1}[0]^{(1)}_{1/2}$ representation is a free hypermultiplet, consisting of a free Lorentz scalar and a free spin-$\half$ fermion, both of which are~$\frak{su}(2)_R$ doublets:
\begin{equation}
\xymatrix  @R=1pc {
 *++[F=]{B_{1}[0]^{(1)}_{\half} } \ar[r]^-Q&  *++[F]{[1]_{1}^{(1)}} }
 \label{3dn3fh}
\end{equation}
\item  The~$B_{1}[0]^{(2)}_1$ representation is an~$8+8$ flavor current multiplet (see section~\ref{sec:fcm}): 
 \begin{equation}
\xymatrix  @R=1pc {
 *++[F=]{B_{1}[0]^{(2)}_1} \ar[r]^-Q&  *++[F]{[1]_{\frac{3}{2}}^{(0)\oplus (2)}} \ar[r]^-Q&  *++[F]{[2]_{2}^{(0)}\oplus [0]_{2}^{(2)}} }
 \label{3dn3gc}
\end{equation}
\item The~$A_{2}[0]^{(0)}_1$ multiplet is an~$8+8$ extra SUSY-current multiplet (see section~\ref{sec:escm}):
 \begin{equation}
\xymatrix  @R=1pc {
 *++[F=]{A_{2}[0]^{(0)}_1} \ar[r]^-Q&  *++[F]{[1]_{\frac{3}{2}}^{(2)}} \ar[r]^-Q&  *++[F]{[2]_{2}^{(2)}\oplus [0]_{2}^{(0)}} \ar[r]^-Q&  *++[F]{[3]_{\frac{5}{2}}^{(0)}}}
 \label{3dn3sc}
 \end{equation}
\item $A_{1}[j]^{(0)}_{\half j +1}$ multiplets  have the following operator content:
 \begin{equation}
\xymatrix  @R=1pc {
 *++[F=]{A_{1}[j]^{(0)}_{\half j +1}} \ar[r]^-Q&  *++[F]{[j+1]_{\half j+\frac{3}{2}}^{(2)}} \ar[r]^-Q&  *++[F]{[j+2]_{\half j+2}^{(2)}} \ar[r]^-Q&  *++[F]{[j+3]_{\half j+\frac{5}{2}}^{(0)}}}
 \label{3dn3em}
 \end{equation}
The case~$j = 1$ is a stress tensor multiplet (see section~\ref{sec:stm}, as well as section 2 of~\cite{Bergshoeff:2010ui}), while all multiplets with~$j \geq 2$ contain higher-spin currents. After subtracting conservation laws, all of these multiplets contain~$8+8$ bosonic and fermionic operators.
\end{itemize}

\subsubsection{$d=3$,  $\mathcal{N}=4$}
 
\label{sec:3dn4curr} 
 
If we apply the results around~\eqref{genncurcons} for~$\CN=4$ (see section~\ref{sec:3dN4defs}), we find that currents can only reside in~$A_{1,2}$-multiplets with~$R_1 = R_2 = 0$ and~$B_1$-multiplets with~$R_1 + R_2 \leq 2$. We will now examine these possibilities in detail:
\begin{itemize}
\item The~$B_1[0]^{(0;0)}_0$ multiplet consists of the unit operator. 
 \item The~$B_{1}[0]^{(1;0)}_{1/2}$ hypermultiplet and~$B_{1}[0]^{(0;1)}_{1/2}$ twisted hypermultiplet, which are exchanged by the mirror automorphism~$M$ of the~$\CN=4$ superconformal algebra (see section~\ref{sec:3dN4defs}), both contain free scalars and spin-$\half$ fermions,
\begin{equation}
\xymatrix  @R=1pc {
 *++[F=]{B_{1}[0]^{(1;0)}_{\half}} \ar[r]^-Q&  *++[F]{[1]_{1}^{(0;1)}} }
 \label{3dn4fh1}
\end{equation}
\begin{equation}
\xymatrix  @R=1pc {
 *++[F=]{B_{1}[0]^{(0;1)}_{\half}} \ar[r]^-Q&  *++[F]{[1]_{1}^{(1;0)}}}
 \label{3dn4fh2}
\end{equation}

\item Both~$B_{1}[0]^{(2;0)}_1$ and~$B_1[0]^{(0;2)}_1$ are~$8+8$ flavor current multiplets (see section~\ref{sec:fcm}), which are exchanged by the mirror automorphism~$M$ (see section~\ref{sec:3dN4defs}): 
\begin{equation}
\xymatrix  @R=1pc {
 *++[F=]{B_{1}[0]^{(2;0)}_1} \ar[r]^-Q&  *++[F]{[1]_{\frac{3}{2}}^{(1;1)}} \ar[r]^-Q&  *++[F]{[2]_{2}^{(0;0)}\oplus [0]_{2}^{(0;2)}}}
 \label{3dn4gc1}
\end{equation}
\begin{equation}
\xymatrix  @R=1pc {
 *++[F=]{B_1[0]^{(0;2)}_1} \ar[r]^-Q&  *++[F]{[1]_{\frac{3}{2}}^{(1;1)}} \ar[r]^-Q&  *++[F]{[2]_{2}^{(0;0)}\oplus [0]_{2}^{(2;0)}} }
 \label{3dn4gc2}
\end{equation}

\item $B_{1}[0]^{(1;1)}_1$ is a~$16+16$ extra SUSY-current multiplet (see section~\ref{sec:escm}):
\begin{equation}
\xymatrix  @R=1pc {
 *++[F=]{B_{1}[0]^{(1;1)}_1} \ar[r]^-Q&  *++[F]{[1]_{\frac{3}{2}}^{(0;0)\oplus (2;0)\oplus (0;2)}} \ar[r]^-Q&  *++[F]{[2]_{2}^{(1;1)}\oplus [0]_{2}^{(1;1)}}  \ar[r]^-Q&  *++[F]{[3]_{\frac{5}{2}}^{(0;0)}} }
 \label{3dn4sc}
\end{equation}

\item $A_2[0]^{(0;0)}_1$ is a~$16+16$ stress tensor multiplet (see section~\ref{sec:stm}, as well as section 2 of~\cite{Bergshoeff:2010ui}):
 \begin{equation}
\xymatrix  @R=1pc {
 *++[F=]{A_2[0]^{(0;0)}_1} \ar[r]^-Q&  *++[F]{[1]_{\frac{3}{2}}^{(1;1)}} \ar[r]^-Q&  *++[F]{[2]_{2}^{(2;0)} \oplus [2]_{2}^{(0;2)}\oplus [0]_{2}^{(0;0)}} \ar[r]^-Q&  *++[F]{[3]_{\frac{5}{2}}^{(1;1)}}\ar[r]^-Q&  *++[F]{[4]_{3}^{(0;0)}} } \phantom{YYY}
 \label{3dn4em}
 \end{equation}

\item $A_{1}[j]^{(0;0)}_{j/2+1}$ multiplets have the following operator content:  \begin{equation}
\xymatrix  @R=1pc {
 *++[F=]{A_{1}[j \geq 1]^{(0;0)}_{{j\over 2}+1}} \ar[r]^-Q&  *++[F]{[j+1]_{\half j+\frac{3}{2}}^{(1;1)}} \ar[r]^-Q&  *++[F]{[j+2]_{\half j+2}^{(2;0)\oplus (0;2)}} \ar[ddll]_Q 
 \\
 \\
*++[F]{[j+3]_{\half j+\frac{5}{2}}^{(1;1)}}\ar[r]^-Q&  *++[F]{[j+4]_{\half j+3}^{(0;0)}} }
 \label{3dn4hs}
 \end{equation}
All of these multiplets contain higher-spin currents. Modulo conservation laws, they contain~$16+16$ operators. 
 \end{itemize}

 \subsubsection{$d=3$, $\mathcal{N}=5$}
 
\label{sec:3dn5curr} 
 
The criterion around~\eqref{genncurcons}, applied to~$\CN=5$ (see section~\ref{sec:3dN5defs}), states that currents must reside in~$A_{1,2}$-multiplets with~$R_1 = R_2 = 0$ or~$B_1$-multiplets with~$R_1 + \half R_2 \leq 1$. Explicitly:
\begin{itemize}
\item The~$B_1[0]_0^{(0,0)}$ multiplets consists of the identity operator. 
\item  $B_{1}[0]^{(0,1)}_{1/2}$ is a free hypermultiplet:
 \begin{equation}
\xymatrix  @R=1pc {
 *++[F=]{B_{1}[0]^{(0,1)}_\half} \ar[r]^-Q&  *++[F]{[1]_{1}^{(0,1)}}}
 \label{3dn5free}
 \end{equation}
\item $B_{1}[0]^{(0,2)}_1$ is a~$32+32$ extra SUSY-current multiplet (see section~\ref{sec:escm}):
 \begin{equation}
\xymatrix  @R=1pc {
 *++[F=]{B_{1}[0]^{(0,2)}_1} \ar[r]^-Q&  *++[F]{[1]_{\frac{3}{2}}^{(0,2) \oplus (1,0)}} \ar[r]^-Q&  *++[F]{[0]_{2}^{(0,2)} \oplus [2]_{2}^{(0,0) \oplus (1,0)} } \ar[r]^-Q&  *++[F]{[3]_{\frac{5}{2}}^{(0,0)}}  }
 \label{3dn5sc}
 \end{equation}
 Note, however, that the~$R$-singlet spin-$1$ current~$[2]_2^{(0,0)}$ at level two is a top component, since its tensor product with the supercharge~$Q = [1]_{1/2}^{(1,0)}$ does not contain an~$R$-symmetry singlet. Therefore this operator is a genuine flavor current, and according to the definition  of  section~\ref{sec:fcm} we can also regard~$B_{1}[0]^{(0,2)}_1$ as a flavor current multiplet. Since this flavor current resides in the same multiplet as an extra SUSY-current, it only exists when supersymmetry is enhanced beyond~$\CN = 5$. 
 
\item The $B_{1}[0]^{(1,0)}_1$ representation is a~$32+32$ stress tensor multiplet (see section~\ref{sec:stm}, as well as section 2 of~\cite{Bergshoeff:2010ui}):
 \begin{equation}
\xymatrix  @R=1pc {
 *++[F=]{B_{1}[0]^{(1,0)}_1} \ar[r]^-Q&  *++[F]{[1]_{\frac{3}{2}}^{(0,0)\oplus(0,2)}} \ar[r]^-Q&  *++[F]{[0]_{2}^{(1,0)} \oplus [2]_{2}^{(0,2)}} \ar[r]^-Q&  *++[F]{[3]_{\frac{5}{2}}^{(1,0)}}\ar[r]^-Q&  *++[F]{[4]_{3}^{(0,0)}} }
 \label{3dn5em}
 \end{equation}
\item The~$A_{2}[0]^{(0,0)}_1$ multiplet (with~$32+32$ operators) contains higher-spin currents:
 \begin{equation}
\xymatrix  @R=1pc {
 *++[F=]{A_{2}[0]^{(0,0)}_1} \ar[r]^-Q&  *++[F]{[1]_{\frac{3}{2}}^{(1,0)}} \ar[r]^-Q&  *++[F]{[0]_{2}^{(0,0)} \oplus [2]_{2}^{(0,2)} } \ar[ddll]_Q \\
\\
*++[F]{[3]_{\frac{5}{2}}^{(0,2)}} \ar[r]^-Q&
*++[F]{[4]_{3}^{(1,0)}}\ar[r]^-Q&  *++[F]{[5]_{\frac{7}{2}}^{(0,0)}} }
 \label{3dn5hs1}
 \end{equation}
\item All~$A_{1}[j]^{(0,0)}_{j/2+1}$ multiplets, which contain~$32+32$ bosonic and fermionic operators, harbor higher-spin currents: 
 \begin{equation}
\xymatrix  @R=1pc {
 *++[F=]{A_{1}[j]^{(0,0)}_{\half j+1}} \ar[r]^-Q&  *++[F]{[j+1]_{\half j+\frac{3}{2}}^{(1,0)}} \ar[r]^-Q&  *++[F]{[j+2]_{\half j+2}^{(0,2)}} \ar[ddll]_Q \\
 \\
 *++[F]{[j+3]_{\half j+\frac{5}{2}}^{(0,2)}}\ar[r]^-Q&  *++[F]{[j+4]_{\half j+3}^{(1,0)}}\ar[r]^-Q&  *++[F]{[j+5]_{\half j+\frac{7}{2}}^{(0,0)}} }
 \label{3dn5hs}
 \end{equation}
\end{itemize}

 \subsubsection{$d=3$,  $\mathcal{N}=6$}

\label{sec:3dn6curr}

In addition to~$A_{1,2}$-multiplets with~$R_1 = R_2 = R_3 =  0$, the criterion~\eqref{genncurcons} applied to~$\CN=6$ theories (see section~\ref{sec:3dn6defs}) states that currents can reside in~$B_1$-multiplets with $R_1 + \half \left(R_2 + R_3\right) \leq 1$. We will now discuss these options in turn:
\begin{itemize}
\item The~$B_1[0]^{(0,0,0)}_0$ multiplet consists of the unit operator.
\item The multiplets~$B_{1}[0]^{(0,1,0)}_{1/2}$  and $B_1[0]^{(0,0,1)}_{1/2}$ are free hypermultiplets, which are exchanged by complex conjugation: 
 \begin{equation}
\xymatrix  @R=1pc {
 *++[F=]{B_{1}[0]^{(0,1,0)}_{\half}} \ar[r]^-Q&  *++[F]{[1]_{1}^{(0,0,1)}}}
 \label{3dn6free1}
 \end{equation}
 \begin{equation}
\xymatrix  @R=1pc {
 *++[F=]{B_1[0]^{(0,0,1)}_{\half}} \ar[r]^-Q&  *++[F]{[1]_{1}^{(0,1,0)}}}
 \label{3dn6free2}
 \end{equation}
 
 \item Both~$B_{1}[0]^{(0,0,2)}_1$  and  $B_1[0]^{(0,2,0)}_1$ are~$32+32$ extra SUSY-current multiplets (see section~\ref{sec:escm}), which are exchanged by complex conjugation:
 \begin{equation}
\xymatrix  @R=1pc {
 *++[F=]{B_{1}[0]^{(0,0,2)}_1} \ar[r]^-Q&  *++[F]{[1]_{\frac{3}{2}}^{(0,1,1)}}\ar[r]^-Q&  *++[F]{[0]_{2}^{(0,2,0)} \oplus [2]_{2}^{(1,0,0)} } \ar[r]^-Q&  *++[F]{[3]_{\frac{5}{2}}^{(0,0,0)}}}
 \label{3dn6super1}
 \end{equation}
 \begin{equation}
\xymatrix  @R=1pc {
 *++[F=]{B_1[0]^{(0,2,0)}_1} \ar[r]^-Q&  *++[F]{[1]_{\frac{3}{2}}^{(0,1,1)}}\ar[r]^-Q&  *++[F]{ [0]_{2}^{(0,0,2)} \oplus [2]_{2}^{(1,0,0)}} \ar[r]^-Q&  *++[F]{[3]_{\frac{5}{2}}^{(0,0,0)}}}
 \label{3dn6super2}
 \end{equation}

\item $B_{1}[0]^{(0,1,1)}_1$ is a~$64+64$ stress tensor multiplet (see section~\ref{sec:stm} and section 2 of~\cite{Bergshoeff:2010ui}):
  \begin{equation}
\xymatrix  @R=1pc {
 *++[F=]{B_{1}[0]^{(0,1,1)}_1} \ar[r]^-Q&  *++[F]{[1]_{\frac{3}{2}}^{(0,0,2)\oplus(0,2,0)\oplus(1,0,0) }} \ar[ddl]_Q\\
 \\
  *++[F]{[0]_{2}^{(0,1,1)} \oplus [2]_{2}^{(0,0,0)\oplus (0,1,1)}} \ar[r]^-Q&  *++[F]{[3]_{\frac{5}{2}}^{(1,0,0)}}\ar[r]^-Q&  *++[F]{[4]_{3}^{(0,0,0)}} }
 \label{3dn6em}
 \end{equation}
Note that the~$R$-singlet spin-$1$ current~$[2]_2^{(0,0,0)}$ at level two is a top component, even though it is not forbidden from mapping to the supersymmetry current~$[3]_{5/2}^{(1,0,0)}$ at level three by quantum numbers. It is therefore an accidental top component and hence a flavor current. One way to see this is to decompose the multiplet in~\eqref{3dn6em} under an~$\hat \CN = 5$ subalgebra. The flavor current then resides in an extra SUSY-current multiplet~\eqref{3dn5sc}, where it is a manifest top component with respect to the~$\hat \CN = 5$ supercharges. Together with the full~$\frak{so}(6)_R$-symmetry of the parent theory, this allows us to conclude that it is in fact a full~$\CN=6$ top component. Therefore, the~$\CN=6$ stress tensor multiplet always contains a flavor current, as was pointed out in~\cite{Bashkirov:2011fr}.

\item The~$64+64$ multiplet~$B_{1}[0]^{(1,0,0)}_1$ contains higher-spin currents:
 \begin{equation}
\xymatrix  @R=1pc {
 *++[F=]{B_{1}[0]^{(1,0,0)}_1} \ar[r]^-Q&  *++[F]{[1]_{\frac{3}{2}}^{(0,0,0)\oplus (0,1,1)}} \ar[r]^-Q&  *++[F]{[0]_{2}^{(1,0,0)} \oplus [2]_{2}^{(0,0,2) \oplus (0,2,0) }  } \ar[ddll]_Q \\
 \\
 *++[F]{[3]_{\frac{5}{2}}^{(0,1,1)}}\ar[r]^-Q& *++[F]{[4]_{3}^{(1,0,0)}} \ar[r]^-Q&  *++[F]{[5]_{\frac{7}{2}}^{(0,0,0)}}}
 \label{3dn6hs1}
 \end{equation} 

\item The~$64+64$ multiplet~$A_{2}[0]^{(0,0,0)}_1$ also contains higher-spin currents:
 \begin{equation}
\xymatrix  @R=1pc {
 *++[F=]{A_{2}[0]^{(0,0,0)}_1} \ar[r]^-Q&  *++[F]{[1]_{\frac{3}{2}}^{(1,0,0)}} \ar[r]^-Q&  *++[F]{[0]_2^{(0,0,0)} \oplus [2]_{2}^{ (0,1,1)}} \ar[ddll]^-Q  \\
 \\
 *++[F]{[3]_{\frac{5}{2}}^{(0,0,2) \oplus(0,2,0)}} \ar[r]^-Q& *++[F]{[4]_{3}^{(0,1,1)}} \ar[r]^-Q& *++[F]{[5]_{\frac{7}{2}}^{(1,0,0)}}\ar[r]^-Q&*++[F]{[6]_{4}^{(0,0,0)}}
 }
 \label{3dn6hs2}
 \end{equation}

\item All~$A_{1}[j]^{(0,0,0)}_{j/2+1}$ multiplets contain higher-spin currents:
 \begin{equation}
\xymatrix  @R=1pc {
 *++[F=]{A_{1}[j \geq 1]^{(0,0,0)}_{{j \over 2}+1}} \ar[r]^-Q&  *++[F]{[j+1]_{\half j+\frac{3}{2}}^{(1,0,0)}} \ar[r]^-Q&  *++[F]{[j+2]_{\half j+2}^{(0,1,1)}} \ar[ddll]_Q &\\
 \\
 *++[F]{[j+3]_{\half j+\frac{5}{2}}^{(0,0,2) \oplus(0,2,0)}} \ar[r]^-Q &  *++[F]{[j+4]_{\half j+3}^{(0,1,1)}} \ar[r]^-Q &   *++[F]{[j+5]_{\half j+\frac{7}{2}}^{(1,0,0)}}\ar[r]^-Q&*++[F]{[j+6]_{\half j+4}^{(0,0,0)}}}
 \label{3dn6hs3}
 \end{equation}
 After subtracting conservation laws, all of these multiplets contain~$64+64$ bosonic and fermionic operators. 
\end{itemize}

\subsubsection{$d=3$,  $\mathcal{N}=7$}

\label{sec:3dn7stm}

As in previous subsections, it is straightforward to enumerate all~$\CN=7$ current multiplets by exploiting the constraints around~\eqref{genncurcons}, which allows currents to reside in~$A_{1,2}$-multiplets with~$R_1 = R_2 = R_3= 0$ or~$B_1$-multiplets with~$R_1 + R_2 + \half R_3 \leq 1$. The upshot of this analysis is that there is a unique stress-tensor multiplet (see section~\ref{sec:stm}), with~$128+128$ bosonic and fermionic operators (see also section 2 of~\cite{Bergshoeff:2010ui}):
  \begin{equation}
\xymatrix  @R=1pc {
 *++[F=]{B_{1}[0]^{(0,0,2)}_1} \ar[r]^-Q&  *++[F]{[1]_{\frac{3}{2}}^{(0,0,2)\oplus(0,1,0) }} \ar[ddl]_Q\\
 \\
  *++[F]{[0]_{2}^{(0,0,2)} \oplus [2]_{2}^{(0,1,0)\oplus (1,0,0)}} \ar[r]^-Q&  *++[F]{[3]_{\frac{5}{2}}^{(0,0,0) \oplus (1,0,0)}}\ar[r]^-Q&  *++[F]{[4]_{3}^{(0,0,0)}} }
 \label{3dn7em}
 \end{equation}
In addition to the~$\CN=7$ supersymmetry currents~$[3]^{(1,0,0)}_{5/2}$ at level three, there is on additional spin-$3 \over 2$ current~$[3]^{(0,0,0)}_{5/2}$ at this level. Since it is an~$R$-symmetry singlet, it cannot map into the stress tensor under the action of the supercharges~$Q = [1]_{1/2}^{(1,0,0)}$, and hence it is a manifest top component. We conclude that this operator is an extra SUSY current, and that the stress tensor multiplet is simultaneously also an extra SUSY-current multiplet (see section~\ref{sec:escm}). Since the stress tensor multiplet must be present in any~$\CN=7$ SCFT, the same is true for the extra SUSY current, and hence supersymmetry is necessarily enhanced beyond~$\CN=7$. In other words, there are no genuine~$\CN=7$ SCFTs. It can be checked that~\eqref{3dn7em} is nothing but the~$\frak{so}(8) \rightarrow \frak{so}(7)$~$R$-symmetry decomposition of the~$\CN=8$ stress tensor multiplets~\eqref{3dn8em1} and~\eqref{3dn8em2} discussed below.  Similar observations were previously made in~\cite{Bashkirov:2011fr}. 

\subsubsection{$d=3$,  $\mathcal{N}=8$}

\label{sec:3dn8curr}

According to the discussion around~\eqref{genncurcons}, conserved currents in~$\CN=8$ theories (see section~\ref{sec:d3n8defs}) can reside in~$A_{1,2}$-multiplets with~$R_1 = R_2 = R_3 =  R_4 = 0$ or~$B_1$-multiplets with~$R_1 + R_2+ \half \left(R_3 + R_4\right) \leq 1$. We will new discuss these candidate multiplets in detail:

\begin{itemize}
\item The~$B_1[0]^{(0,0,0,0)}_0$ multiplet consists of the unit operator.

\item  Both~$B_{1}[0]_{1/2}^{(0,0,1,0)}$ and~$B_1[0]_{1/2}^{(0,0,0,1)}$ are free hypermultiplets. They are exchanged by the~$\Z_2$ triality subgroup~$T$ described in section~\ref{sec:d3n8defs}, which fixes the~$Q$-supercharges in the vector representation~$(1,0,0,0)$ of~$\frak{so}(8)_R$, and exchanges the two spinor representations~$(0,0,1,0)$ and~$(0,0,0,1)$:
 \begin{equation}
\xymatrix  @R=1pc {
 *++[F=]{B_{1}[0]_{\half}^{(0,0,1,0)}} \ar[r]^-Q&  *++[F]{[1]_{1}^{(0,0,0,1)}}}
 \label{3dn8free1}
 \end{equation}
 \begin{equation}
\xymatrix  @R=1pc {
 *++[F=]{B_1[0]_{\half}^{(0,0,0,1)}} \ar[r]^-Q&  *++[F]{[1]_{1}^{(0,0,1,0)}}}
 \label{3dn8free2}
 \end{equation}

\item Both~$B_{1}[0]^{(0,0,2,0)}_1$  and $B_1[0]_1^{(0,0,0,2)}$ are~$128+128$ stress tensor multiplets (see section~\ref{sec:stm}, as well as section 2 of~\cite{Bergshoeff:2010ui}); they are exchanged by the same~$\Z_2$ triality subgroup~$T$ (see section~\ref{sec:d3n8defs}) that exchanges the two hypermultiplets in~\eqref{3dn8free1} and~\eqref{3dn8free2}:
 \begin{equation}
\xymatrix  @R=1pc {
 *++[F=]{B_{1}[0]^{(0,0,2,0)}_1} \ar[r]^-Q&  *++[F]{[1]_{\frac{3}{2}}^{(0,0,1,1)}} \ar[r]^-Q&  *++[F]{[0]_{2}^{(0,0,0,2)} \oplus [2]_{2}^{(0,1,0,0)} } \ar[r]^-Q&  *++[F]{[3]_{\frac{5}{2}}^{(1,0,0,0)}} \ar[r]^-Q&  *++[F]{[4]_{3}^{(0,0,0,0)}}}
 \label{3dn8em1}
 \end{equation}
 \begin{equation}
\xymatrix  @R=1pc {
 *++[F=]{B_1[0]_1^{(0,0,0,2)}} \ar[r]^-Q&  *++[F]{[1]_{\frac{3}{2}}^{(0,0,1,1)}} \ar[r]^-Q&  *++[F]{ [0]_{2}^{(0,0,2,0)} \oplus [2]_{2}^{(0,1,0,0)}} \ar[r]^-Q&  *++[F]{[3]_{\frac{5}{2}}^{(1,0,0,0)}} \ar[r]^-Q&  *++[F]{[4]_{3}^{(0,0,0,0)}}}
 \label{3dn8em2}
 \end{equation}
An irreducible quantum field theory (i.e.~a theory without locally decoupled subsectors) is expected to have a unique stress-tensor (see for instance~\cite{Maldacena:2011jn}). If this is the case, specifying the stress tensor multiplet also fixes a triality frame.   

\item The $B_{1}[0]^{(0,0,1,1)_1}$ multiplet contains higher-spin currents:
\begin{equation}
\xymatrix  @R=1pc {
 *++[F=]{B_{1}[0]^{(0,0,1,1)}_1} \ar[r]^-Q&  *++[F]{[1]_{\frac{3}{2}}^{(0,0,0,2)\oplus(0,0,2,0)\oplus(0,1,0,0)}} \ar[r]^-Q&  *++[F]{[0]_{2}^{(0,0,1,1)} \oplus [2]_{2}^{(0,0,1,1) \oplus(1,0,0,0)} } \ar[ddll]_Q\\
  \\
 *++[F]{[3]_{\frac{5}{2}}^{(0,0,0,0)\oplus(0,1,0,0)}}\ar[r]^-Q&   *++[F]{[4]_{3}^{(1,0,0,0)}} \ar[r]^-Q & *++[F]{[5]_{\frac{7}{2}}^{(0,0,0,0)}}}
 \label{3dn8hs1}
 \end{equation}
 After taking into account conservation laws, it contains~$256+256$ bosonic and fermionic operators.

\item $B_{1}[0]_1^{(1,0,0,0)}$ multiplets, with~$256+256$ operators, also harbor higher-spin currents:
 \begin{equation}
\xymatrix  @R=1pc {
 *++[F=]{B_{1}[0]_1^{(1,0,0,0)}} \ar[r]^-Q&  *++[F]{[1]_{\frac{3}{2}}^{(0,0,0,0)\oplus(0,1,0,0)}} \ar[r]^-Q&  *++[F]{[0]_{2}^{(1,0,0,0)} \oplus [2]_{2}^{(0,0,1,1)}} \ar[r]^-Q&  *++[F]{[3]_{\frac{5}{2}}^{(0,0,0,2) \oplus (0,0,2,0)}}\ar[dddlll]_Q\\
  \\
  \\
  *++[F]{[4]_{3}^{(0,0,1,1)}}\ar[r]^-Q & *++[F]{[5]_{\frac{7}{2}}^{(0,1,0,0)}} \ar[r]^-Q&  *++[F]{[6]_{4}^{(1,0,0,0)}}\ar[r]^-Q& *++[F]{[7]_{\frac{9}{2}}^{(0,0,0,0)}}  \phantom{qqq}}
 \label{3dn8hs2}
 \end{equation}

\item The operator content of a~$B_{1}[0]_1^{(0,1,0,0)}$ multiplet is given by
 \begin{equation}
\xymatrix  @R=1pc {
 *++[F=]{B_{1}[0]_1^{(0,1,0,0)}} \ar[r]^-Q&  *++[F]{[1]_{\frac{3}{2}}^{(0,0,1,1)\oplus(1,0,0,0)}} \ar[r]^-Q&  *++[F]{[0]_{2}^{(0,1,0,0)} \oplus [2]_{2}^{(0,0,0,0)\oplus (0,0,0,2)\oplus(0,0,2,0)}} \ar[ddll]_{Q}  \\
 \\
  *++[F]{[3]_{\frac{5}{2}}^{(0,0,1,1)}}\ar[r]^-Q&  *++[F]{[4]_{3}^{(0,1,0,0)}} \ar[r]^-Q&  *++[F]{[5]_{\frac{7}{2}}^{(1,0,0,0)}} \ar[dd]^-{Q}  \\ \\
&& *++[F]{[6]_{4}^{(0,0,0,0)}}  }
 \label{3dn8hs3}
 \end{equation}
Modulo conservation laws, this multiplet contains~$256+256$ operators. Note that the~$R$-neutral spin-1 current~$[2]_2^{(0,0,0,0)}$ at level two is a manifest top component, and hence a flavor current. Therefore~$B_{1}[0]_1^{(0,1,0,0)}$ is also a flavor current multiplet (see section~\ref{sec:fcm}), albeit one that contains higher-spin currents.

\item The~$256+256$ multiplet~$A_{2}[0]_1^{(0,0,0,0)}$ contains higher-spin currents:
 \begin{equation}
\xymatrix  @R=1pc {
*++[F=]{A_{2}[0]_1^{(0,0,0,0)}} \ar[r]^-Q&  *++[F]{[1]_{\frac{3}{2}}^{(1,0,0,0)}} \ar[r]^-Q&  *++[F]{ [0]_{2}^{(0,0,0,0)} \oplus [2]_{2}^{(0,1,0,0)}} \ar[r]^-Q&  *++[F]{[3]_{\frac{5}{2}}^{(0,0,1,1)}}\ar[dddlll]_{Q}\\
  \\
  \\
*++[F]{[4]_{3}^{(0,0,0,2)\oplus (0,0,2,0)}}\ar[r]^-Q&  *++[F]{[5]_{\frac{7}{2}}^{(0,0,1,1)}} \ar[r]^-Q  & *++[F]{[6]_{4}^{(0,1,0,0)}}\ar[r]^-Q&*++[F]{[7]_{\frac{9}{2}}^{(1,0,0,0)}}\ar[dd]^-{Q}\\
  \\
& &  & *++[F]{[8]_{5}^{(0,0,0,0)}}   }
 \label{3dn8hs4}
 \end{equation}

\item The operator content of~$A_{1}[j]^{(0,0,0,0)}_{j/2+1}$ multiplets, which always include higher-spin currents, is given by
 \begin{equation}
\xymatrix  @R=1pc {
 *++[F=]{A_{1}[j \geq 1]^{(0,0,0,0)}_{\half j +1}} \ar[r]^-Q&  *++[F]{[j+1]_{\half j+\frac{3}{2}}^{(1,0,0,0)}} \ar[r]^-Q&  *++[F]{[j+2]_{\half j+2}^{(0,1,0,0)}} \ar[dddll]_{Q}\\
 \\
 \\
 *++[F]{[j+3]_{\half j+\frac{5}{2}}^{(0,0,1,1)}}\ar[r]^-Q &*++[F]{[j+4]_{\half j+3}^{(0,0,0,2)\oplus (0,0,2,0)}}\ar[r]^-Q& *++[F]{[j+5]_{\half j+\frac{7}{2}}^{(0,0,1,1)}}\ar[dddll]_{Q} \\
 \\
 \\
  *++[F]{[j+6]_{\half j+4}^{(0,1,0,0)}}\ar[r]^-Q& *++[F]{[j+7]_{\half j+\frac{9}{2}}^{(1,0,0,0)}}\ar[r]^-Q&*++[F]{[j+8]_{\half j+5}^{(0,0,0,0)}} }
 \label{3dn8hs5}
 \end{equation}
 After subtracting conservation laws, these multiplets contain~$256+256$ bosonic and fermionic operators.
\end{itemize}

\subsection{Superconformal Current Multiplets:  $d=4$}

\label{sec:d4curr}

Here we tabulate superconformal current multiplets in four dimensions, which contain short representations of the conformal algebra~$\frak{so}(4,2)$. The unitary representations of this algebra are summarized in table~\ref{tab:4Dcurrents} (see also page~10 of~\cite{Minwalla:1997ka}), which indicates the Lorentz quantum numbers of the CP, unitarity bounds on its scaling dimension, and the quantum numbers of the primary null state that results when these bounds are saturated (see section~\ref{sec:4dmults} for a summary of our conventions in~$d = 4$). We also indicate whether the CP is a free field annihilated by~$\square$. 

\smallskip

\renewcommand{\arraystretch}{1.6}
\renewcommand\tabcolsep{6pt}
\begin{table}[H]
  \centering
  \begin{tabular}{ | l | l |l |c| }
\hline
\multicolumn{1}{|c|}{\bf Primary} &  \multicolumn{1}{|c|}{\bf Unitarity Bound} & \multicolumn{1}{|c|}{\bf Null State } & \multicolumn{1}{|c|}{\bf Comments}\\
\hline
\hline
$ [j ; \overline{j}]_{\Delta}~,~j ,\overline{j} \geq 1 $ & $\Delta \geq \half \left(j+\overline{j}\right)  + 2 $ & $[j-1;\overline{j}-1]_{\Delta+1}$ & -\\
\hline
\hline
$ [j; 0]_{\Delta}~,~j\geq 1 $ & $\Delta \geq \half j+ 1$ & $[j-1;1]_{\Delta+1}$ & free field\\
\hline
$ [0; \overline{j}]_{\Delta}~,~\overline{j}\geq 1 $ & $\Delta \geq  \half \overline{j} + 1$ & $[1;\overline{j}-1]_{\Delta+1}$ & free field \\
\hline
$ [0; 0]_{\Delta}~~ $ & $\Delta \geq 1$ & $[0;0]_{\Delta+2}$ & free field \\
\hline
\hline 
$ [0;0]_{\Delta}~~ $ & $\Delta = 0 $ & $[1;1]_{\Delta+1}$ & unit operator\\
\hline
\end{tabular}
  \caption{Unitary representations of the conformal algebra in~$d=4$.  }
   \label{tab:4Dcurrents}
\end{table}

\smallskip

\noindent In general, we count currents or free fields modulo conservation laws or field equations. Applying this to the operators in table~\ref{tab:4Dcurrents}, and recalling that the dimension of the Lorentz representation~$[j; \b j]$ is~$(j+1)(\b j+1)$, we conclude that conserved currents~$[j ; \b j]_{j/2 + \b j/2 + 2}$ (with~$j, \b j \geq 1$) always contain~$(j+1)(\b j+1) - j \b j = j + \b j + 1$ independent operators. By contrast, the free fields~$[j; 0]_{j/2+1}$ (and similarly~$[0; \b j]_{\b j/2 + 1}$) do not contain any independent operators. To see this, note that the null state~$[j-1, 1]_{j/2 + 2}$ is itself a conserved current, with~$j+1$ degrees of freedom. 

As before, we follow the method described in section~\ref{sec:findsccm} to identify all superconformal current multiplets in~$d = 4$. We must therefore substitute~$h_1 = \half (j + \b j)$ (see~\eqref{hdlambda} in appendix~\ref{app:liealg}) and~$d = 4$ into the general bound~\eqref{primaryh} that must be satisfied by the SCP~$\CV$ of any such superconformal multiplet, 
\begin{equation}\label{primaryhii}
\Delta_\CV \leq \half\left(j + \b j\right) + 2~.
\end{equation}
Together with the superconformal unitarity constraints on~$\CV$ summarized in section~\ref{sec:4dmults}, we will use this bound to identify and analyze all candidate superconformal current multiplets. Below, we will do this explicitly for~$1 \leq \CN \leq 4$; as discussed in section~\ref{sec:maxsusy}, there are no quantum field theories with~$\CN \geq 5$.  It is straightforward to enumerate the candidate current multiplets for all values of~$\CN$ using the form of the superconformal unitarity bounds in appendix~\ref{app:genN}: 
\begin{itemize}
\item $A_{\ell}\overline{A}_{\b \ell}$-multiplets (with~$\ell, \b \ell = 1,2$) whose SCP is neutral under~$\frak{su}(\CN)_R$ and whose~$\frak{u}(1)_{R}$ charge~$r$ is given by
\begin{equation}
\left(\frac{4-\mathcal{N}}{\mathcal{N}}\right)r = j-\overline{j}~.
\end{equation}
When~$\mathcal{N}=4$, there is no~$\frak{u}(1)_R$-symmetry, so that~$r = 0$ and hence~$j=\overline{j}$.

\item $A_{\ell}\overline{B}_{1}$-multiplets (with~$\ell = 1,2$) whose primary has vanishing~$\frak{su}(\CN)_R$ quantum numbers and~$\frak{u}(1)_{R}$ charge~$r$ given by
\begin{equation}\label{u1rabmultcond}
\left(\frac{4-\mathcal{N}}{\mathcal{N}}\right)r = j+2~.
\end{equation}
There are also conjugate~$B_{1}\overline{A}_{\b \ell}$ multiplets, which are~$\frak{su}(\CN)_R$-neutral and have~$\frak{u}(1)_R$ charge~$r$ given by 
\begin{equation}\label{u1rabmultcondconj}
\left(\frac{4-\mathcal{N}}{\mathcal{N}}\right)(-r) = \overline{j}+2~.
\end{equation}
When~$\CN = 4$ we must set~$r = 0$, so that~\eqref{u1rabmultcond} and~\eqref{u1rabmultcondconj} become inconsistent. Hence these multiplets do not exist when~$\CN=4$. 

\item  $A_{\ell}\overline{B}_{1}$-multiplets (with~$\ell = 1,2$) whose~$\frak{su}(\CN)_R$ Dynkin labels~$R_i~(i = 1, \ldots, \CN-1)$ vanish for all values of~$i$, except for a single~$i = \hat i$ for which~$R_{\hat i} = 1$. (This corresponds to a fundamental weight of~$\frak{su}(\CN)_R$.) The~$\frak{u}(1)_{R}$ charge of such multiplets is given by
\begin{equation}
\left(\frac{4-\mathcal{N}}{\mathcal{N}}\right)r = j+4\left(\frac{\mathcal{N}-\hat i}{\mathcal{N}}\right)~.
\end{equation}
The conjugate~$B_{1}\overline{A}_{\b \ell}$ multiplets instead satisfy
\begin{equation}
\left(\frac{4-\mathcal{N}}{\mathcal{N}}\right)(-r) = \overline{j}+4\left(\frac{\hat i}{\mathcal{N}}\right)~.
\end{equation}
Note that these multiplets do not exist for~$\CN=1$, because the~$R$-symmetry is abelian, or for~$\CN=4$, because there is no~$\frak{u}(1)_R$ symmetry and we must set~$r = 0$.

\item $B_{1}\overline{B}_{1}$-multiplets whose~$\frak{su}(\CN)_R$ Dynkin labels~$R_i$ satisfy the bound
\begin{equation}\label{sundynkbound}
\sum_{i = 1}^{\CN-1} R_i \leq 2~,
\end{equation}
and whose~$\frak{u}(1)_R$ charge~$r$ is given by
\begin{equation}\label{u1b1b1barfix}
\left({4-\CN \over \CN}\right) r = {2 \over \CN} \sum_{i = 1}^{\CN-1} \left(\CN - 2i \right) R_i~.
\end{equation}
Note that for~$\CN=1$ the right-hand side vanishes, because the~$R$-symmetry is abelian, while for~$\CN=4$ the left-hand side vanishes, because the~$\frak{u}(1)_R$ symmetry is absent. 
\end{itemize}
As was the case for~$d = 3, \CN=2$ theories (see the beginning of section~\ref{sec:3dn2curr}), some multiplets of the form~$A_{1,2}\b L$ or~$B_1 \b L$ (as well as their conjugates) may satisfy the bound~\eqref{primaryhii} for certain values of the~$\frak{u}(1)_R$ charge, but it can be checked that these multiplets do not contain conserved currents or free fields. 

\subsubsection{$d=4$,  $\mathcal{N}=1$}

\label{sec:d4n1curr}

If we apply the constraints above to~$\CN=1$ (see section~\ref{sec:4dn1defs}), we find the following conserved current multiplets:
\begin{itemize}
\item The~$B_1 \b B_1[0;0]_0^{(0)}$ multiplet consists of the unit operator. 
\item The multiplets~$A_\ell \b B_1[j;0]_{j/2+1}^{({1 \over 3}(j+2))}$ (with~$\ell = 1$ if~$j \geq 1$ and~$\ell = 2$ if~$j = 0$) are chiral free fields of spin-$j \over 2$. The case~$j = 0$ is a standard Lorentz-scalar chiral free field of~$R$-charge~$r = {2 \over 3}$, while~$j = 1$ describes a free vector multiplet of~$R$-charge~$r = 1$. The cases~$j \geq 2$ describe higher-spin free fields:
 \begin{equation}
\xymatrix  @R=2pc {
 *++[F=]{ {[j;0]}_{\half j+1}^{({1 \over 3}(j+2))}} \ar[r]^-Q&  *++[F]{{[j+1;0]}_{\half j+\frac{3}{2}}^{({1 \over 3}(j - 1))}}}
 \label{4dn1freev}
 \end{equation}
The complex conjugate anti-chiral free fields are~$B_1 \b A_{\b \ell} [0;\b j]_{\b j/2+1}^{(-{1 \over 3}(\b j+2))}$; we do not tabulate them explicitly. 

\item The operator content of~$A_\ell \b A_{\b \ell}[j; \b j]^{({1 \over 3}(j - \b j))}_{j/2 + \b j/2 + 2}$ multiplets (with~$\ell =1$ if~$j \geq 1$, $\ell = 2$ if~$j = 0$, and likewise for~$\b \ell, \b j$) is given by
\begin{equation}   \label{4dn1a1a1bar}
 \xymatrix @C=7.5pc @R=7.5pc @!0 @dr {
*++[F=]{A_\ell \b A_{\b \ell}[j; \b j]^{(r = {1 \over 3}(j - \b j))}_{\Delta = \half(j + \b j) + 2}} \ar[r]|--{{~\b Q~}} \ar[d]|--{~Q~} & *++[F]{ [j;\b j + 1]^{(r+1)}_{\Delta + \half}}  \ar[d]|--{~Q~} & \\
*++[F]{[j+1;\b j]_{\Delta + \half}^{(r-1)}}   \ar[r]|--{{~\b Q~}}  &  *++[F]{[j+1;\b j+1]_{\Delta + 1}^{(r)}} 
}
\end{equation}
Accounting for conservation laws, there are~$2(j + \b j) + 4$ bosonic (and equally many fermionic) operators. The~$A_2 \b A_2$-multiplet with~$j = \b j = 0$ is a flavor current multiplet (see section~\ref{sec:fcm}), while the~$A_1 \b A_2$-multiplet with~$j = 1, \b j = 0$ (and its conjugate~$A_2 \b A_1$-multiplet with $j = 0, \b j = 1$) are extra SUSY-current multiplets (see section~\ref{sec:escm}). The~$A_1 \b A_1$-multiplet with~$j = \b j = 1$ is a stress tensor multiplet (see section~\ref{sec:stm}). All other~$A_\ell \b A_{\b \ell}$ current multiplets contain higher spin currents. 
\end{itemize}

\subsubsection{$d=4$,  $\mathcal{N}=2$}

\label{sec:d4n2curr}

The constraints summarized at the beginning of section~\ref{sec:d4curr} lead to the following conserved current multiplets in~$\CN=2$ theories (see section~\ref{sec:d4n2defs}); these multiplets have also been analyzed in~\cite{Dolan:2002zh}:

\begin{itemize}
\item $B_1 \b B_1$-multiplets with~$R \leq 2$ and~$r = 0$:
\begin{itemize}
\item[$\star$] The~$B_1 \b B_1[0;0]_0^{(0;0)}$ multiplet consists of the unit operator.
\item[$\star$] $B_{1}\b{B}_{1}[0;0]_1^{(1; 0)}$ is a free hypermultiplet:
\begin{equation}   \label{4dn2hyper}
 \xymatrix @C=6pc @R=6pc @!0 @dr {
*++[F=]{B_{1}\b{B}_{1}[0;0]_1^{(1; 0)}} \ar[r]|--{{~\b Q~}} \ar[d]|--{~Q~} & *++[F]{ [0;1]^{(1;1)}_{3\over2}}   & \\
*++[F]{[1;0]_{3 \over 2}^{(1;-1)}}  &  &
}
\end{equation}

\item[$\star$] $B_{1}\b{B}_{1}[0;0]_2^{(2; 0)}$ is an~$8+8$ flavor current multiplet:
\begin{equation}\label{4dn2flavcurr}
 \xymatrix @C=6pc @R=6pc @!0 @dr {
*++[F=]{B_{1}\b{B}_{1}[0;0]_2^{(2; 0)}} \ar[r]|--{{~\b Q~}} \ar[d]|--{~Q~} & *++[F]{ [0;1]^{(1;1)}_{5\over2}}   \ar[d]|--{{~Q~}} \ar[r]|--{{~\b Q~}}  & 
*++[F]{[0;0]_{3 }^{(0;2)}}  &  & \\
*++[F]{[1;0]_{5 \over 2}^{(1;-1)}}   \ar[d]|--{{~Q~}} \ar[r]|--{{~\b Q~}} & *++[F]{[1;1]_{3}^{(0;0)}} & \\
 *++[F]{ [0;0]^{(0;-2)}_{3}}
}
\end{equation}

\end{itemize}

\item $A_\ell \b B_1[j; 0]^{(0;j+2)}_{j/2+1}$ multiplets (with~$\ell = 1$ if~$j \geq 1$ and~$\ell = 2$ if~$j = 0$) and their complex conjugate~$B_1 \b A_{\b \ell}[0; \b j]^{(0;-\b j -2)}_{\b j/2+1}$ multiplets (which we do not tabulate) contain free fields:
 \begin{equation}
\xymatrix  @R=1pc {
 *++[F=]{A_\ell \b B_1[j; 0]^{(0;r = j+2)}_{\Delta = \half j+1}} \ar[r]^-Q&  *++[F]{ {[j+1;0]}_{\Delta+\half}^{(1;r-1)}}\ar[r]^-Q&  *++[F]{{[j+2;0]}_{\Delta+1}^{(0;r-2)}}
 }
 \label{4dn2freev}
 \end{equation}
The case~$j=0$ describes a free vector multiplet, while multiplets with~$j \geq 1$ contain higher-spin free fields. 

\item $A_{2}\overline{B}_{1}[0;0]_2^{(1; 2)}$ is a~$16+16$ extra-SUSY current multiplet (see section~\ref{sec:escm}):
\begin{align}\label{d4n2escm}
  & \hskip-43pt \xymatrix @C=7.3pc @R=7.3pc @!0 @dr {
*++[F=]{A_{2}\overline{B}_{1}[0;0]_2^{(1; 2)}} \ar[r]|--{{~\b Q~}} \ar[d]|--{~Q~} & *++[F]{ [0;1]^{(0;3)}_{5\over2}} \ar[d]|--{~Q~} 
\\
*++[F]{[1;0]_{5 \over 2}^{(0;1) \oplus (2;1)}}  \ar[d]|--{~Q~} \ar[r]|--{~\b Q~} & *++[F]{[1;1]_{3}^{(1;2)}} \ar[d]|--{~Q~} & \\
*++[F]{\begin{aligned} & [0;0]_{3}^{(1;0)}~\\
& [2;0]_{3}^{(1;0)}
\end{aligned}
} \ar[r]|--{~\b Q~} \ar[d]|--{~Q~} & *++[F]{[2;1]_{7 \over 2}^{(0;1)}}   & 
\\
*++[F]{[1;0]_{7 \over 2}^{(0;-1)}}
}
\end{align}

\item The~$A_{1}\overline{B}_{1}[j;0]^{(1; j+2)}_{\half j + 2}$ multiplets contain higher-spin currents:
\begin{align}  \label{4dn2hs1}
  & \hskip-43pt \xymatrix @C=7.3pc @R=7.3pc @!0 @dr {
*++[F=]{A_{1}\overline{B}_{1}[j\geq 1;0]^{(1; r = j+2)}_{\Delta = \half j + 2}} \ar[r]|--{{~\b Q~}} \ar[d]|--{~Q~} & *++[F]{ [j;1]^{(0;r+1)}_{\Delta + \half}} \ar[d]|--{~Q~} 
\\
*++[F]{\begin{aligned} & [j-1;0]_{\Delta+\half}^{(0;r-1)} \\[-6pt]
& [j+1;0]_{\Delta+\half}^{(0;r-1) \oplus (2;r-1)}
\end{aligned}
}  \ar[d]|--{~Q~} \ar[r]|--{~\b Q~} & *++[F]{[j+1;1]_{\Delta+1}^{(1;r)}} \ar[d]|--{~Q~} & \\
*++[F]{\begin{aligned} & [j;0]_{\Delta+1}^{(1;r-2)}~\\
& [j+2;0]_{\Delta+1}^{(1;r-2)}
\end{aligned}
} \ar[r]|--{~\b Q~} \ar[d]|--{~Q~} & *++[F]{[j+2;1]_{\Delta+{3 \over 2}}^{(0;r-1)}}   & 
\\
*++[F]{[j+1;0]_{\Delta+{3 \over 2}}^{(0;r-3)}}
}
\end{align}
After taking into account conservation laws, this multiplet contains~$8j + 16$ bosonic, and equally many fermionic, operators. 

\item The operator content of~$A_{\ell}\overline{A}_{\b \ell}[j; \b j]^{(0; j-\overline{j})}_{j/2+\b j/2 +2}$ multiplets (with~$\ell = 1$ if~$j \geq 1$, $\ell =2$ if~$j = 0$, and similarly for~$\b \ell, \b j$) is given by
\begin{align}   \label{4dn2hs}
  & \hskip-43pt \xymatrix @C=7.3pc @R=7.3pc @!0 @dr {
*++[F=]{A_{\ell}\overline{A}_{\b \ell}[j; \b j]^{(0; r = j-\overline{j})}_{\Delta = \half (j + \b j) +2}} \ar[r]|--{{~\b Q~}} \ar[d]|--{~Q~} & *++[F]{ [j;\b j + 1]^{(1;r+1)}_{\Delta + \half}} \ar[d]|--{~Q~} \ar[r]|--{{~\b Q~}} \ar[d]|--{~Q~} & *++[F]{ [j;\b j + 2]^{(0;r+2)}_{\Delta + 1}} \ar[d]|--{~Q~}
\\
*++[F]{  [j+1;\b j]_{\Delta+\half}^{(1;r-1)}
}  \ar[d]|--{~Q~} \ar[r]|--{~\b Q~} & *++[F]{[j+1;\b j + 1]_{\Delta+1}^{(0;r) \oplus (2;r) }} \ar[d]|--{~Q~} \ar[r]|--{{~\b Q~}} & *++[F]{ [j+1;\b j + 2]^{(1;r+1)}_{\Delta + {3 \over 2}}} \ar[d]|--{~Q~}
 \\
*++[F]{  [j+2;\b j]_{\Delta+1}^{(0;r-2)}
} \ar[r]|--{~\b Q~} & *++[F]{[j+2;\b j + 1]_{\Delta+{3 \over 2}}^{(1;r-1)}}   \ar[r]|--{{~\b Q~}}  & *++[F]{[j+2;\b j + 2]_{\Delta+2}^{(0;r)}}
}
\end{align}
The case~$j = \b j = 0$ (with~$\ell = \b \ell = 2$) is a stress tensor multiplets (see section~\ref{sec:stm}), while all other cases describe multiplets containing higher-spin currents. After subtracting conservation laws, these multiplets contain~$8(j+ \b j) + 24$ bosonic (and the same number of fermionic) operators. 

\end{itemize}

\subsubsection{$d=4$, $\mathcal{N}=3$}

\label{sec:d4n3curr}

If we apply the constraints spelled out at the beginning of section~\ref{sec:d4curr} to~$\CN=3$ (see section~\ref{sec:d4n3defs}), we find the following conserved current multiplets:
\begin{itemize}
\item $B_1 \b B_1$-multiplets with~$R_1 + R_2 \leq 2$ and~$r = 2(R_1-R_2)$:
\begin{itemize}
\item[$\star$] The~$B_1 \b B_1[0;0]_0^{(0,0;0)}$ multiplet consists of the unit operator.
\item[$\star$] The multiplet~$B_1 \b B_1[0;0]_1^{(1,0;2)}$ and its complex conjugate~$B_1 \b B_1[0;0]_1^{(0,1;-2)}$, which we do not explicitly tabulate, describe a free~$\CN=3$ vector multiplet:
\begin{equation}\label{freed4n3vecmul}
 \xymatrix @C=6pc @R=6pc @!0 @dr {
*++[F=]{B_1 \b B_1[0;0]_1^{(1,0;2)}} \ar[r]|--{{~\b Q~}} \ar[d]|--{~Q~} & *++[F]{ [0;1]^{(0,0;3)}_{3\over2}}   & \\
*++[F]{[1;0]_{3 \over 2}^{(0,1;1)}} \ar[d]|--{~Q~} &  & \\
*++[F]{[2;0]_2^{(0,0;0)}} & & 
}
\end{equation}
\item[$\star$] The multiplet~$B_1 \b B_1[0;0]_2^{(2,0;4)}$ and its complex conjugate~$B_1 \b B_1[0;0]_2^{(0,2;-4)}$, which we do not tabulate, are~$32+32$ extra SUSY-current multiplets (see section~\ref{sec:escm}):
\begin{align}\label{d4n3escm}
  & \hskip-43pt \xymatrix @C=7pc @R=7pc @!0 @dr {
*++[F=]{B_1 \b B_1[0;0]_2^{(2,0;4)}} \ar[r]|--{{~\b Q~}} \ar[d]|--{~Q~} & *++[F]{ [0;1]^{(1,0;5)}_{5\over2}} \ar[d]|--{~Q~} \ar[r]|--{{~\b Q~}}  & *++[F]{ [0;0]^{(0,0;6)}_{3}} 
\\
*++[F]{[1;0]_{5 \over 2}^{(1,1;3)}}  \ar[d]|--{~Q~} \ar[r]|--{~\b Q~} & *++[F]{[1;1]_{3}^{(0,1;4)}} \ar[d]|--{~Q~} & \\
*++[F]{\begin{aligned} & [0;0]_{3}^{(0,2;2)}~\\
& [2;0]_{3}^{(1,0;2)}
\end{aligned}
} \ar[r]|--{~\b Q~} \ar[d]|--{~Q~} & *++[F]{[2;1]_{7 \over 2}^{(0,0;3)}}   & 
\\
*++[F]{[1;0]_{7 \over 2}^{(0,1;1)}} \ar[d]|--{~Q~}
\\
*++[F]{[0;0]_{4}^{(0,0;0)}} 
}
\end{align}

\item[$\star$] $B_1 \b B_1[0;0]_2^{(1,1;0)}$ is a~$64+64$ stress tensor multiplet:
\begin{align}\label{d4n3stm}
  & \hskip-43pt \xymatrix @C=7pc @R=7pc @!0 @dr {
*++[F=]{B_1 \b B_1[0;0]_2^{(1,1;0)}} \ar[r]|--{{~\b Q~}} \ar[d]|--{~Q~} & *++[F]{ [0;1]^{(0,1;1) \oplus (2,0;1)}_{5\over2}} \ar[d]|--{~Q~} \ar[r]|--{{~\b Q~}}  & *++[F]{ \begin{aligned} & [0;0]^{(1,0;2)}_{3} \\
& [0;2]^{(1,0;2)}_{3}
\end{aligned}}  \ar[d]|--{~Q~} \ar[r]|--{{~\b Q~}}  & *++[F]{ [0;1]^{(0,0;3)}_{7\over2}}
\\
*++[F]{[1;0]_{5 \over 2}^{(0,2;-1) \oplus (1,0;-1)}}  \ar[d]|--{~Q~} \ar[r]|--{~\b Q~} & *++[F]{[1;1]_{3}^{(0,0;0) \oplus (1,1;0)}} \ar[d]|--{~Q~} \ar[r]|--{~\b Q~} &  *++[F]{[1;2]_{7 \over 2}^{(0,1;1)}} \ar[d]|--{~Q~}
 \\
*++[F]{\begin{aligned} & [0;0]_{3}^{(0,1;-2)}~\\
& [2;0]_{3}^{(0,1;-2)}
\end{aligned}
} \ar[r]|--{~\b Q~} \ar[d]|--{~Q~} & *++[F]{[2;1]_{7 \over 2}^{(1,0;-1)}} \ar[r]|--{~\b Q~}   & *++[F]{[2;2]_{4}^{(0,0;0)}}
\\
*++[F]{[1;0]_{7 \over 2}^{(0,0;-3)}} 
}
\end{align}
\end{itemize}
\item $A_\ell \b B_1[j; 0]^{(0,0;3j+6)}_{j/2+1}$ multiplets (with~$\ell = 1$ if~$j \geq 1$ and~$\ell = 2$ if~$j = 0$) and their complex conjugate~$B_1 \b A_{\b \ell}[0; \b j]^{(0,0;-3\b j-6)}_{\b j/2+1}$ multiplets (which we do not tabulate) contain higher-spin free fields:
\begin{equation}
 \xymatrix @C=5.8pc @R=5.8pc @!0 @dr {
*++[F=]{A_\ell \b B_1[j; 0]^{(0,0;3j+6)}_{\half j+1}} \ar[d]|--{~Q~} &    & \\
*++[F]{[j+1; 0]^{(1,0;3j+5)}_{\half j+{3 \over 2}}} \ar[d]|--{~Q~} &  & \\
*++[F]{[j+2; 0]^{(0,4;3j+4)}_{\half j+2}} \ar[d]|--{~Q~} & & \\
*++[F]{[j+3; 0]^{(0,0;3j+3)}_{\half j+{5 \over 2}}}
}
\end{equation}
\item The multiplets~$A_\ell \b B_1[j; 0]^{(1,0;3j+8)}_{j/2+2}$ and~$A_\ell \b B_1[j; 0]^{(0,1;3j+4)}_{j/2+2}$ (with~$\ell = 1$ if~$j \geq 1$ and~$\ell = 2$ if~$j = 0$), as well as their complex conjugates~$B_1 \b A_{\b \ell}[0; \b j]^{(0,1;-3\b j-8)}_{\b j/2+2}$ and~$B_1 \b A_{\b \ell}[0; \b j]^{(1,0;-3\b j-4)}_{\b j/2+2}$ (which we do not tabulate), contain higher-spin currents. After taking into account conservation laws, $A_\ell \b B_1[j; 0]^{(1,0;3j+8)}_{j/2+2}$ multiplets harbor~$32j + 64$ bosonic (and equally many fermionic) operators, while~$A_\ell \b B_1[j; 0]^{(0,1;3j+4)}_{j/2+2}$ multiplets harbor~$32j+96$. 
\begin{align}
  & \hskip-43pt
\xymatrix @C=7pc @R=7pc @!0 @dr {
*++[F=]{A_\ell \b B_1[j; 0]^{(1,0;r = 3j+8)}_{\Delta = \half j+2}} 
\ar[d]|--{~Q~} \ar[r]|--{{~\b Q~}}
& *+[F]{\scriptstyle [j; 1]_{\Delta + \half}^{(0,0;r+1)} } 
\ar[d]|--{~Q~} 
\\
*+[F]{ \scriptstyle [j-1; 0]^{(0,1;r-1)}_{\Delta+\half}~,~[j+1;0]_{\Delta + \half}^{(0,1;r-1) \oplus (2,0;r-1)}} \ar[d]|--{~Q~} \ar[r]|--{{~\b Q~}} & *+[F]{ \scriptstyle [j+1; 1]_{\Delta+1}^{(1,0;r) } } \ar[d]|--{~Q~} 
\\
*+[F]{\begin{aligned} & \scriptstyle [j-2;0]_{\Delta+1}^{ (0,0;r-2)}~,~[j;0]_{\Delta+1}^{ (0,0;r-2) \oplus (1,1;r-2)} \\[-6pt]
&\scriptstyle [j+2 ; 0]^{(0,0;r-2) \oplus (1,1;r-2)}_{\Delta+1}
\end{aligned}} \ar[r]|--{{~\b Q~}} \ar[d]|--{~Q~} 
& *+[F]{ \scriptstyle [j+2; 1]_{\Delta+{3 \over 2}}^{(0,1;r-1) }}  \ar[d]|--{~Q~} 
\\
*+[F]{\begin{aligned} & \scriptstyle [j-1;0]_{\Delta+{3 \over 2}}^{ (1,0;r-3)}~,~[j+3; 0]^{(1,0;r-3)}_{\Delta+{3 \over 2}} \\[-6pt]
&\scriptstyle [j+1;0]_{\Delta+{3\over 2}}^{ (0,2;r-3) \oplus (1,0;r-3)} 
\end{aligned}
}\ar[r]|--{{~\b Q~}} \ar[d]|--{{~Q~}} 
& *+[F]{ \scriptstyle [j+3;1]_{\Delta+2}^{(0,0;r-2)}} 
\\ 
*+[F]{\begin{aligned} & \scriptstyle [j;0]_{\Delta+2}^{ (0,1;r-4)}~,~[j+2;0]^{(0,1;r-4)}_{\Delta+2} 
\end{aligned}}  \ar[d]|--{{~Q~}} & 
\\
*+[F]{ \scriptstyle [j+1;0]_{\Delta+{5 \over 2}}^{ (0,0;r-5)} }
}
\end{align}


\begin{align}
  & \hskip-43pt
\xymatrix @C=7pc @R=7pc @!0 @dr {
*++[F=]{A_\ell \b B_1[j; 0]^{(0,1;r = 3j+4)}_{\Delta = \half j+2}} 
\ar[d]|--{~Q~} \ar[r]|--{{~\b Q~}}
& *+[F]{\scriptstyle [j; 1]_{\Delta + \half}^{(1,0;r+1)} } 
\ar[d]|--{~Q~} \ar[r]|--{{~\b Q~}}
& *+[F]{\scriptstyle [j; 2]_{\Delta + 1}^{(0,0;r+2)} } \ar[d]|--{~Q~}
\\
*+[F]{ \scriptstyle [j-1; 0]^{(0,0;r-1)}_{\Delta+\half}~,~[j+1;0]_{\Delta + \half}^{(0,0;r-1) \oplus (1,1;r-1)}} \ar[d]|--{~Q~} \ar[r]|--{{~\b Q~}} & *+[F]{ \scriptstyle [j+1; 1]_{\Delta+1}^{(0,1;r) \oplus (2,0;r) } } \ar[d]|--{~Q~} \ar[r]|--{{~\b Q~}}
& *+[F]{\scriptstyle [j+1; 2]_{\Delta + {3 \over 2}}^{(1,0;r+1)} } \ar[d]|--{~Q~}
\\
*+[F]{\begin{aligned} & \scriptstyle [j;0]_{\Delta+1}^{ (1,0;r-2)} \\[-6pt]
&\scriptstyle [j+2; 0]^{(0,2;r-2) \oplus (1,0;r-2)}_{\Delta+1}
\end{aligned}} \ar[r]|--{{~\b Q~}} \ar[d]|--{~Q~} 
& *+[F]{ \scriptstyle [j+2; 1]_{\Delta+{3 \over 2}}^{(0,0;r-1) \oplus (1,1;r-1) }}  \ar[d]|--{~Q~} \ar[r]|--{{~\b Q~}} & *+[F]{ \scriptstyle [j+2; 2]_{\Delta+2}^{(0,1;r)} } \ar[d]|--{~Q~}
\\
*+[F]{\begin{aligned} & \scriptstyle [j+1;0]_{\Delta+{3 \over 2}}^{ (0,1;r-3)}\\[-6pt]
& \scriptstyle [j+3; 0]^{(0,1;r-3)}_{\Delta+{3 \over 2}} 
\end{aligned}
}\ar[r]|--{{~\b Q~}} \ar[d]|--{{~Q~}} 
& *+[F]{ \scriptstyle [j+3;1]_{\Delta+2}^{(1,0;r-2)}} \ar[r]|--{{~\b Q~}} & *+[F]{ \scriptstyle [j+3;2]_{\Delta+{5 \over 2}}^{(0,0;r-1)}} 
\\ 
*+[F]{ \scriptstyle [j+2;0]^{(0,0;r-4)}_{\Delta+2} }   
}
\end{align}

\item $A_\ell \b A_{\b \ell}[j; \b j]_{j/2+\b j/2 + 2}^{(0,0; 3(j - \b j))}$ multiplets (with~$\ell = 1$ if~$j \geq 1$, $\ell = 2$ if~$j = 0$, and similarly for~$\b \ell$ and~$\b j$) also contain higher-spin currents:
\begin{align}
  & \hskip-43pt
\xymatrix @C=7pc @R=7pc @!0 @dr {
*++[F=]{A_\ell \b A_{\b \ell}[j; \b j]_{\Delta = \half \left(j + \b j\right) + 2}^{(0,0; r = 3(j - \b j))}} 
\ar[d]|--{~Q~} \ar[r]|--{{~\b Q~}}
& *+[F]{\scriptstyle [j;\b j+1]_{\Delta + \half}^{(0,1;r+1)} } 
\ar[d]|--{~Q~} \ar[r]|--{{~\b Q~}}
& *+[F]{ \scriptstyle [j;\b j+2]_{\Delta + 1}^{(1,0; r + 2)}} \ar[d]|--{~Q~} \ar[r]|--{{~\b Q~}} 
& *+[F]{ \scriptstyle [j;\b j+3]_{\Delta + {3 \over 2}}^{(0,0;r+3)} }  \ar[d]|--{{~ Q~}} 
\\
*+[F]{ \scriptstyle [j+1;\b j]_{\Delta + \half}^{(1,0;r-1)}} \ar[d]|--{~Q~} \ar[r]|--{{~\b Q~}} & *+[F]{ \scriptstyle [j+1;\b j+1]_{\Delta+1}^{(0,0;r) \oplus (1,1;r) } } \ar[d]|--{~Q~} \ar[r]|--{{~\b Q~}} 
& *+[F]{ \scriptstyle [j+1;\b j+2]_{\Delta + {3 \over 2}}^{(0,1;r+1) \oplus (2,0;r+1)}}  \ar[d]|--{~Q~} \ar[r]|--{~\b Q~}
&  *+[F]{ \scriptstyle [j+1;\b j+3]_{\Delta+2}^{(1,0;r+2)}}\ar[d]|--{~Q~} 
\\
*+[F]{\scriptstyle [j+2;\b j]_{\Delta+1}^{ (0,1;r-2)}} \ar[r]|--{{~\b Q~}} \ar[d]|--{~Q~} 
& *+[F]{ \scriptstyle [j+2;\b j+1]_{\Delta+{3 \over 2}}^{(0,2;r-1) \oplus (1,0;r-1)}}  \ar[d]|--{~Q~} \ar[r]|--{{~\b Q~}} 
& *+[F]{ \scriptstyle [j+2;\b j+2]_{\Delta+2}^{(0,0;r) \oplus (1,1;r)}}  \ar[d]|--{~Q~} \ar[r]|--{{~\b Q~}} 
& *+[F]{ \scriptstyle [j+2;\b j+3]_{\Delta+{5 \over 2}}^{(0,1;r+1)}} \ar[d]|--{~Q~}
\\
*+[F]{\scriptstyle [j+3;\b j]_{\Delta+{3 \over 2}}^{(0,0;r-3)}
}\ar[r]|--{{~\b Q~}} 
& *+[F]{ \scriptstyle [j+3;\b j+1]_{\Delta+2}^{(0,1;r-2)}} \ar[r]|--{{~\b Q~}}  &  *+[F]{ \scriptstyle [j+3;\b j+2]_{\Delta+{5 \over 2}}^{(1,0;r-1)}} \ar[r]|--{{~\b Q~}} & *+[F]{ \scriptstyle [j+3;\b j+3]_{\Delta+3}^{(0,0;r)}} 
}
\end{align}
After taking into account conservation laws, these multiplets contain~$32(j + \b j) + 128$ bosonic, and as many fermionic, operators. 
\end{itemize} 

\subsubsection{$d=4$,  $\mathcal{N}=4$}

\label{sec:d4n4curr}

In order to apply the general constraints discussed at the beginning of section~\ref{sec:d4curr} to the case~$\CN=4$, we must set~$r = 0$ in all formulas since the~$\frak{u}(1)_R$ symmetry is absent. It follows that conserved currents can only reside in~$A_\ell \b A_{ \ell}$-multiplets (with~$\ell = 1,2$) with~$R_1 = R_2 = R_3 = 0$ and~$j = \b j$, or in~$B_1 \b B_1$-multiplets with~$R_1 + R_2 + R_3 \leq 2$ and~$R_1 = R_3$. We will now examine this possibilities in turn (see also~\cite{Dolan:2002zh} for a detailed discussion):

\begin{itemize}

\item The~$B_1[0;0]_0^{(0,0,0)}$ multiplet consists of the unit operator. 

\item $B_1 \b B_1[0;0]_1^{(0,1,0)}$ is a free vector multiplet:  
\begin{equation}\label{freed4n4vecmul}
 \xymatrix @C=6pc @R=6pc @!0 @dr {
*++[F=]{B_1 \b B_1[0; 0]^{(0,2,0)}_1} \ar[r]|--{{~\b Q~}} \ar[d]|--{~Q~} & *++[F]{ [0;1]^{(1,0,0)}_{3\over2}} \ar[r]|--{{~\b Q~}}  & *++[F]{[0;2]_2^{(0,0,0)}} \\
*++[F]{[1;0]_{3 \over 2}^{(0,0,0)}} \ar[d]|--{~Q~} &  & \\
*++[F]{[2;0]_2^{(0,0,0)}} & & 
}
\end{equation}

\item $B_1 \b B_1[0;0]_2^{(0,2,0)}$ is a~$128+128$ stress tensor multiplet (see section~\ref{sec:stm}): 
\begin{align}\label{d4n4stm}
  & \hskip-43pt
\xymatrix @C=5.8pc @R=5.8pc @!0 @dr {
*++[F=]{B_1\b B_1[0;0]_2^{(0,2,0)} } 
\ar[d]|--{~Q~} \ar[r]|--{{~\b Q~}}
& *+[F]{\scriptstyle [0;1]_{5 \over 2}^{ (1,1,0)} } 
\ar[d]|--{~Q~} \ar[r]|--{{~\b Q~}}
& *+[F]{\begin{aligned} & \scriptstyle [0;0]_3^{(2,0,0)}~\\[-6pt]
& \scriptstyle [0;2]_3^{(0,1,0) }
~\end{aligned}} \ar[d]|--{~Q~} \ar[r]|--{{~\b Q~}} 
& *+[F]{\scriptstyle [0;1]_{7 \over 2}^{(1,0,0)}}  \ar[r]|--{{~\b Q~}} 
& *+[F]{\scriptstyle [0;0]_{4}^{(0,0,0)}}
\\
*+[F]{ \scriptstyle [1;0]_{5 \over 2}^{(0,1,1)}} \ar[d]|--{~Q~} \ar[r]|--{{~\b Q~}} & *+[F]{ \scriptstyle [1;1]_3^{(1,0,1) } } \ar[d]|--{~Q~} \ar[r]|--{{~\b Q~}} 
& *+[F]{ \scriptstyle [1;2]_{7 \over 2}^{(0,0,1)}}  \ar[d]|--{~Q~}
& 
\\
*+[F]{\begin{aligned} & \scriptstyle [0;0]_3^{(0,0,2)}~\\[-6pt]
& \scriptstyle [2;0]_3^{(0,1,0)}~
\end{aligned}} \ar[r]|--{{~\b Q~}} \ar[d]|--{~Q~} 
& *+[F]{ \scriptstyle [2;1]_{7 \over 2}^{(1,0,0)}} \ar[r]|--{{~\b Q~}} 
& *+[F]{ \scriptstyle [2;2]_4^{(0,0,0)}}
& 
\\
*+[F]{\scriptstyle [1;0]_{7 \over 2}^{(0,0,1)}} \ar[d]|--{{~Q~}} 
& & & \\
*+[F]{\scriptstyle [0;0]_{4}^{(0,0,0)}} 
}
\end{align}

\item $B_1 \b B_1[0;0]_2^{(1,0,1)}$ is a~$384+384$ multiplet that contains higher-spin currents:
\begin{align}\label{d4n4hs1}
  & \hskip-43pt
\xymatrix @C=6.4pc @R=6.4pc @!0 @dr {
*++[F=]{B_1 \b B_1[0;0]_2^{(1,0,1)} } 
\ar[d]|--{~Q~} \ar[r]|--{{~\b Q~}}
& *+[F]{\scriptstyle [0;1]_{5 \over 2}^{(0,0,1) \oplus (1,1,0)} } 
\ar[d]|--{~Q~} \ar[r]|--{{~\b Q~}}
& *+[F]{\begin{aligned} & \scriptstyle [0;0]_3^{(0,1,0)}~\\[-6pt]
& \scriptstyle [0;2]_3^{(0,1,0) \oplus (2,0,0) }
~\end{aligned}} \ar[d]|--{~Q~} \ar[r]|--{{~\b Q~}} 
& *+[F]{\begin{aligned} & \scriptstyle [0;1]_{7 \over 2}^{(1,0,0)} ~\\[-6pt]
& \scriptstyle [0;3]_{7 \over 2}^{(1,0,0) }
~\end{aligned}}  \ar[d]|--{{~ Q~}} \ar[r]|--{{~\b Q~}} 
& *+[F]{\scriptstyle [0;2]_{4}^{(0,0,0)}}
\\
*+[F]{ \scriptstyle [1;0]_{5 \over 2}^{(0,1,1) \oplus (1,0,0)}} \ar[d]|--{~Q~} \ar[r]|--{{~\b Q~}} & *+[F]{ \scriptstyle [1;1]_3^{(0,0,0) \oplus (0,2,0) \oplus (1,0,1) } } \ar[d]|--{~Q~} \ar[r]|--{{~\b Q~}} 
& *+[F]{ \scriptstyle [1;2]_{7 \over 2}^{(0,0,1) \oplus (1,1,0)}}  \ar[d]|--{~Q~} \ar[r]|--{~\b Q~}
&  *+[F]{ \scriptstyle [1;3]_{4}^{(0,1,0)}}\ar[d]|--{~Q~}
\\
*+[F]{\begin{aligned} & \scriptstyle [0;0]_3^{(0,1,0)}~\\[-6pt]
& \scriptstyle [2;0]_3^{(0,0,2) \oplus (0,1,0)}~
\end{aligned}} \ar[r]|--{{~\b Q~}} \ar[d]|--{~Q~} 
& *+[F]{ \scriptstyle [2;1]_{7 \over 2}^{(0,1,1) \oplus (1,0,0)}}  \ar[d]|--{~Q~} \ar[r]|--{{~\b Q~}} 
& *+[F]{ \scriptstyle [2;2]_4^{(0,0,0) \oplus (1,0,1)}}  \ar[d]|--{~Q~} \ar[r]|--{{~\b Q~}} 
& *+[F]{ \scriptstyle [2;3]_{9 \over 2}^{(0,0,1)}} \ar[d]|--{~Q~} 
\\
*+[F]{\begin{aligned} & \scriptstyle [1;0]_{7 \over 2}^{(0,0,1)}~\\[-6pt]
& \scriptstyle [3;0]_{7 \over 2}^{(0,0,1)}~
\end{aligned}
} \ar[d]|--{{~Q~}} \ar[r]|--{{~\b Q~}} 
& *+[F]{ \scriptstyle [3;1]_{4}^{(0,1,0)}} \ar[r]|--{{~\b Q~}}  &  *+[F]{ \scriptstyle [3;2]_{9 \over 2}^{(1,0,0)}} \ar[r]|--{{~\b Q~}} & *+[F]{ \scriptstyle [3;3]_{5}^{(0,0,0)}}
\\
*+[F]{\scriptstyle [2;0]_{4}^{(0,0,0)}} 
}
\end{align}

\item $A_\ell \b A_{\ell}[j; j]^{(0,0,0)}_{j + 2}$ multiplets (with~$\ell = 1$ if~$j \geq 1$ and~$\ell = 2$ if~$j = 0$) also contain higher-spin currents; the case $\ell = 2, j = 0$ is the Konishi multiplet:\footnote{~In standard~$\CN=4$ gauge theories, the Konishi multiplet~$\CK$ is a long multiplet~$\CK = L \b L[0;0]^{(0,0,0)}_{\Delta}$, whose anomalous dimension~$\Delta$ depends on the gauge coupling~$g$. When~$g \rightarrow 0$ the theory becomes free and~$\Delta \rightarrow 2$, where~$\CK$ breaks apart into several short multiplets according to the last equation in~\eqref{4dn4recomb}. In the free theory, the primary of~$\CK$ resides in an~$A_2 \b A_2[0;0]^{(0,0,0)}_2$ short multiplet, which is tabulated in~\eqref{konishi}.} 
\begin{align}\label{konishi}
  & \hskip-43pt
\xymatrix @C=6.4pc @R=6.4pc @!0 @dr {
*++[F=]{A_\ell \b A_{\ell}[j; j]^{(0,0,0)}_{j + 2}} 
\ar[d]|--{~Q~} \ar[r]|--{{~\b Q~}}
& *+[F]{\scriptstyle [j;j+1]_{j+{5 \over 2}}^{(0,0,1)} } 
\ar[d]|--{~Q~} \ar[r]|--{{~\b Q~}}
& *+[F]{ \scriptstyle [j;j+2]_{j+3}^{(0,1,0)}} \ar[d]|--{~Q~} \ar[r]|--{{~\b Q~}} 
& *+[F]{ \scriptstyle [j;j+3]_{j+{7 \over 2}}^{(1,0,0)} }  \ar[d]|--{{~ Q~}} \ar[r]|--{{~\b Q~}} 
& *+[F]{\scriptstyle [j;j+4]_{j+4}^{(0,0,0)}}\ar[d]|--{{~ Q~}} 
\\
*+[F]{ \scriptstyle [j+1;j]_{j+{5 \over 2}}^{(1,0,0)}} \ar[d]|--{~Q~} \ar[r]|--{{~\b Q~}} & *+[F]{ \scriptstyle [j+1;j+1]_{j+3}^{(0,0,0) \oplus (1,0,1) } } \ar[d]|--{~Q~} \ar[r]|--{{~\b Q~}} 
& *+[F]{ \scriptstyle [j+1;j+2]_{j+{7 \over 2}}^{(0,0,1) \oplus (1,1,0)}}  \ar[d]|--{~Q~} \ar[r]|--{~\b Q~}
&  *+[F]{ \scriptstyle [j+1;j+3]_{j+4}^{(0,1,0) \oplus (2,0,0)}}\ar[d]|--{~Q~} \ar[r]|--{~\b Q~} & *+[F]{ \scriptstyle [j+1;j+4]_{j+{9 \over 2}}^{(1,0,0)}} \ar[d]|--{~Q~} 
\\
*+[F]{\scriptstyle [j+2;j]_{j+3}^{ (0,1,0)}} \ar[r]|--{{~\b Q~}} \ar[d]|--{~Q~} 
& *+[F]{ \scriptstyle [j+2;j+1]_{j+{7 \over 2}}^{(0,1,1) \oplus (1,0,0)}}  \ar[d]|--{~Q~} \ar[r]|--{{~\b Q~}} 
& *+[F]{ \scriptstyle [j+2;j+2]_{j+4}^{(0,0,0) \oplus (0,2,0) \oplus (1,0,1)}}  \ar[d]|--{~Q~} \ar[r]|--{{~\b Q~}} 
& *+[F]{ \scriptstyle [j+2;j+3]_{j+{9 \over 2}}^{(0,0,1) \oplus (1,1,0)}} \ar[d]|--{~Q~} \ar[r]|--{~\b Q~} & 
*+[F]{ \scriptstyle [j+2;j+4]_{j+5}^{(0,1,0)}} \ar[d]|--{~Q~}
\\
*+[F]{\scriptstyle [j+3;j]_{j+{7 \over 2}}^{(0,0,1)}
} \ar[d]|--{{~Q~}} \ar[r]|--{{~\b Q~}} 
& *+[F]{ \scriptstyle [j+3;j+1]_{j+4}^{(0,0,2) \oplus (0,1,0)}} \ar[d]|--{~Q~} \ar[r]|--{{~\b Q~}}  &  *+[F]{ \scriptstyle [j+3;j+2]_{j+{9 \over 2}}^{(0,1,1) \oplus (1,0,0)}} \ar[d]|--{~Q~} \ar[r]|--{{~\b Q~}} & *+[F]{ \scriptstyle [j+3;j+3]_{j+5}^{(0,0,0) \oplus (1,0,1)}} \ar[d]|--{~Q~} \ar[r]|--{{~\b Q~}} & *+[F]{ \scriptstyle [j+3;j+4]_{j+{11\over 2}}^{(0,0,1)}} \ar[d]|--{~Q~}
\\
*+[F]{\scriptstyle [j+4;j]_{j+4}^{(0,0,0)}}  \ar[r]|--{{~\b Q~}} & *+[F]{ \scriptstyle [j+4;j+1]_{j+{9\over 2}}^{(0,0,1)}} \ar[r]|--{{~\b Q~}} & *+[F]{ \scriptstyle [j+4;j+2]_{j+5}^{(0,1,0)}} \ar[r]|--{{~\b Q~}} & *+[F]{ \scriptstyle [j+4;j+3]_{j+{11\over 2}}^{(1,0,0)}}  \ar[r]|--{{~\b Q~}} & *+[F]{ \scriptstyle [j+4;j+4]_{j+6}^{(0,0,0)}}
}
\end{align}
Taking into account conservation laws, this multiplet contains~$256j + 640$ bosonic (and as many fermionic) operators. 
\end{itemize}

\subsection{Superconformal Current Multiplets: $d=5$}

\label{sec:5dcurrents}

In this subsection we enumerate all superconformal current multiplets in five dimensions, which contain short representations of the conformal algebra~$\frak{so}(5,2)$. The unitary representations of this algebra are summarized in table~\ref{tab:5Dcurrents} (see also page~10 of~\cite{Minwalla:1997ka}), which indicates the Lorentz quantum numbers of the CP, unitarity bounds on its scaling dimension, and the quantum numbers of the primary null state that results when these bounds are saturated (see section~\ref{sec:5dusm} for a summary of our conventions in~$d = 5$). We also indicate whether the CP is a free field annihilated by~$\square$. 

\smallskip

\renewcommand{\arraystretch}{1.6}
\renewcommand\tabcolsep{6pt}
\begin{table}[H]
  \centering
  \begin{tabular}{ | l | l |l |c| }
\hline
\multicolumn{1}{|c|}{\bf Primary} &  \multicolumn{1}{|c|}{\bf Unitarity Bound} & \multicolumn{1}{|c|}{\bf Null State } & \multicolumn{1}{|c|}{\bf Comments}\\
\hline
\hline
$ [j_1, j_2]_{\Delta}~,~j_2 \geq 1 $ & $\Delta \geq \half \left(j_1 + 2 j_2 \right)  + 3 $ & $[j_1,j_2-1]_{\Delta+1}$ & -\\
\hline
$ [j_1, 0]_{\Delta}~,~j_1 \geq 2 $ & $\Delta \geq \half j_1+ 2$ & $[j_1-2,1]_{\Delta+1}$ & -\\
\hline
\hline
$ [1,0]_{\Delta}~~ $ & $\Delta \geq  2$ & $[1,0]_{\Delta+1}$ & free field \\
\hline
$ [0, 0]_{\Delta}~~ $ & $\Delta \geq {3 \over 2}$ & $[0,0]_{\Delta+2}$ & free field \\
\hline
\hline 
$ [0,0]_{\Delta}~~ $ & $\Delta = 0 $ & $[0,1]_{\Delta+1}$ & unit operator\\
\hline
\end{tabular}
  \caption{Unitary representations of the conformal algebra in~$d=5$.  }
   \label{tab:5Dcurrents}
\end{table}

\smallskip

\noindent In general, we count currents or free fields modulo conservation laws or field equations. Here we need the fact that the dimension of a general~$\frak{sp}(4)$ Lorentz representation~$[j_1, j_2]$ is
\begin{equation}
\dim [j_1, j_2] = {1 \over 6} \left(j_1+1\right)\left(j_2+1\right) \left(j_1+j_2 + 2\right) \left(j_1 + 2 j_2 + 3\right)~.
\end{equation}
Explicitly, we find the following operator counts for the currents and free fields in table~\ref{tab:5Dcurrents}:
\begin{itemize}
\item Conserved currents~$[j_1, j_2 \geq 1]_{j_1/2 + j_2 + 3}$ contain~$\dim [j_1, j_2] - \dim [j_1, j_2-1]$ operators. 
\item Currents of the form~$[j_1 \geq 2, 0]_{j_1/2 + 2}$ contain~$\dim [j_1, 0] - \dim[j_1-2, 1] + \dim[j_1 -2, 0] = 2(j_1+1)$ operators, because the null state~$[j_1-2, 1]_{j_1/2 + 3}$ is itself a conserved current.  
\item The free fields~$[1,0]_{2}, [0,0]_{3/2}$ contain no operator degrees of freedom. 
\end{itemize}

\smallskip

We can now follow the method of section~\ref{sec:findsccm} and substitute~$h_1 = \half (j_1 + 2 j_2)$ (see~\eqref{hdlambda} in appendix~\ref{app:liealg}) and~$d = 5$ into the general bound~\eqref{primaryh} that must be satisfied by the SCP~$\CV$ of any superconformal current multiplet, 
\begin{equation}\label{primaryhii}
\Delta_\CV \leq \half\left(j_1 + 2 j_2 \right) + 3~.
\end{equation}
Together with the superconformal unitarity constraints on~$\CV$ summarized in section~\ref{sec:5dusm}, we can use this bound to enumerate all superconformal current multiplets in~$d = 5$, together with their operator content. We also count the number of bosonic and fermionic CPs, modulo conservation laws.

One upshot of this analysis is that there are several short conformal multiplets that never arise in SCFTs, because they are not embedded in any superconformal current multiplet. This is the case for currents of the form~$[j_{1} \geq 2,0]_{j_{1}/2+2}$ and~$[j_1 \geq 3,j_2]_{j_{1}/2+j_{2}+3}$ in table~\ref{tab:5Dcurrents}.

We now examine the superconformal current multiplets in detail:
 
\begin{itemize}
\item The~$C_1^[0,0]_0^{(0)}$ multiplet consists of the unit operator.
\item $C_1[0,0]_{3/2}^{(1)}$ is a free hypermultiplet: 
\begin{equation}
\xymatrix  @R=1pc {
 *++[F=]{[0,0]_{\frac{3}{2}}^{(1)}} \ar[r]^-Q&  *++[F]{[1,0]_{2}^{(0)}} }
 \label{5dH}
\end{equation}
\item $C_1[0,0]_{3}^{(2)}$ is an~$8+8$ flavor-current multiplet (see section~\ref{sec:fcm}):
\begin{equation}
\xymatrix  @R=1pc {
 *++[F=]{[0,0]_{3}^{(2)}} \ar[r]^-Q&  *++[F]{[1,0]_{\frac{7}{2}}^{(1)}} \ar[r]^-Q&  *++[F]{[0,0]_{4}^{(0)} \oplus [0,1]_{4}^{(0)} }}
 \label{5dG}
\end{equation}
\item $B_2[0,0]_3^{(0)}$ is a stress tensor multiplet (see section~\ref{sec:stm}), with~$32+32$ operators:
\begin{equation}
\xymatrix  @R=1pc {
 *++[F=]{[0,0]_{3}^{(0)}} \ar[r]^-Q&  *++[F]{[1,0]_{\frac{7}{2}}^{(1)}} \ar[r]^-Q&  *++[F]{[0,1]_{4}^{(2)}\oplus[2,0]_{4}^{(0)}}\ar[r]^-Q&  *++[F]{[1,1]_{\frac{9}{2}}^{(1)}}\ar[r]^-Q&  *++[F]{[0,2]_{5}^{(0)}}}
 \label{5dEM}
\end{equation}
\item All~$B_1[0, j_2]^{(0)}_{j_2 + 3}$ multiplets contain higher-spin currents:
 \begin{equation}
\xymatrix  @R=2pc {
 *++[F=]{B_1[0, j_2 \geq 1]^{(0)}_{j_2 + 3}} \ar[r]^-Q &  *++[F]{[1,j_2]_{j_2+\frac{7}{2}}^{(1)}} \ar[r]^-Q &  *++[F]{[0,j_2+1]_{j_2+4}^{(2)}\oplus [2,j_2]_{j_2+4}^{(0)}}  \ar[ld]_Q  \\ 
&  *++[F]{[1,j_2+1]_{j_2+\frac{9}{2}}^{(1)} }\ar[r]^-Q &  *++[F]{[0,j_2+2]_{j_2+5}^{(0)}}
  } 
 \label{3dn8hs3}
 \end{equation}
After taking into account conservation laws, this multiplet contains~$8 (j_2+2)^2$ bosonic and the same number of fermionic operators. 
\end{itemize} 

\subsection{Superconformal Current Multiplets: $d=6$}

\label{sec:d6cur}

Here we tabulate all superconformal current multiplets in six dimensions, which contain short representations of the conformal algebra~$\frak{so}(6,2)$. The unitary representations of this algebra are summarized in table~\ref{tab:6Dcurrents} (see also page~10 of~\cite{Minwalla:1997ka}), which indicates the Lorentz quantum numbers of the CP, unitarity bounds on its scaling dimension, and the quantum numbers of the primary null state that results when these bounds are saturated (see section~\ref{sec:6dusm} for a summary of our conventions in~$d = 6$). We also indicate whether the CP is a free field annihilated by~$\square$. 

\smallskip

\renewcommand{\arraystretch}{1.6}
\renewcommand\tabcolsep{6pt}
\begin{table}[H]
  \centering
  \begin{tabular}{ | l | l |l |c| }
\hline
\multicolumn{1}{|c|}{\bf Primary} &  \multicolumn{1}{|c|}{\bf Unitarity Bound} & \multicolumn{1}{|c|}{\bf Null State } & \multicolumn{1}{|c|}{\bf Comments}\\
\hline
\hline
$ [j_1, j_2,j_3]_{\Delta}~,~j_2 \geq 1 $ & $\Delta \geq \half \left(j_1 + 2 j_2 + j_3 \right)  + 4 $ & $[j_1,j_2-1,j_3]_{\Delta+1}$ & -\\
\hline
$ [j_1, 0,j_3]_{\Delta}~,~j_1, j_3 \geq 1 $ & $\Delta \geq \half \left(j_1 + j_3\right)+ 3$ & $[j_1-1,1,j_3-1]_{\Delta+1}$ & -\\
\hline
\hline
$ [j_1,0,0]_{\Delta}~,~~j_1 \geq 1$ & $\Delta \geq  \half j_1 + 2$ & $[j_1-1,0,1]_{\Delta+1}$ & free field \\
\hline
$ [0,0,j_3]_{\Delta}~,~~j_3 \geq 1$ & $\Delta \geq  \half j_3 + 2$ & $[1,0,j_3-1]_{\Delta+1}$ & free field \\
\hline
$ [0, 0,0]_{\Delta}~~ $ & $\Delta \geq {2}$ & $[0,0,0]_{\Delta+2}$ & free field \\
\hline
\hline 
$ [0,0,0]_{\Delta}~~ $ & $\Delta = 0 $ & $[0,1,0]_{\Delta+1}$ & unit operator\\
\hline
\end{tabular}
  \caption{Unitary representations of the conformal algebra in~$d=6$.  }
  \label{tab:6Dcurrents}
\end{table}

\smallskip

\noindent In general, we count currents or free fields modulo conservation laws or field equations. Here we need the fact that the dimension of a general~$\frak{su}(4)$ Lorentz representation~$[j_1, j_2, j_3]$ is
\begin{equation}
\dim [j_1, j_2, j_3] = {1 \over 12} \left(j_1+1\right)\left(j_2+1\right)\left(j_3+1\right) \left(j_1+j_2 + 2\right)\left(j_2+j_3 + 2\right) \left(j_1 + j_2 + j_3+ 3\right)~.
\end{equation}
This leads to the following operator counts for the currents and free fields in table~\ref{tab:6Dcurrents}:
\begin{itemize}
\item Conserved currents~$[j_1, j_2 \geq 1, j_3]_{j_1/2 + j_2 + j_3/2+ 4}$ contain~$\dim [j_1, j_2, j_3] - \dim [j_1, j_2-1, j_3]$ operators. 
\item Currents of the form~$[j_1 \geq 1, 0, j_3 \geq 1]_{j_1/2 + j_3/2+ 3}$ contain
\begin{align}
\dim [j_1, 0, j_3]  - \dim[j_1-1, & 1,j_3-1]  + \dim[j_1 -1, 0,j_3-1]  \\
& = {1 \over 6} (j_1 + j_3+1)(j_1 + j_3+2)(j_1 + j_3+3)
\end{align}
operators, because the null state~$[j_1-1, 1,j_3-1]_{j_1/2 + j_3/2 + 4}$ is itself a conserved current.  
\item The free fields~$[j_1,0,0]_{j_1/2+2}$ (and similarly~$[0,0,j_3]_{j_3/2+2}$) contain no operator degrees of freedom, because their null state~$[j_1-1, 0, 1]_{j_1/2 + 3}$ is itself a conserved current containing~${1 \over 6} (j_1+1)(j_1+2)(j_1+3)$ operators.
\end{itemize}

\smallskip

We follow section~\ref{sec:findsccm} and substitute~$h_1 = \half (j_1 + 2 j_2 + j_3)$ (see~\eqref{hdlambda} in appendix~\ref{app:liealg}), $d = 6$ into the general bound~\eqref{primaryh}, which must be satisfied by the SCP~$\CV$ of any superconformal current multiplet,
\begin{equation}\label{primaryhii}
\Delta_\CV \leq \half (j_1 + 2 j_2 + j_3) + 4~.
\end{equation}
Together with the superconformal unitarity constraints on~$\CV$ summarized in section~\ref{sec:6dusm},  this bound allows us to identify and analyze all candidate superconformal current multiplets. Below, we will do this explicitly for~$1 \leq \CN \leq 2$, i.e.~for~$(1,0)$ and~$(2,0)$ theories; as discussed in section~\ref{sec:maxsusy}, there are no~$(\CN,0)$ SCFTs with~$\CN \geq 3$. We will also count the number of bosonic and fermionic CPs in these multiplets, modulo conservation laws.  It is straightforward to enumerate the candidate current multiplets for all values of~$\CN$ using the form of the superconformal unitarity bounds in appendix~\ref{app:genN}: 
\begin{itemize}
\item $B_{1,2,3}$-multiplets that are neutral under the~$\frak{sp}(2\CN)_R$ symmetry, i.e.~all~$R$-symmetry Dynkin labels vanish, $R_i  = 0~(i = 1, \ldots, \CN)$.
\item $C_{1,2}$-multiplets with at most one non-vanishing~$R$-symmetry Dynkin index, i.e.
\begin{equation}
\sum_{i = 1}^\CN R_i \leq 1~. 
\end{equation} 
\item $D_{1}$-multiplets whose~$R$-symmetry Dynkin labels satisfy the bound
\begin{equation}\label{6dcurrsymbd}
\sum_{i = 1}^\CN R_i \leq 2~. 
\end{equation} 
\end{itemize}

One general conclusion of our analysis is that certain short representation of the~$\frak{so}(6,2)$ conformal algebra do cannot occur in six-dimensional SCFTs, because they do not reside in any unitary superconformal multiplet. This is the case for all currents of the form 
\begin{equation}
 [j_1 \geq 1,0,j_2 \geq 1]_{\Delta = \frac{1}{2}(j_1+j_3)+3}~. 
\end{equation}
When~$j_1 = j_3$, this representation describes a conserved two-form current 
\begin{equation}
j_{\mu\nu} = j_{[\mu\nu]}~, \qquad \d^\mu j_{\mu\nu} = 0~, \qquad \Delta_{j_{\mu\nu}} = 4~.
\end{equation}
Such a current naturally couples two a two-form gauge field~$B_{\mu\nu}$ via a marginal interaction,\begin{equation}
\Delta \SL = B^{\mu\nu} j_{\mu\nu} + \cdots~,
\end{equation}
where the ellipsis denotes possible higher-order seagull terms that may be needed to ensure gauge invariance. The well-known analogue of this operation in~$d = 4$ is the marginal gauging of a flavor current~$j_\mu$ by an ordinary vector gauge field~$A_\mu$, which plays a crucial role in copious examples. By contrast, the fact that the two-form current~$j_{\mu\nu}$ does not exist in~$d = 6$ SCFTs means that there is no straightforward notion of gauging a two-form symmetry in these theories, despite the fact that the dynamics of these theories is believed to involve two-form gauge fields in some way. The absence of two-form currents also has implications for anomaly matching in six-dimensional SCFTs~\cite{6danom}.

\subsubsection{$d = 6, \CN = (1,0)$}

\label{sec:d6n1curr}

If we apply the general constraints around~\eqref{6dcurrsymbd} to~$\CN=1$ (see section~\ref{sec:d6n1defs}), we find the following current multiplets:
\begin{itemize}
\item The multiplet~$D_1[0,0,0]_0^{(0)}$ consists of the unit operator.
\item $D[0,0,0]_2^{(1)}$ is a free hypermultiplet: 
\begin{equation}
\xymatrix  @R=1pc {
 *++[F=]{D[0,0,0]_2^{(1)}} \ar[r]^-Q&  *++[F]{[1,0,0]_{\frac{5}{2}}^{(0)}} }
 \label{6dH}
\end{equation}

\item $D_{1}[0,0,0]_4^{(2)}$ is an~$8+8$ flavor current multiplet (see section~\ref{sec:fcm}):
\begin{equation}
\xymatrix  @R=1pc {
 *++[F=]{[0,0,0]_{4}^{(2)}} \ar[r]^-Q&  *++[F]{[1,0,0]_{\frac{9}{2}}^{(1)}} \ar[r]^-Q&  *++[F]{[0,1,0]_{5}^{(0)}}}
 \label{6dG}
\end{equation}
\item $C_{2}[0,0,0]^{(0)}_2$ is a free tensor multiplet: 
\begin{equation}
\xymatrix  @R=1pc {
 *++[F=]{[0,0,0]_{2}^{(0)}} \ar[r]^-Q&  *++[F]{[1,0,0]_{\frac{5}{2}}^{(1)}} \ar[r]^-Q&  *++[F]{[2,0,0]_{3}^{(0)}}}
 \label{6dT1}
\end{equation}
\item $C_{2}[0,0,0]_4^{(1)}$ is an extra SUSY-current multiplet (see section~\ref{sec:escm}) with~$32+32$ bosonic and fermionic CPs:
\begin{equation}
\xymatrix  @R=1pc {
 *++[F=]{[0,0,0]_{4}^{(1)}} \ar[r]^-Q&  *++[F]{[1,0,0]_{\frac{9}{2}}^{(0)\oplus (2)}} \ar[r]^-Q&  *++[F]{[0,1,0]_{5}^{(1)}\oplus [2,0,0]_{5}^{(1)}}\ar[r]^-Q&  *++[F]{[1,1,0]_{\frac{11}{2}}^{(0)}}}
 \label{6dSC}
 \end{equation}
\item The multiplet~$C_1[j_1, 0,0]^{(0)}_{j_1/2 + 2}$ contains higher-spin free fields: 
\begin{equation}
\xymatrix  @R=1pc {
 *++[F=]{C_1[j_1 \geq 1, 0,0]^{(0)}_{\half j_1 + 2}} \ar[r]^-Q&  *++[F]{[j_1+1,0,0]_{\frac{1}{2}j_1+\frac{5}{2}}^{(1)}} \ar[r]^-Q&  *++[F]{[j_1+2,0,0]_{\frac{1}{2}j_1+3}^{(0)}} }
 \label{6dC1}
\end{equation}
\item $C_{1}[j_1, 0,0]^{(1)}_{j_1/2+4}$ contains higher-spin currents:
\begin{equation}
\xymatrix  @R=1pc {
 *++[F=]{C_{1}[j_1\geq 1, 0,0]^{(1)}_{\half j_1+4}} \ar[r]^-Q &  *++[F]{[j_1-1,1,0]_{\frac{1}{2}j_1+\frac{9}{2}}^{(0)}~,~[j_1+1,0,0]_{\frac{1}{2}j_1+\frac{9}{2}}^{ (0) \oplus(2) } } \ar[dl]_Q\\
*++[F]{[j_1,1,0]_{\frac{1}{2}j_1+5}^{(1)}~,~[j_1+2,0,0]_{\frac{1}{2}j_1+5}^{(1)} } \ar[r]^-Q &
   *++[F]{[j_1+1,1,0]_{\frac{1}{2}j_1+\frac{11}{2}}^{(0)}}}
 \label{6dC2}
\end{equation}
After accounting for conservation laws this multiplet contains~${4 \over 3} (j_1+2)(j_1+3)(j_1+4)$ bosonic, and equally many fermionic, CPs. 

\item $B_3[0,0,0]_4^{(0)}$ is a~$40+40$ stress tensor multiplet (see section~\ref{sec:stm}):
\begin{equation}
\xymatrix  @R=1pc {
 *++[F=]{B_3[0,0,0]_4^{(0)}} \ar[r]^-Q&  *++[F]{[1,0,0]_{\frac{9}{2}}^{(1)}} \ar[r]^-Q&  *++[F]{[0,1,0]_{5}^{(2)}\oplus[2,0,0]_{5}^{(0)}}\ar[ddl]_Q \\ \\
& *++[F]{[1,1,0]_{\frac{11}{2}}^{(1)}}\ar[r]^-Q&  *++[F]{[0,2,0]_{6}^{(0)}}}
 \label{6dEM1}
\end{equation}

\item The~$B_\ell[j_1, j_2, 0]^{(0)}_{j_1/2 + j_2 + 4}$ multiplets, with~$\ell =1$ if~$j_2 \geq 1$ and~$\ell = 2$ if~$j_1 \geq 1, j_2 = 0$, all contain higher spin currents:
\begin{equation}
\xymatrix  @R=1pc {
 *++[F=]{B_\ell[j_1, j_2, 0]^{(0)}_{\Delta = \half j_1 + j_2 + 4}} \ar[r]^-Q &  *++[F]{\begin{aligned} & [j_1-1,j_2+1,0]_{\Delta + \half}^{(1)} \\[-2pt]
 & [j_1+1,j_2,0]_{\Delta + \half}^{(1)}
 \end{aligned}} \ar[ddl]_Q \\ \\ 
*++[F]{\begin{aligned}
& [j_1-2, j_2+2, 0]_{\Delta+1}^{(0)} \\[2pt]
& [j_1, j_2+1, 0]_{\Delta+1}^{(0) \oplus (2)} \\[2pt]
& [j_1+2, j_2, 0]_{\Delta+1}^{(0)}
\end{aligned} } \ar[r]^-Q& *++[F]{\begin{aligned}
& [j_1-1, j_2+2, 0]^{(1)}_{\Delta + {3 \over 2}} \\[-2pt]
& [j_1+1, j_2+1, 0]^{(1)}_{\Delta + {3 \over 2}}
\end{aligned}}\ar[r]^-Q&  *++[F]{
[j_1, j_2+2, 0]^{(0)}_{\Delta + 2}
}}
 \label{6dEM1}
\end{equation}
Accounting for conservation laws, there are~${4 \over 3} \left(j_1+1\right)\left(j_2+2\right) \left(j_1 + j_2 + 3\right)\left(j_1 + 2 j_2 + 5\right)$ bosonic, and as many fermionic, operators. When~$j_1 = 0$, the operators $[j_1-1, j_2+1,0]^{(1)}_{\Delta + \half}$ and $[j_1-1, j_2+2, 0]^{(1)}_{\Delta + {3 \over 2}}$ disappear, while $[j_1-2, j_2+2, 0]^{(0)}_{\Delta+1}$ and $[j_1, j_2+1, 0]^{(0)}_{\Delta+1}$ cancel. When~$j_1 = 1$, we only drop the operator $[j_1-2, j_2+2, 0]^{(0)}_{\Delta+1}$. 
\end{itemize}

\subsubsection{$d = 6, \CN = (2,0)$}

\label{sec:currd6n2}

\label{sec:d6n2curr}

The constraints around~\eqref{6dcurrsymbd}, applied to~$\CN=2$ (see section~\ref{sec:d6n2defs}), state that currents can only reside in~$B_{1,2,3}$-multiplets with~$R_1 = R_2 = 0$, $C_{1,2}$-multiplets with~$R_1 + R_2 \leq 1$, and~$D_1$-multiplets with~$R_1+R_2 \leq 2$. We will now consider these in more detail: 
\begin{itemize}
\item The multiplet~$D_1[0,0,0]_0^{(0,0)}$ consists of the unit operator.
\item $D_1[0,0,0]_2^{(0,1)}$ is a free tensor multiplet: 
\begin{equation}
\xymatrix  @R=1pc {
 *++[F=]{D_1[0,0,0]_{2}^{(0,1)}} \ar[r]^-Q&  *++[F]{[1,0,0]_{\frac{5}{2}}^{(1,0)}} \ar[r]^-Q&  *++[F]{[2,0,0]_{3}^{(0,0)}}}
 \label{6dT2}
\end{equation}
\item $D_1[0,0,0]_2^{(1,0)}$ consists of free fields, some of which carry higher spin: 
\begin{equation}
\xymatrix  @R=1pc {
 *++[F=]{D_1[0,0,0]_{2}^{(1,0)}} \ar[r]^-Q&  *++[F]{[1,0,0]_{\frac{5}{2}}^{(0,0) \oplus (0,1)}} \ar[r]^-Q&  *++[F]{[2,0,0]_{3}^{(1,0)}} \ar[r]^-Q&  *++[F]{[3,0,0]_{7 \over 2}^{(0,0)}}  }
 \label{6dT2}
\end{equation}
\item $D_1[0,0,0]_4^{(0,2)}$ is a~$128+128$ stress tensor multiplet (see section~\ref{sec:stm}):
\begin{equation}
\xymatrix  @R=1pc {
 *++[F=]{
 D_1[0,0,0]_4^{(0,2)}
 } \ar[r]^-Q&  *++[F]{[1,0,0]_{\frac{9}{2}}^{(1,1)}} \ar[r]^-Q&  *++[F]{[0,1,0]_{5}^{(2,0)}\oplus [2,0,0]_{5}^{(0,1)}} \ar[ddl]_Q \\ \\
& *++[F]{[1,1,0]_{\frac{11}{2}}^{(1,0)}}\ar[r]^-Q&  *++[F]{[0,2,0]_{6}^{(0,0)}}}
 \label{6dEM2}
\end{equation}

\item The multiplets~$D_1[0,0,0]_4^{(1,1)}$ (with~$512+512$ bosonic and fermionic operators) and $D_1[0,0,0]_4^{(2,0)}$ (with~$640+640$ operators)  both contain higher-spin currents:
\begin{equation}
\xymatrix  @R=1pc {
 *++[F=]{
 D_1[0,0,0]_4^{(1,1)}
 } \ar[r]^-Q&  *++[F]{[1,0,0]_{\frac{9}{2}}^{(0,1) \oplus (0,2) \oplus (2,0) }} \ar[r]^-Q&  *++[F]{\begin{aligned} & [0,1,0]_{5}^{(1,0) \oplus (1,1)} \\[-2pt]
 & [2,0,0]_{5}^{(1,0) \oplus (1,1)}
 \end{aligned} } \ar[dddll]_Q \\ \\ \\
*++[F]{\begin{aligned} & [1,1,0]_{\frac{11}{2}}^{(0,0) \oplus (0,1) \oplus (2,0) } \\[-2pt]
& [3,0,0]_{11 \over 2}^{(0,1)}
\end{aligned} }\ar[r]^-Q&  *++[F]{\begin{aligned} & [0,2,0]_{6}^{(1,0)} \\[-2pt]
& [2,1,0]_6^{(1,0)}
\end{aligned} }  \ar[r]^-Q& *++[F]{[1,2,0]^{(0,0)}_{13 \over 2}}  }
 \label{6dn2hsd2}
\end{equation}
\smallskip
\begin{equation}
\xymatrix  @R=1pc {
 *++[F=]{
 D_1[0,0,0]_4^{(2,0)}
 } \ar[r]^-Q&  *++[F]{[1,0,0]_{\frac{9}{2}}^{(1,0) \oplus (1,1)}} \ar[r]^-Q&  *++[F]{\begin{aligned} & [0,1,0]_{5}^{(0,0) \oplus (0,1) \oplus (0,2)} \\[-2pt]
 & [2,0,0]_{5}^{(0,1) \oplus (2,0)}
 \end{aligned} } \ar[dddll]_Q \\ \\ \\
*++[F]{\begin{aligned} & [1,1,0]_{\frac{11}{2}}^{(1,0) \oplus (1,1)} \\[-2pt]
& [3,0,0]_{11 \over 2}^{(1,0)}
\end{aligned} }\ar[r]^-Q&  *++[F]{\begin{aligned} & [0,2,0]_{6}^{(2,0)} \\[-2pt]
& [2,1,0]_6^{(0,0) \oplus (0,1)}
\end{aligned} }  \ar[r]^-Q& *++[F]{[1,2,0]^{(1,0)}_{13 \over 2}} \ar[dd]^-QQ \\ \\
& & *++[F]{[0,3,0]_7^{(0,0)}
} }
 \label{6dn2hsd1}
\end{equation}

\item The multiplet~$C_\ell[j_1,0,0]^{(0,0)}_{j_1/2 + 2}$ (with~$\ell = 1$ if~$j_1 \geq 1$ and~$\ell = 2$ if~$j_1 = 0$) contains higher-spin free fields:
 \begin{equation}
\xymatrix  @R=1pc {
 *++[F=]{
 C_\ell[j_1,0,0]^{(0,0)}_{\half j_1 + 2}
 } \ar[r]^-Q&  *++[F]{[j_1 + 1,0,0]_{\half j_1 + {5 \over 2} }^{(1,0) }} \ar[r]^-Q& *++[F]{[j_1 + 2,0,0]_{\half j_1 + 3 }^{(0,0) \oplus (0,1)}}  \ar[ddl]_Q \\ \\ 
&  *++[F]{[j_1 + 3,0,0]_{\half j_1 + {7 \over 2} }^{(1,0) }}  \ar[r]^-Q& *++[F]{[j_1 + 4,0,0]_{\half j_1 + 4 }^{(0,0) }}  
}
 \label{6dn2c1r1r2zeroff}
\end{equation}
\item The multiplet~$C_\ell[j_1,0,0]^{(1,0)}_{j_1/2 + 4}$ (with~$\ell = 1$ if~$j_1 \geq 1$ and~$\ell = 2$ if~$j_1 = 0$) contains higher-spin currents:
\begin{equation}
\xymatrix  @R=1pc {
 *++[F=]{
 C_\ell[j_1,0,0]^{(1,0)}_{\Delta = \half j_1 + 4}
 } \ar[r]^-Q&  *++[F]{\begin{aligned} & \scriptstyle [j_1-1, 1,0]_{\Delta + \half}^{(0,0) \oplus (0,1)} \\[-2pt]
 & \scriptstyle [j_1 + 1,0,0]_{\Delta + \half }^{(0,0) \oplus (0,1) \oplus (2,0) } 
 \end{aligned} } \ar[r]^-Q& *++[F]{\begin{aligned} &  \scriptstyle [j_1-2, 2, 0]_{\Delta+1}^{(1,0)}  \\[-2pt]
 &\scriptstyle [j_1, 1, 0]_{\Delta + 1}^{2 (1,0) \oplus (1,1)} \\[-2pt]
 & \scriptstyle [j_1 + 2,0,0]_{\Delta + 1}^{2 (1,0) \oplus (1,1)}
 \end{aligned}}  \ar[dddll]_Q \\ \\ \\
 *++[F]{\begin{aligned} & \scriptstyle [j_1-3, 3, 0]^{(0,0)}_{\Delta + {3 \over 2}} \\[-2pt]
 & \scriptstyle [j_1-1, 2, 0]_{\Delta + {3 \over 2}}^{(0,0) \oplus (0,1) \oplus (2,0)} \\[-2pt]
 & \scriptstyle [j_1+1, 1, 0]^{2 (0,0) \oplus 2(0,1) \oplus (0,2) \oplus (2,0)} \\[-2pt]
 & \scriptstyle  [j_1+3, 0,0]_{\Delta + {3 \over 2}}^{(0,0) \oplus (0,1) \oplus (2,0)}
 \end{aligned}}  \ar[r]^-Q& *++[F]{
\begin{aligned}  & \scriptstyle  [j_1-2, 3, 0]^{(1,0)}_{\Delta + 2} \\[-2pt]
 & \scriptstyle  [j_1, 2, 0]^{2(1,0) \oplus (1,1)}_{\Delta + 2} \\[-2pt]
 &\scriptstyle [j_1 + 2, 1, 0]^{2(1,0) \oplus (1,1)}_{\Delta + 2}  \\[-2pt]
& \scriptstyle [j_1+4, 0,0]^{(1,0)}_{\Delta + 2}
 \end{aligned} 
}   
\ar[r]^-Q& *++[F]{
\begin{aligned}  & \scriptstyle  [j_1-1, 3, 0]^{(0,0) \oplus (0,1) }_{\Delta + {5 \over 2}} \\[-2pt]
 & \scriptstyle  [j_1+1, 2, 0]^{(0,0) \oplus (0,1) \oplus (2,0) }_{\Delta + {5 \over 2}} \\[-2pt]
 &\scriptstyle [j_1+3, 1, 0]^{(0,0) \oplus (0,1) }_{\Delta + {5 \over 2}} 
 \end{aligned} 
}
\ar[dddl]_Q \\ \\ \\
&  *++[F]{\begin{aligned} & \scriptstyle [j_1, 3, 0]^{(1,0)}_{\Delta + 3} \\[-2pt]
 & \scriptstyle  [j_1+2 , 2, 0]^{(1,0)}_{\Delta + 3}
 \end{aligned}} 
 \ar[r]^-Q& *++[F]{\scriptstyle
 [j_1+1, 3, 0]^{(0,0)}_{\Delta + {7 \over 2}}
 }   
}
 \label{6dn2c1r1oner2zerohs}
\end{equation}
Taking into account conservation laws, this multiplet contains~${128 \over 3} \left(j_1+2\right)\left(j_1 + 4\right) \left(j_1 + 6\right)$ bosonic, and the same number of fermionic, operators. When~$j_1 = 2$, we must drop $[j_1 -3, 3, 0]^{(0,0)}_{\Delta + {3 / 2}}$. When~$j_1 = 1$ we drop $[j_1-2, 2, 0]^{(1,0)}_{\Delta + 1}$ and~$[j_1-2, 3, 0]^{(1,0)}_{\Delta+2}$, while $[j_1-3, 3, 0]^{(0,0)}_{\Delta + {3 / 2}}$ and $[j_1-1, 2, 0]^{(0,0)}_{\Delta+{3 / 2}}$ cancel. Finally, when $j_1 = 0$, we must omit $[j_1-1, 1, 0]^{(0,0) \oplus (0,1)}_{\Delta + 1/2}$ and $[j_1-1, 3, 0]^{(0,0) \oplus (0,1)}_{\Delta + {5 / 2}}$, while $[j_1-2, 2, 0]^{(1,0)}_{\Delta + 1}$ cancels $[j_1, 1, 0]^{(1,0)}_{\Delta+1}$, $[j_1-3, 3, 0]^{(0,0)}_{\Delta + {3 / 2}}$ cancels $[j_1+1, 1, 0]^{(0,0)}_{\Delta+{3 / 2}}$, and $[j_1-2, 3, 0]^{(1,0)}_{\Delta+2}$ cancels $[j_1, 2, 0]^{(1,0)}_{\Delta+2}$.

\item $C_\ell[j_1,0,0]^{(0,1)}_{j_1/2 + 4}$ (with~$\ell = 1$ if~$j_1 \geq 1$ and~$\ell = 2$ if~$j_1 = 0$) also contains higher-spin currents:
\begin{equation}
\xymatrix  @R=1pc {
 *++[F=]{
 C_\ell[j_1,0,0]^{(0,1)}_{\half j_1 + 4}
 } \ar[r]^-Q&  *++[F]{\begin{aligned} & \scriptstyle [j_1-1, 1,0]_{\Delta + \half}^{(1,0)} \\[-2pt]
 & \scriptstyle [j_1 + 1,0,0]_{\Delta + \half }^{(1,0) \oplus (1,1)} 
 \end{aligned} } \ar[r]^-Q& *++[F]{\begin{aligned} &  \scriptstyle [j_1-2, 2, 0]_{\Delta+1}^{(0,0)}  \\[-2pt]
 &\scriptstyle [j_1, 1, 0]_{\Delta + 1}^{(0,0) \oplus (0,1) \oplus (2,0)  } \\[-2pt]
 & \scriptstyle [j_1 + 2,0,0]_{\Delta + 1}^{(0,0) \oplus (0,1) \oplus (0,2) \oplus (2,0)}
 \end{aligned}}  \ar[ddddll]_Q \\ \\ \\ \\
 *++[F]{\begin{aligned} & \scriptstyle [j_1-1, 2, 0]_{\Delta + {3 \over 2}}^{(1,0)} \\[-2pt]
 & \scriptstyle [j_1+1, 1, 0]^{2 (1,0) \oplus (1,1)}_{\Delta + {3 \over 2}} \\[-2pt]
 & \scriptstyle  [j_1+3, 0,0]_{\Delta + {3 \over 2}}^{(1,0) \oplus (1,1)}
 \end{aligned}}  \ar[r]^-Q& *++[F]{
\begin{aligned}  & \scriptstyle  [j_1, 2, 0]^{(0,0) \oplus (0,1)}_{\Delta + 2} \\[-2pt]
 &\scriptstyle [j_1 + 2, 1, 0]^{(0,0) \oplus (0,1) \oplus (2,0)}_{\Delta + 2}  \\[-2pt]
& \scriptstyle [j_1+4, 0,0]^{(0,1)}_{\Delta + 2}
 \end{aligned} 
}   
\ar[r]^-Q& *++[F]{
\begin{aligned}   & \scriptstyle  [j_1+1, 2, 0]^{(1,0)}_{\Delta + {5 \over 2}} \\[-2pt]
 &\scriptstyle [j_1+3, 1, 0]^{(1,0)}_{\Delta + {5 \over 2}} 
 \end{aligned} 
}
\ar[dd]^-Q \\ \\
& &  *++[F]{\begin{aligned} & \scriptstyle  [j_1+2 , 2, 0]^{(0,0)}_{\Delta + 3}
 \end{aligned}} 
}
 \label{6dn2c1r1zeror2onehs}
\end{equation}
After subtracting conservation laws, this multiplet contains~${64 \over 3} \left(j_1 + 3\right)\left(j_1+4\right)\left(j_1+5\right)$ bosonic, and the same number of fermionic, operators. When~$j_1 = 1$, we must drop $[j_1-2, 2, 0]_{\Delta+1}^{(0,0)}$. Similarly, when~$j_1 = 0$, we drop $[j_1-1, 1, 0]^{(1,0)}_{\Delta+1/2}$ and $[j_1-1, 2, 0]^{(1,0)}_{\Delta + 3/2}$, while~$[j_1-2, 2, 0]^{(0,0)}_{\Delta+1}$ cancels $[j_1, 1, 0]^{(0,0)}_{\Delta+1}$. 

\item All multiplets of the form~$B_\ell[j_1,j_2,0]^{(0,0)}_{j_1/2 + j_2 + 4}$ (with~$\ell = 1$ if~$j_2 \geq 1$, $\ell = 2$ if~$j_2 = 0$, $j_1 \geq 1$, and~$\ell = 3$ if~$j_1 = j_2 = 0$) contain higher-spin currents:
\begin{equation}
\xymatrix  @R=1pc {
 *++[F=]{
 B_\ell[j_1,j_2,0]^{(0,0)}_{\Delta = \half j_1 + j_2 + 4}
 } \ar[r]^-Q&  *++[F]{\begin{aligned} & \scriptstyle [j_1-1, j_2+1,0]_{\Delta + \half}^{(1,0)} \\[-2pt]
 & \scriptstyle [j_1 + 1,j_2,0]_{\Delta + \half }^{(1,0)} 
 \end{aligned} } \ar[r]^-Q& *++[F]{\begin{aligned} &  \scriptstyle [j_1-2, j_2+2, 0]_{\Delta+1}^{(0,0) \oplus (0,1) }  \\[-2pt] 
 &\scriptstyle [j_1, j_2 + 1, 0]_{\Delta + 1}^{(0,0) \oplus (0,1) \oplus (2,0)  } \\[-2pt]
 & \scriptstyle [j_1 + 2,j_2,0]_{\Delta + 1}^{(0,0) \oplus (0,1)}
 \end{aligned}}  \ar[dddll]_Q \\ \\ \\ 
 *++[F]{\begin{aligned} & \scriptstyle [j_1 -3, j_2 + 3, 0]^{(1,0)}_{\Delta + {3 \over 2}} \\[-2pt]
 & \scriptstyle [j_1-1, j_2+2, 0]_{\Delta + {3 \over 2}}^{2(1,0) \oplus (1,1)} \\[-2pt]
 & \scriptstyle [j_1+1, j_2 + 1, 0]^{2 (1,0) \oplus (1,1)}_{\Delta + {3 \over 2}} \\[-2pt]
 & \scriptstyle  [j_1+3, j_2,0]_{\Delta + {3 \over 2}}^{(1,0)}
 \end{aligned}}  \ar[r]^-Q& *++[F]{
\begin{aligned}  & \scriptstyle [j_1-4, j_2 + 4, 0]^{(0,0)}_{\Delta + 2} \\[-2pt]
& \scriptstyle [j_1-2, j_2 + 3, 0]_{\Delta + 2}^{(0,0) \oplus (0,1) \oplus (2,0)} \\[-2pt]
& \scriptstyle  [j_1, j_2+2, 0]^{2 (0,0) \oplus 2 (0,1) \oplus (0,2) \oplus (2,0)}_{\Delta + 2} \\[-2pt]
 &\scriptstyle [j_1 + 2, j_2 + 1, 0]^{(0,0) \oplus (0,1) \oplus (2,0)}_{\Delta + 2}  \\[-2pt]
& \scriptstyle [j_1+4, j_2,0]^{(0,0)}_{\Delta + 2}
 \end{aligned} 
}   
\ar[r]^-Q& *++[F]{
\begin{aligned}   & \scriptstyle [j_1-3, j_2+4, 0]^{(1,0)}_{\Delta + {5 \over 2}} \\[-2pt]
& \scriptstyle [j_1 -1, j_2+3, 0]_{\Delta + {5 \over 2}}^{2(1,0) \oplus (1,1)} \\[-2pt]
& \scriptstyle  [j_1+1, j_2+2, 0]^{2(1,0) \oplus (1,1)}_{\Delta + {5 \over 2}} \\[-2pt]
 &\scriptstyle [j_1+3, j_2+1, 0]^{(1,0)}_{\Delta + {5 \over 2}} 
 \end{aligned} 
}
\ar[dddll]^Q \\ \\ \\
 *++[F]{
\begin{aligned}   & \scriptstyle [j_1-2, j_2+4, 0]^{(0,0) \oplus (0,1) }_{\Delta + 3} \\[-2pt]
& \scriptstyle [j_1, j_2+3, 0]^{(0,0) \oplus (0,1) \oplus (2,0) }_{\Delta + 3} \\[-2pt]
& \scriptstyle  [j_1+2, j_2+2, 0]^{(0,0) \oplus (0,1) }_{\Delta + 3} 
 \end{aligned} 
} \ar[r]^-Q & *++[F]{
\begin{aligned}   & \scriptstyle [j_1-1, j_2+4, 0]^{(1,0) }_{\Delta + {7 \over 2}} \\[-2pt]
& \scriptstyle [j_1+1, j_2+3, 0]^{(1,0) }_{\Delta + {7 \over 2}}
 \end{aligned} 
} \ar[r]^-Q &  *++[F]{ \scriptstyle  [j_1 ,j_2 + 4, 0]^{(0,0)}_{\Delta + 4}
 } 
}
 \label{6dn2b1r1r2zerohs}
\end{equation}
Accounting for conservation laws, there are ${64 \over 3} \left(j_1 + 1\right)\left(j_2+3\right) \left(j_1 + j_2 + 4\right) \left(j_1 + 2 j_2 + 7\right)$ bosonic (and equally many fermionic) operators. For small values of~$j_1$, we must remove certain operators:
\begin{itemize}
\item[$\star$] When~$j_1 = 3$, we drop~$[j_1-4, j_2 + 4, 0]^{(0,0)}_{\Delta+2}$.
\item[$\star$] When~$j_1 = 2$, we drop~$[j_1-3, j_2+3, 0]^{(1,0)}_{\Delta + {3/2}}$ and~$[j_1-3, j_2+4, 0]^{(1,0)}_{\Delta + {5 /2}}$. We also cancel~$[j_1-4, j_2 + 4, 0]^{(0,0)}_{\Delta + 2}$ against~$[j_1-2, j_2 + 3, 0]^{(0,0)}_{\Delta + 2}$.
\item[$\star$] When~$j_1 = 1$, we drop~$[j_1-2, j_2+2, 0]^{(0,0) \oplus (0,1)}_{\Delta+1}$, $[j_1-2, j_2+3, 0]^{(0,0) \oplus (0,1) \oplus (2,0)}_{\Delta+2}$, and~$[j_1-2, j_2 + 4, 0]^{(0,0) \oplus (0,1)}_{\Delta + 3}$. We also cancel $[j_1 - 3, j_2+3, 0]^{(1,0)}_{\Delta + 3/2}$ against $[j_1-1, j_2+2, 0]^{(1,0)}_{\Delta + 3/2}$, $[j_1 - 4, j_2+4, 0]^{(0,0)}_{\Delta + 2}$ against $[j_1, j_2+2, 0]^{(0,0)}_{\Delta + 2}$, and $[j_1 - 3, j_2+4, 0]^{(1,0)}_{\Delta + 5/2}$ against $[j_1-1, j_2+3, 0]^{(1,0)}_{\Delta + 5/2}$.
\item[$\star$] When~$j_1 = 0$, we drop all operators of the form~$[j_1-1, \cdots]$. We also cancel $[j_1-2, j_2+2, 0]^{(0,0) \oplus (0,1)}_{\Delta + 1}$ against $[j_1, j_2+1, 0]^{(0,0) \oplus (0,1)}_{\Delta + 1}$, $[j_1-3, j_2+3, 0]^{(1,0)}_{\Delta + 3/2}$ against $[j_1+1, j_2+1, 0]^{(1,0)}_{\Delta + 3/2}$, $[j_1-4, j_2+4, 0]^{(0,0)}_{\Delta+2}$ against $[j_1+2, j_2+1,0]^{(0,0)}_{\Delta + 2}$, $[j_1-2, j_2+3, 0]^{(0,0)\oplus (0,1) \oplus (2,0)}_{\Delta+2}$ against $[j_1, j_2+2, 0]^{(0,0) \oplus (0,1) \oplus (2,0)}_{\Delta+2}$, $[j_1-3, j_2+4, 0]^{(1,0)}_{\Delta+5/2}$ against $[j_1 + 1, j_2 + 2, 0]^{(1,0)}_{\Delta + 5/2}$, and $[j_1 - 2, j_2+4, 0]^{(0,0) \oplus (0,1)}_{\Delta + 3}$ against $[j_1, j_2+3, 0]^{(0,0) \oplus (0,1)}_{\Delta + 3}$.  
\end{itemize} 
\end{itemize}

\section{Decompositions of Superconformal Multiplets under Subalgebras}

\label{sec:subalgdec}

It is occasionally useful to view SCFTs with~$\CN$-extended supersymmetry as special examples of theories with~$\widehat \CN <\CN$ supersymmetry.  As was already discussed in section~\ref{sec:escm}, the~$Q$-supercharges of~$\CN$-supersymmetry then decompose as in~\eqref{qdecomp},
\begin{equation}\label{qdecomp2}
Q~\rightarrow~\hat Q \oplus Q'~,
\end{equation}
where~$\hat Q$ are the~$\hat \CN$-supercharges, while the remaining~$Q'$ supercharges enhance~$\hat \CN$ to~$\CN$. Similarly, the~$R$-symmetry algebra decomposes as in~\eqref{rdecomp},
\begin{equation}\label{rdecomp2}
\frak R~\rightarrow~\frak {\hat R} \oplus \frak F \oplus \CR_\text{off-diag.}~.
\end{equation}
Here~$\frak {\hat R}$ is the~$\hat R$-subalgebra corresponding to~$\hat \CN$, while~$\frak F$ is its commutant inside~$\frak R$. The~$\hat Q$-supercharges in~\eqref{qdecomp} are charged under~$\frak{\hat R}$, but not under~$\frak F$, so that~$\frak F$ is a flavor symmetry from the point of view of~$\hat \CN$-supersymmetry.  

Irreducible multiplets of~$\CN$-superconformal symmetry decompose into finitely many irreducible~$\hat \CN$-multiplets, each of which may carry a representation of the flavor symmetry~$\frak F$. Such decompositions can be useful for a variety of purposes, see e.g.~sections~\ref{sec:escm} and~\ref{sec:maxsusy} for some applications. They also provide an effective way to organize and present the operator content of multiplets in theories with~$N_Q>8$. These were not tabulated in section~\ref{sec:tables} because they contain too many operators.  

It is straightforward to decompose a long~$\CN$-multiplet: its SCP~$\CV$ decomposes into a finite sum of~$\hat \CN$-SCPs~$\hat \CV$, according to the decomposition of its~$R$-symmetry representation into~$\frak{R} \oplus \frak{F}$ representations,
\begin{equation}\label{cvdeccva}
\CV = \bigoplus_{\hat \CV} \hat \CV \qquad (\text{finite sum})~.
\end{equation}
Each~$\hat \CV$ is the SCP of a long~$\hat \CN$-multiplet. Acting on them with the additional~$Q'$ supercharges generates new~$\hat \CN$-SCPs at higher levels, which also give rise to long~$\hat \CN$-multiplets. 

The decomposition of short multiplets is more intricate, e.g.~some~$Q'$-descendants may vanish, or they may be related to~$\hat Q$-descendants, because of~$\CN$-null states that cannot be understood as null states of the~$\hat \CN$-subalgebra. Even the decomposition~\eqref{cvdeccva} of the~$\CN$-SCP~$\CV$ into~$\hat \CN$-SCPs~$\hat \CV$ often gives rise to a rich variety of~$\hat \CN$-multiplets, some of which can be less short, or even long. This can be seen from the superconformal unitarity bounds in~\eqref{scabcdboundintro}, 
\begin{equation}\label{unitaritygeneral}
\Delta_\CV \geq f(L_\CV)+g( R_\CV) + \delta_A \equiv \Delta_A~~\text{or}~~\Delta_\CV = f(L_\CV)+g( R_\CV) + \delta_{B, C, D} \equiv \Delta_{B,C,D}~.
\end{equation}
The function~$f(L_\CV)$ of the Lorentz symmetry Dynkin labels, the available shortening types ($L$, $A$,~$B$,~etc.), and the shifts~$\delta _A>\delta _B>\delta_C > \delta_D$ are the same for the~$\CN$ and~$\hat \CN$ superconformal algebras. The functions~$g(R_\CV)$ and~$\hat g(\hat R_{\hat \CV})$ are different, but they are related in a simple way: $\hat g$ is the pullback of~$g$ via the embedding~$\hat{\frak{R}} \subset \frak{R}$. It follows that any~$\hat \CV$ in~\eqref{cvdeccva} will satisfy the bound~$\hat g(\hat R_{\hat{ \CV}_A}) \leq g(R_\CV)$. If this bound is saturated, then~$\hat \CV$ obeys the same type of shortening condition as~$\CV$; if the bound is strict, then~$\hat \CV$ gives rise to a longer (or even a long)~$\hat \CN$-multiplet. 

We will now illustrate some of these features in simple examples.

\subsection{$d = 6, \, \CN =(2,0) \rightarrow \widehat \CN =(1,0)$}


Our conventions for superconformal multiplets in six dimensions are summarized in section~\ref{sec:6dusm}. We now decompose~$\CN=(2,0)$ (see section~~\ref{sec:d6n2defs}) under~$\hat \CN = (1,0)$ (see section~\ref{sec:d6n1defs}). Following~\eqref{rdecomp2}, the~$(2,0)$~$R$-symmetry~$\frak{R} = \frak{sp}(4)_R$ decomposes into a flavor symmetry~$\frak F$ and the~$(1,0)$~$\hat R$-symmetry~$\hat{\frak{R}}$,
\begin{equation}
\frak{F} = \frak{su}(2)_L~, \qquad \hat{\frak R} = \frak{su}(2)_{R}~.
\end{equation}
We will write~$\frak{su}(2)_L \oplus \frak{su}(2)_R$ weights as~$(n_L; n_R)$, where~$n_{L, R} \in \Z_{\geq 0}$ are~$\frak{su}(2)$ Dynkin indices. The supercharge decomposition~\eqref{qdecomp2} then takes the following form:
\begin{equation}\label{twozeroqs}
Q \in [1,0,0]_{\half}^{(1,0)} \quad \longrightarrow \quad \hat{Q} \in [1,0,0]_{\half}^{(0; 1)} \quad \oplus \quad  Q' \in [1,0,0]_{\half}^{(1; 0)}~.
\end{equation}

An irreducible~$\frak{sp}(4)_{R}$ representation~$(R_1, R_2)$ decomposes into~$(R_1+1)(R_2+1)$ irreducible representations of~$\frak{su}(2)_L \oplus \frak{su}(2)_R$,
\begin{equation}
\frak{sp}(4)_R\to \frak{su}(2)_L \oplus \frak{su}(2)_R: \qquad (R_1, R_2)\to \bigoplus _{n_1=0}^{R_1}\bigoplus _{n_2=0}^{R_2} (R_1-n_1+n_2; n_1+n_2)~.\label{sptosofour}
\end{equation}
Note that the right-hand side is symmetric under the exchange~$\frak{su}(2)_L\leftrightarrow \frak{su}(2)_R$, because this exchange can be brought about by an~$\frak{sp}(4)_R$ transformation. The decomposition~\eqref{sptosofour} is particularly simple when~$R_1 = 0$, corresponding to symmetric, traceless~$\frak{so}(5)_R \simeq \frak{sp}(4)_R$ tensors, 
\begin{equation}\label{Spdecompzero}
(0, R)\to \bigoplus_{n=0}^{R}(n;n)~.
\end{equation}

We would like to decompose an~$\CN = (2,0)$ multiplet $X[j_1, j_2, j_3]^{(R_1, R_2)}$ into $\hat \CN = (1,0)$ multiplets of the form $\hat Y[j_1, j_2, j_3]^{(n_L; n_R)}$. Here~$X, \hat Y \in \{L, A, B, C, D\}$ refer to the possible shortening types in six dimensions. In the notation of~\eqref{unitaritygeneral}, the~$(2,0)$ unitarity bounds in table~\ref{tab:6DN2} and the~$(1,0)$ unitarity bounds in table~\ref{tab:6DN1} can be expressed as follows:
\begin{align}\label{6dgs}
& f(j_1, j_2, j_3) = \half (j_1 + 2 j_2 + 3j_3)~,\cr
& g(R_1, R_2) = 2 (R_1 + R_2)~, \qquad \hat g(n_R) = 2 n_R~,\cr
& \delta_A = 6~, \qquad \delta_B = 4~, \qquad \delta_C = 2~, \qquad \delta_D = 0~.
\end{align}
It follows from~\eqref{sptosofour} that decomposing the SCP~$\CV$ of the~$X$-multiplet as in~\eqref{cvdeccva} always leads to a SCP~$\hat \CV$ with~$n_{R}=R_{1}+R_{2}$ and~$n_L = R_2$. Comparing with~\eqref{6dgs}, we see that~$g(R_1, R_2) = \hat g(n_R)$. Therefore~$\hat \CV$ is the SCP of a~$\hat \CN=(1,0)$ multiplet whose shortening type~$\hat Y$ is the same as that of the~$X$-multiplet, i.e.~$\hat Y = X$. All other~$(1,0)$ SCPs that appear in the decomposition of~$\CV$ have~$n_R < R_1 +R_2$, and hence~$g(R_1, R_2)$ differs from~$\hat g(n_R)$ by a strictly positive, even integer. It follows that these SCPs give rise to~$\hat \CN= (1,0)$ multiplets with a larger~$\delta$-offset, whose shortening type~$\hat Y$ is strictly longer than that of~$X$, i.e.~$\hat Y > X$. Schematically, we can therefore write the decomposition of the~$X$-multiplet as follows:
\begin{equation}\label{schematicXXhat}
X[j_{1},j_{2},j_{3}]^{(R_{1},R_{2})}_{\Delta} \quad \rightarrow \quad \hat X[j_{1},j_{2},j_{3}]^{(R_{2} ; R_{1}+R_{2})}_{\Delta} \oplus \big\{\hat Y_{\ell = 0} > X\big\} \oplus \{\ell > 0\}~.
\end{equation}
Here~$\big\{\hat Y_{\ell = 0} > X\big\}$ denotes~$(1,0)$ multiplets whose SCPs arise from decomposing the SCP~$\CV$ of the~$X$-multiplet (at level~$\ell = 0$), but whose shortening type~$\hat Y$ is longer that that of~$X$. The term~$\{\ell > 0\}$ indicates additional~$(1,0)$ multiplets whose SCPs reside at higher level. Such multiplets can arise by acting with the additional~$Q'$ supercharges on~$\CV$. 

In general, determining the independent~$Q'$ descendants is essentially as difficult as constructing the full~$(2,0)$ multiplet directly, because null states involving the~$Q$-supercharges relate~$\hat Q$ and~$Q'$ descendants. A class of $(2,0)$ multiplets whose~$(1,0)$ decomposition is particularly simple consists of~$\half$-BPS~$D_1$-multiplets with~$R_1 = 0$ and arbitrary~$R_2 = R$. Their~$(1,0)$-decomposition takes the following form: 
\begin{align}
D_{1}[0,0,0]^{(0,R)}_{2R} \quad \rightarrow \quad & \widehat{D}_{1}[0,0,0]^{(R;R)}_{2R} \oplus \widehat{C}_{2}[0,0,0]^{(R-1;R-1)}_{2R}\oplus \widehat{B}_{3}[0,0,0]^{(R-2;R-2)}_{2R} \nonumber \\
&\oplus \widehat{A}_{4}[0,0,0]^{(R-3;R-3)}_{2R} \bigoplus_{n=4}^{R} \widehat{L}[0,0,0]^{(R-n;R-n)}_{2R}~.\label{simpledecomp}
\end{align}
If~$R \leq 3$, we must omit all multiplets on the right-hand side whose flavor or~$R$-symmetry Dynkin indices are negative. By comparing with~\eqref{Spdecompzero}, we see that all of these terms arise from decomposing the $(2,0)$ SCP under~$\frak{su}(2)_L \oplus \frak{su}(2)_R$. In the notation of~\eqref{schematicXXhat}, $X = D_1$, while all other multiplets that appear on the right-hand side of~\eqref{simpledecomp} belong to~$\big\{\hat Y_{\ell = 0} > X \big\}$. The terms~$\left\{\ell > 0\right\}$ in~\eqref{schematicXXhat} are absent.\footnote{~This can be verified by counting the total number of operators that appear on the two sides of~\eqref{simpledecomp}, e.g.~using the {\tt Mathematica} package in~\cite{tdmm}. There are~$\frac{64}{3}R(R-1)(2R-1)$ bosonic operators, and equally many fermionic ones, on both sides. } This amounts to the statement that all~$Q'$-descendants are related to~$\hat Q$-descendants by the~$D_1$ shortening condition. 

If we set~$R = 2$, the decomposition~\eqref{simpledecomp} reduces to 
\begin{equation}
D_{1}[0,0,0]^{(0,2)}_{4} \quad \rightarrow \quad \widehat{D}_{1}[0,0,0]^{(2;2)}_{4}\oplus \widehat{C}_{2}[0,0,0]^{(1;1)}_{4}\oplus \widehat{B}_{3}[0,0,0]^{(0;0)}_{4}~.
\end{equation}
Each multiplet in this equation has a simple interpretation: the left-hand side is the~$(2,0)$ stress tensor multiplet in~\eqref{6dEM2}, while the right-hand side consists of the following~$(1,0)$ multiplets:
\begin{itemize}
\item A~$\widehat{B}_{3}[0,0,0]^{(0;0)}_{4}$ stress tensor multiplet (tabulated in~\eqref{6dEM1}), which is neutral under the~$\frak{su}(2)_L$ flavor symmetry.
\item A~$\widehat{C}_{2}[0,0,0]^{(1;1)}_{4}$ extra SUSY-current multiplet (tabulated in~\eqref{6dSC}); it gives rise to the~$Q'$ supercharges, which are~$\frak{su}(2)_L$ doublets (see~\eqref{twozeroqs}). 
\item A~$\widehat{D}_{1}[0,0,0]^{(2;2)}_{4}$ flavor current multiplet (tabulated in~\eqref{6dG}) that contains the~$\frak{su}(2)_L$ currents. 
\end{itemize}

The simplest example of a decomposition rule that goes beyond~\eqref{simpledecomp} involves a $(2,0)$ $D_{1}[0,0,0]^{(1,2)}_{6}$ multiplet, which was discussed in section \ref{sec:leftovernegs}. Its decomposition into~$(1,0)$ multiplets takes the following form:
\begin{align}
D_{1}&[0,0,0]^{(1,2)}_{6} \quad \rightarrow \quad \left[ \widehat{D}_{1}[0,0,0]^{(2;3)}_{6}\oplus \widehat{C}_{2}[0,0,0]^{(3;2)}_{6}\oplus \widehat{C}_{2}[0,0,0]^{(1;2)}_{6} \oplus \widehat{B}_{3}[0,0,0]^{(2;1)}_{6} \right.  \cr
 &  \oplus \left. \widehat{B}_{3}[0,0,0]^{(0;1)}_{6} \oplus \widehat{A}_{4}[0,0,0]^{(1;0)}_{6}\right]_{\ell = 0} \oplus \left[\widehat{C}_{1}[1,0,0]^{(2;2)}_{13\over 2}\oplus \widehat{B}_{2}[1,0,0]^{(1;1)}_{13\over 2}\oplus \widehat{A}_{3}[1,0,0]^{(0;0)}_{13\over 2}\right]_{\ell = 1}~.\nonumber
\end{align}
Here the SCPs of the~$(1,0)$ multiplets inside~$[\cdots]_{\ell = 0}$ arise by decomposing the~$(2,0)$ SCP~$\CV$ under~$\frak{su}(2)_L \oplus \frak{su}(2)_R$, while the~$(1,0)$ multiplets inside~$[\cdots]_{\ell = 1}$ are non-trivial~$Q'$ descendants of~$\CV$ at level~$\ell=1$. The latter were schematically denoted by~$\{\ell > 1\}$ in~\eqref{schematicXXhat}.

\subsection{$d = 4, \, \CN =3 \rightarrow \widehat \CN =2$}

Our conventions for superconformal multiplets in four dimensions are summarized in section~\ref{sec:4dmults}. We now decompose~$\CN=3$ (see section~~\ref{sec:d4n3defs}) under~$\hat \CN = 2$ (see section~\ref{sec:d4n2defs}). Following~\eqref{rdecomp2}, the~$\CN=3$~$R$-symmetry~$\frak R=\frak{su}(3)_R\oplus \frak{u}(1)_{r}$ decomposes into a flavor symmetry~$\frak F$ and the~$\hat \CN = 2$~$R$-symmetry~$\hat{\frak{R}}$,
\begin{equation}
\frak{F} = \frak{u}(1)_f~, \qquad \hat{\frak R} = \frak{su}(2)_{\hat R} \oplus u(1)_{\hat r}~.
\end{equation}
We will label~$\hat{\frak{R}} \oplus \frak{F}$ representations by~$(\hat R; \hat r ; f)$. The generators of~$\frak{u}(1)_r$, $\frak{u}(1)_{\hat r}$, and~$\frak{u}(1)_f$ are related as follows:
\begin{equation}
r=\hat r+2 f~.
\end{equation}
The combination~$f-\hat r$ is a generator of~$\frak{su}(3)_R$; it is quantized in integral units. The chiral~$\CN=3$ supercharges decompose as follows (the antichiral supercharges can be obtained by complex conjugation): 
\begin{equation}
Q \in [1;0]_\half^{(1,0;-1)} \quad \rightarrow \quad \hat{Q} \in [1;0]_\half^{(1;-1;0)} \quad \oplus \quad Q' \in [1;0]_\half^{(0;1;-1)}~.
\end{equation}

More generally, a representation~$(R_{1},R_{2};r)$ of the~$\mathcal{N}=3$ $R$-symmetry decomposes into the following sum of~$(\hat R;\hat r;f)$ representations:
\begin{eqnarray}
\label{su3su2decomp}
(R_{1},R_{2};r) & = & \sideset{}{~'}\bigoplus_{\hat R=|R_{1}-R_{2}|}^{R_{1}+R_{2}} \left(\hat R\,; \frac{1}{3}(r-2R_{1}+2R_{2})\, ;\, \frac{1}{3}(r+R_{1}-R_{2})\right) \nonumber \\&  & \bigoplus_{n=0}^{R_{1}-1} ~~\sideset{}{~'}\bigoplus_{\hat R=|R_{2}-n|}^{R_{2}+n} \left(\hat R\,;\frac{1}{3}(r+4R_{1}+2R_{2})-2n \, ; \, n-\frac{1}{3}(2R_{1}+R_{2}-r)\right) \\
&& \bigoplus_{n=0}^{R_{2}-1} ~~\sideset{}{~'}\bigoplus_{\hat R=|R_{1}-n|}^{R_{1}+n} \left(\hat R\, ;2n-\frac{1}{3}(2R_{1}+4R_{2}-r)\, ;\, \frac{1}{3}(r+R_{1}+2R_{2})-n\right)~.\nonumber 
\end{eqnarray}
Here~$\oplus'_{\hat R}$ indicates a sum over~$\hat R$ in steps of~$2$, and the second or third line must be omitted if~$R_{1} = 0$ or $R_{2} = 0$, respectively. 

The chiral~$\CN = 3$, $\hat \CN=2$ unitarity bounds in tables~\ref{tab:4DN3C}, \ref{tab:4DN2C} are of the form~\eqref{unitaritygeneral}, with 
\begin{equation}\label{nthreegs}
g(R_1, R_2, r) =\frac{2}{3}(2R_1+R_2)-\frac{1}{6}r~, \qquad
\hat g(\hat R, \hat r) = \hat R-\frac{1}{2} \hat r~.
\end{equation}
The terms with~$\hat R=R_1+n~(0\leq n\leq R_{2})$ in the decomposition~\eqref{su3su2decomp} satisfy~$g = \hat g$. If these terms arise from decomposing an~$\CN=3$ SCP of chiral shortening type~$X$, then they obey the same shortening type as~$\hat \CN=2$ SCPs. The other terms in~\eqref{su3su2decomp} lead to less short (or long)~$\hat \CN =2$ multiplets. Similarly, the antichiral~$\CN=3$, $\hat \CN=2$ unitarity bounds in tables~\ref{4DN3AC}, \ref{4DN2AC} are of the form~\eqref{unitaritygeneral}, with 
\begin{equation}
\bar g(R_1, R_2;r) =\frac{2}{3}(R_1+2R_2)+\frac{1}{6}r~, \qquad  
\hat{\bar g} (\hat R, \hat r) =\hat R +\frac{1}{2}\hat r~.
\end{equation}
Therefore~$\bar g \geq \hat{\bar g}$, with equality for the terms with~$\hat R=R_{2}+n~(0\leq n \leq R_{1})$ in~\eqref{su3su2decomp}.

If~$R_{1} = 0$ or~$R_{2} = 0$, then all terms in~\eqref{su3su2decomp} satisfy either~$g = \hat g$ or~$\bar g = \hat{\bar g}$. In this case there is a simple decomposition rule for the corresponding short multiplets. For instance, a~$B_{1}\bar {X}[0; \bar j]^{(0, R_2; r)}$ multiplet, with~$X \in \{L, A, B\}$ and~$R_2 \geq 1$, decomposes as follows:
\begin{equation}\label{NthreeRzerotwo}
 B_{1}\b  X[0; \bar j] ^{(0,R_2 ; r)} \quad \to \quad \bigoplus_{\hat R=0}^{R_2} \, \widehat{B}_{1}\widehat{\bar {Y}}[0;\bar j]^{(\hat R\,; \, 2\hat R+\frac{1}{3}r-\frac{4}{3}R_2\,;\,\frac{1}{3}r+\frac{2}{3}R_{2}-\hat R)}~.
 \end{equation}
Here the antichiral~$\hat \CN=2$ shortening type ~$\hat{\b Y}$ on the right-hand side can vary over the sum; it is determined by matching the quantum numbers of the operators to the last row of table~\ref{tab:4DN2CAC}. Every~$\hat \CN=2$ multiplet in~\eqref{NthreeRzerotwo} arises by decomposing the SCP of the~$\CN=3$ multiplet according to~\eqref{su3su2decomp}. Hence, there are no independent~$Q'$ or~$\b Q'$ descendants at higher levels. 
 
Another example where the decomposition of the  $\CN =3$ SCP yields all the $\hat \CN=2$ SCPs is the~$\mathcal{N}=3$ stress tensor multiplet in~\eqref{d4n3stm}. It decomposes according to the following rule: 
\begin{align}
B_1\bar B_1[0;0]^{(1,1 ;  0)}_2 \quad \to   \quad & \widehat{B}_1\widehat{\bar {B}}_1[0;0]^{(2 ;  0 ; 0)}_2 \oplus \widehat{B}_1\widehat{\bar{ A}}_2[0;0]^{(1  ;  -2 ; 1)}_2 \cr
& \oplus  \widehat{A}_2\widehat{\bar {B}}_1[0;0]^{(1 ; 2 ; -1)}_2 \oplus \widehat{A}_2\widehat{\bar {A}}_2[0;0]^{(0 ; 0 ; 0)}_2~.\label{tthreetotwo}
\end{align}
Here~$\widehat{B}_1\widehat{\bar {B}}_1[0;0]^{(2 ;  0 ; 0)}_2$ is the~$\frak{u}(1)_f$ flavor current multiplet (see~\eqref{4dn2flavcurr}); $\widehat{B}_1\widehat{\bar{ A}}_2[0;0]^{(1  ;  -2 ; 1)}_2$ and its complex conjugate~$\widehat{A}_2\widehat{\bar {B}}_1[0;0]^{(1 ; 2 ; -1)}_2$ are extra SUSY-current multiplets (see~\eqref{d4n2escm}); and $\widehat{A}_2\widehat{\bar {A}}_2[0;0]^{(0 ; 0 ; 0)}_2$ is the~$\hat \CN=2$ stress tensor multiplet (see~\eqref{4dn2hs}).

In general, decomposing the~$\CN=3$ SCP does not yield the full decomposition rule: there can be non-trivial~$Q'$ or~$\b Q'$ descendants that give rise to new~$\hat \CN=2$ SCPs. 

\bigskip

\section*{Acknowledgements}\noindent We are grateful to C.~Beem, V.~Dobrev, J.~Maldacena, S.~Minwalla, S.~Raju, N.~Seiberg, C.~Vafa, and~B.~van Rees for helpful discussions. The work of CC is supported by a Martin and Helen Chooljian membership at the Institute for Advanced Study and DOE grant DE-SC0009988. TD is supported by the Fundamental Laws Initiative at Harvard University, as well as DOE grant DE-SC0007870 and NSF grant PHY-1067976.  KI is supported by DOE grant DE-SC0009919 and the Dan Broida Chair.  

\appendix

\section{Lie Algebras and Representations}

\label{app:liealg}

In this appendix we review some aspects of Lie algebra representation theory that are needed throughout the paper. The material is standard, see for instance~\cite{FultonHarris, Fuchs:1997jv,DiFrancesco:1997nk}. For computations, we used the LieART {\tt Mathematica} package~\cite{Feger:2012bs}. 

\subsection{Weights}

Let~${\frak g}_r$ be a complex simple Lie algebra of rank~$r \in \Z_{\geq 1}$. All states in a representation of the algebra~${\frak g}_r$ possess a weight~$\lambda \in \Lambda _w$, where~$\Lambda _w \in \R^r$ is the weight lattice of~${\frak g}_r$. There are several convenient bases for the~$r$-dimensional vector space spanned by~$\Lambda_w$. We will mostly use a basis of fundamental weights~$\omega_i$, in which the expansion coefficients~$\lambda_i$ are integers known as Dynkin labels,
\begin{equation}
\lambda = \sum_{i =1}^r \lambda_i \omega_i~, \qquad \lambda_i \in \Z~.
\end{equation}
Weights are labeled by~$r$-tuples of Dynkin lables; throughout this paper we write either~$[\lambda_1, \ldots, \lambda_r]$ for Lorentz weights or~$(\lambda_1, \ldots, \lambda_r)$ for~$R$-symmetry weights.\footnote{~This differs from some conventions in the literature, which use~$[\cdots]$, $(\cdots)$, etc.~to distinguish different bases for the weight space of a given Lie algebra. Unless we explicitly say otherwise, we use Dynkin labels.} 

Another basis for the weight space is furnished by the simple roots~$\alpha_i$ of the algebra,
\begin{equation}\label{alphabasis}
\lambda = \sum_{i = 1}^r x_i \alpha_i~.
\end{equation}
In order to change between the~$\alpha$-basis and the~$\omega$-basis, we use the integer-valued Cartan matrix~$A_{ij}$ that characterizes the algebra~${\frak g}_r$,
\begin{equation}\label{alphomtrans}
\alpha_i = \sum_{j = 1}^r A_{ij} \omega_j~.
\end{equation}
The~$i^{\text{th}}$ row of the Cartan matrix thus specifies the Dynkin labels of the simple root~$\alpha_i$. (The simple roots are themselves weights in the adjoint representation of~${\frak g_r}$.) Comparing~\eqref{alphabasis} and~\eqref{alphomtrans}, we see that the expansion coefficients in the two bases are related by
\begin{equation}
x_i = \sum_{j = 1}^r \lambda_j \left(A^{-1}\right)_{ji}~.
\end{equation}
Note that the~$x_i$ are not in general integers. Below, we will introduce yet a third basis for the weight space of the special orthogonal algebras~$\frak{b}_r$ and~$\frak{d}_r$. 

Every finite-dimensional, irreducible representation of~${\frak g}_r$ contains a unique state whose weight~$\lambda$ is such that every other weight~$\mu$  in the representation can be expressed as
\begin{equation}\label{hwdef}
\mu = \lambda - \sum_{i = 1}^r n_i \alpha_i~, \qquad n_i \in \Z_{\geq 0}~.
\end{equation}
We call~$\lambda$ the highest weight of the representation. Two finite-dimensional irreducible representations are equivalent if and only if their highest weights coincide. We can therefore label such representations by their highest weights. We can rephrase~\eqref{hwdef} by saying that every weight~$\mu$ is obtained by lowering the highest weight~$\lambda$ a finite number of times using the simple roots~$\alpha_i$. In this context, it is standard to introduce a lowering operator~$E_{-\alpha_i}$ for every simple root, which acts as~$E_{-\alpha_i} \lambda = \lambda - \alpha_i$. It is a useful fact (which entered the discussion following~\eqref{moreqstars}) that~$E_{-\alpha_i}$ annihilates a highest weight~$\lambda$ if and only if the~$i^\text{th}$ Dynkin label vanishes, $\lambda_i = 0$. This can, for instance, be shown by introducing a positive-definite inner product~$(\mu, \mu')$ (and hence a norm) on weights, such that~$(\omega _i, \alpha _j)\sim \delta _{ij}$. Then the norm of~$E_{-\alpha_i} \lambda$ satisfies~$\big|E_{-\alpha _{i}}\lambda\big |^2 \sim (\lambda, \alpha _{i})\sim \lambda _{i}$. (In the preceding formulas~$\sim$ indicates that we have omitted various positive factors.) Therefore~$E_{-\alpha_i} \lambda = 0$ if and only if~$\lambda_i = 0$. 

The expression~\eqref{hwdef} implies several useful facts about the~$\alpha$-basis~\eqref{alphabasis}. If we express $\lambda = \sum_{i = 1}^r x_i \alpha_i$  and~$\mu = \sum_{i = 1}^r y_i \alpha_i$, then~$y_i \leq x_i$. Therefore, given two irreducible representations with highest weights~$\lambda^{(1)} = \sum_{i = 1}^r x_i^{(1)} \alpha_i$ and~$\lambda^{(2)} = \sum_{i = 1}^r x_i^{(2)} \alpha_i$, the highest weights~$\lambda = \sum_{i = 1}^r x_i \alpha_i$ of the irreducible representation that appear in the tensor product decomposition~$\lambda^{(1)} \otimes \lambda^{(2)} = \oplus_\lambda \lambda$ must satisfy
\begin{equation}
x_i\leq x_i^{(1)}+x_i^{(2)}~.   
\end{equation}
This follows from the fact that every~$\lambda$ can be obtained by lowering~$\lambda^{(1)} + \lambda^{(2)}$ using the~$E_{-\alpha_i}$. 

\subsection{The Classical Lie Algebras} 

In this paper we only encounter the classical Lie algebras~$\frak{a}_r = \frak{su}(r+1)$, $\frak{b}_r = \frak{so}(2r+1)$, $\frak{c}_r = \frak{sp}(2r)$, and~$\frak{d}_r = \frak{so}(2r)$. Note the following exceptional isomorphisms at low rank,
\begin{subequations}
\begin{align}
& \frak{su}(2) = \frak{so}(3) = \frak{sp}(2)~,\\
& \frak{so}(4) = \frak{su}(2) \oplus  \frak{su}(2)~,\\
& \frak{so}(5) = \frak{sp}(4)~,\\
& \frak{so}(6) = \frak{su}(4)~.
\end{align}
\end{subequations}

\subsubsection{Unitarity Algebras}

The unitary algebras are~$\frak{a}_r=\frak{su}(r+1)$. We denote their weights~$[\lambda _1, \ldots, \lambda _r]$ using Dynkin labels $\lambda_i \in \Z_{\geq 0}$. A representation with highest weight~$[\lambda _1, \ldots, \lambda _r]$ corresponds to a Young tableaux with~$\lambda_i$ columns of height~$i$; the total number of boxes is~$\sum _{i=1}^r i \lambda_i$. The highest weight of the complex conjugate representation is obtained by exchaning~$\lambda_i \to \lambda_{r+1-i}$. 

\subsubsection{Orthogonal Algebras}

The orthogonal algebras are~$\frak{b}_r = \frak{so}(2r + 1)~(r \geq 1)$ and~$\frak{d}_r =\frak{so}(2r)~(r \geq 2)$. We will denote their weights~$[\lambda_1, \ldots, \lambda_r]$ using Dynkin labels~$\lambda_i \in \Z_{\geq 0}$. Note that the Dynkin labels of the isomorphic algebras~$\frak{so}(6)$ and~$\frak{su}(4)$ are related as follows:
\begin{equation}
\lambda _{1,2}^{\frak{so}(6)}=\lambda _{2,1}^{\frak{su}(4)}~, \qquad \lambda _3^{\frak{so}(6)}=\lambda _3^{\frak{su}(4)}~.
\end{equation}

It is sometimes convenient to switch to an orthogonal basis of weights~$[h_1, \ldots, h_r]$. The~$h_i$ are eigenvalues under rotations in~$r$ mutually orthogonal two-planes, and hence~$h_i \in \half \Z$. They always satisfy
\begin{equation}
h_i = h_{i + 1} + \lambda_i \qquad (1 \leq i \leq r-1)~.
\end{equation}
Since the~$\lambda_i$ are non-negative integers, it follows that~$h_1 \geq h_2 \geq \cdots \geq h_r$, and that the~$h_i$ are either all integral (corresponding to tensor weights) or all half-integral (corresponding to spinor weights). 

The precise relation between the orthogonal weights~$h_i$ and the Dynkin labels~$\lambda_i$ is different for the odd and even orthogonal algebras:
\begin{itemize}
\item In the odd case~$\frak{b}_r = \frak{so}(2r+1)$, we have
\begin{equation}\label{hblambda}
h_{i} = \lambda_i + \lambda_{i + 1} + \cdots + \lambda_{r-1} + {\lambda_r \over 2} \quad \left(1 \leq i \leq r-1\right)~, \qquad h_r = {\lambda_r \over 2}~.
\end{equation}
Therefore all~$h_i \geq 0$. The~$(2r+1)$-dimensional vector representation of~$\frak{so}(2r+1)$ corresponds to~$h_1 = \lambda_1 = 1$, with all other~$h_i = \lambda_i = 0$. A~$2^r$-dimensional Dirac spinor has~$\lambda_r = 1$, with all other~$\lambda_i = 0$, and hence all orthogonal weights take the value~$h_i = \half$. 

\item  In the even case~$\frak{d}_r = \frak{so}(2r)$, we have
\begin{subequations}\label{hdlambda}
\begin{align}
\label{lineone} & h_i = \lambda_i + \lambda_{i + 1} + \cdots + \lambda_{r-2} + {\lambda_{r-1} + \lambda_r \over 2} \qquad (1 \leq i \leq r-2)~, \\
& h_{r-1} = {\lambda_{r-1} + \lambda_r \over 2}~, \qquad h_r = {\lambda_{r-1} - \lambda_r \over 2}~.
\end{align}
\end{subequations}
Therefore~$h_1 \geq h_2 \geq \cdots \geq h_{r-1} \geq |h_r| \geq 0$, but~$h_r$ can be negative. A~$2r$-dimensional vector of~$\frak{so}(2r)$ always has~$h_1 = 0$, with all other~$h_i = 0$. Generically, this corresponds to~$\lambda_1 = 0$, with all other~$\lambda_i = 0$. The case~$r = 2$ is exceptional, because~$\frak{so}(4) = \frak{su}(2) \oplus \b{\frak{su}(2)}$ is not simple. The~$\frak{so}(4)$ Dynkin labels~$\lambda_1 = j$ and~$\lambda_2 = \b j$ correspond to the Dynkin labels of the two~$\frak{su}(2)$'s, so that a four-dimensional vector of~$\frak{so}(4)$ has highest weight $[\lambda_1 = j, \lambda_2 = \b j] = [1,1]$. 

The orthogonal weights of a~$2^r$-dimensional, left-handed chiral spinor of~$\frak{so}(2r)$ are given by~$[h_1, \ldots, h_r] = [\half, \ldots, \half]$, corresponding to~$[\lambda_1, \ldots, \lambda_r] = [0,\ldots, 0, 1, 0]$. Similarly, a~$2^r$-dimensional, right-handed chiral spinor has orthogonal weights~$[h_1, \ldots, h_r] = [\half, \ldots, -\half]$, corresponding to~$[\lambda_1, \ldots, \lambda_r] = [0,\ldots, 0, 1]$. In~$d = 2r = 4$ dimensions, this is the familiar fact that left- and right-handed chiral spinors transform as~$[1;0]$ and~$[0;1]$ of~$\frak{su}(2) \oplus \b {\frak{su}(2)}$. 

Similarly, the orthogonal weights~$[h_1, \cdots, h_r] = [1, \ldots, 1,  \pm 1]$, which correspond to Dynkin labels~$[\lambda_1, \ldots, \lambda_r] = [0, \cdots, 0,2,0]$ (for~$+$) and~$[0, \cdots, 0,2]$ (for~$-$), describe self-dual and anti-self-dual~$r$-forms in~$d = 2r$ dimensions.
\end{itemize}

\subsubsection{Symplectic Algebras}

The symplectic algebras are~$\frak{c}_r=\frak{sp}(2r)$. As before, we denote their weights~$[\lambda _1, \ldots \lambda _r]$ by Dynkin labels~$\lambda _i\in \Z _{\geq 0}$.  The~$n$-dimensional fundamental representation is~$[1,0,\ldots, 0]$, while~$[0,1,0,\ldots, 0]$ is a two-index, symplectic-traceless antisymmetric tensor. Note that the Dynkin labels of the isomorphic algebras~$\frak{sp}(4)$ and~$\frak{so}(5)$ are related by
\begin{equation}
\lambda ^{\frak{sp}(4)}_{1,2}=\lambda ^{\frak{so}(5)}_{2,1}~.
\end{equation}

\subsection{The Racah-Speiser Algorithm}

\label{app:RSalg}

Here we briefly review the Racah-Speiser (RS) algorithm, which allows for the efficient decomposition of tensor product representations. See~\cite{Fuchs:1997jv,Dolan:2002zh,Kinney:2005ej} for additional details. 

Given two irreducible representations of a Lie algebra~$\frak{g}_r$, with highest weights~$\lambda^{(1)}$ and~$\lambda^{(2)}$, the RS algorithm produces the highest weights~$\lambda$ of the irreducible representations that appear in the decomposition of the tensor product~$\lambda^{(1)} \otimes \lambda^{(2)} = \oplus_\lambda \lambda$. It does so by adding to the highest weight~$\lambda^{(1)}$ the entire weight system~$\big\{\mu_a^{(2)}\big\}~(a = 1, \ldots, \dim \lambda^{(2)})$ of the~$\lambda^{(2)}$-representation (note that~$\mu^{(2)}_{a = 1} = \lambda^{(2)}$), or vice versa:
\begin{equation}\label{RSterms}
\lambda^{(1)} \otimes \lambda^{(2)} = \bigoplus _{a=1}^{\dim \lambda^{(2)} } \left( \lambda^{(1)}+ \mu^{(2)}_a\right)\Big|_\text{RS} = \bigoplus _{a=1}^{\dim \lambda^{(1)} } \left( \mu^{(1)}_a + \lambda^{(2)}\right)\Big|_\text{RS}~. 
\end{equation}
The two direct sums in this equation are the same, since the tensor product is symmetric under the exchange~$\lambda ^{(1)} \leftrightarrow \lambda^{(2)}$, although this is not manifest from either expression. In what follows we will focus on the first sum, which runs over the weights of the~$\lambda^{(2)}$-representation. 

The symbol~$\big|_\text{RS}$ in~\eqref{RSterms} indicates a crucial aspect of the RS algorithm: 
\begin{itemize}
\item If the weight~$\lambda^{(1)}+ \mu^{(2)}_a$ is dominant (i.e.~all of its Dynkin labels are non-negative), and hence a valid highest weight, then~$ \big( \lambda^{(1)}+ \mu^{(2)}_a\big)\big|_\text{RS}$ is simply the representation with that highest weight.
\item If~$\lambda^{(1)}+ \mu^{(2)}_a$ is not a valid highest weight, because some of its Dynkin labels are negative, then~$\big( \lambda^{(1)}+ \mu^{(2)}_a\big)\big|_\text{RS}$ is obtained by subjecting~$\lambda^{(1)}+ \mu^{(2)}_a$ to a sequence of (shifted) Weyl reflections~$\sigma_i$ that map it to the fundamental Weyl chamber. This leads to a valid highest weight~$\t \lambda$, which contributes to the direct sum in~\eqref{RSterms} with a sign factor~$\chi$ that is determined by the~$\sigma_i$ (see below). Explicitly,
\begin{equation}\label{chifactordef}
\big( \lambda^{(1)}+ \mu^{(2)}_a\big)\big|_\text{RS} = \chi\left(\lambda^{(1)}+ \mu^{(2)}_a\right) \, \t \lambda~, \qquad \chi = \pm 1, 0~.
\end{equation}
This leads to three possibilities:
\begin{itemize}
\item[$\star$] If~$\chi =1$, we add a representation with highest weight~$\t \lambda$ to the tensor product decomposition in~\eqref{RSterms}.
\item[$\star$] If~$\chi = -1$, we remove a representation with highest weight~$\t \lambda$ from the decomposition. Whenever this case arises, such a representation is guaranteed to be supplied by a different term in the direct sum.
\item[$\star$] If~$\chi = 0$, we simply drop the term~$\big( \lambda^{(1)}+ \mu^{(2)}_a\big)\big|_\text{RS}$ in~\eqref{RSterms}.
\end{itemize}
\end{itemize}

Given any weight~$\lambda$, the sign factor~$\chi(\lambda)$ that appears in~\eqref{chifactordef} receives a contribution of~$-1$ for every Weyl reflection:
\begin{equation}\label{Weylminus}
\left[\lambda_1, \ldots, \lambda_r\right]^{\sigma_i} = -\left[\lambda_1, \ldots, \lambda_r\right]~,
\end{equation}
where
\begin{equation}\label{Weylreflect}
\left[\lambda_1, \ldots, \lambda_r\right]^{\sigma_i} = \sigma_i \left(\left[\lambda_1, \ldots, \lambda_r\right] + \rho\right) - \rho~.
\end{equation}
Here~$\rho = [1, \ldots, 1]$ is the Weyl vector, while the~$\sigma_i~(i = 1, \ldots, r)$ are a basis of Weyl reflections:
\begin{equation}
\sigma_i \left(\left[\lambda_1, \ldots, \lambda_r\right]\right) = \left[\lambda_1 - \lambda_i A_{i1}, \ldots, \lambda_j - \lambda_i A_{ij}, \ldots, \lambda_r - \lambda_i A_{ir}\right]~.
\end{equation}
In this formula~$A_{ij}$ is the Cartan matrix of the Lie algebra, and the repeated~$i$ index is not summed.  

In general~\eqref{Weylminus} implies that~$\chi(\lambda) = (-1)^n$, where~$n$ is the number of reflections~\eqref{Weylreflect} that are needed to map~$\lambda$ to the fundamental Weyl chamber. (This number is well defined mod$\, 2$.) For some weights~$\lambda$ this is impossible, because they satisfy~$\lambda^{\sigma_i} = \lambda$. For such weights, consistency with~\eqref{Weylminus} requires us to assign~$\chi(\lambda) = 0$. 

As a simple illustration of the RS algorithm, consider the case~$\frak{g}_r=\frak{su}(2)$. The tensor product of two~$\frak{su}(2)$ representations with Dynkin labels~$n_{1,2} \in \Z_{\geq 0}$ decomposes as follows:
\begin{align}\label{su2RS}
[n_1]\otimes [n_2]~&=[n_1-n_2]\Big|_\text{RS}\oplus [n_1-n_2 + 2]\Big|_\text{RS} \oplus \cdots \oplus [n_1+n_2]\Big|_\text{RS} \cr
& = [|n_1 - n_2|] \oplus [|n_1 - n_2| + 2] \oplus \cdots \oplus [n_1 + n_2]~.
\end{align}
Note that the true tensor product decomposition has~$\min(n_1+1, n_2+1)$ terms, while the RS algorithm involves a sum over~$n_2+1$ terms. If $n_2>n_1$, the extra terms cancel. This can be checked by applying~\eqref{Weylminus} and~\eqref{Weylreflect} to conclude that~$[n] = -[n-2]$. For instance, $[-1] = -[-1]$, so that~$\chi\left([-1]\right) = 0$ and~$[-1]\big|_\text{RS}$ does not contribute. 

For convenience, we collect the following RS reflection and sign rules:
\begin{align}
& \frak{su}(2)~:~[-\lambda_1] = -[\lambda_1-2]~,\cr
& \frak{su}(3)~:~[\lambda _1, \lambda _2] = -[-\lambda _1-2, \lambda _1+\lambda _2+1]=-[\lambda _1+\lambda +2+1, -\lambda _2-2]~,\cr
& \frak{sp}(4)~:~ [\lambda _1, \lambda _2] = -[-\lambda _1-2, \lambda _1+\lambda _2+1]=-[\lambda _1+2\lambda +2+2, -\lambda _2-2]~,\cr
& \frak{su}(4)~:~ [\lambda _1, \lambda _2, \lambda _3] =  -[-\lambda _1-2, \lambda _1+\lambda _2+1, \lambda _3] = -[\lambda _1, \lambda _2+\lambda _3+1, -\lambda _3-2] \cr
& \hskip97pt =-[\lambda _1+\lambda +1+2, -\lambda _2-2, \lambda _3+\lambda _2+1]~.
\end{align}

\section{Unitary Superconformal Multiplets for General~$\mathcal{N}$}

\label{app:genN}

In section~\ref{sec:usm} we discussed the complete classification of unitary superconformal multiplets for each value of $\mathcal{N}$ with~$N_Q \leq 16$. It is sometimes useful that the representation theory for different values of $\mathcal{N}$ is quite uniform. In this appendix we present the unitarity bounds in $d=3,4,6$ accordingly. 

\subsection{$d=3$}

\label{sec:d3app}

The superconformal algebra is $\frak{osp}(\mathcal{N}|4)$.  The $R$-symmetry is $\frak{so}(\mathcal{N})$; its representations are denoted by Dynkin labels~$R_i$. The supercharges carry the following quantum numbers, in the vector representation of $\frak{so}(\CN)$:
\begin{equation}
Q_{\alpha}\hspace{.5in} \mathrm {transforms \ as}\hspace{.5in}[j]_{\Delta}^{(R_i)}=[1]^{(1,0,\cdots,0)}_{1/2}~.
\end{equation}
For all $\mathcal{N}\geq 3,$ unitary representations are classified by table \ref{tab:3DNgeneral} where $h_1$ is given in terms of the $R_i$, as in~\eqref{hblambda} or~\eqref{hdlambda}, by $h_1=R_1+\cdots +R_{\half (\CN -1)-1}+\half R_{\half (\CN -1)}$ for odd $\CN$, or $h_1=R_1+\cdots + \half (R_{\half \CN -1}+R_{\half \CN})$ for even $\CN$. 

\renewcommand{\arraystretch}{1.5}
\begin{table}[H]
  \centering
  \begin{tabular}{ |c|lr| l|l| }
\hline
{\bf Name} &  \multicolumn{2}{|c|}{\bf Primary} &  \multicolumn{1}{|c|}{\bf Unitarity Bound} & {\bf Null State } \\
\hline
\hline
$L$ & $[j]_{\Delta}^{(R_{i})}$&  &$\Delta>\frac{1}{2} \,j+h_{1}+1$ & \multicolumn{1}{|c|}{$-$} \\
\hline
$A_{1}$ & $[j]_{\Delta}^{(R_{i})}~,$& $(j\geq1)$ &$\Delta=\frac{1}{2} \, j+h_{1}+1$ & $[j-1]_{\Delta+1/2}^{(R_{1}+1,R_{2}, \cdots, R_{[\mathcal{N}/2]})}$ \\
\hline 
$A_{2}$ & $[0]_{\Delta}^{(R_{i})}$& $$ &$\Delta=h_{1}+1$ & $[0]_{\Delta+1}^{(R_{1}+2,R_{2}, \cdots, R_{[\mathcal{N}/2]})}$ \\
\hline
$B_{1}$ & $[0]_{\Delta}^{(R_{i})}$& $$ &$\Delta=h_{1}$ & $[1]_{\Delta+1/2}^{(R_{1}+1,R_{2}, \cdots, R_{[\mathcal{N}/2]})}$ \\
\hline
\end{tabular}
  \caption{Unitary representations of the $3d,$ $\mathcal{N}\geq3$ superconformal algebra.}
  \label{tab:3DNgeneral}
\end{table}

\subsection{$d=4$}

Operators are labelled by $[j, \bar j]_{\Delta}^{(R_1, \dots , R_{\CN -1}; r)}$ where $[j, \bar j]$ are the integer $\frak{su}(2)\oplus \overline{\frak{su}(2)}$ Lorentz Dynkin indices, $(R_1, \dots R_{\CN-1})$ are the $\frak{su}(\CN)$ Dynkin labels, and the $\frak{u}(1)_R$ charge is $r$ (it is not present for $\CN =4)$.  The unitarity bounds and relations involve the  quantities 
\begin{eqnarray}
S\equiv \frac{2}{\mathcal{N}}\left(\sum_{\ell=1}^{\mathcal{N}-1}(\mathcal{N}-\ell)R_{\ell}\right)-\left(\frac{4-\mathcal{N}}{2\mathcal{N}}\right)r & = & \begin{cases}- \left(\frac{3}{2}\right)r~, & \mathcal{N}=1 \\ R-\left(\frac{1}{2}\right)r~, & \mathcal{N}=2 \\ \frac{1}{2}\left(3R_1+2R_2+R_3\right)~, & \mathcal{N}=4\end{cases}
\end{eqnarray}
and
\begin{eqnarray}
\overline{S}=\frac{2}{\mathcal{N}}\left(\sum_{\ell=1}^{\mathcal{N}-1}(\ell)R_{\ell}\right)+\left(\frac{4-\mathcal{N}}{2\mathcal{N}}\right)r & = & \begin{cases}\left(\frac{3}{2}\right)r~, & \mathcal{N}=1 \\ R+\left(\frac{1}{2}\right)r~, & \mathcal{N}=2 \\ \frac{1}{2}\left(R_1+2R_2+3R_3\right)~, & \mathcal{N}=4\end{cases}
\end{eqnarray}
The representations are classified by tables~\ref{tab:4DC} and~\ref{tab:4DAC}

\renewcommand{\arraystretch}{1.5}
\begin{table}[H]
  \centering
  \begin{tabular}{ |c|lr| l|l| }
\hline
{\bf Name} &  \multicolumn{2}{|c|}{\bf Primary} &  \multicolumn{1}{|c|}{\bf Unitarity Bound} & {\bf Null State } \\
\hline
\hline
$L$ & $[j,\overline{j}]_{\Delta}^{(R_{i},r)}$&  &$\Delta>j+S+2$ & \multicolumn{1}{|c|}{$-$} \\
\hline
$A_{1}$ & $[j,\overline{j}]_{\Delta}^{(R_{i},r)}~,$& $(j\geq1)$ &$\Delta= j+S+2$ & $[j-1,\overline{j}]_{\Delta+1/2}^{(R_{1}+1,R_{2},\cdots,R_{\mathcal{N}-1},r-1)}$ \\
\hline 
$A_{2}$ & $[0,\overline{j}]_{\Delta}^{(R_{i},r)}$& $$ &$\Delta=S+2$ & $[0,\overline{j}]_{\Delta+1}^{(R_{1}+2,R_{2}, \cdots, R_{\mathcal{N}-1},r-2)}$ \\
\hline
$B_{1}$ & $[0,\overline{j}]_{\Delta}^{(R_{i},r)}$& $$ &$\Delta=S$ & $[1,\overline{j}]_{\Delta+1/2}^{(R_{1}+1, R_{2}, \cdots, R_{\mathcal{N}-1},r-1)}$ \\
\hline
\end{tabular}
  \caption{The chiral half of unitary representations of the $4d$ superconformal algebras.  (In the special case $\mathcal{N}=4,$ the charge $r$ is absent.)}
  \label{tab:4DC}
\end{table}

\renewcommand{\arraystretch}{1.5}
\begin{table}[H]
  \centering
  \begin{tabular}{ |c|lr| l|l| }
\hline
{\bf Name} &  \multicolumn{2}{|c|}{\bf Primary} &  \multicolumn{1}{|c|}{\bf Unitarity Bound} & {\bf Null State } \\
\hline
\hline
$\overline{L}$ & $[j,\overline{j}]_{\Delta}^{(R_{i},r)}$&  &$\Delta>\overline{j}+\overline{S}+2$ & \multicolumn{1}{|c|}{$-$} \\
\hline
$\overline{A}_{1}$ & $[j,\overline{j}]_{\Delta}^{(R_{i},r)}~,$& $(\overline{j}\geq1)$ &$\Delta= \overline{j}+\overline{S}+2$ & $[j,\overline{j}-1]_{\Delta+1/2}^{(R_{1},R_{2},\cdots,R_{\mathcal{N}-1}+1,r+1)}$ \\
\hline 
$\overline{A}_{2}$ & $[j,0]_{\Delta}^{(R_{i},r)}$& $$ &$\Delta=\overline{S}+2$ & $[j,0]_{\Delta+1}^{(R_{1},R_{2}, \cdots, R_{\mathcal{N}-1}+2,r+2)}$ \\
\hline
$\overline{B}_{1}$ & $[j,0]_{\Delta}^{(R_{i},r)}$& $$ &$\Delta=\overline{S}$ & $[j,1]_{\Delta+1/2}^{(R_{1}, R_{2}, \cdots, R_{\mathcal{N}-1}+1,r+1)}$ \\
\hline
\end{tabular}
  \caption{The antichiral half of unitary representations of the $4d$ superconformal algebras.  (In the special case $\mathcal{N}=4,$ the charge $r$ is absent.)}
  \label{tab:4DAC}
\end{table}

\subsection{$d=6$}

\label{sec:d6app}

The superconformal algebra in six dimensions is a real form of $osp(8|2\mathcal{N});$  the $R$-symmetry is $\frak{sp}(2\mathcal{N}).$ 
Operators are labelled by the quantum numbers $[j_1, j_2, j_3]_\Delta ^{(R_1, \dots, R_\CN)}$ where $[j_1, j_2,j_3]$ are Lorentz $\frak{su}(4)\cong\frak{su}(6)$ Dynkin labels, and $(R_1, \dots, R_\CN)$ are  $\frak{sp}(2\mathcal{N})$ Dynkin labels.  The supercharges are $Q\in [1,0,0]_{1/2}^{(1,0, \dots, 0)}$.    The unitarity relations depend on the $R_i$ only via 
\begin{equation}
R\equiv R_{1}+R_{2}+\cdots +R_{\mathcal{N}}~.
\end{equation}
The list of unitary irreducible representations is given in table \ref{tab:6D}.

\renewcommand{\arraystretch}{1.5}
\begin{table}[H]
  \centering
  \begin{tabular}{ |c|lr| l|l| }
\hline
{\bf Name} &  \multicolumn{2}{|c|}{\bf Primary} &  \multicolumn{1}{|c|}{\bf Unitarity Bound} & {\bf Null State } \\
\hline
\hline
$L$ & $[j_1,j_2,j_3]_{\Delta}^{(R_{i})}$&  &$\Delta>\frac{1}{2}\left(j_1+2j_2+3j_3\right)+2R+6$ & \multicolumn{1}{|c|}{$-$} \\
\hline
$A_{1}$ &$[j_1,j_2,j_3]_{\Delta}^{(R_{i})}~,$& $ j_3\geq1)$ &$\Delta=\frac{1}{2}\left(j_1+2j_2+3j_3\right)+2R+6$ & $[j_1,j_2,j_3-1]_{\Delta+1/2}^{(R_{1}+1, R_{2},\cdots,R_{\mathcal{N}})}$ \\
\hline 
$A_{2}$ &$[j_1,j_2,0]_{\Delta}^{(R_{i})}~,$& $(j_2\geq1)$ &$\Delta=\frac{1}{2}\left(j_1+2j_2\right)+2R+6$ & $[j_1,j_2-1,0]_{\Delta+1}^{(R_{1}+2, R_{2},\cdots,R_{\mathcal{N}})}$ \\
\hline
$A_{3}$ &$[j_1,0,0]_{\Delta}^{(R_{i})}~,$& $(j_1\geq1)$ &$\Delta=\frac{1}{2}j_1+2R+6$ & $[j_1-1,0,0]_{\Delta+3/2}^{(R_{1}+3, R_{2},\cdots,R_{\mathcal{N}})}$ \\
\hline
$A_{4}$ &$[0,0,0]_{\Delta}^{(R_{i})}$& $$ &$\Delta=2R+6$ & $[0,0,0]_{\Delta+2}^{(R_{1}+4, R_{2},\cdots,R_{\mathcal{N}})}$ \\
\hline
$B_{1}$ &$[j_1,j_2,0]_{\Delta}^{(R_{i})}~,$& $(j_2\geq1)$ &$\Delta=\frac{1}{2}\left(j_1+2j_2\right)+2R+4$ & $[j_1,j_2-1,1]_{\Delta+1/2}^{(R_{1}+1, R_{2},\cdots,R_{\mathcal{N}})}$ \\
\hline
$B_{2}$ &$[j_1,0,0]_{\Delta}^{(R_{i})}~,$& $(j_1\geq1)$ &$\Delta=\frac{1}{2}j_1+2R+4$ & $[j_1-1,0,1]_{\Delta+1}^{(R_{1}+2, R_{2},\cdots,R_{\mathcal{N}})}$ \\
\hline
$B_{3}$ &$[0,0,0]_{\Delta}^{(R_{i})}$& $$ &$\Delta=2R+4$ & $[0,0,1]_{\Delta+3/2}^{(R_{1}+3, R_{2},\cdots,R_{\mathcal{N}})}$ \\
\hline
$C_{1}$ &$[j_1,0,0]_{\Delta}^{(R_{i})}~,$& $(j_1\geq1)$ &$\Delta=\frac{1}{2}j_1+2R+2$ & $[j_1-1,1,0]_{\Delta+1/2}^{(R_{1}+1, R_{2},\cdots,R_{\mathcal{N}})}$ \\
\hline
$C_{2}$ &$[0,0,0]_{\Delta}^{(R_{i})}$& $$ &$\Delta=2R+2$ & $[0,1,0]_{\Delta+1}^{(R_{1}+2, R_{2},\cdots,R_{\mathcal{N}})}$ \\
\hline
$D_{1}$ &$[0,0,0]_{\Delta}^{(R_{i})}$& $$ &$\Delta=2R$ & $[1,0,0]_{\Delta+1/2}^{(R_{1}+1, R_{2},\cdots,R_{\mathcal{N}})}$ \\
\hline
\end{tabular}
  \caption{Unitary representations of the $6d$ superconformal algebras.  }
  \label{tab:6D}
\end{table}


\bibliographystyle{utphys}
\bibliography{multiplets}
\end{document}